\documentclass[preprint,12pt]{elsarticle}
\usepackage{hyperref}
\hypersetup{
    colorlinks=true,
    allcolors=black,
    pdftitle={SMC_parallel_version},
}

\usepackage{amssymb}
\usepackage{amsmath}
\usepackage{booktabs}
\usepackage{makecell}
\usepackage{multirow}
\usepackage{subcaption}
\usepackage{float}
\newcolumntype{T}{>{\tiny}c}
\newcolumntype{S}{>{\scriptsize}c}
\newcolumntype{F}{>{\footnotesize}c}
\newcolumntype{Q}{>{\scriptsize}l}
\newcolumntype{N}{>{\normalsize}l}

\journal{Computers in Biology and Medicine}

\begin{document}

\begin{frontmatter}



\title{Accelerated 3D-3D rigid registration of echocardiographic images obtained from apical window using particle filter}


\author[raddi]{Thanuja Uruththirakodeeswaran\corref{cor1}}
\ead{uruththi@ualberta.ca}
\cortext[cor1]{Corresponding author}
\author[maz]{Harald Becher}
\author[raddi]{Michelle Noga}
\author[raddi]{Lawrence H. Le}
\author[comsci]{Pierre Boulanger}
\author[maz]{Jonathan Windram}
\author[raddi,comsci]{Kumaradevan Punithakumar} 

\affiliation[raddi]{organization={Department of Radiology and Diagnostic Imaging, University of Alberta},
            city={Edmonton},
            postcode={T6G 2T4}, 
            state={Alberta},
            country={Canada}}
\affiliation[maz]{organization={Division of Cardiology, Mazankowski Alberta Heart Institute},
            city={Edmonton},
            postcode={T6G 2B7}, 
            state={Alberta},
            country={Canada}}
\affiliation[comsci]{organization={Department of Computing Science, University of Alberta},
            city={Edmonton},
            postcode={T6G 2E1}, 
            state={Alberta},
            country={Canada}}

\begin{abstract}
The perfect alignment of 3D echocardiographic images captured from various angles has improved image quality and broadened the field of view. This study proposes an accelerated sequential Monte Carlo (SMC) algorithm for 3D-3D rigid registration of transthoracic echocardiographic images with significant and limited overlap taken from apical window that is robust to the noise and intensity variation in ultrasound images. The algorithm estimates the translational and rotational components of the rigid transform through an iterative process and requires an initial approximation of the rotation and translation limits. We perform registration in two ways: the image-based registration computes the transform to align the end-diastolic frame of the apical nonstandard image to the apical standard image and applies the same transform to all frames of the cardiac cycle, whereas the mask-based registration approach uses the binary masks of the left ventricle in the same way. The SMC and exhaustive search (EX) algorithms were evaluated for 4D temporal sequences recorded from 7 volunteers who participated in a study conducted at the Mazankowski Alberta Heart Institute. The evaluations demonstrate that the mask-based approach of the accelerated SMC yielded a Dice score value of 0.819 ± 0.045 for the left ventricle and gained $16.7 \times$ speedup compared to the CPU version of the SMC algorithm.
\end{abstract}

\begin{graphicalabstract}
\begin{figure*}[h!]
    \centering
    \includegraphics[width=0.6\textwidth]{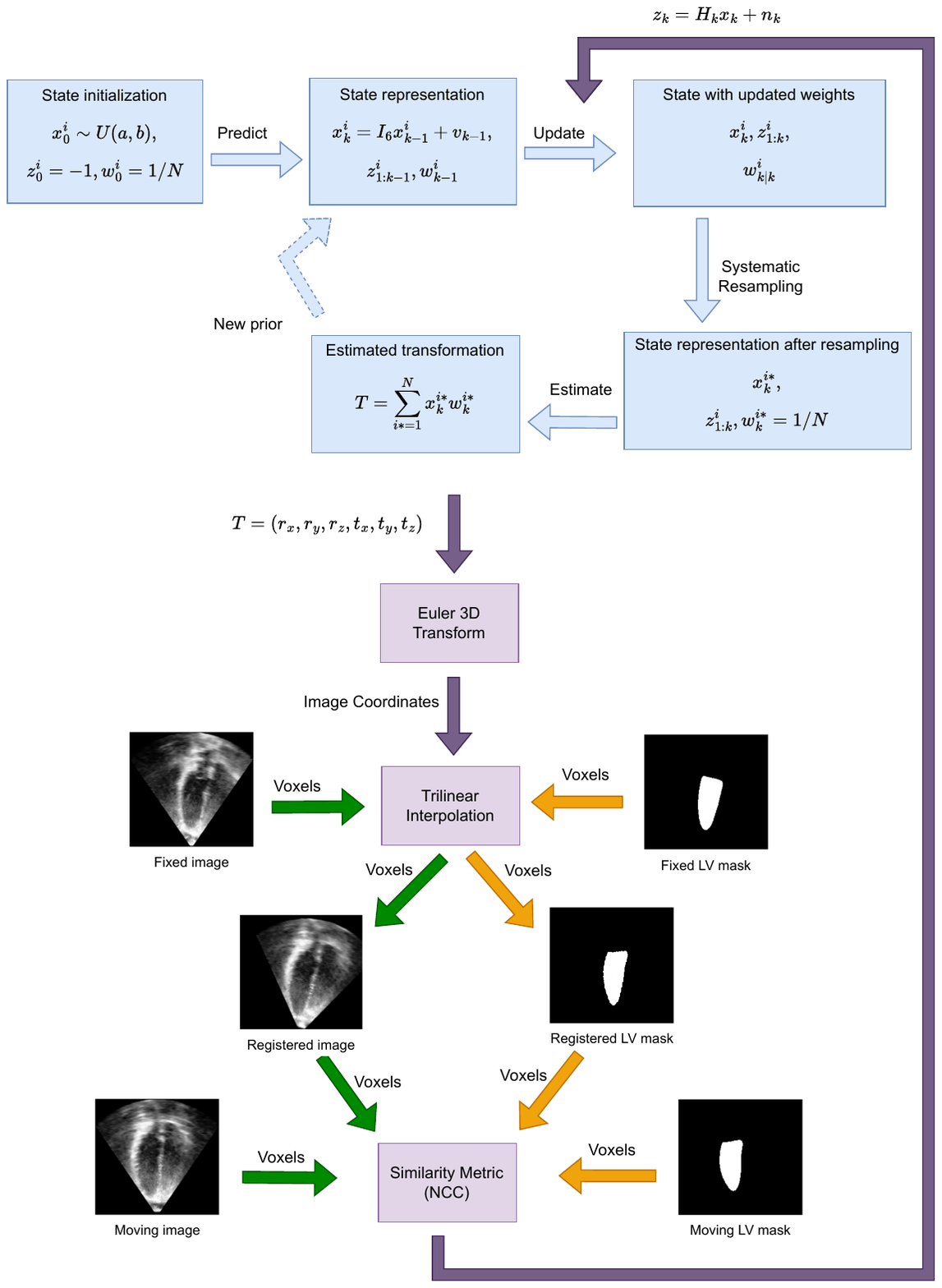}
    \label{fig:fig_pf_method_ga}
    \caption{The 3D-3D rigid registration using the particle filter (PF) algorithm, where the state of the system is represented by N particles ($x^i$) and its associated weights ($w^i$) are highlighted along with the measurement values ($z^i$). The estimated transform by the PF algorithm is used to register the source image/LV mask with the target image/LV mask for image-based (directed with green arrows) and mask-based (directed with orange arrows) registration, respectively. The 3D image and LV masks are denoted using the coronal view in this diagram.}
\end{figure*} 
\end{graphicalabstract}

\begin{highlights}
\item This study proposes a novel accelerated rigid registration for 3D TTE images with significant and limited overlap using particle filter that use GPU hardware.
\item The mask-based registration approach can overcome the intensity variation between different views of TTE images taken from the apical window.
\item The speedup gained between parallel and non-parallel versions of the particle filter demonstrates the efficiency of the algorithm.
\end{highlights}


\end{frontmatter}



\section{Introduction}
Cardiovascular diseases caused nearly 19.91 million deaths worldwide in 2021 \cite{martin_2024_2024}. Different cardiac imaging modalities are used for diagnosis, prognosis, planning and guiding treatment, monitoring disease progression and other follow-up processes, and scientific studies. Ultrasound (US) is widely used to analyze the structure and function of the heart due to its high temporal resolution, portability, accessibility, affordability, and non-ionizing examination. Three-dimensional echocardiography (3DE) is preferred over two-dimensional echocardiography (2DE) for cardiac chamber volume quantification because of the lack of geometrical assumptions about the shape and elimination of apical view shortening \cite{corbett_role_2024, franca_my_2024, henry_three-dimensional_2022}. In addition, it provides a realistic view of the dynamic functional anatomy of the heart to assess heart valves, congenital defects, and surrounding structures\cite{surkova_current_2016}. 

Three-dimensional transesophageal echocardiography (TEE) is used to assess heart valve anatomy and function, while 3D transthoracic echocardiography (TTE) is used to evaluate improved measurement of cardiac chambers, strain imaging with 3D speckle tracking, and valvular disease \cite{masoy_aberration_2024, surkova_current_2016}. The 3D TTE recordings obtained from the apical window on the chest are used to quantify the LV volumes and ejection fraction compared to other views \cite{franca_my_2024, lang_eaease_2012, ostenfeld_comparison_2008}.  In clinical practice, acquiring 3D images in which most of the LV walls can be visualized is standard. However, the image quality is often not optimal in all walls, or parts of the walls are obscured. Therefore, additional non-standard views are recorded. These non-standard recordings are obtained from different transducer positions and orientations to get improved visualization of those segments that are suboptimally displayed in the standard recording. This study focuses on full volume \cite{menciotti_three-dimensional_2024} images obtained from the apical window. The apical standard and non-standard images are used to calculate the volume of the left ventricle (LV) and the ejection fraction (EF) of LV \cite{corbett_role_2024, hung_3d_2007}.

Although the US provides real-time images, analyzing ultrasound images is not easy because of the low imaging quality due to the noise level and presence of artifacts, the high dependence on operator/diagnostician experience, and the high variability across different US manufacturers’ systems \cite{ghorbani_deep_2020, liu_deep_2019, che_ultrasound_2017}. The whole heart cannot be scanned in a single 3DE scan that only captures a portion of the heart. Acquiring 3DE images from various positions that necessitate a method to properly align them for clinical tasks could help overcome the field of view limitation. 

The process of aligning two or more images into a single geometric coordinate system to determine the feature or intensity correlation between them is known as image registration. Because these images are taken at various times from different views using the same or various modalities, they must be correctly aligned for clinical tasks that use computer systems to visualize overlapped images \cite{cao_deep_2024, strittmatter_deep_2024}. Techniques used to register images differ depending on input image modalities, dimensions, region of interest, nature of the information, required transform type, optimization methods, and interaction level \cite{fu_deep_2020, fookes_rigid_2003}. Image intensities and features, including landmarks, can be used to align images depending on the imaging modality and required accuracy. The 3DE image segmentation and fusion may require the registration as a preprocessing step that helps to identify the endocardial border and improve the quality of the image \cite{mao_complete_2023, krishnaswamy_novel_2023, lamb_multi-view_2021, szmigielski_real-time_2010}. The performance of a registration algorithm is critical when using aligned 3D datasets for diagnostic purposes and procedural planning with segmentation and fusion. 

Rigid registration uses translation and rotation to transform one image into another, and it can be used as a global preregistration of images for non-rigid registration. Further, it can be used to align intrapatient images and the registration of multimodal images when considering its application to medical images \cite{fookes_rigid_2003, zhang_ghmm_2024}. Image registration algorithms that use parallel processing with Graphical Processing Unit (GPU) hardware can align the images faster than non-parallel or CPU-based implementations \cite{wang_modetv2_2024, punithakumar_gpu-accelerated_2017, shamonin_fast_2013, shams_survey_2010}.

\section{Related Work}
This section summarizes algorithms used for 3D rigid registration of 3DE images using traditional and machine learning approaches for monomodal registration. A phase calculation was performed to extract features, and it was combined with voxel intensities to perform multiwindow real-time 3D TTE image registration as image intensities may vary within the cardiac structure with factors such as ultrasound beam incidence angle and the depth of tissue \cite{grau_registration_2007}. The study initialized registration using rigid registration with landmarks that find the transform corresponding to the minimum mean squared distance by the singular value decomposition method. Later, speckle filtering was performed using the coherence-enhancing edge detection method to remove artifacts that were present as uncorrelated speckle patterns due to a large movement of the probe. A novel similarity metric was proposed considering the differences in phase and orientation values accumulated over a cardiac cycle. To minimize the calculation time, the end-diastolic (ED) and end-systolic (ES) frames of a cardiac cycle were considered. Rigid registration was performed using a multiresolution framework (scales: 1/4, 1/2, 1) by minimizing the similarity metric using the Powel method. The algorithm and the similarity metric were implemented using Matlab and ITK for CPU hardware, respectively.

Landmark-based rigid registration \cite{mulder_registration_2011} of 3D TTE images was performed in three ways using the normalized cross correlation (NCC) and mutual information metrics: single frame registration, multiframe registration with the ED and ES frames, and multiframe registration with all frames. The Elastix \cite{klein_elastix_2010} toolkit’s multiresolution approach was used to register images with three resolutions. The accuracy was measured using the Euclidean distance between automatic registration and manual registration, and the best result was obtained for multiframe registration with only the ED and ES frames using the NCC metric. The extended version \cite{mulder_atlas-based_2017} of the previous study proposed an atlas-based mosaicing (ABM) for rigid registration of multiview 3D TEE images to create a single volume capturing the left atrium (LA) with an extended FOV that can be used for cardiac interventions. The registration was performed using landmarks by the Elastix toolkit with the NCC metric as a multiresolution approach using three resolutions with a downsampling factor of 2. The TRE values were calculated for pairwise registration and global registration in addition to the registration error calculated for LA segmentations. The accuracy of the ABM approach was compared against regular pairwise registration, and it was concluded that it performed better than the pairwise registration.

The authors \cite{danudibroto_spatiotemporal_2016} performed spatiotemporal registration on multiview 3D apical TTE images using multibeat acquisition for four cardiac cycles with ECG gating and a 65 degree sector angle. A piecewise one-dimensional cubic B-spline method was used for temporal registration, and the accuracy was evaluated using mitral valve opening (MVO) error, resulting in half the time of the temporal resolution and a NCC value indicating improvement. A multiscale iterative Farneback optic flow method performed the spatial registration and yielded better accuracy than Procrustes analysis using manual contouring as a benchmark in terms of Hausdroff distance (HD) and mean absolute distance (MAD) values between LV contours. The proposed method gave low HD and MAD values. A novel principal component analysis (PCA) metric \cite{peressutti_registration_2017} outperformed the state-of-the-art phase-based metric in terms of accuracy, robustness, and execution time when performing rigid registration of multiview 3D TTE temporal sequences. The images were acquired from four healthy patients using apical and modified parasternal windows with electrocardiogram (ECG) gating. Four to five sequences were obtained per participant with an average frame rate of 15. The accuracy was measured using translational (mm) and rotational (degree) errors and target registration error (TRE) (mm) compared to the ground truth transform calculated by tracking the probe movement using optical tracking over 48 image pairs.

 Multiview 3D TEE images were prealigned with feature based SIFT3D \cite{rister_volumetric_2017} algorithm using rigid transformation before aligning the images using the proposed nonrigid registration \cite{mao_complete_2023}. A sequential fusion strategy was used to register and fuse images repeatedly. A new image was considered the fixed image, while the fused image that was reconstructed at the same phase as the fixed image was considered the moving image to align images taken at different cardiac phases and views to enlarge the FOV with motion estimation. The images of four patients were taken as 3D sequences with an initial overlap of 50\% by moving the transducer within [-50, 50] mm and [-10, 10] degrees range, resulting in 29 sequences in total. The algorithm was implemented in Matlab for CPU, and it took approximately one hour to register a pair of images. The authors claimed that fused images had high accuracy and better quality.

The study \cite{carnahan_multi-view_2022} proposed a 3D TEE image fusion method with registration performed using the Elastix toolkit as a preprocessing approach. The volumes of three patients were acquired using two beat acquisitions from a mid-esophageal position and several transgastric positions with ECG gating to match the cardiac phase with the minimum spatial overlap of 80\% between subsequent volumes. The researchers performed rigid registration with the sum of pairwise NCC as the loss function to get global alignment between all volumes using fully simultaneous groupwise registration for the ES frame of each volume. Later, two cycles of semi-simultaneous registration were performed, followed by the nonrigid registration of frames at each cardiac phase in a semi-simultaneous framework to compensate for the difference in frame rate and anatomical variation in time. The semi-simultaneous framework used a multiresolution approach with the same loss function and the adaptive stochastic gradient descent optimizer at four resolution levels with smoothing. A single volume was obtained after fusion with high quality to view the mitral valve and subvalvular structures that cardiac surgeons can use for procedural planning. An echocardiographic expert validated the accuracy of the proposed fusion method for the patient data. A two step rigid and nonrigid registration algorithms \cite{shanmuganathan_two-step_2024} were used to align multiview 3D TTE images that can be later fused and used to evaluate the function and structure of the heart. Three landmarks (the right coronary cusp and the anterior and posterior mitral valve leaflet attachment on the mitral valve annulus) were predicted using a neural network trained using a reinforcement learning algorithm and registered rigidly with the Simple ITK library. The result was improved further by the Simple Elastix (SE) \cite{marstal_simpleelastix_2016} algorithm using a multimetric multiresolution B-spline transform.

With low computing resources, single pairwise image registration can be carried out iteratively using classical image registration techniques that compute the transform model using energy functions \cite{grzech_variational_2022}. The particle filter (PF) approach uses a set of random particles with weights to represent posterior density for implementing a recursive Bayesian filter by Monte Carlo (MC) sampling. Using the Bayesian filtering framework for image registration has the benefit of being robust to noise and handling outliers and local minima convergence \cite{porter_point_2021}. This study proposes a parallel implementation of the sequential MC approach \cite{uruththirakodeeswaran_3d3d_2024} to register the temporal sequence of TTE volumes using intensities of images and masks separately and compare their performance for rigid registration. To the best of the authors’ knowledge, this is the first attempt to use the parallel version of the sequential MC approach for rigid registration even though there are algorithms to perform registration using non-parallel hardware \cite{porter_point_2021, wong_adaptive_2010, arce-santana_affine_2012, mejia-rodriguez_elastic_2011, abdel-basset_feature_2017}.

\section{Methods}
For clinical analysis of the entire heart, precise alignment of 3DE images acquired from multiple views and at different times is essential. The image and binary mask intensity based rigid registration algorithm proposed in this study for 3DE images of different views of the apical window is significant, as there are no algorithms available now for the GPU architecture, and this is the extended work of \cite{uruththirakodeeswaran_3d3d_2024} with more experiments. This study proposes a sequential Monte Carlo based rigid registration of apical standard and nonstandard images in two ways: the image-based registration computes the transform to align the ED frame of the apical nonstandard image to the apical standard image and applies the same transform to all frames of the cardiac cycle, whereas the mask-based registration approach uses the binary masks of the LV in the same way. The mask-based registration is used to compare the accuracy against image-based registration, as it is expected that intensities of voxels may change for different views of 3DE images because of the difference in angle of imaging, image quality and artifacts, and other technical factors related to image acquisition. We expect no intensity changes due to physiological and pathological changes in the heart muscle, as the images were acquired within a short interval.

\subsection{Registration}
\begin{figure*}[t!bp]
    \centering
    \includegraphics[height=15cm,keepaspectratio]{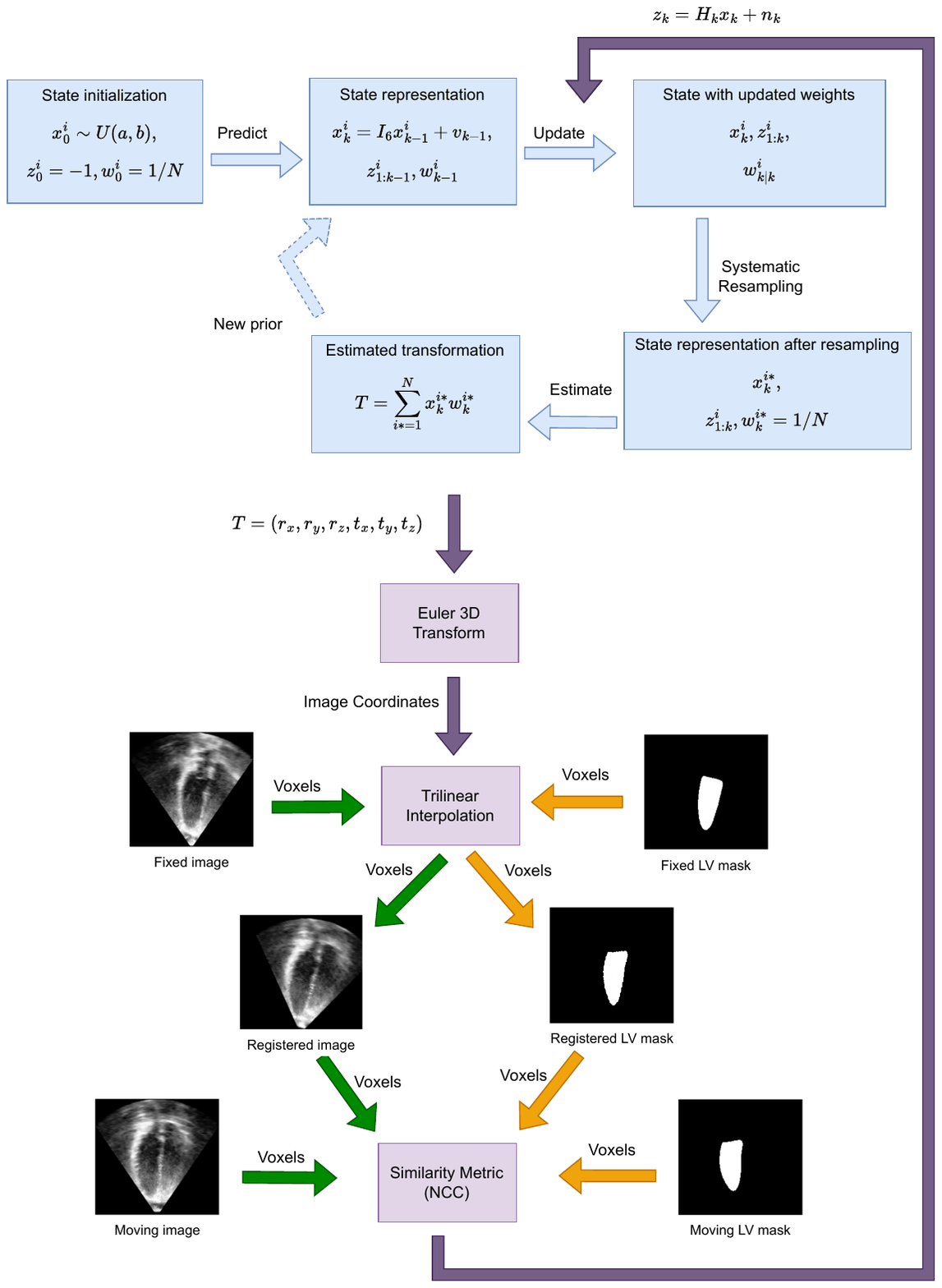}
    \caption{The 3D-3D rigid registration using the PF algorithm, where the state of the system is represented by N particles ($x^i$) and its associated weights ($w^i$) are highlighted along with the measurement values ($z^i$) after prediction, update, resampling, and estimation steps. The estimated transform by the PF algorithm is used to register the source image/LV mask with the target image/LV mask for image-based (directed with green arrows) and mask-based (directed with orange arrows) registration, respectively. The similarity metric value between them is used as the measurement value when updating the weights in the next iteration. The same transform calculates the Dice score value between the LV masks of the source and target images. This figure is adapted from \cite{uruththirakodeeswaran_3d3d_2024}. The 3D image and LV masks are denoted using the coronal view in this diagram.}
    \label{fig:fig_pf_method}
\end{figure*}

The PF algorithm performs rigid registration on 3DE images with significant and limited overlap. The apical standard and non-standard images of the same volunteer are considered to perform registration with translation and rotation. The voxel intensity of the image and LV mask are considered independently to accomplish the task without using landmarks or features, as shown in Figure \ref{fig:fig_pf_method}. The objective of the registration algorithm is to find the best transform ($\hat{\phi}$) that minimizes the dissimilarity value between source ($S$) and target ($T$) images when applying transform $\phi$. Because the similarity metric alone is insufficient to estimate the accuracy of a registration algorithm \cite{rohlfing_image_2012} and due to the intensity variation between different views of 3DE images, the PF algorithm is evaluated using a similarity metric and an overlapping score.

\begin{align} \label{eq:eq_rf}
\hat{\phi}=\arg \min _{\phi} E_D\left(I_T, I_S(\phi)\right)
\end{align}

\subsection{Dataset}\label{subsec3}
The 3D TTE dataset was obtained from a research study conducted at the Mazankowski Alberta Heart Institute and was approved by the Health Research Ethics Board of the University of Alberta. The study used a robotic arm to collect images in a controlled manner, putting less strain on the sonographer, utilizing a Philips X5-1 probe and a Philips EPIQ 7C scanner (Philips Healthcare, Eindhoven, The Netherlands) \cite{punithakumar_force_2024}. There were 66 image pairs from 7 volunteers (gender ratio: 3 male and 4 female, age range: above 18) considered to evaluate the algorithm using 1073 frames with an average frame rate of 20.49. All the other volunteer datasets do not have apical standard and non-standard image pairs except one that was ignored due to poor image quality.

The image acquisition and volume metadata details are listed in Table \ref{tab:table_metadata}. There were 189, 188, 126, 150, 96, 138, and 186 frames considered for each volunteer. The binary masks of the LV area were annotated by a sonographer using Tomtec Arena \footnote{\url{https://www.tomtec.de/products/tomtec-arena}} software. The binary masks were used after the registration to validate the performance of the image-based PF algorithm by applying the transform to the mask of the source image to be considered as the registered mask when calculating the overlapping score after each iteration. The mask-based registration used the binary masks as source and target to perform registration, as shown in Figure \ref{fig:fig_pf_method}, and calculated the overlapping score similar to image-based registration. The final transform estimated by the mask-based registration is applied to the source image to align it with the target image.

\begin{table}[H]
\caption{The metadata of 4D TTE image (temporal) sequences of 7 volunteers. The apical standard and nonstandard sequences are denoted with S and N, respectively. The total number of image pairs and 3D volumes considered are 66 and 1073, respectively.}
\label{tab:table_metadata}
    \begin{tabular*}{\textwidth}{@{\extracolsep\fill}cccccc}
        \toprule
        VID &
        \multicolumn{1}{p{1cm}}{\centering No. \\ of  \\ seq} &
        \multicolumn{1}{p{2cm}}{\centering Acquisition  \\ type} & 
        \multicolumn{1}{p{1cm}}{\centering Frame  \\ rate \\(Hz)}  &
        \multicolumn{1}{p{3cm}}{\centering Volume  \\ size \\ (voxel)} &
        \multicolumn{1}{p{3cm}}{\centering Volume  \\ resolution \\ (mm)}  \\
        \midrule
         V1	& 3S, 3N & HMQ & 22 & ${224 \times 176 \times 208}$ & ${0.87 \times 1.08 \times 0.73}$ \\
         V2 & 3S, 4N & HMQ & 22 & ${224 \times 176 \times 209}$ & ${0.98 \times 1.21 \times 0.82}$ \\
         V3 & 3S, 3N & HMQ & 16 & ${224 \times 208 \times 208}$ & ${1.09 \times 1.30 \times 0.92}$ \\
         V4 & 3S, 3N & HMQ & 21 & ${224 \times 176 \times 208}$ & ${1.09 \times 1.30 \times 0.92}$ \\ 
         V5 & 3S, 3N & HMQ & 16 & ${224 \times 208 \times 208}$ & ${1.04 \times 1.23 \times 0.87}$ \\
         V6 & 3S, 3N & HMQ & 21 & ${224 \times 176 \times 208}$ & ${0.92 \times 1.15 \times 0.78}$ \\
         V7 & 3S, 3N & HMQ & 22 & ${224 \times 176 \times 208}$ & ${0.87 \times 1.08 \times 0.73}$ \\
        \bottomrule
    \end{tabular*}
\end{table}

\subsection{Implementation Details}\label{subsec4}

The parallel (GPU) version of the PF algorithm was implemented using Python (3.11.5) and PyTorch (2.4.1), while the non-parallel (CPU) version of the algorithm was implemented using ITK (5.3.0) for the same Python version. The NCC \eqref{eq:eq_ncc} was used to assess the similarity between the source and target images. The Euler 3D transform was used to calculate new image coordinates after applying the rigid transform ${(r_x, r_y, r_z, t_x, t_y, t_z)}$ defined using the homogeneous coordinates as in \eqref{eq:eq_rigid_coor} where ${r_{ij}}$ is a $(3 \times 3)$ rotational matrix. The source image was aligned with the target image using trilinear interpolation to find voxel values at non-grid positions, recursively reducing dissimilarity between the images.

\begin{align} \label{eq:eq_ncc}
NCC(T, S)=\frac{(\sum_{i=1}^N\left(T_i-\bar{T}\right)\left(S_i-\bar{S}\right))^2}{{\sum_{i=1}^N\left(T_i-\bar{T}\right)^2}\sum_{i=1}^N\left(S_i-\bar{S}\right)^2}
\end{align}

\begin{align}\label{eq:eq_rigid_coor}
    T = \left[\begin{array}{cccc}
    r_{11} & r_{12} & r_{13} & t_{x} \\
    r_{21} & r_{22} & r_{23} & t_{y}\\
    r_{31} & r_{32} & r_{33} & t_{z}\\
    0 & 0 & 0 & 1 \end{array}\right]
\end{align}

The PF algorithm found the rigid transform to register ED frames and applied the same transform to all the other frames in the cardiac cycle. The parameters of the PF algorithm remained the same as mentioned in the previous work \cite{uruththirakodeeswaran_3d3d_2024} for both GPU and CPU versions. The images were normalized with a mean of 0 and a standard deviation of 1 before registration. The nonrigid (SE) registration was implemented with the images and their LV binary masks together using Simple ITK 2.2.1 for the same Python version to improve the accuracy of registration after performing a rigid transform. The previous study used only the images for SE registration. The advanced normalized correlation and transformed bending energy penalty were used as metrics to perform multimetric, multiresolution registration using a recursive image pyramid. The SE registration was performed for 100 iterations for each of the five levels of resolution.

\subsection{Baseline and Evaluation Metrics}\label{subsec5}

The registration accuracy is evaluated using the NCC as the similarity metric and the Dice score (DSC) as the overlapping score. The image similarity is measured on a scale from 0 to 1, where 1 indicates a high correlation between voxel intensities. The DSC between the 3D binary masks of the source and target images $S$ and $T$ is defined as \eqref{eq:eq_ds}, where a value of 0 indicates that there is no overlap between the corresponding images and a value of 1 indicates that the images are perfectly aligned. The exhaustive search (EX) method is used as a baseline method to compare the accuracy of the PF algorithm and was implemented using the same ITK framework's exhaustive optimizer \footnote{\url{https://examples.itk.org/src/numerics/optimizers/exhaustiveoptimizer/documentation}} for the same Python version. The parameters remained the same as in previous work, and registration was performed using the image and LV binary mask intensities separately, similar to the PF algorithm.

\begin{align} \label{eq:eq_ds}
DSC(S,T) = \frac{2 * |S \cap T|}{|S| + |T|}
\end{align}

\section{Results}
\label{sec1}
\begin{figure}[h!tbp]
    \centering
    \includegraphics[width=1.0\textwidth]{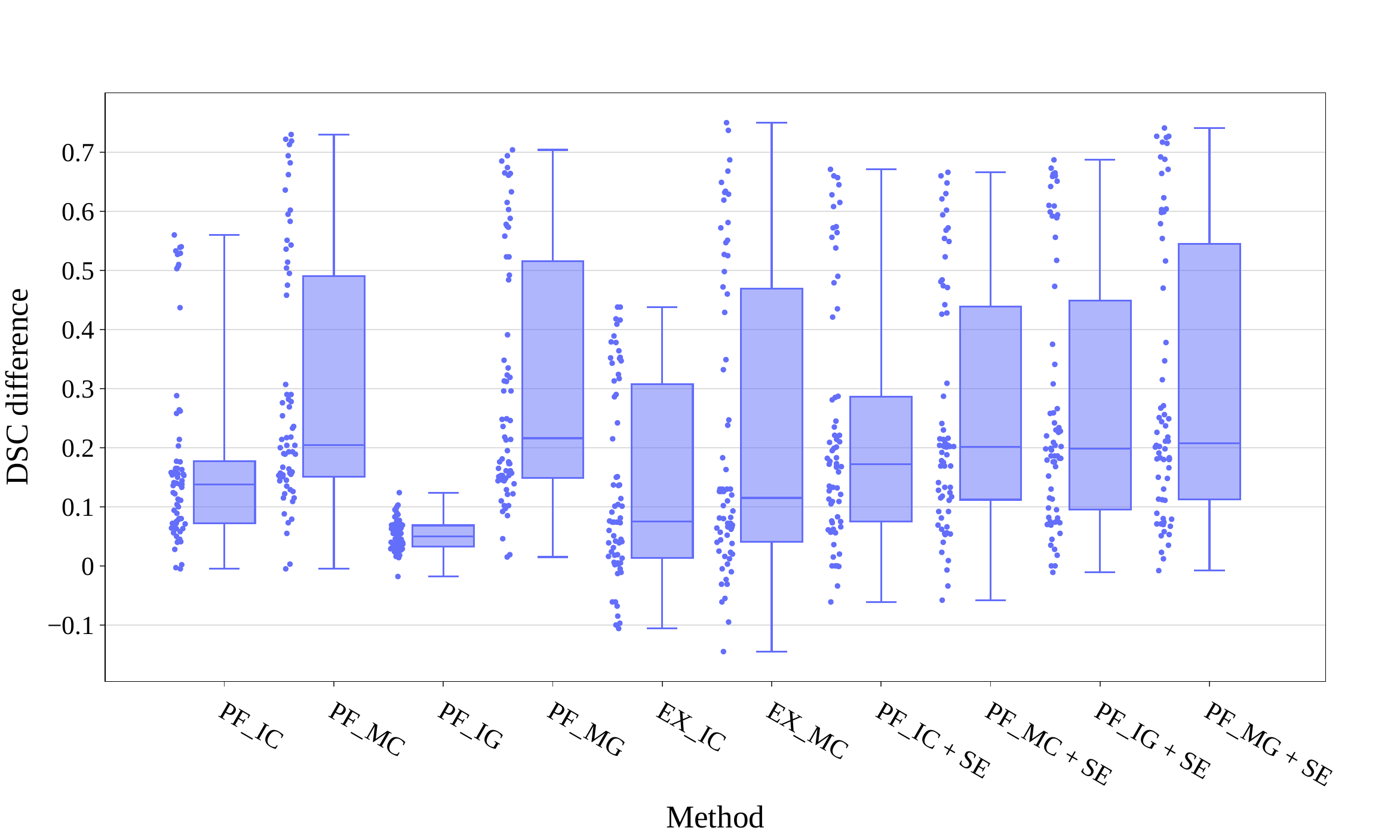}
    \caption{The 3D-3D rigid registration results using the PF and EX algorithms. The figure shows the DSC difference after registration for non-parallel (CPU) and parallel (GPU) versions of the PF and baseline (ES) algorithms for rigid registration using image intensities and LV segmentation masks. The PF results were further improved by applying the nonrigid (SE) algorithm using the images and their corresponding LV segmentation masks. The DSC difference between PF + SE and before registration is shown here. IC and IG refer to image-based registration performed on CPU and GPU hardware, respectively. The MC and MG refer to mask-based registration in the same way.}
    \label{fig:fig_pf_es_se_ncc}
\end{figure}

\begin{figure*}[!ht]
    \centering
    \begin{tabular}{SSSSSSS}
         & \multicolumn{2}{c}{\scriptsize PF (CPU)} & \multicolumn{2}{c}{\scriptsize PF (GPU)} & \multicolumn{2}{c}{\scriptsize EX (CPU)} \\
         & Before & After & Before & After & Before & After \\
        \makecell[b]{Min\vspace{20pt}} & 
        {\includegraphics[width=0.12\textwidth]{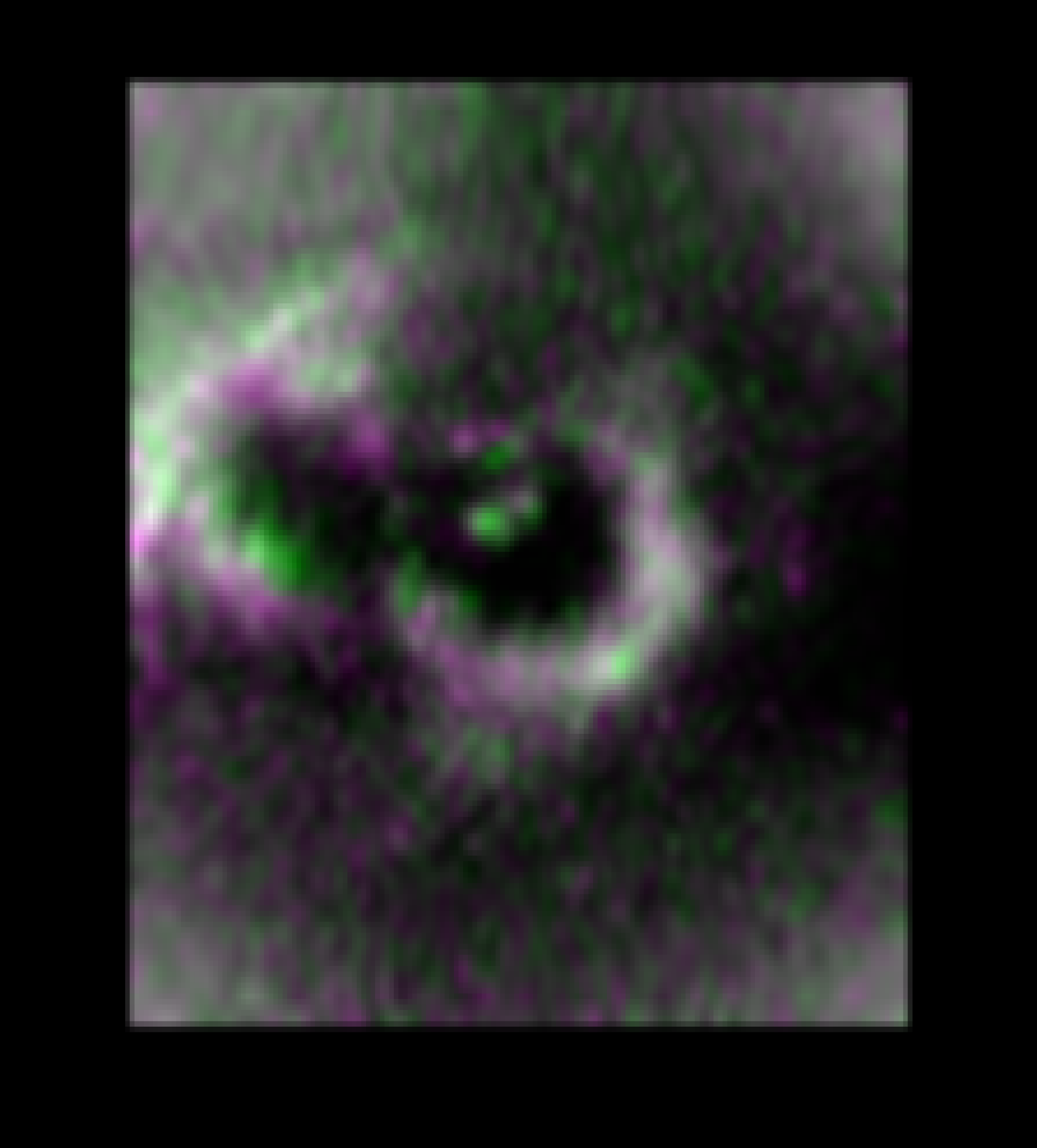}} &
        {\includegraphics[width=0.12\textwidth]{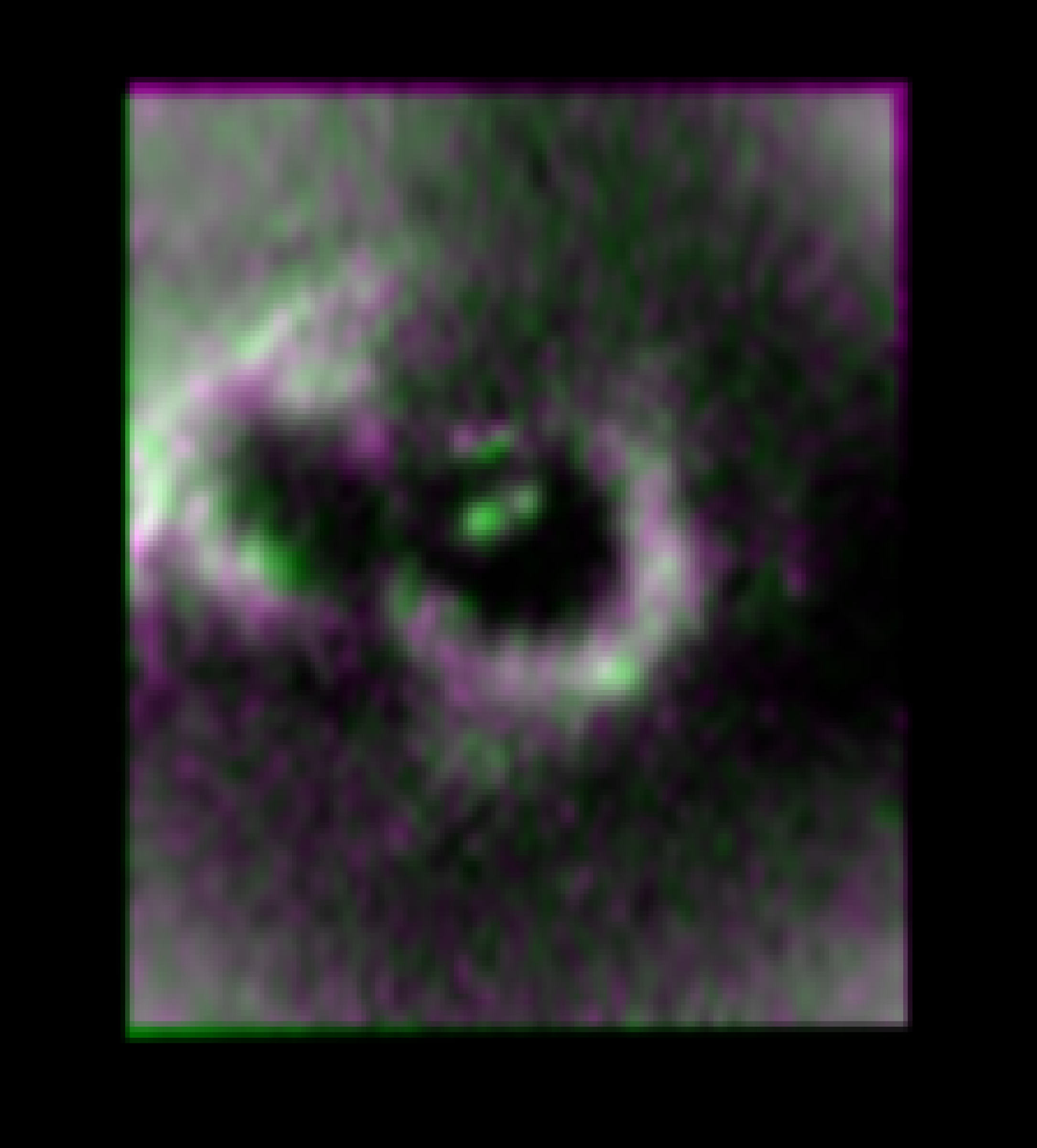}} &
        {\includegraphics[width=0.12\textwidth]{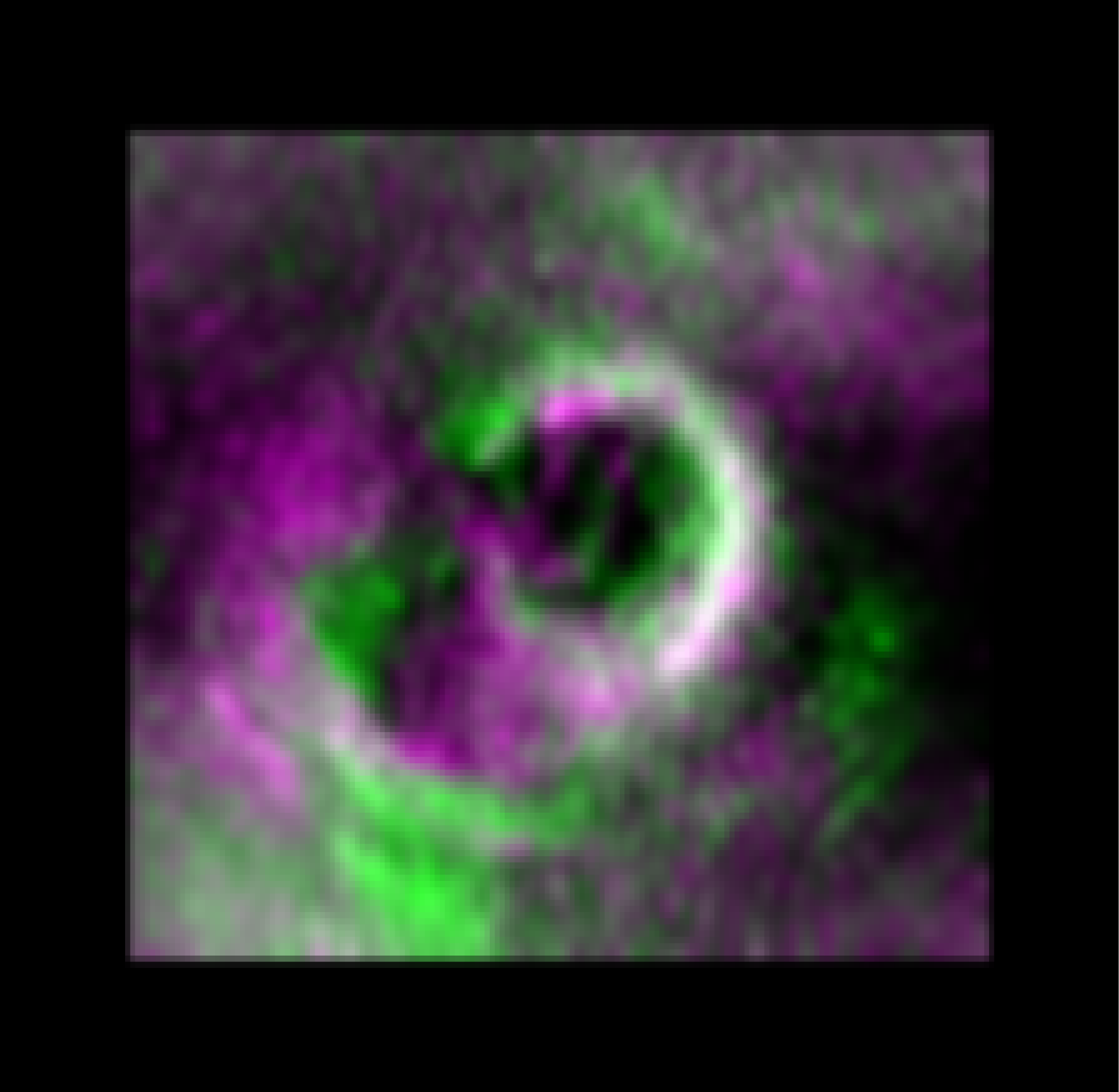}} &
        {\includegraphics[width=0.12\textwidth]{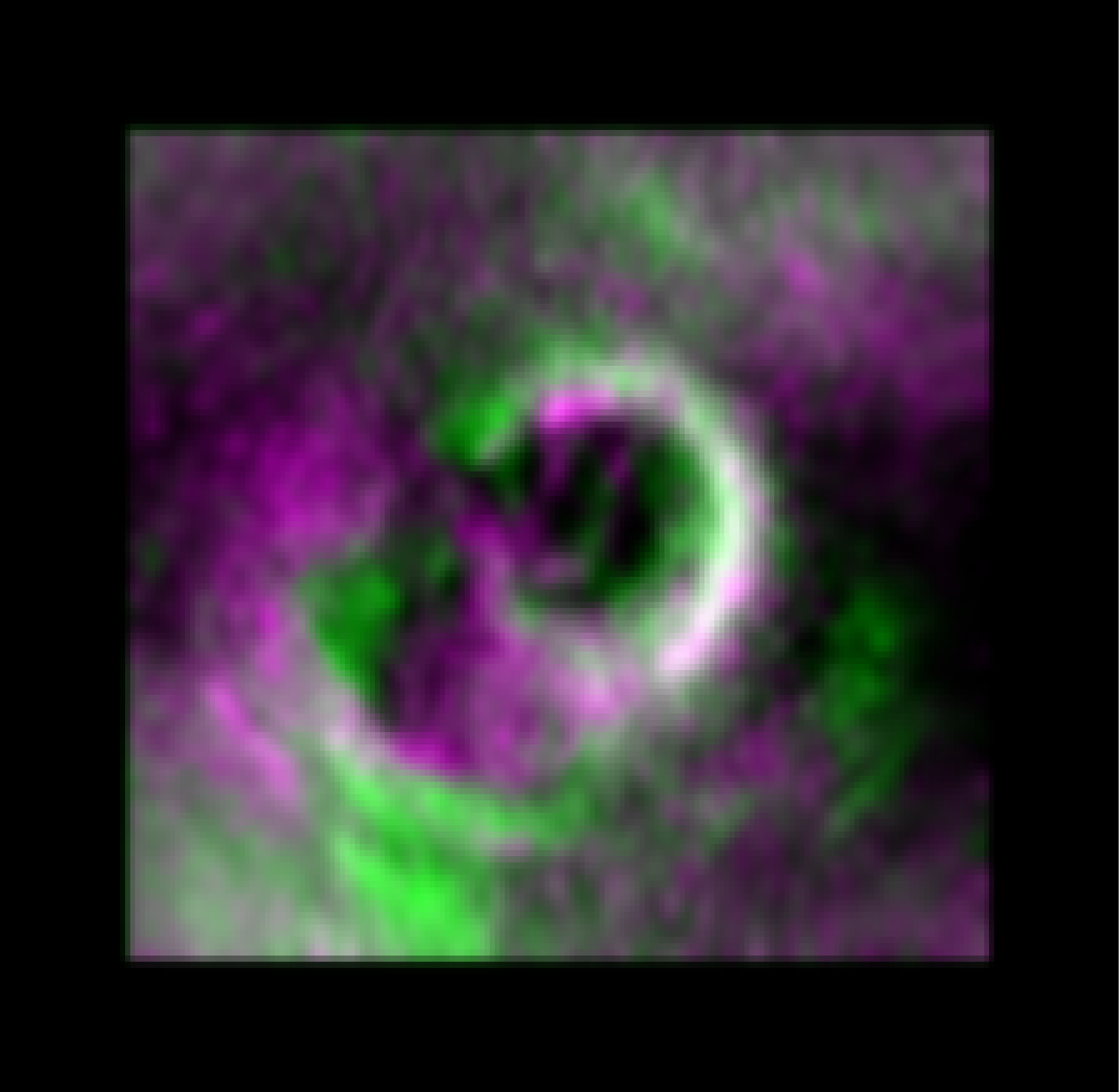}} &
        {\includegraphics[width=0.12\textwidth]{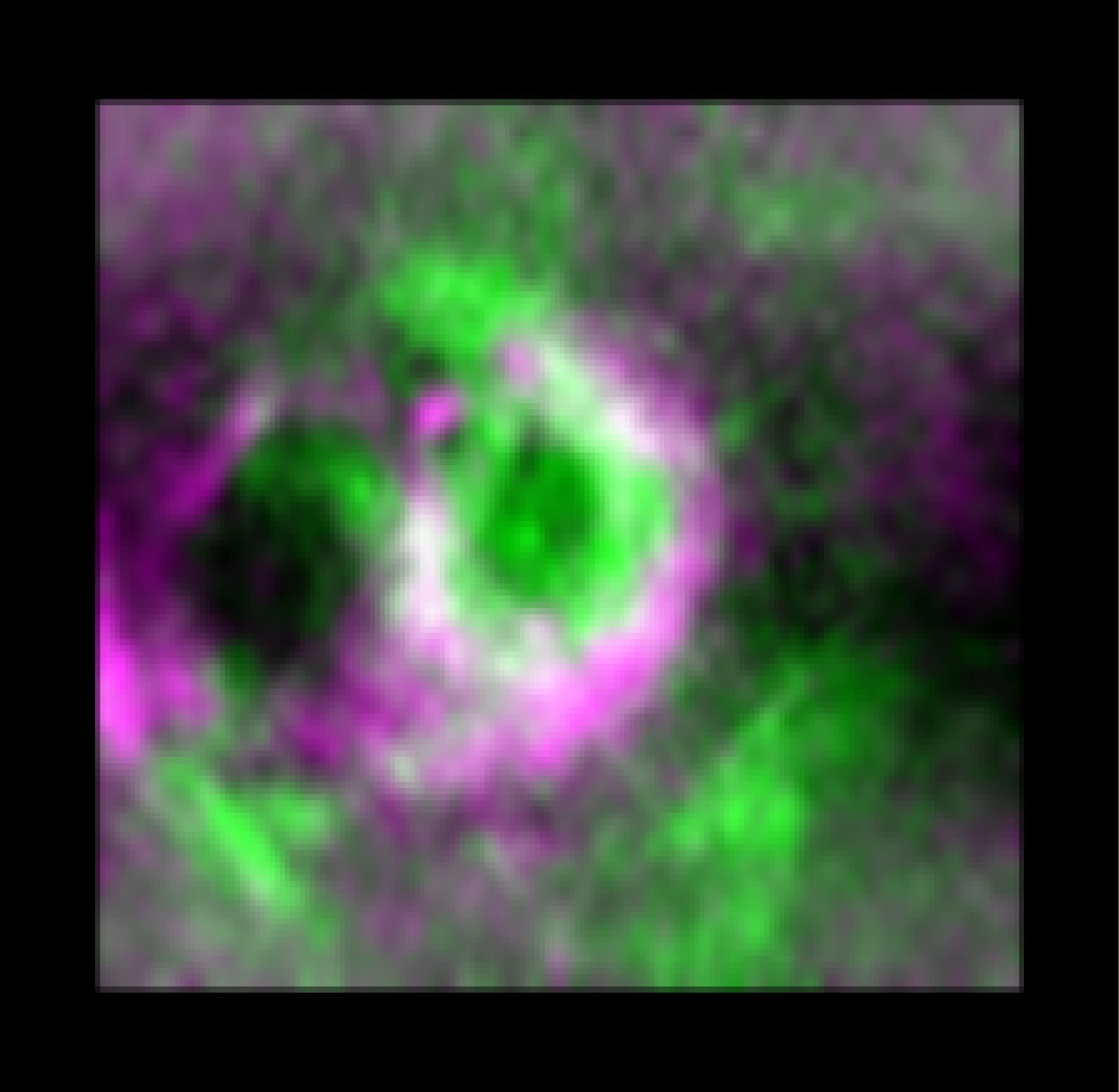}} &
        {\includegraphics[width=0.12\textwidth]{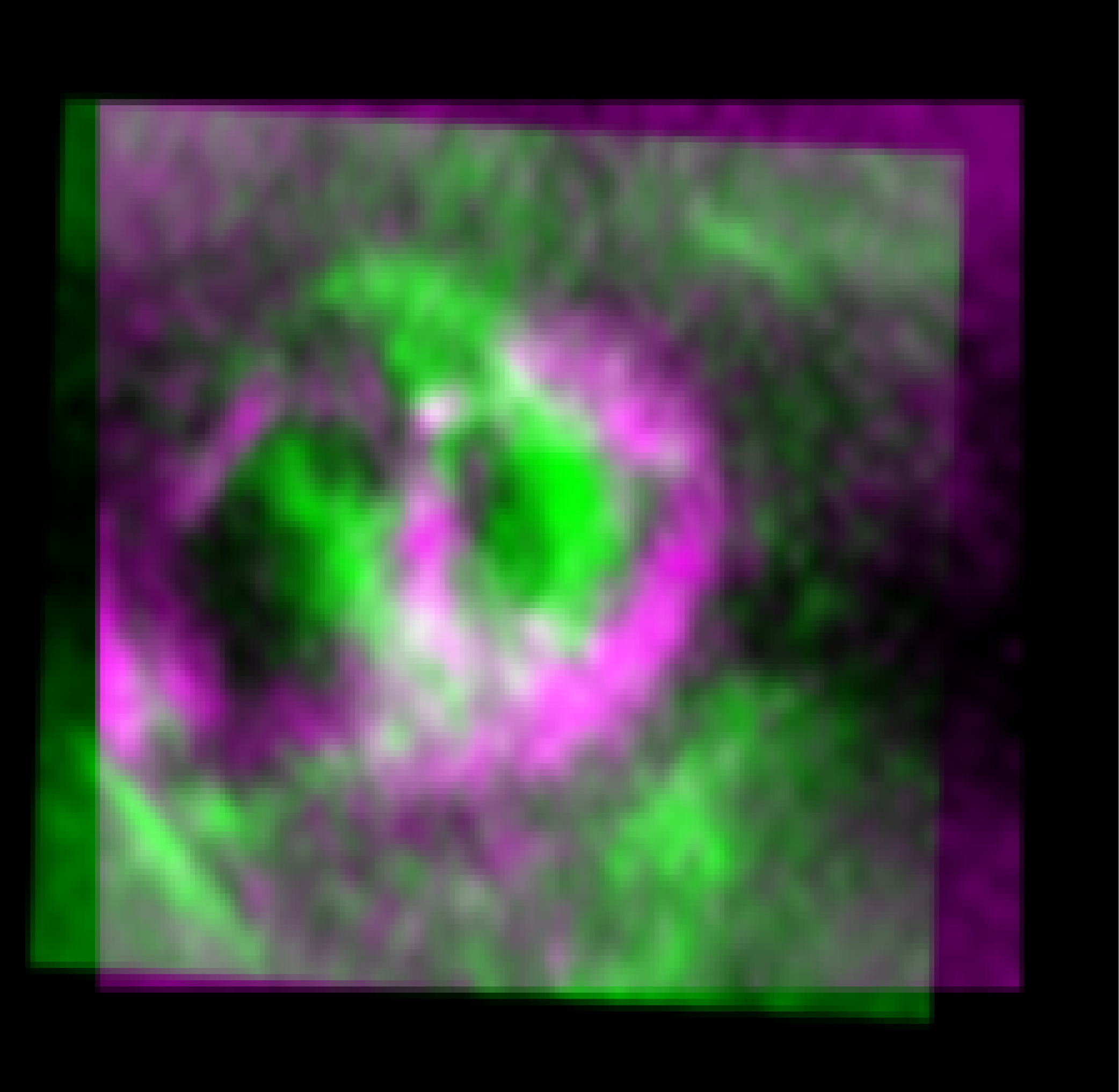}} \\
        \makecell[b]{Q1\vspace{20pt}} & 
        {\includegraphics[width=0.12\textwidth]{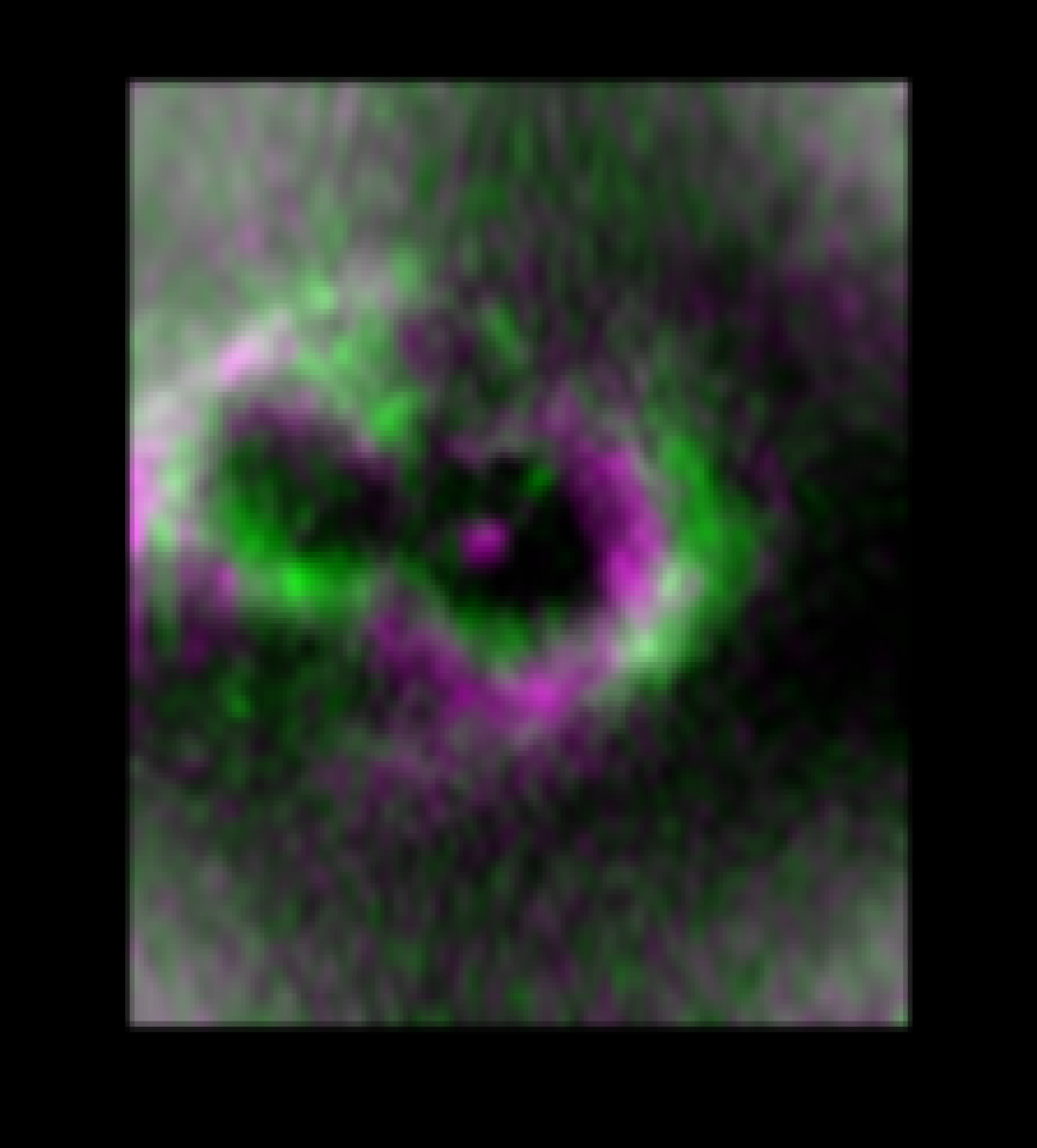}} &
        {\includegraphics[width=0.12\textwidth]{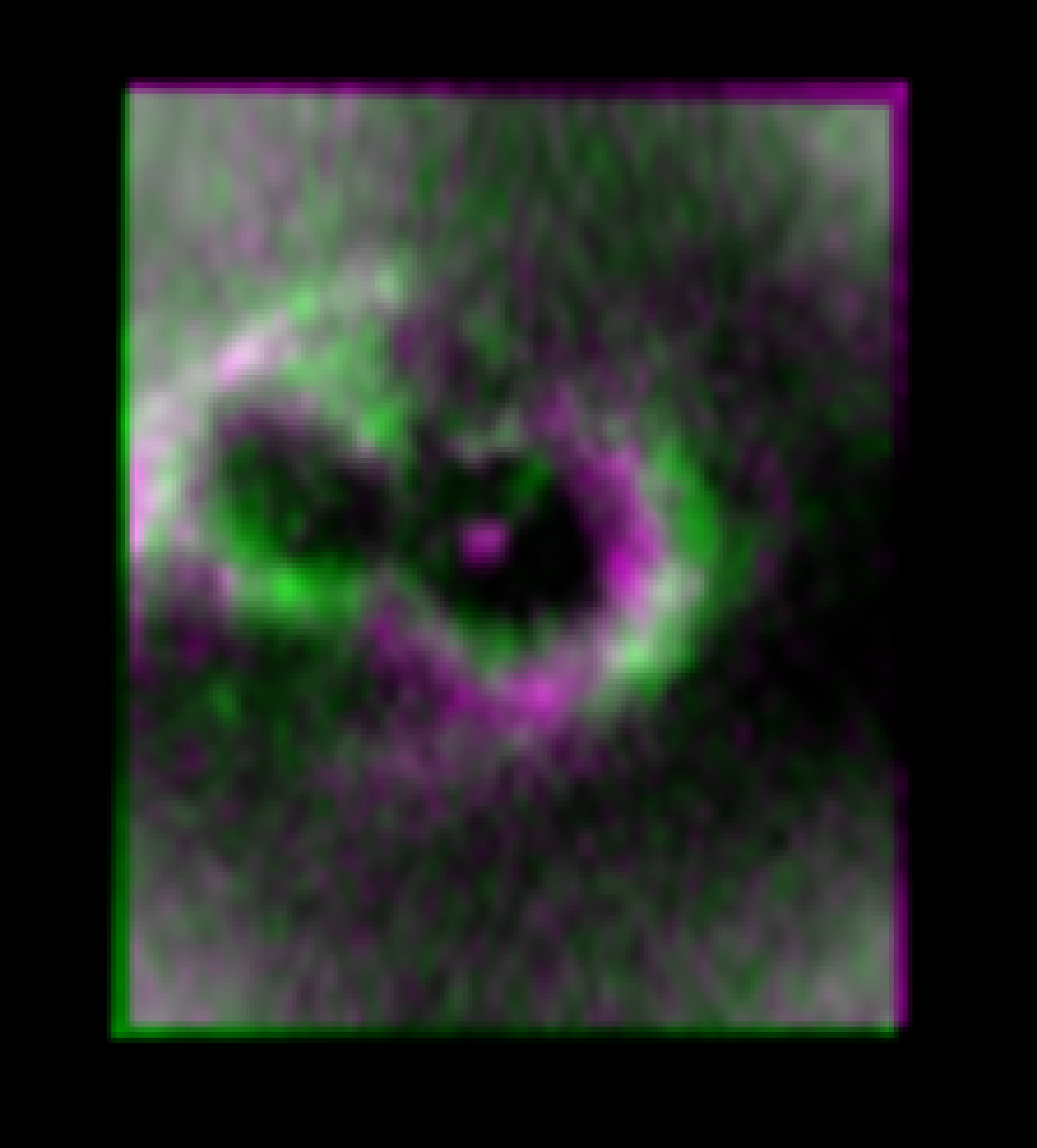}} &
        {\includegraphics[width=0.12\textwidth]{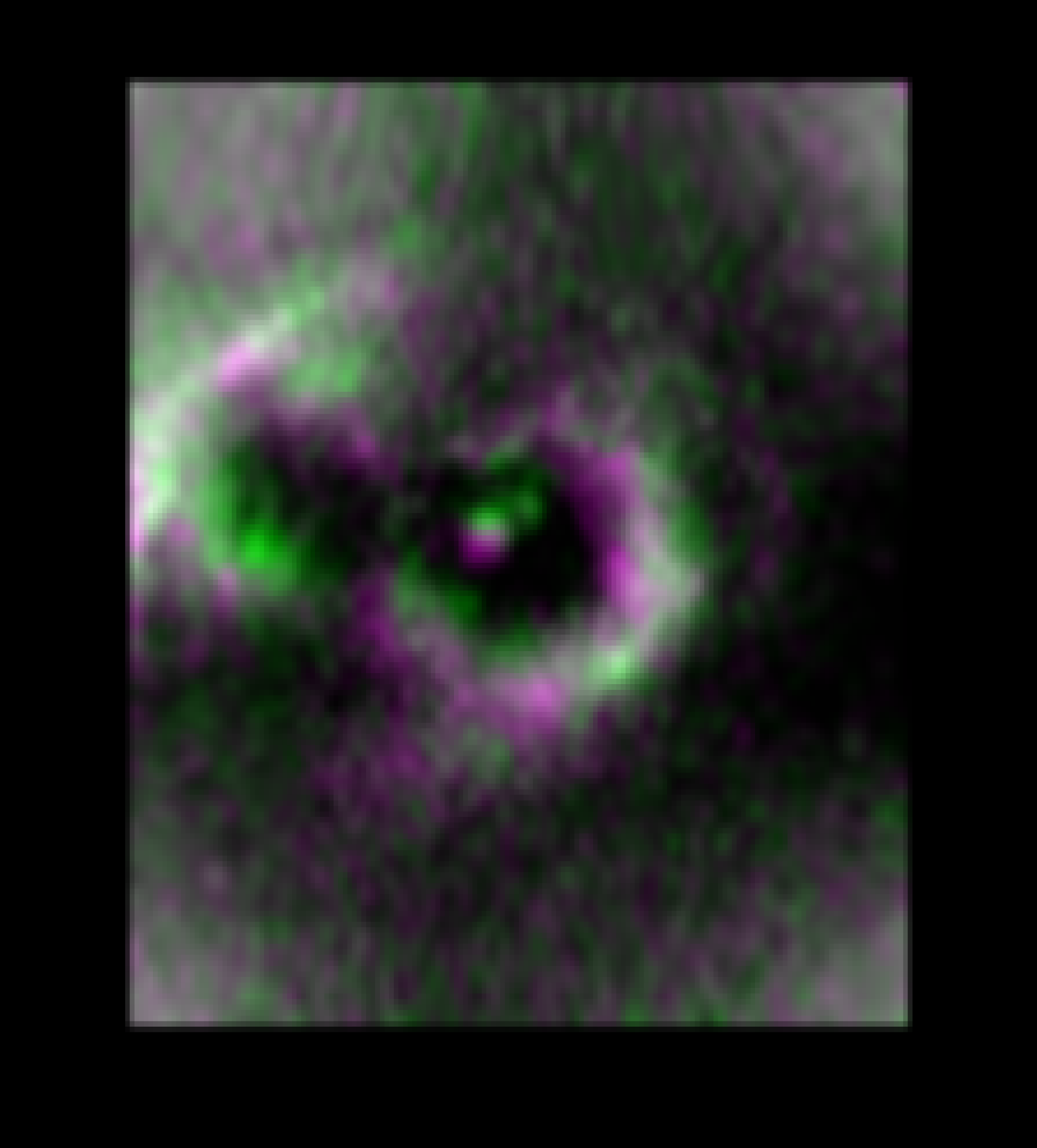}} &
        {\includegraphics[width=0.12\textwidth]{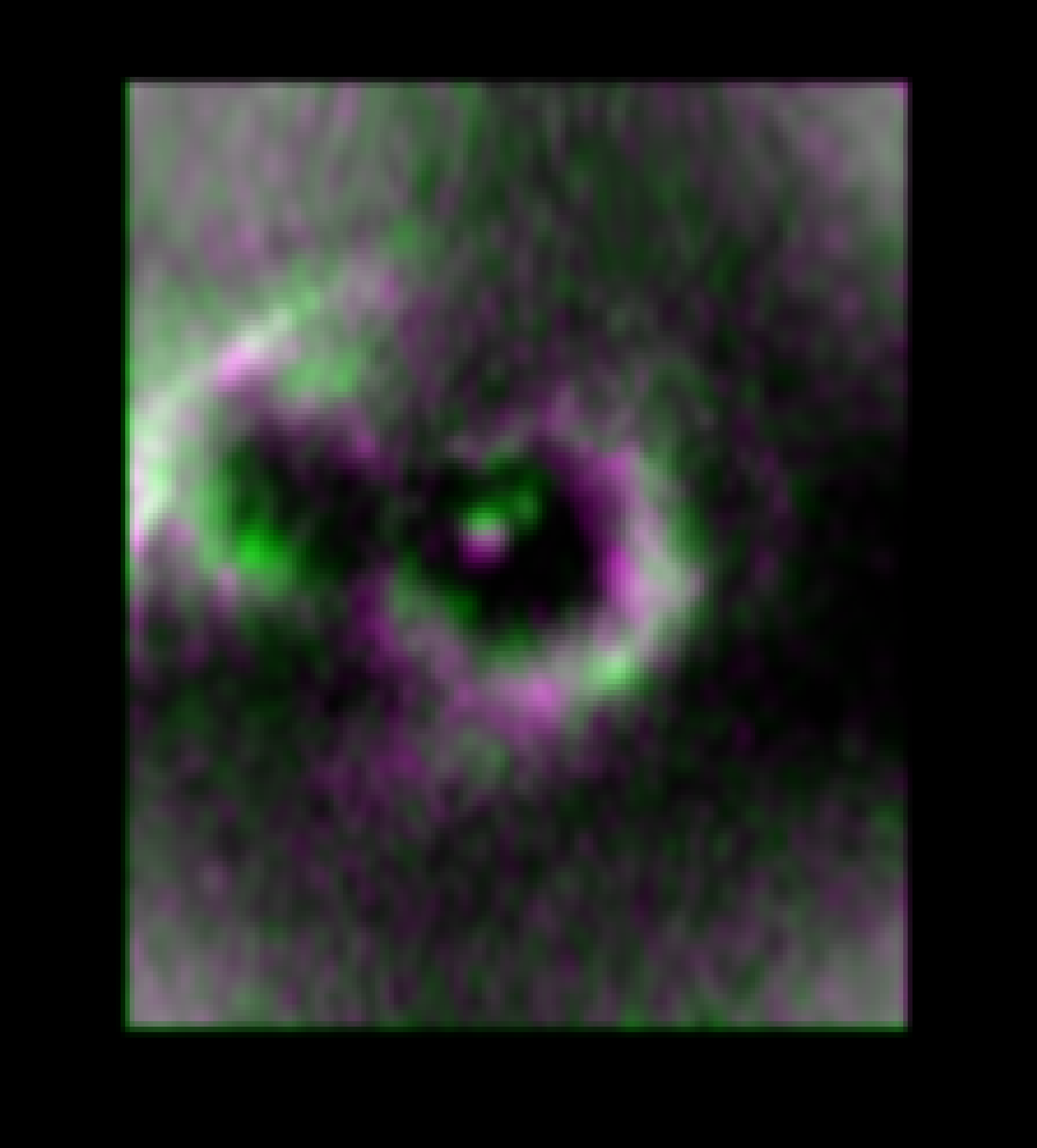}} & 
        {\includegraphics[width=0.12\textwidth]{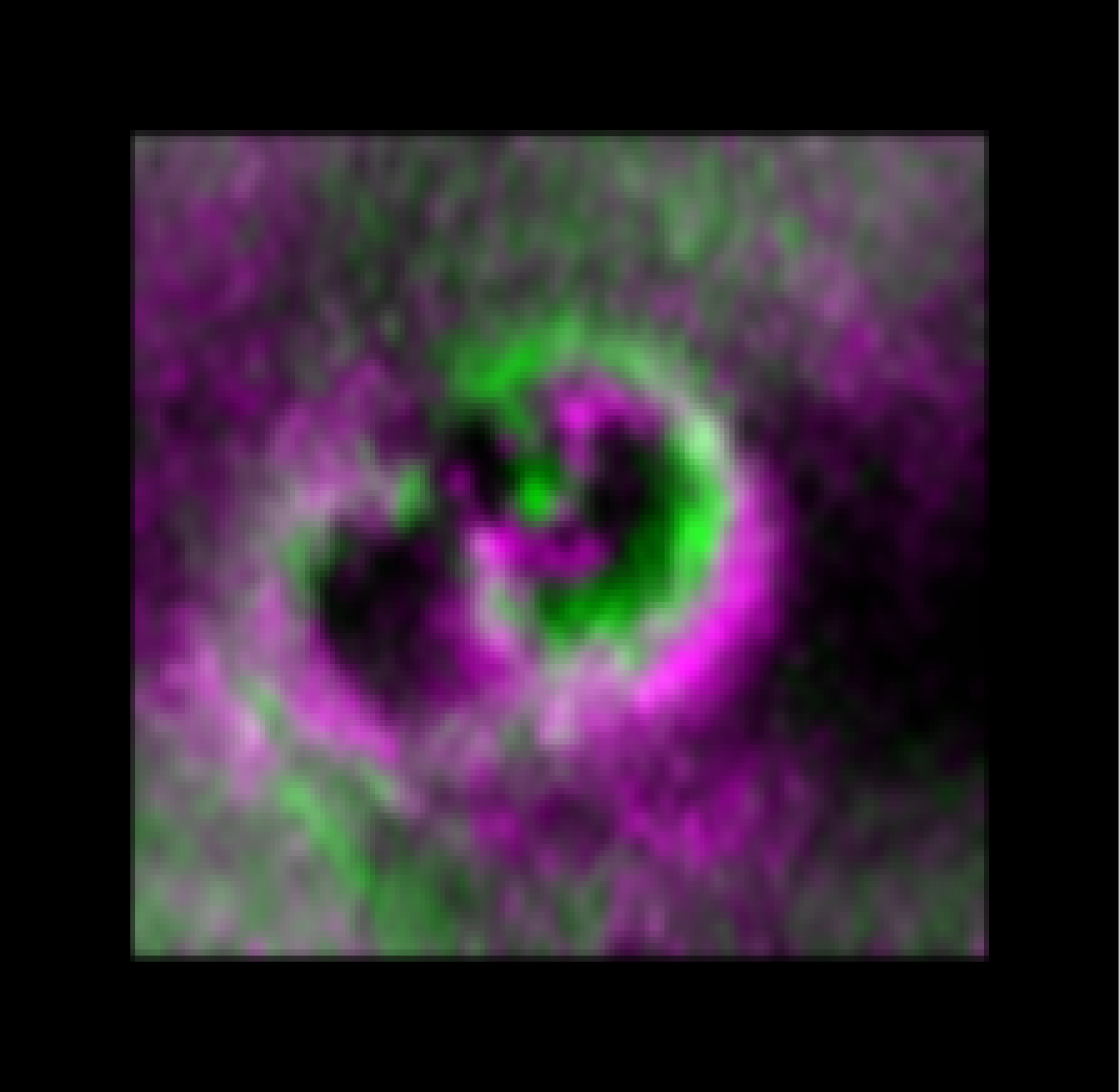}} &
        {\includegraphics[width=0.12\textwidth]{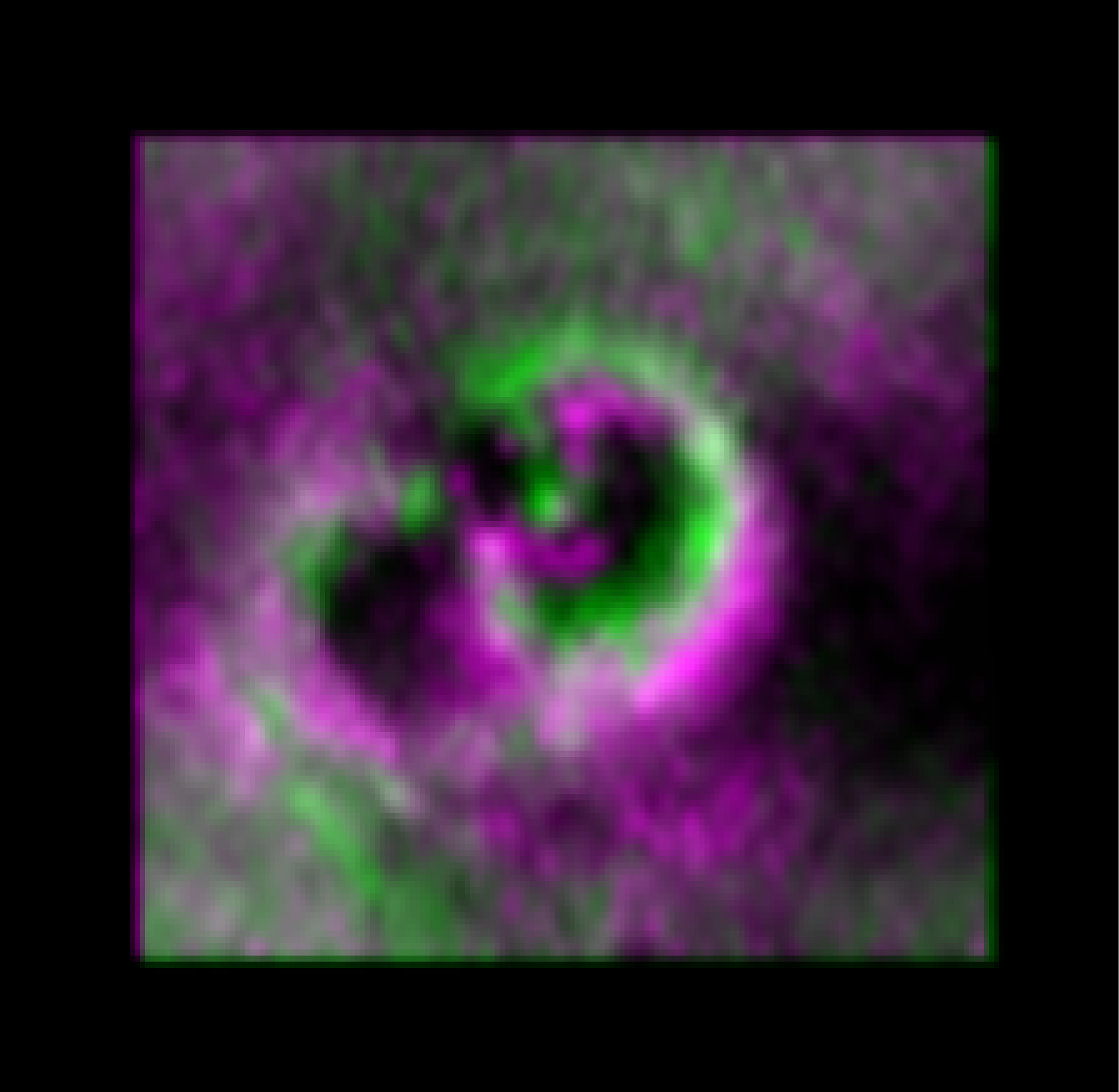}} \\ 
        \makecell[b]{Q2\vspace{20pt}} &
        {\includegraphics[width=0.12\textwidth]{figures/fig_before_Im_019_20230817_104435_3D-Im_008_20230817_103907_3D_Axial_0.pdf}} &
        {\includegraphics[width=0.12\textwidth]{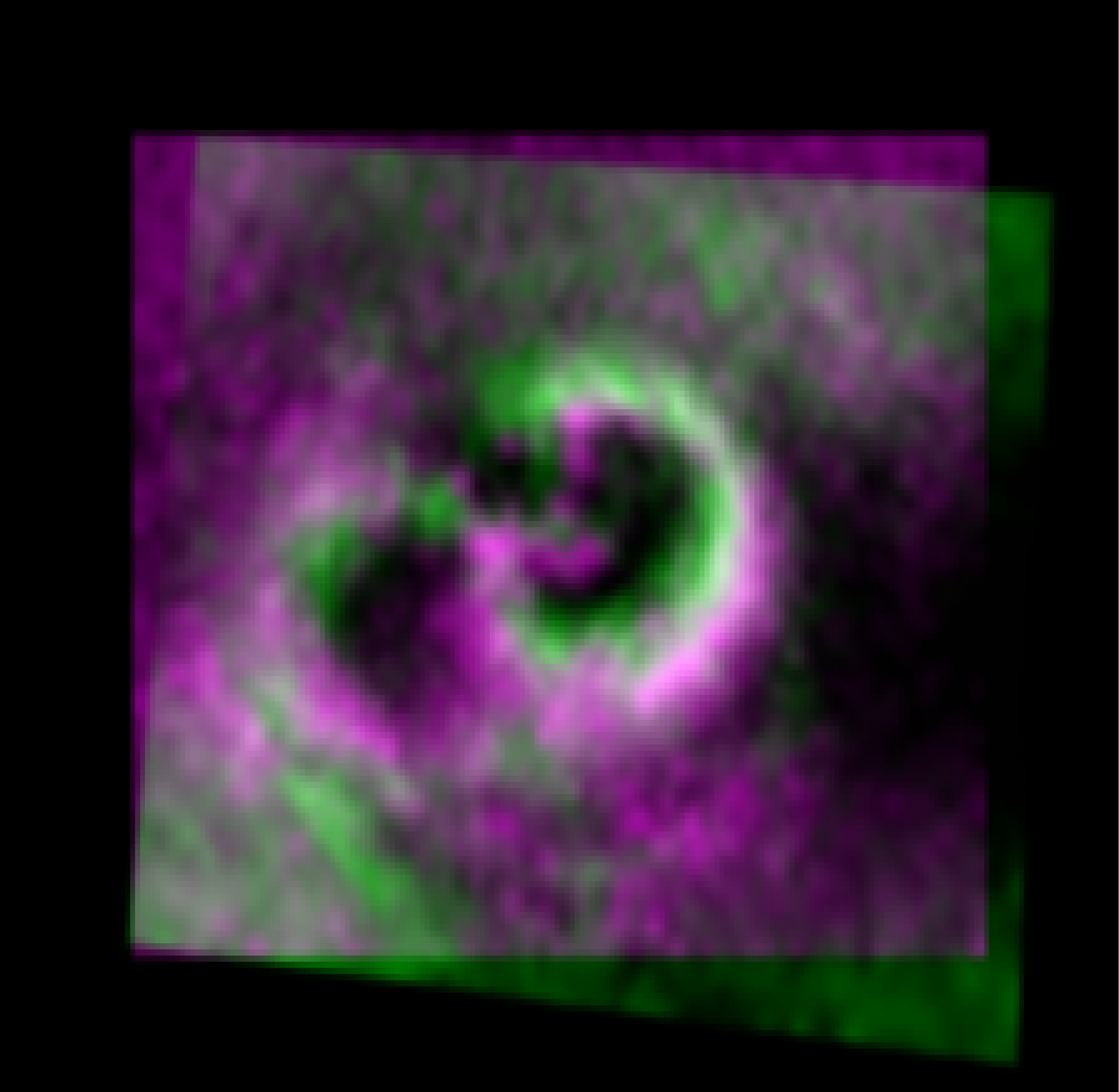}} &
        {\includegraphics[width=0.12\textwidth]{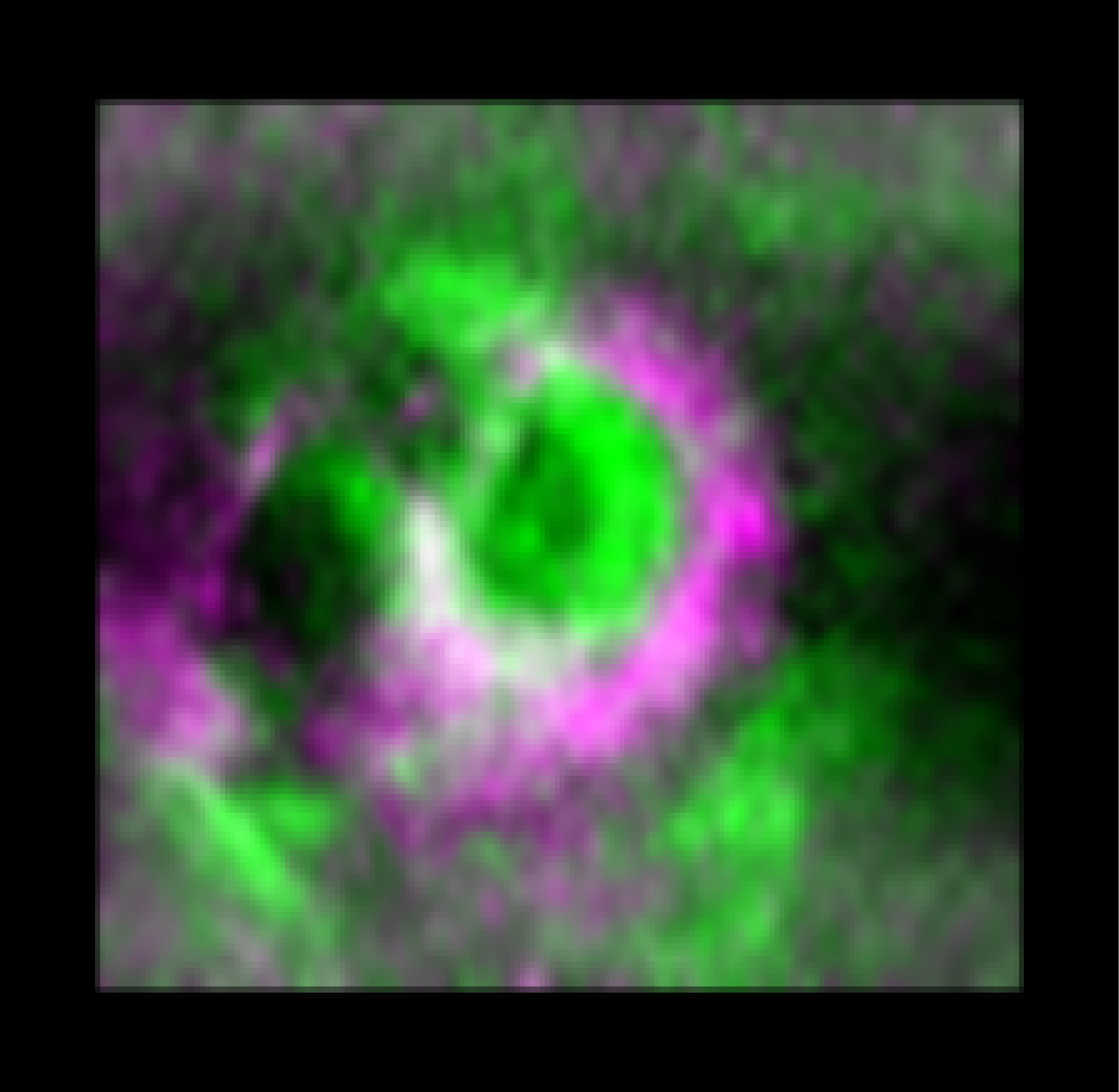}} &
        {\includegraphics[width=0.12\textwidth]{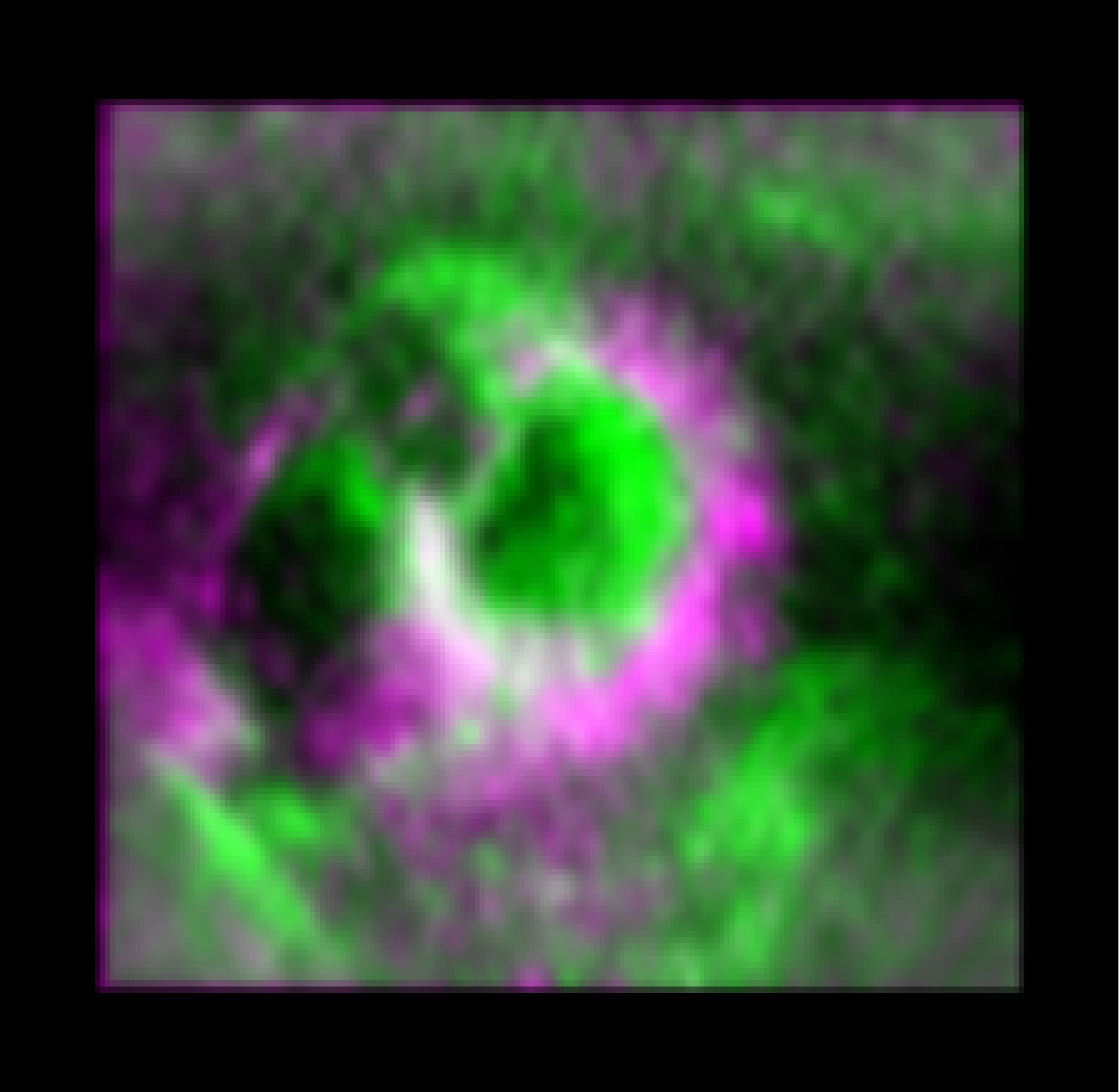}} &
        {\includegraphics[width=0.12\textwidth]{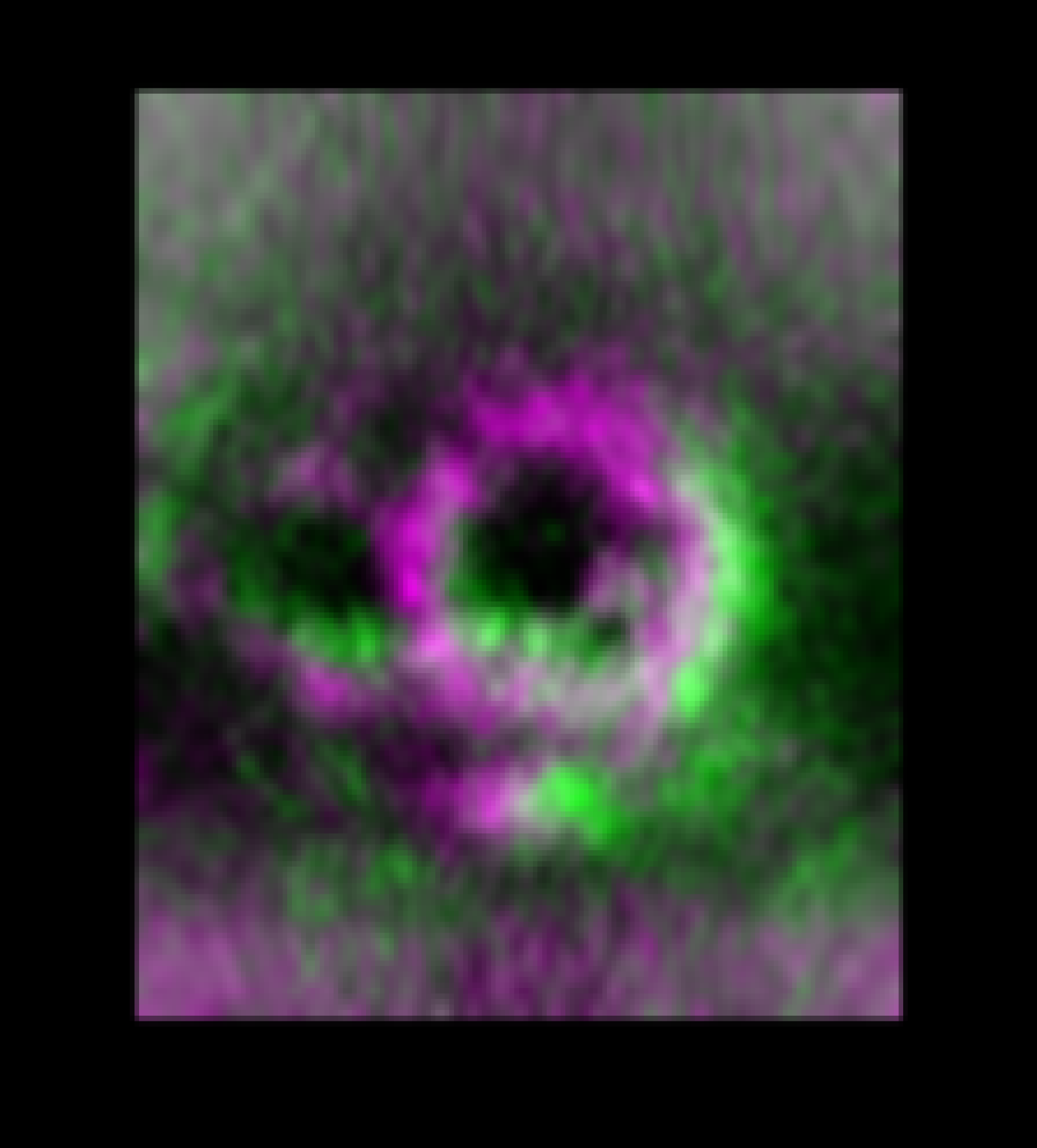}} &
        {\includegraphics[width=0.12\textwidth]{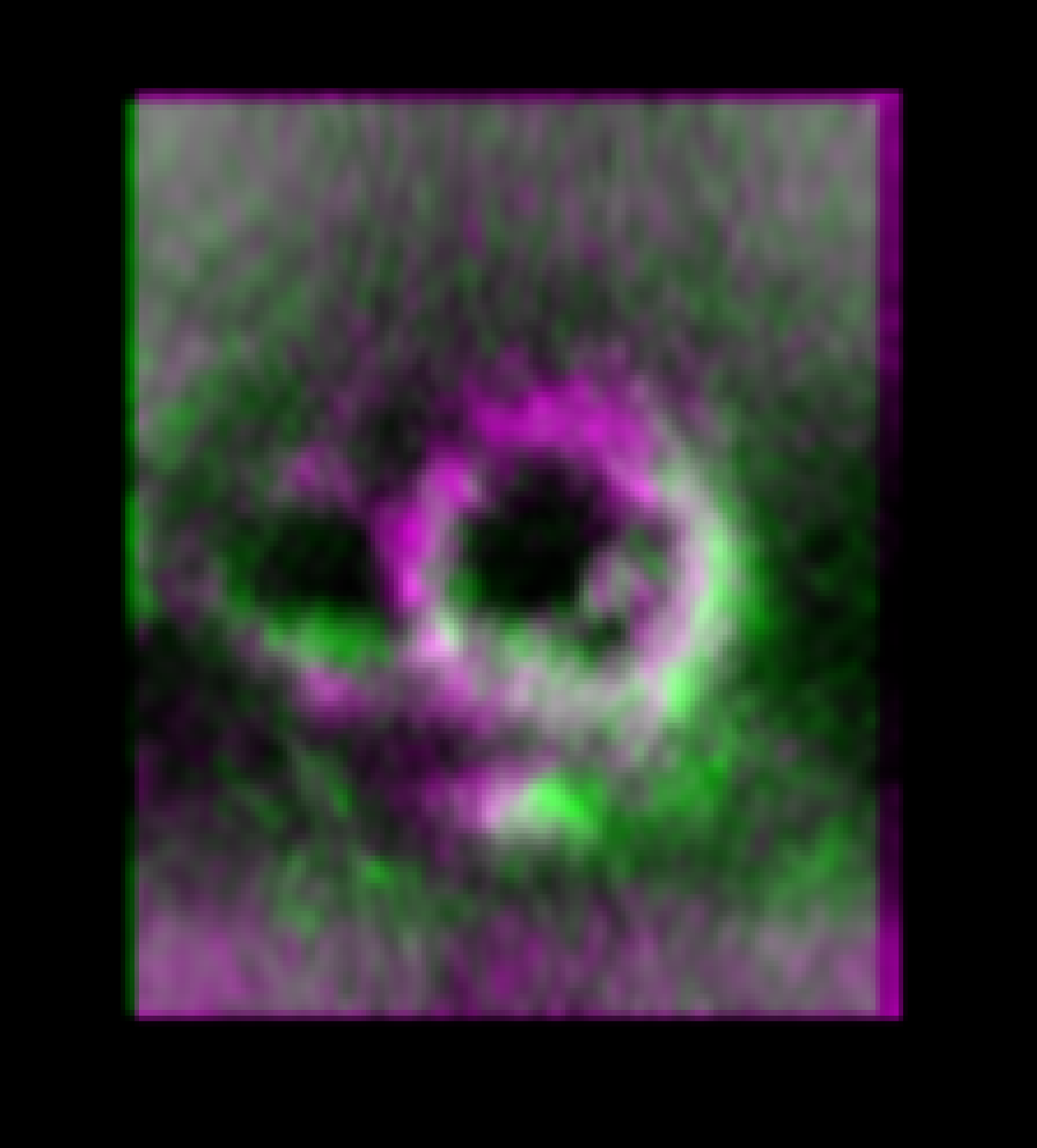}} \\
        \makecell[b]{Q3\vspace{20pt}} &
        {\includegraphics[width=0.12\textwidth]{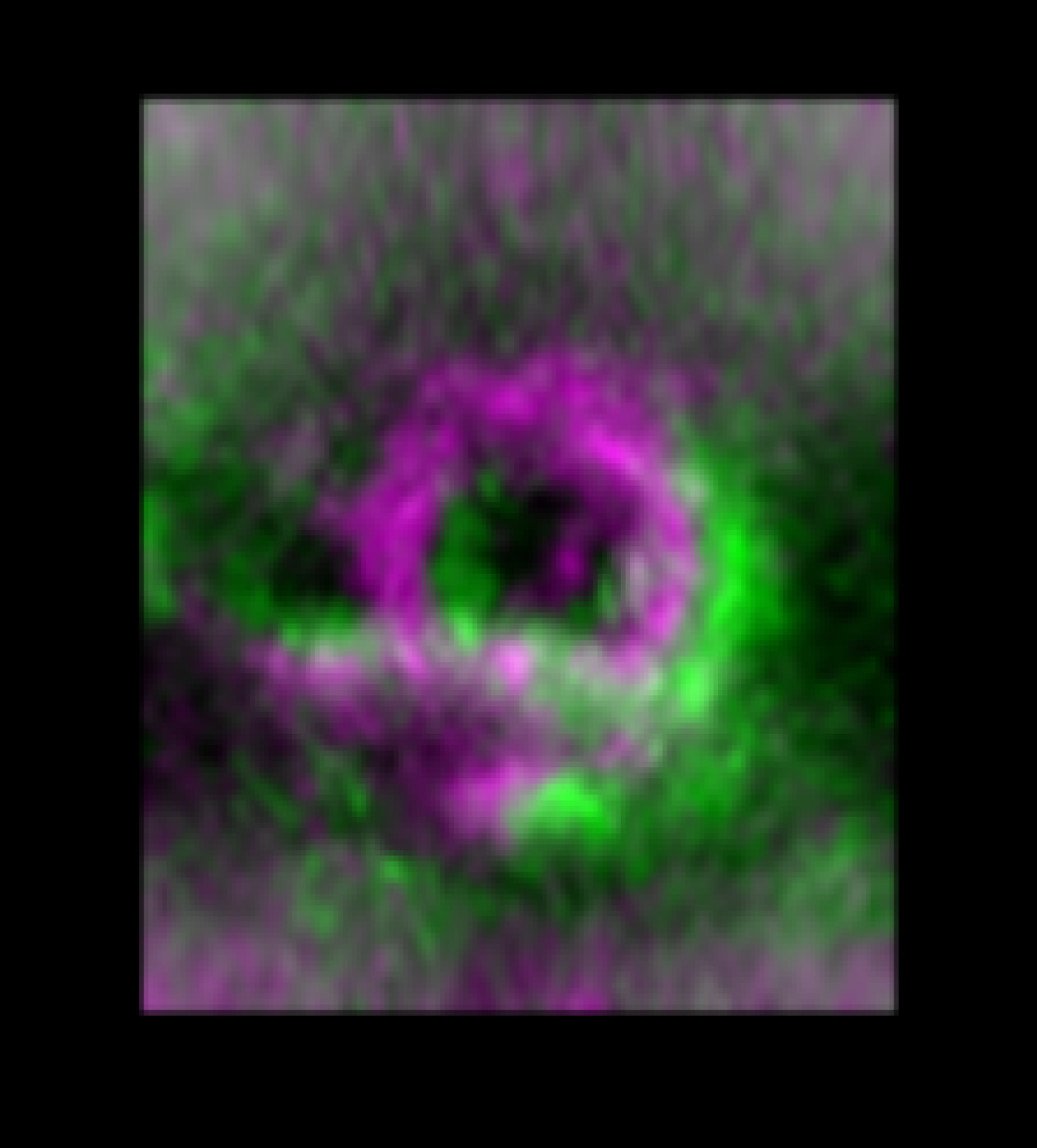}} &
        {\includegraphics[width=0.12\textwidth]{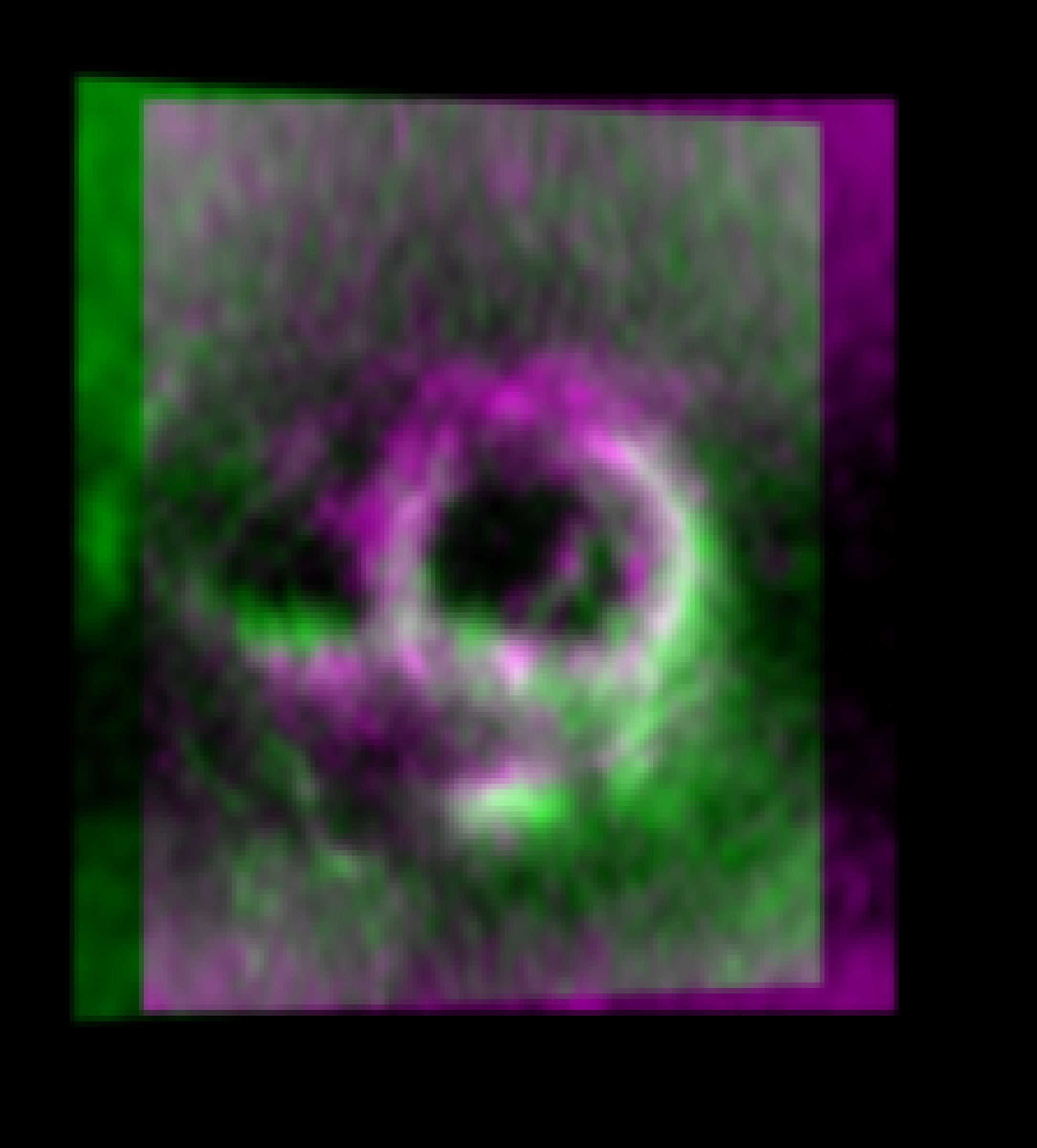}} &
        {\includegraphics[width=0.12\textwidth]{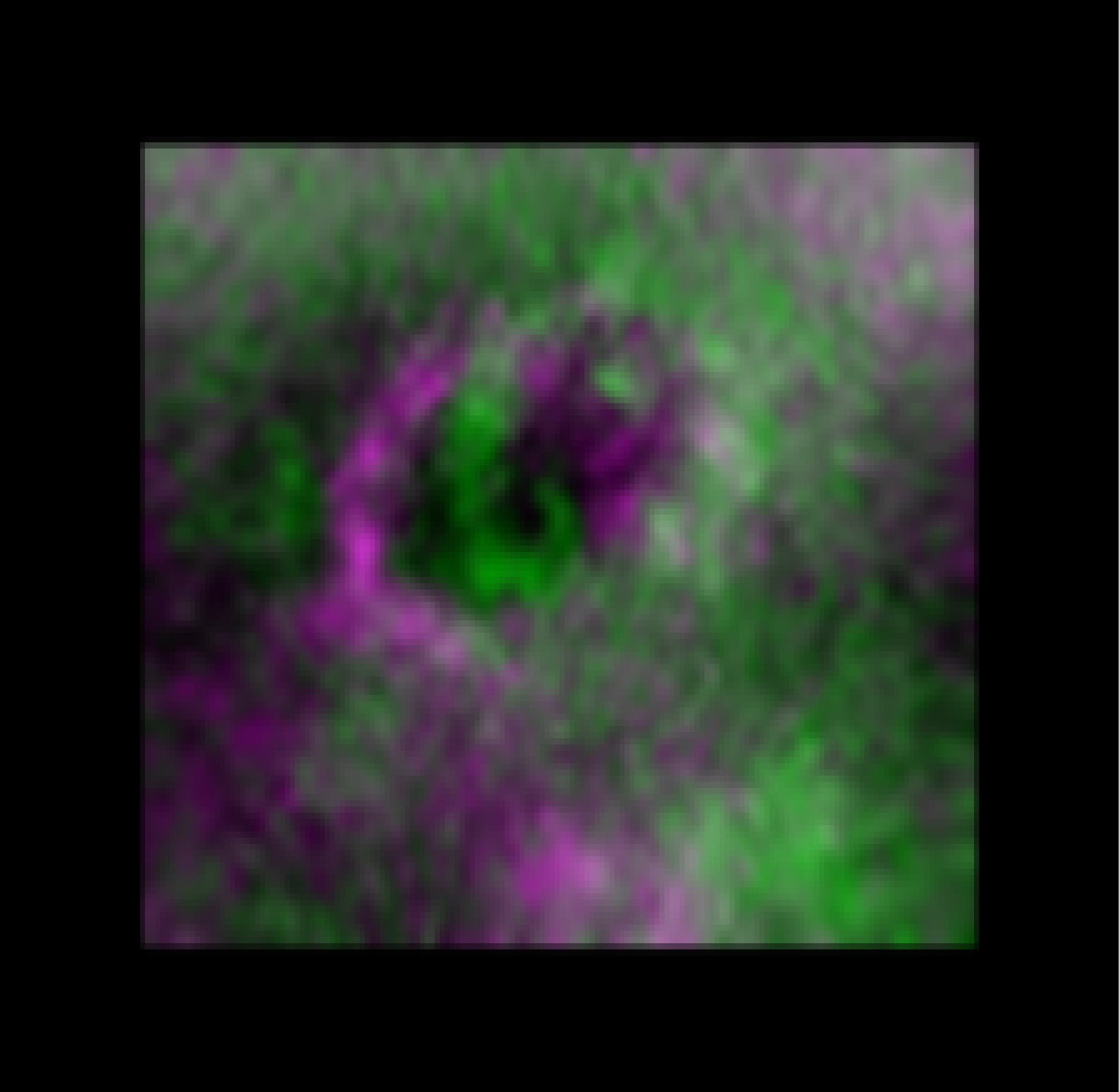}} &
        {\includegraphics[width=0.12\textwidth]{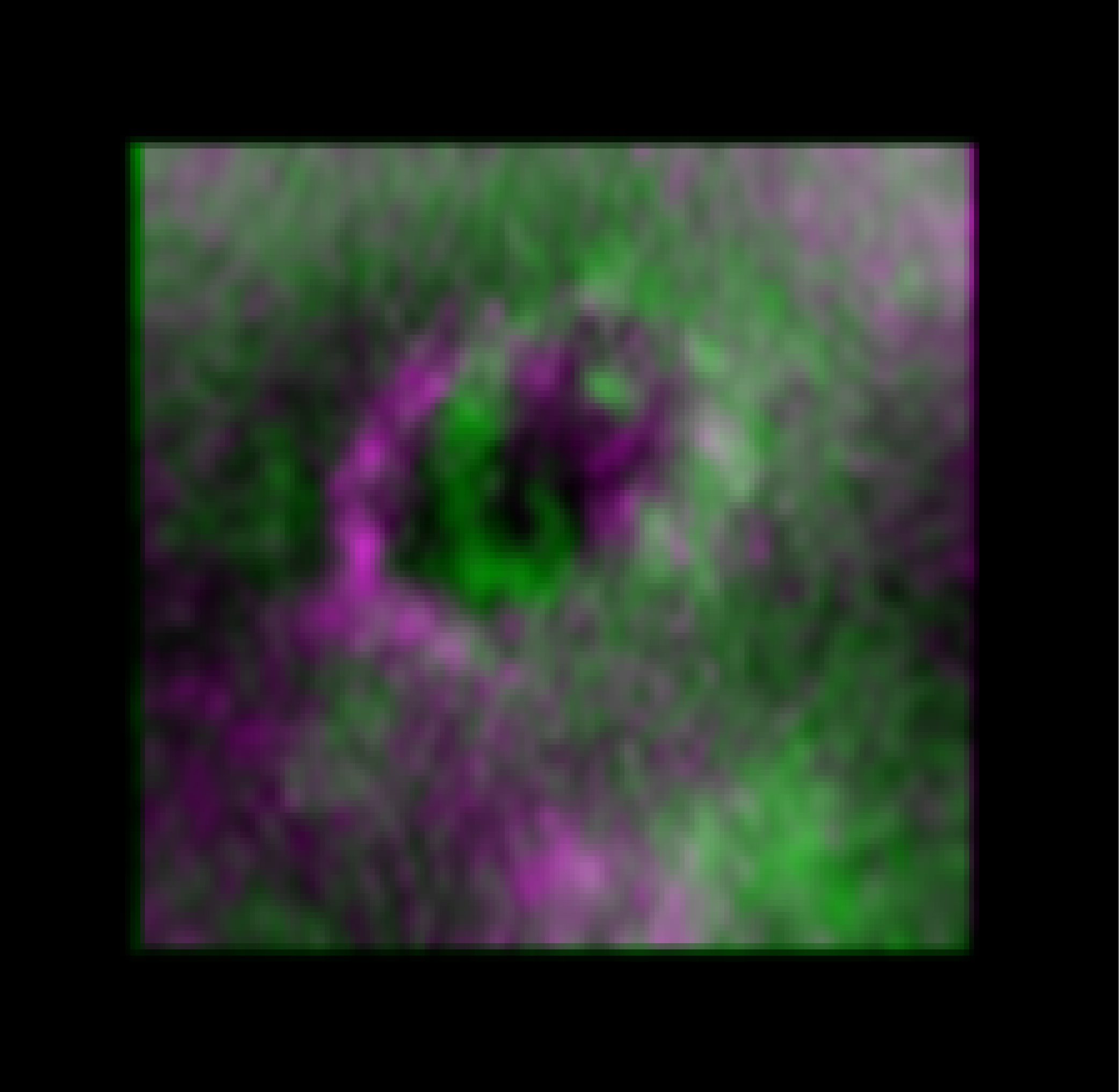}} &
        {\includegraphics[width=0.12\textwidth]{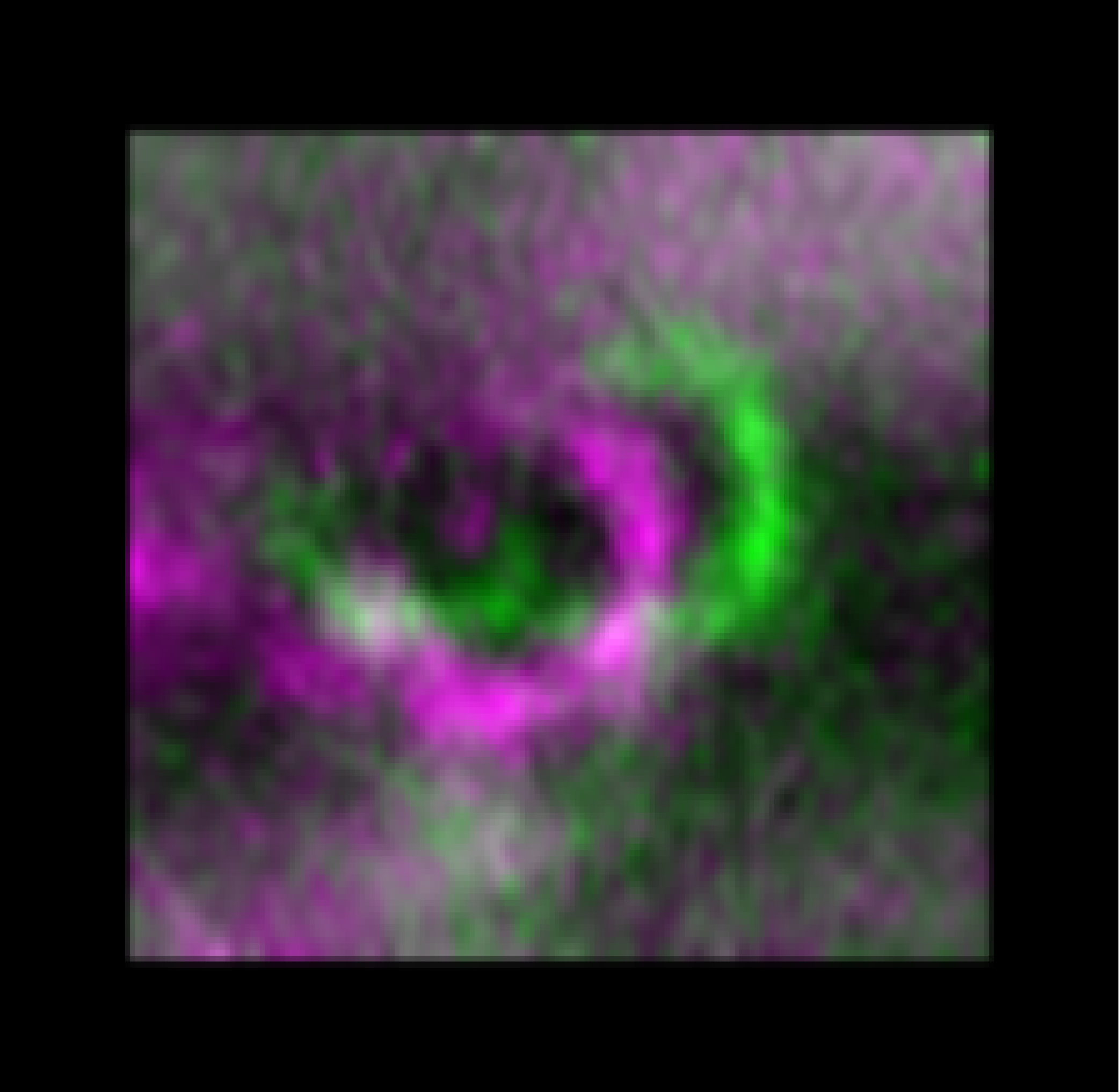}} &
        {\includegraphics[width=0.12\textwidth]{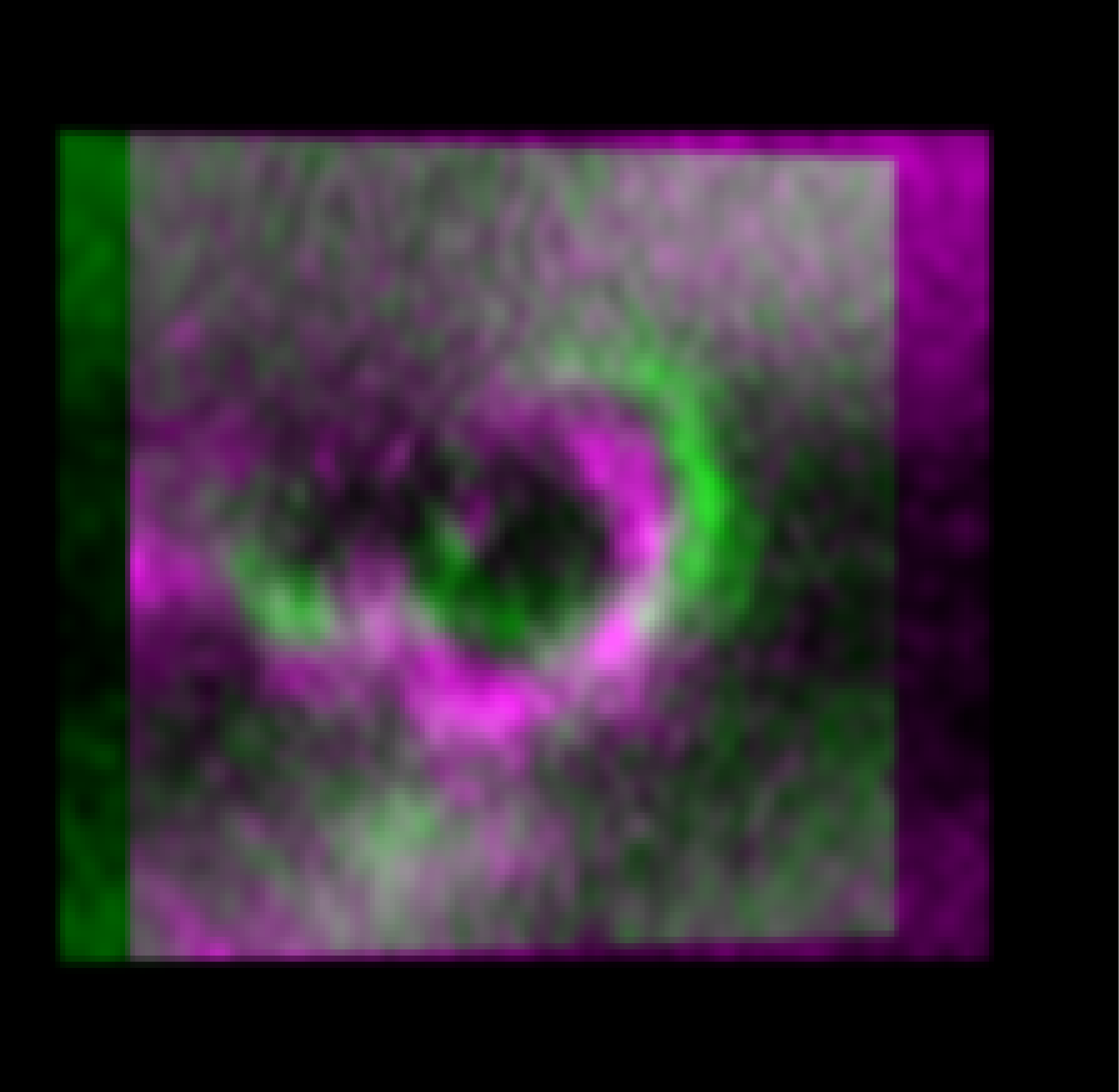}} \\
        \makecell[b]{Max\vspace{20pt}} & 
        {\includegraphics[width=0.12\textwidth]{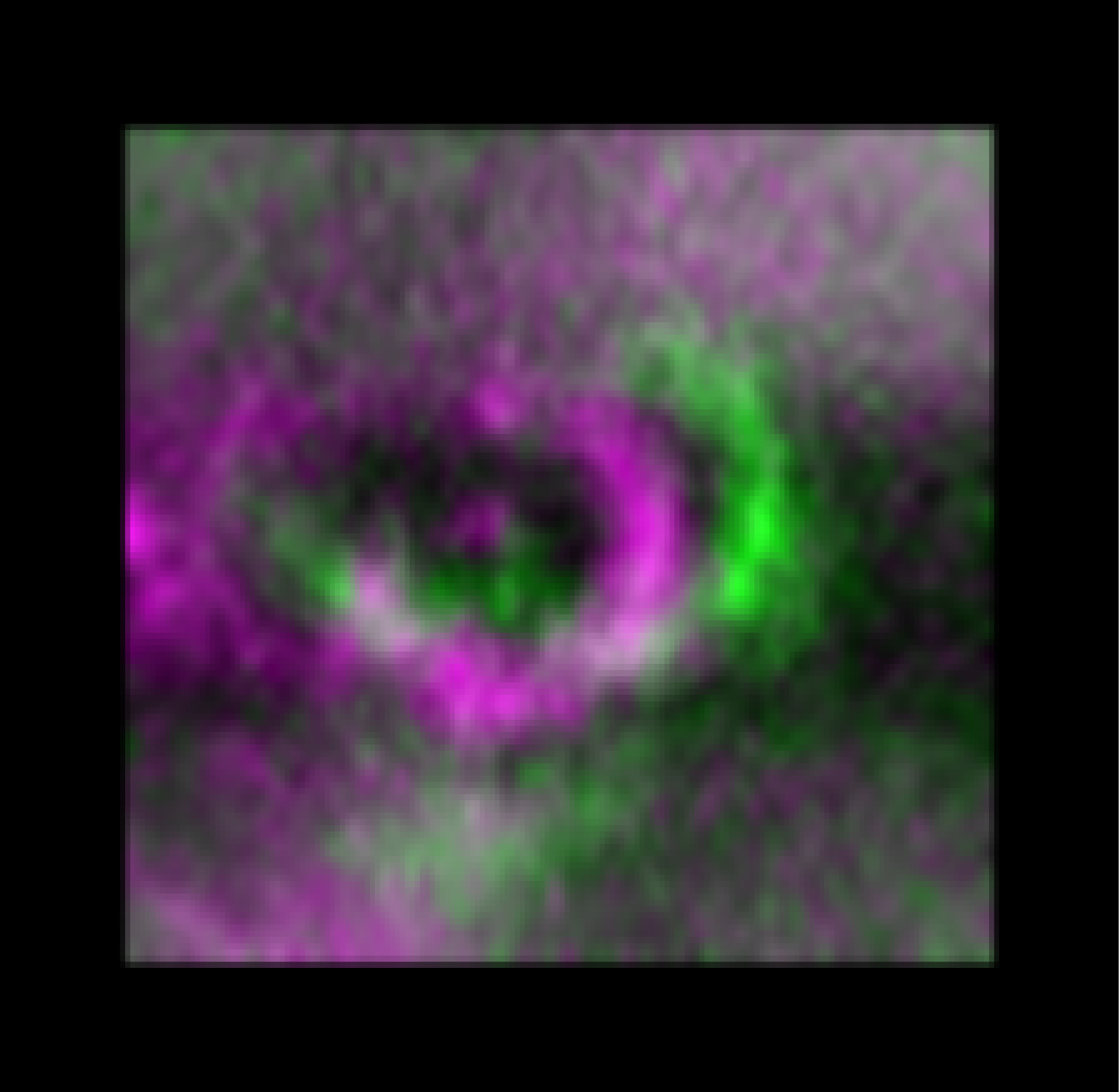}} &
        {\includegraphics[width=0.12\textwidth]{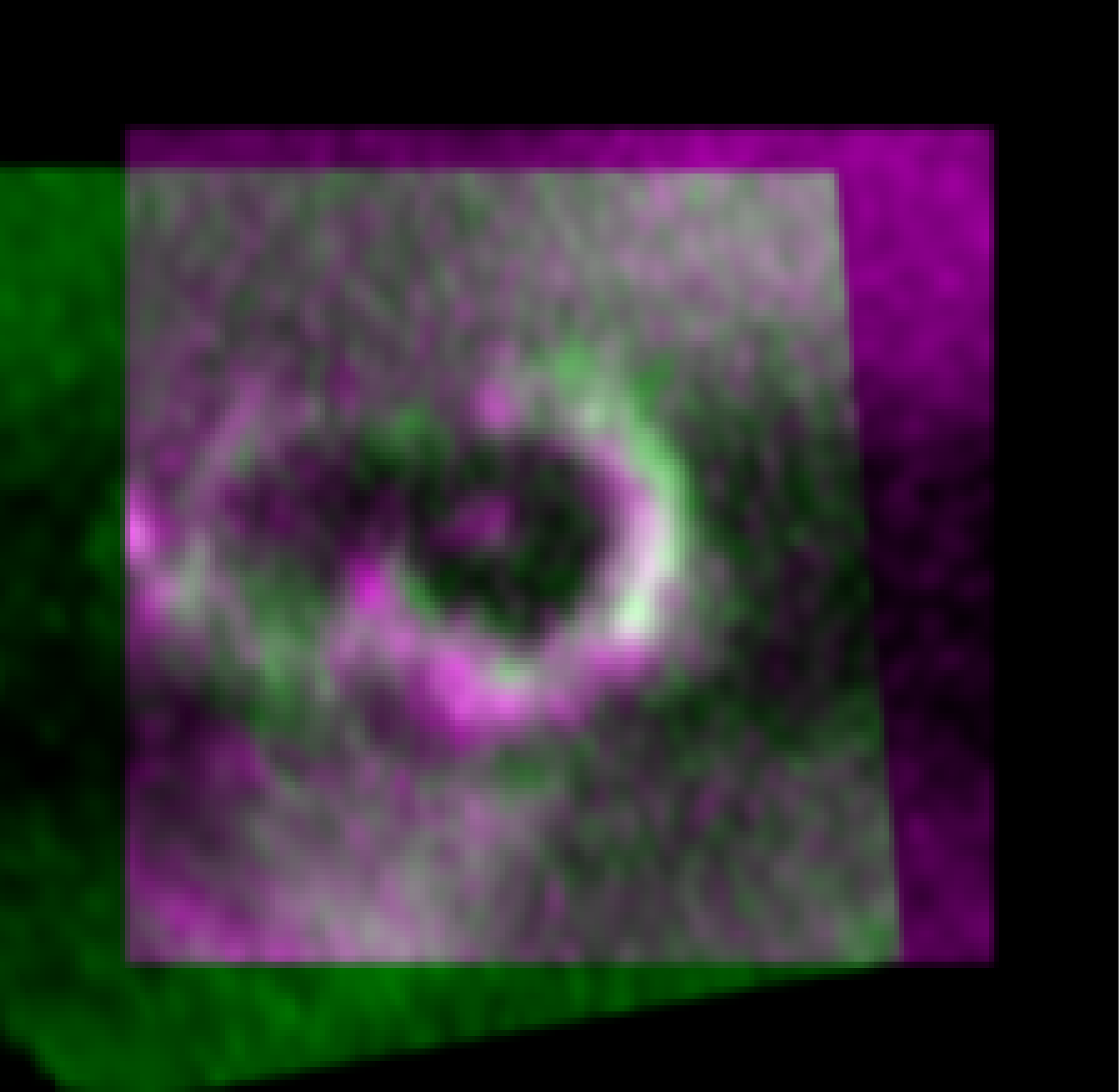}} &
        {\includegraphics[width=0.12\textwidth]{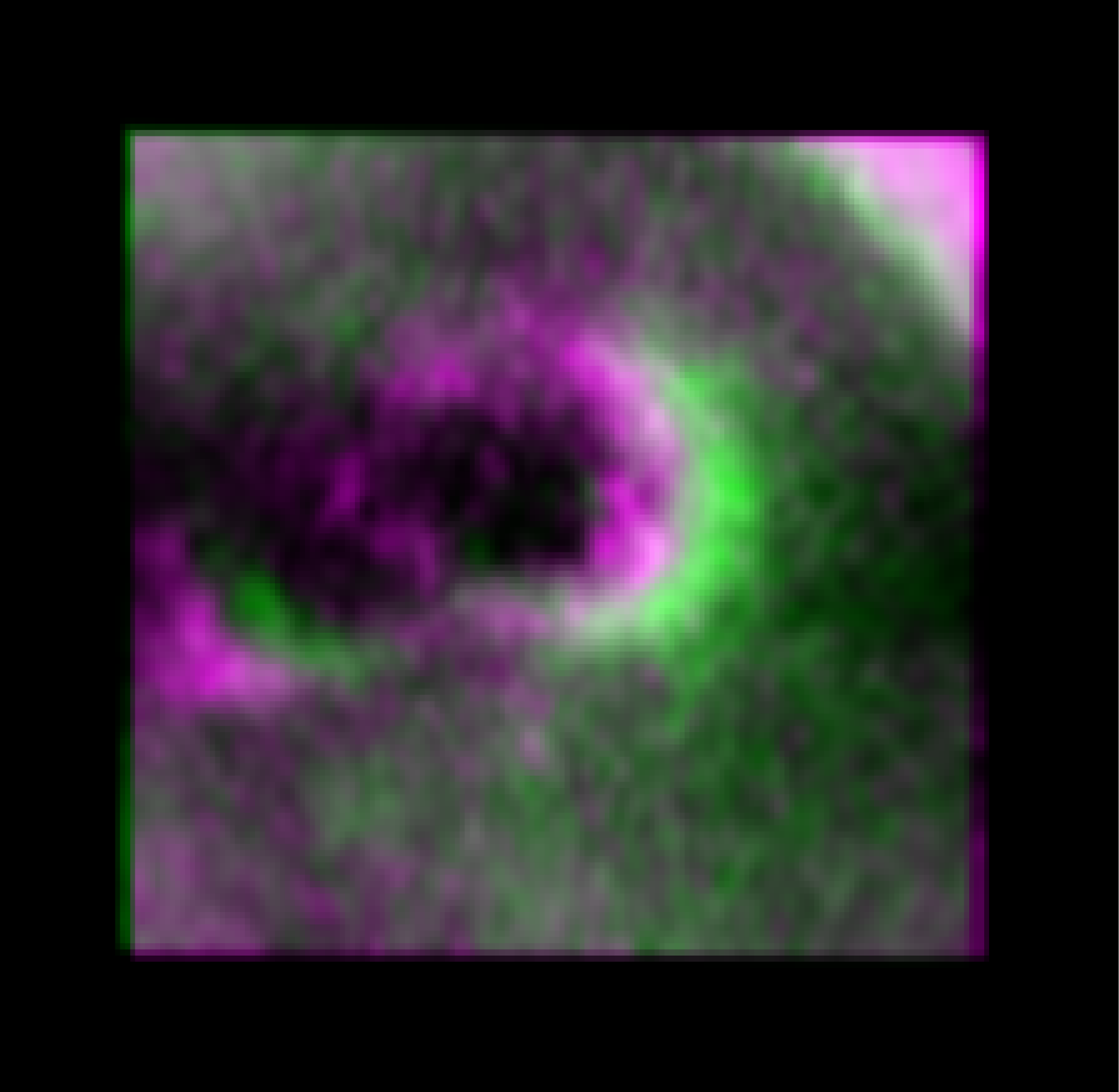}} &
        {\includegraphics[width=0.12\textwidth]{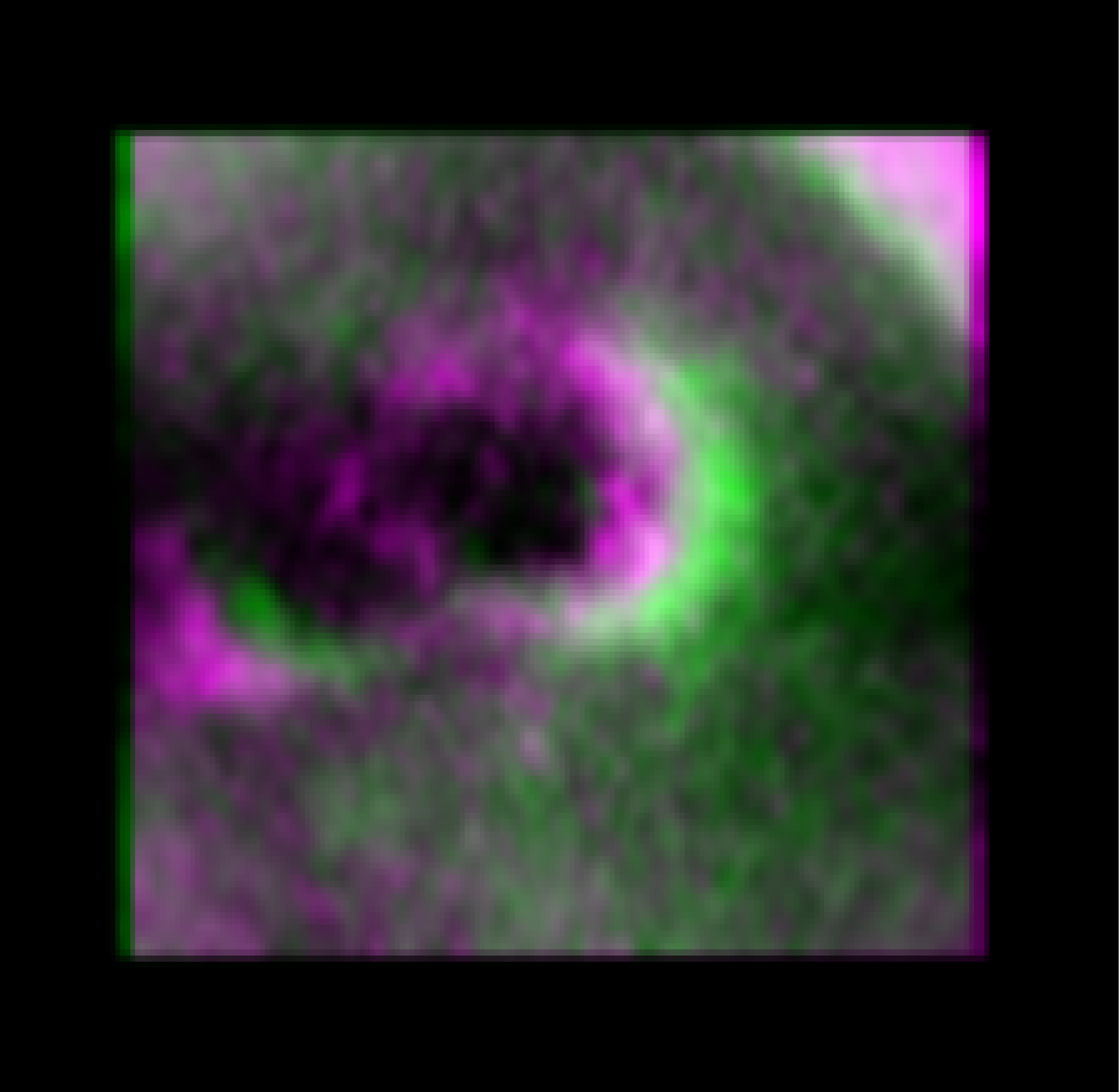}} &
        {\includegraphics[width=0.12\textwidth]{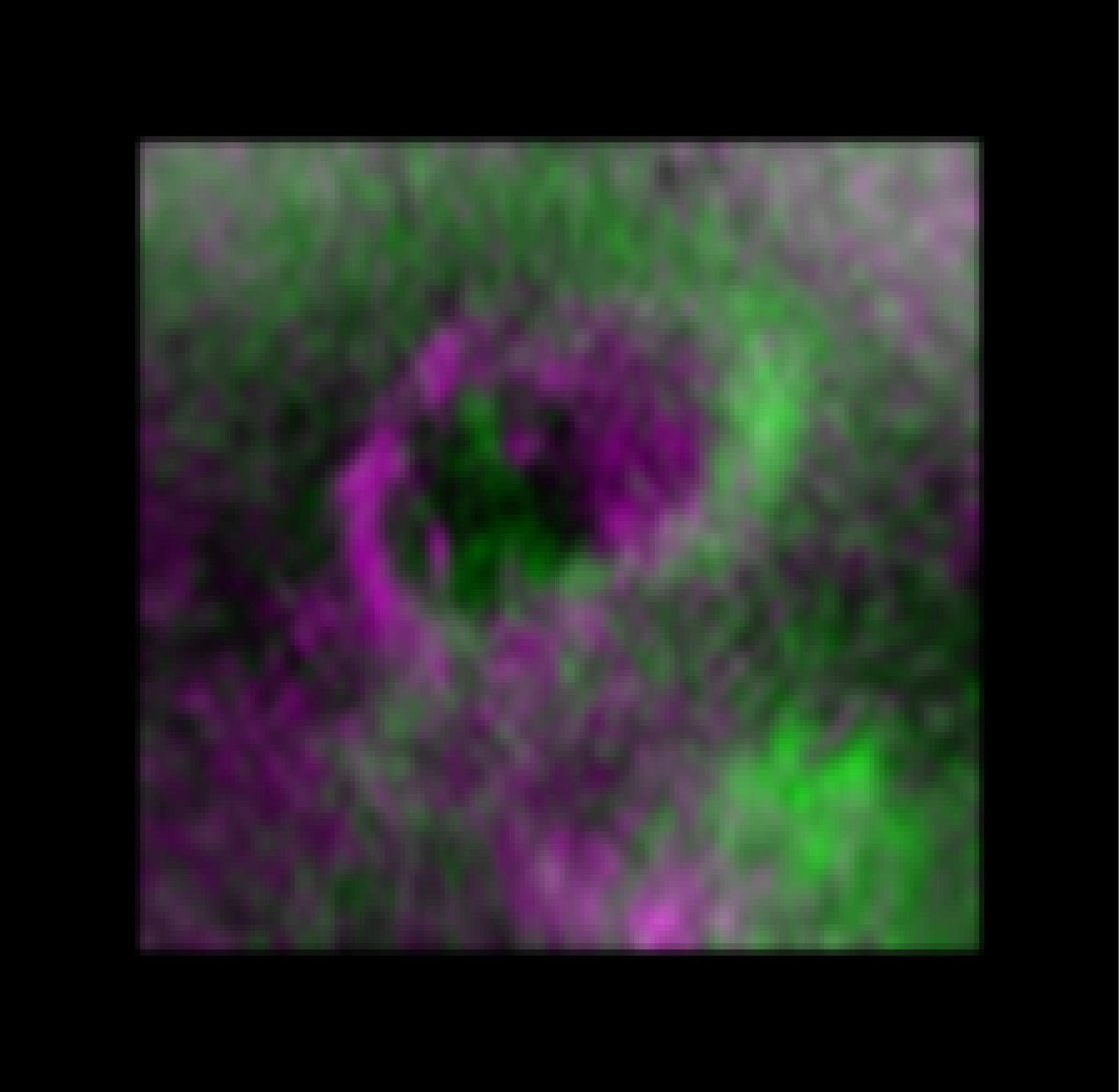}} &
        {\includegraphics[width=0.12\textwidth]{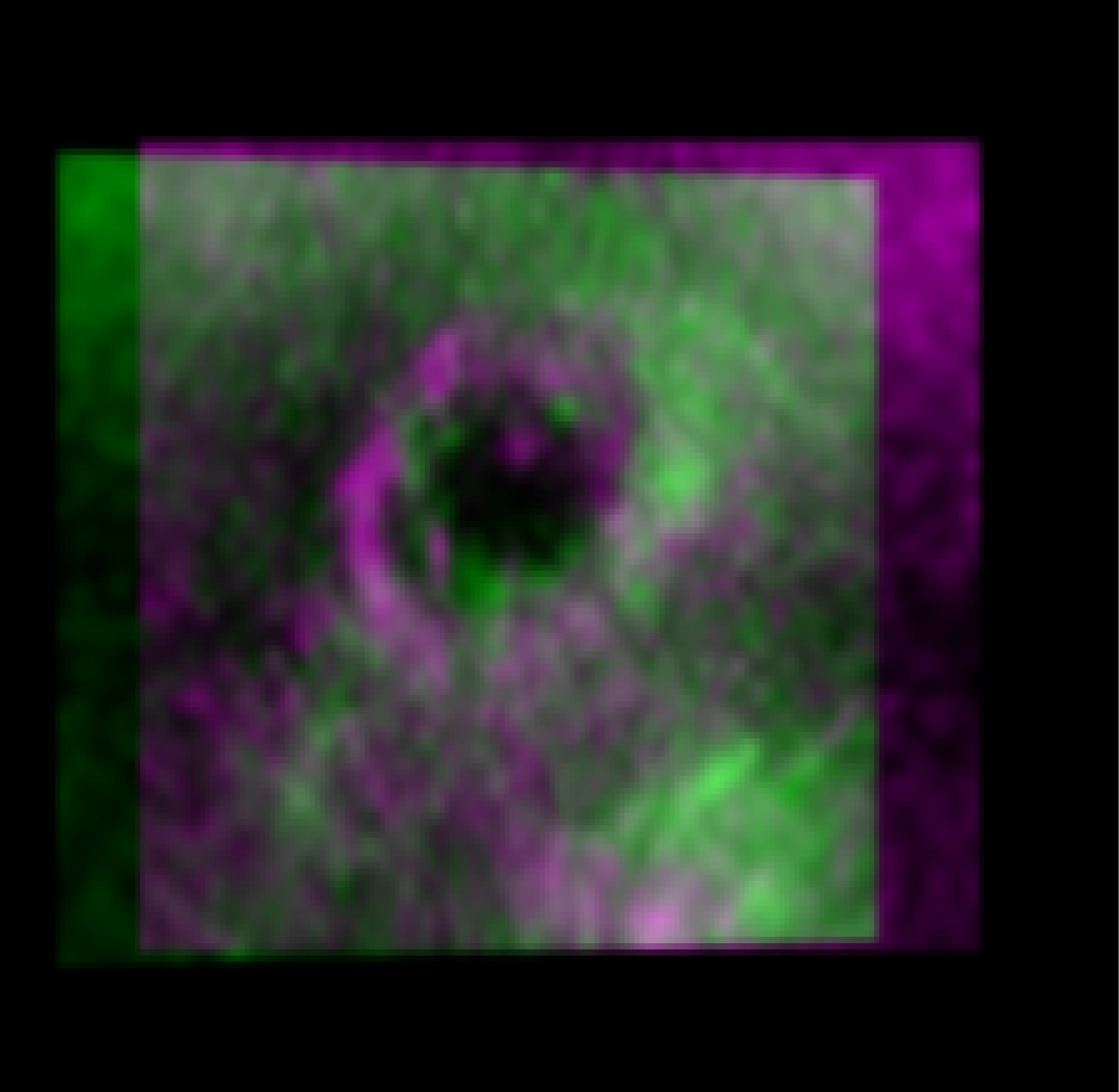}} \\
    \end{tabular}
    \caption{Image-based rigid (PF and EX) registration results of the ED frame of image pairs for percentile DSC difference values of the axial view. Two consecutive columns show images before and after the registration for each category. The source and target images are shown in green and purple colors.}
    \label{fig:fig_perc_images_axial_rigid_img}
\end{figure*}

\begin{figure*}[!ht]
    \centering
    \begin{tabular}{SSSSSSS}
         & \multicolumn{2}{c}{\scriptsize PF (CPU)} & \multicolumn{2}{c}{\scriptsize PF (GPU)} & \multicolumn{2}{c}{\scriptsize EX (CPU)} \\
         & Before & After & Before & After & Before & After \\
         \makecell[b]{Min\vspace{15pt}} & 
        {\includegraphics[width=0.12\textwidth]{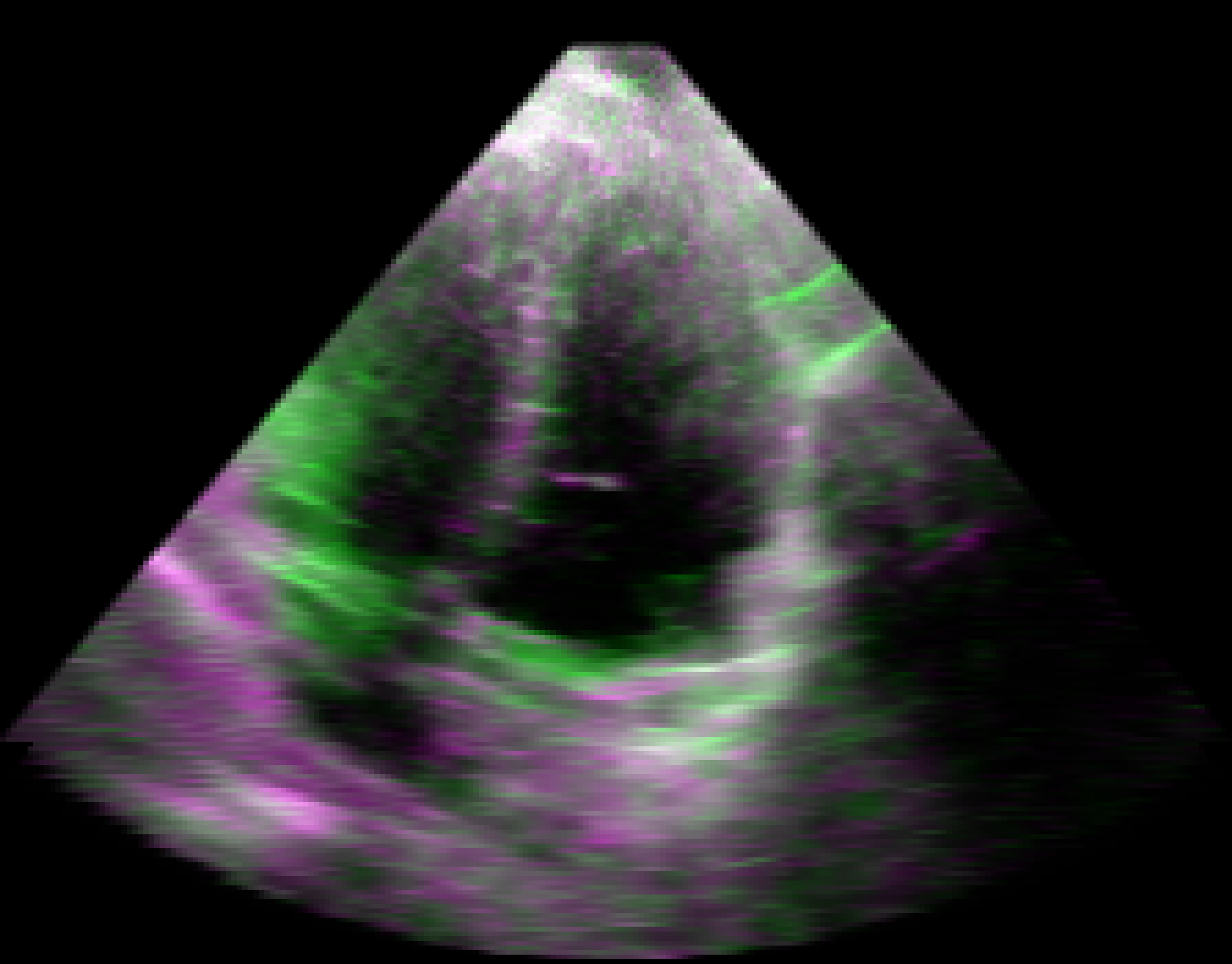}} &
        {\includegraphics[width=0.12\textwidth]{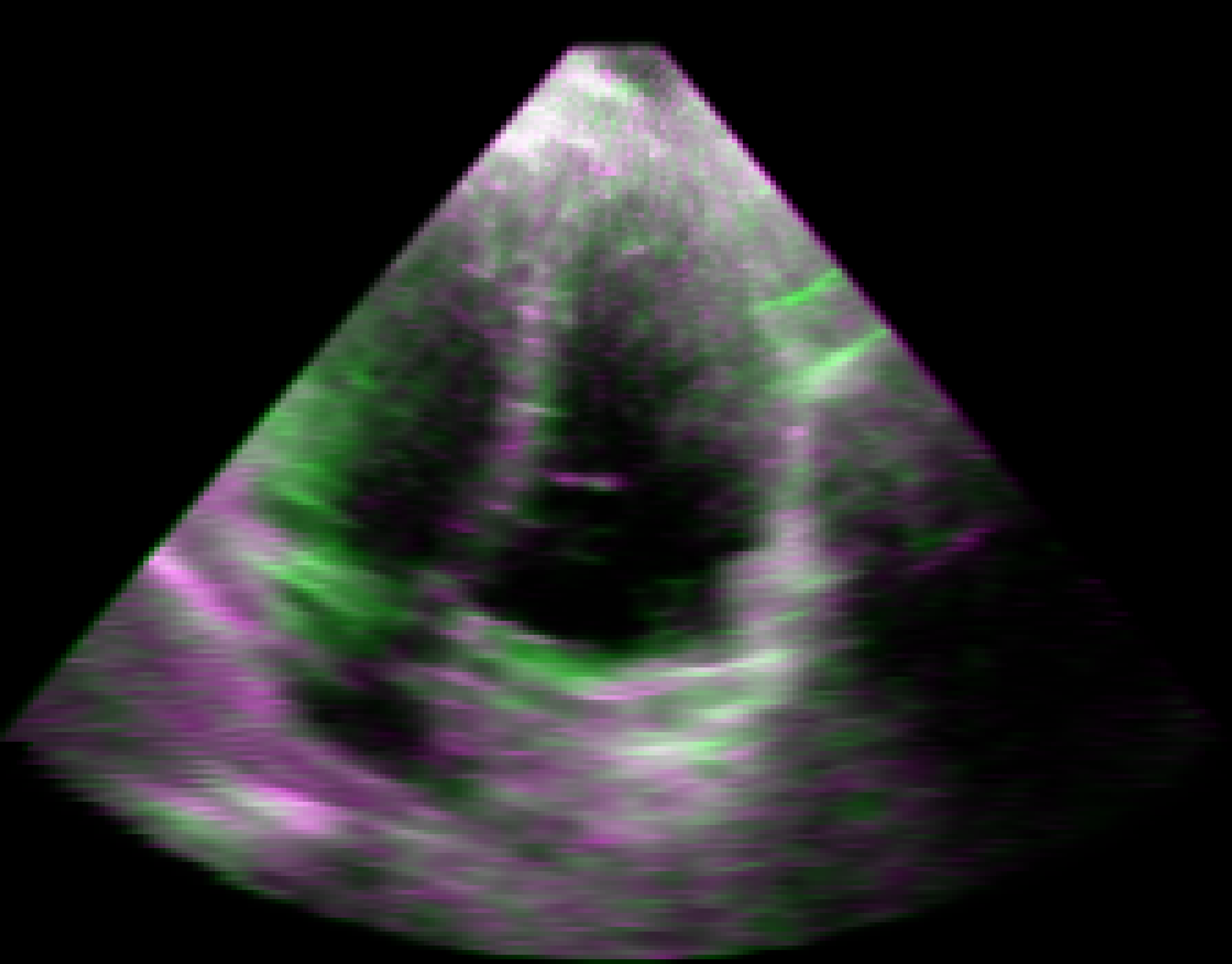}} &
        {\includegraphics[width=0.12\textwidth]{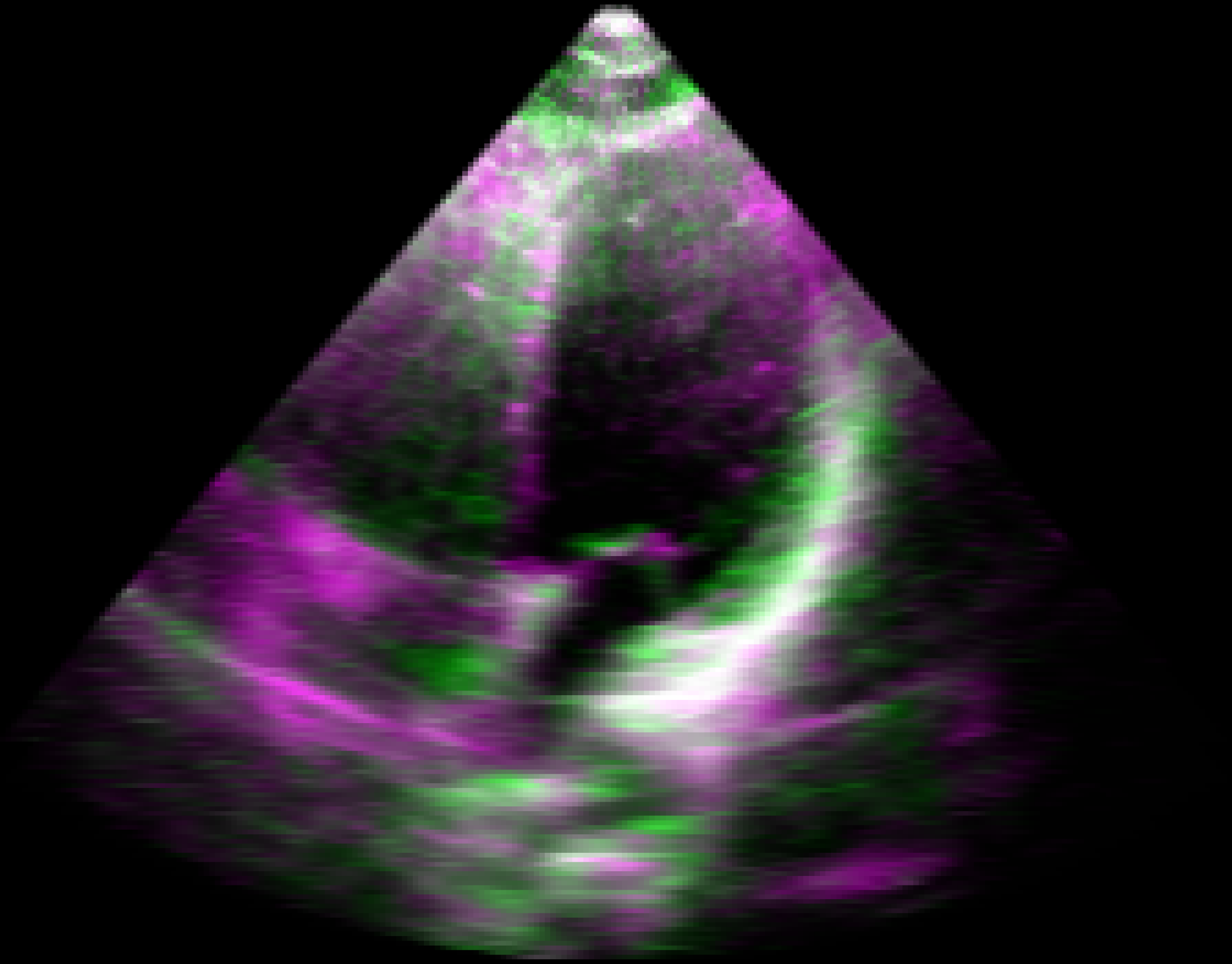}} &
        {\includegraphics[width=0.12\textwidth]{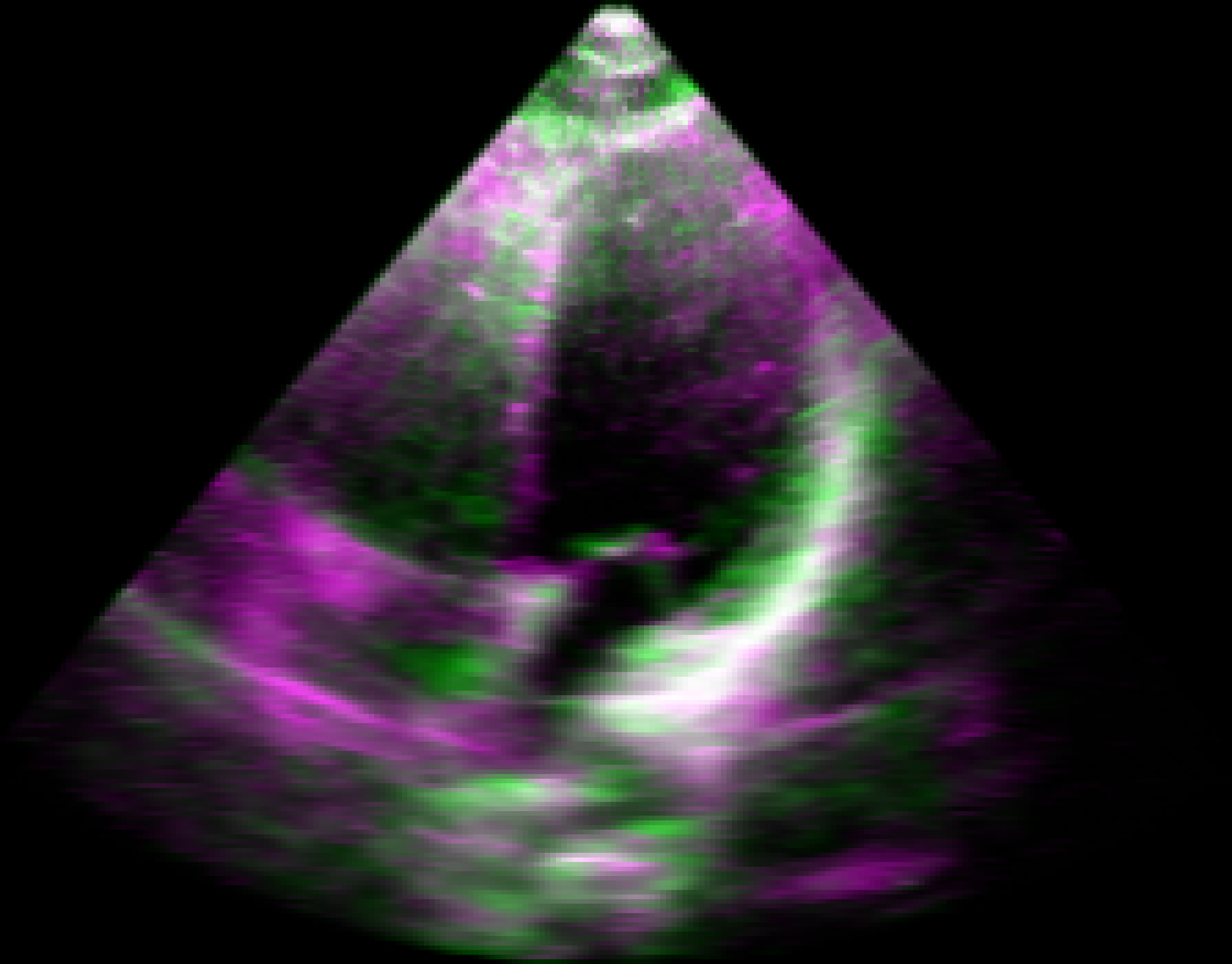}} &
        {\includegraphics[width=0.12\textwidth]{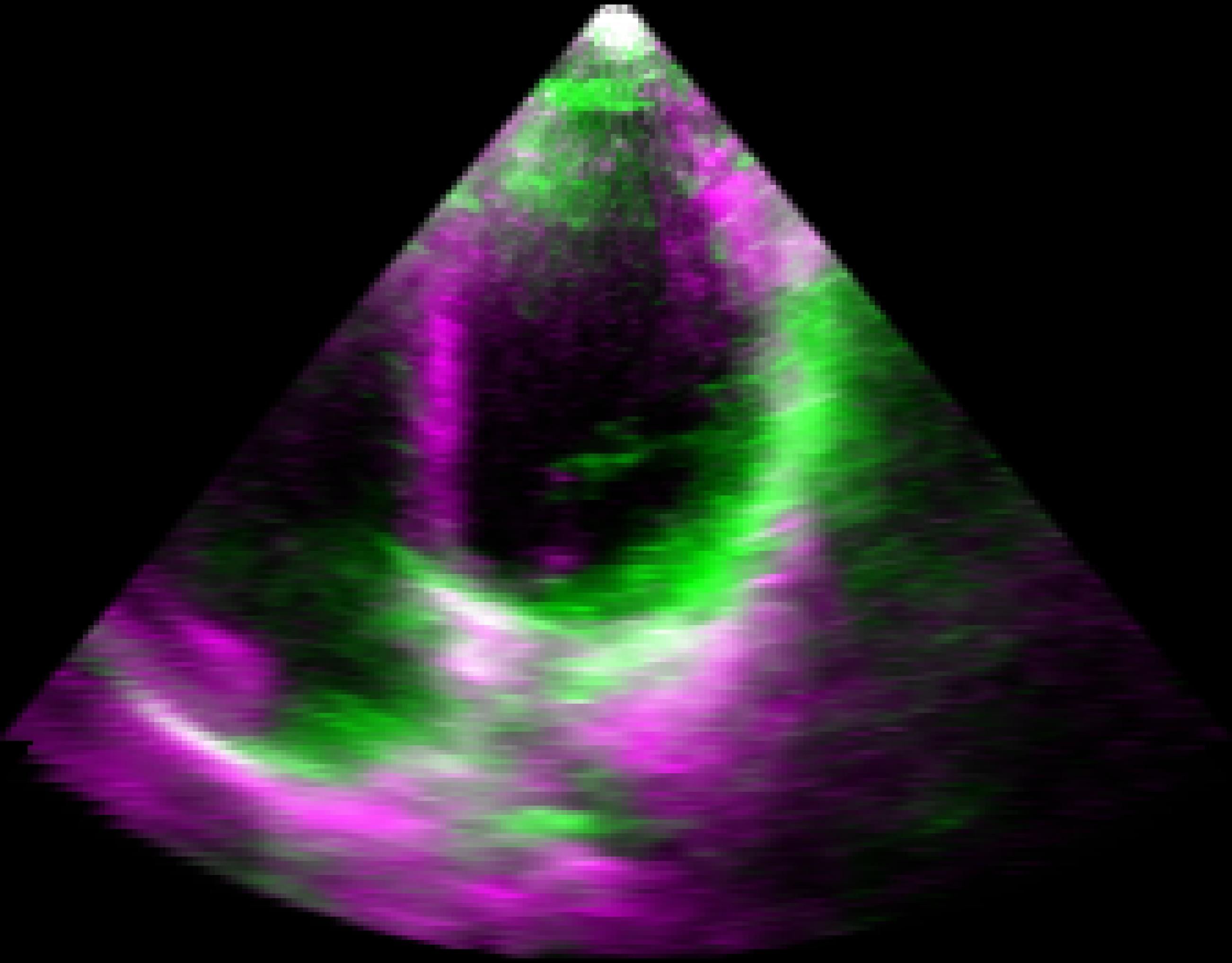}} &
        {\includegraphics[width=0.12\textwidth]{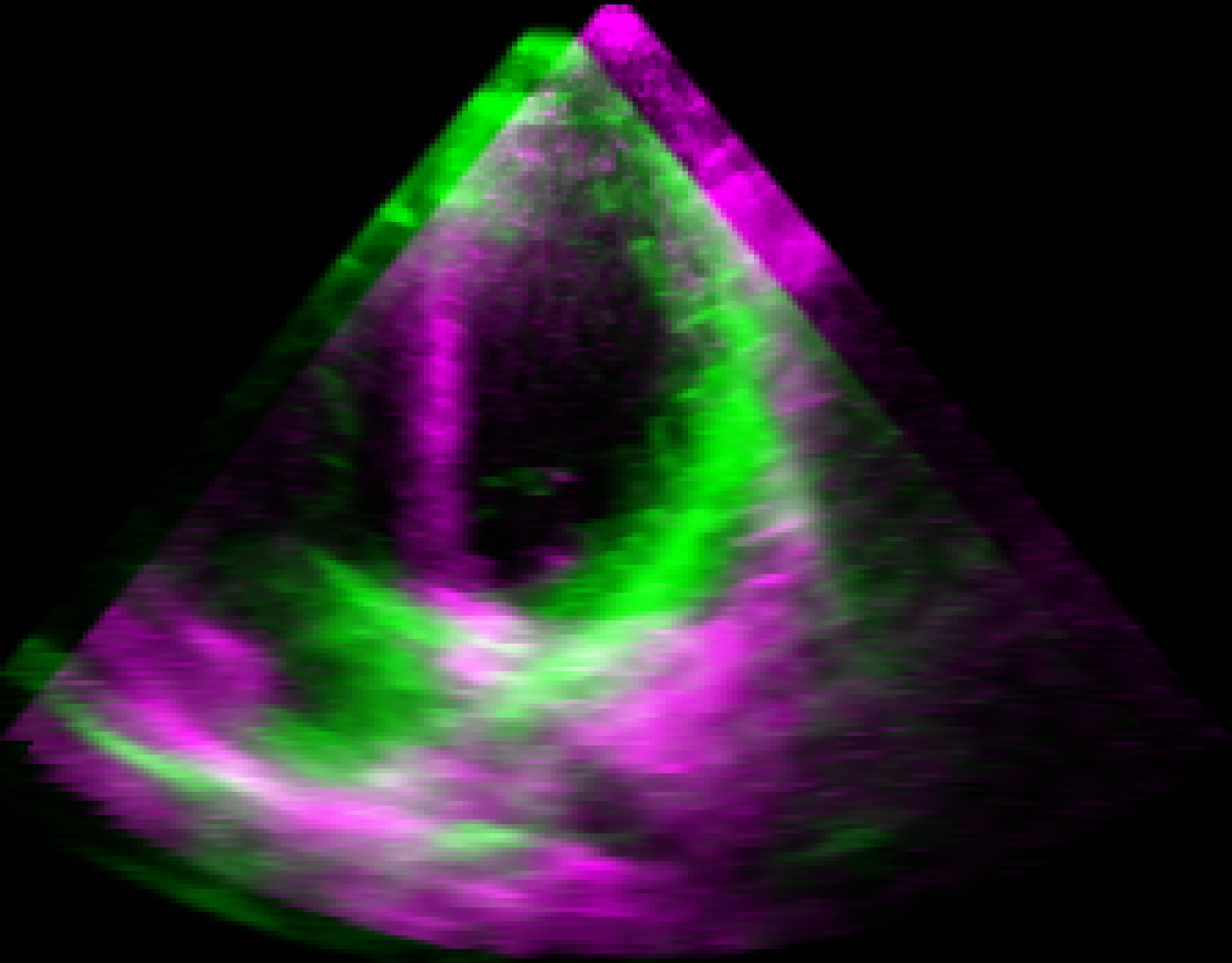}} \\
        \makecell[b]{Q1\vspace{15pt}} & 
        {\includegraphics[width=0.12\textwidth]{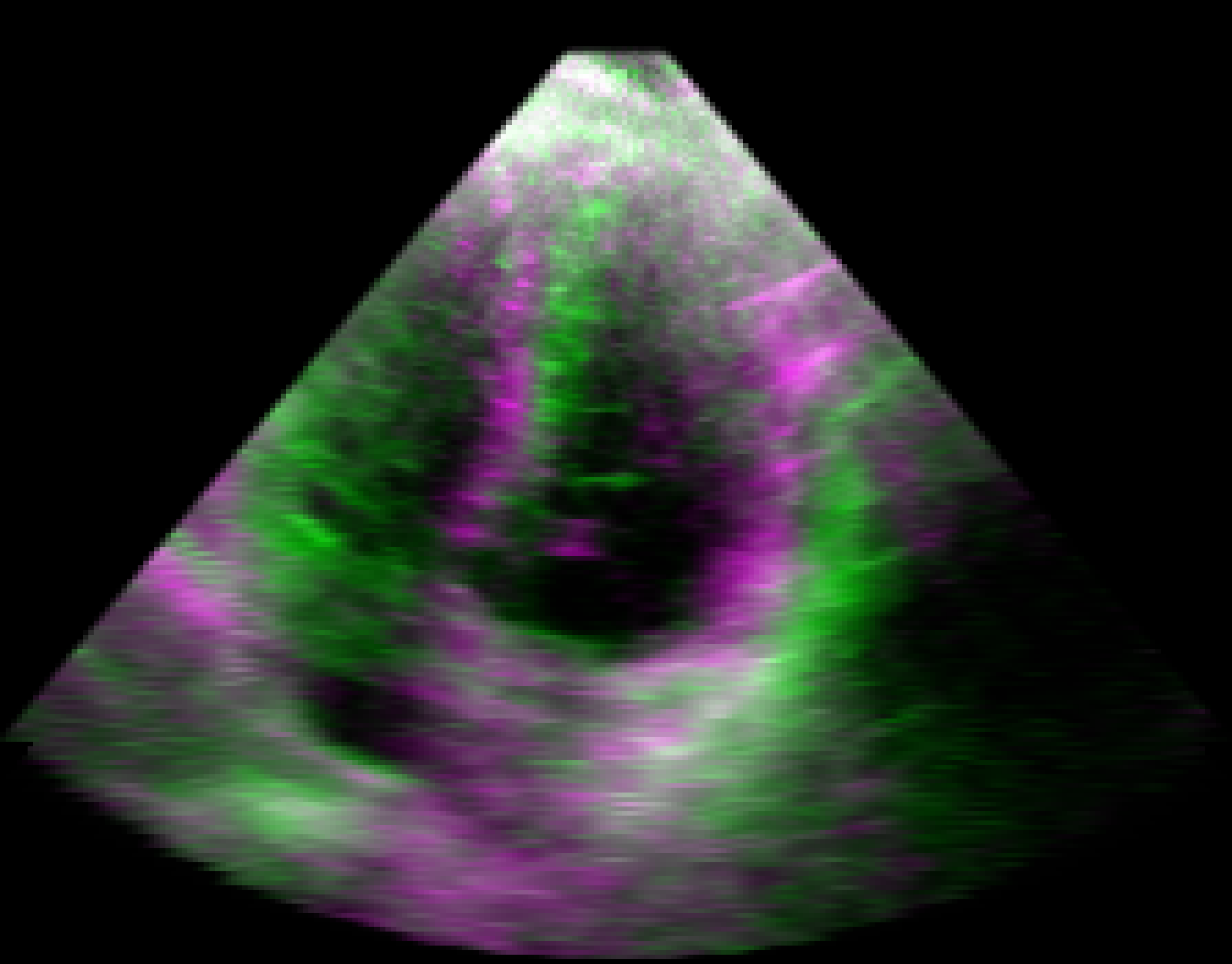}} &
        {\includegraphics[width=0.12\textwidth]{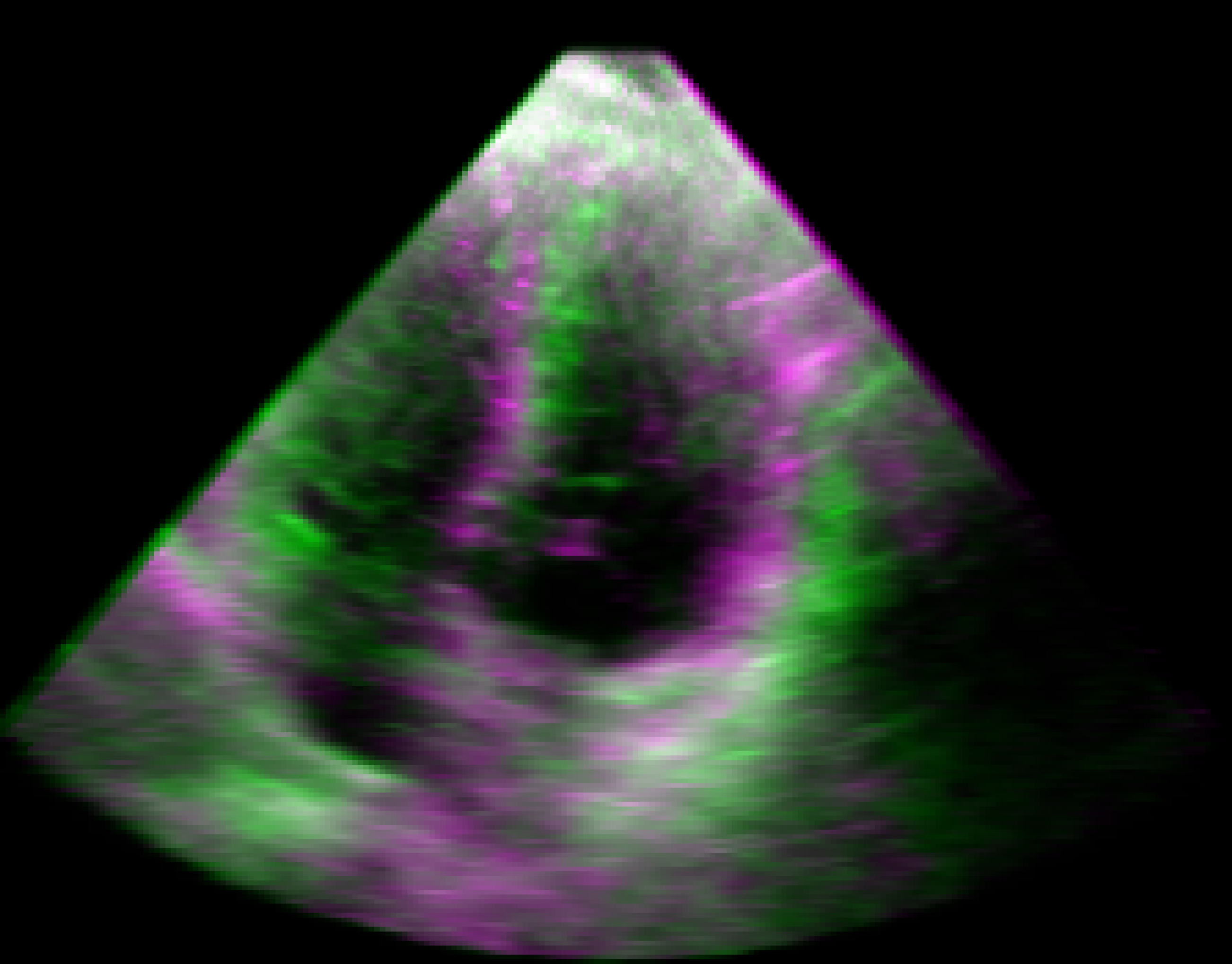}} &
        {\includegraphics[width=0.12\textwidth]{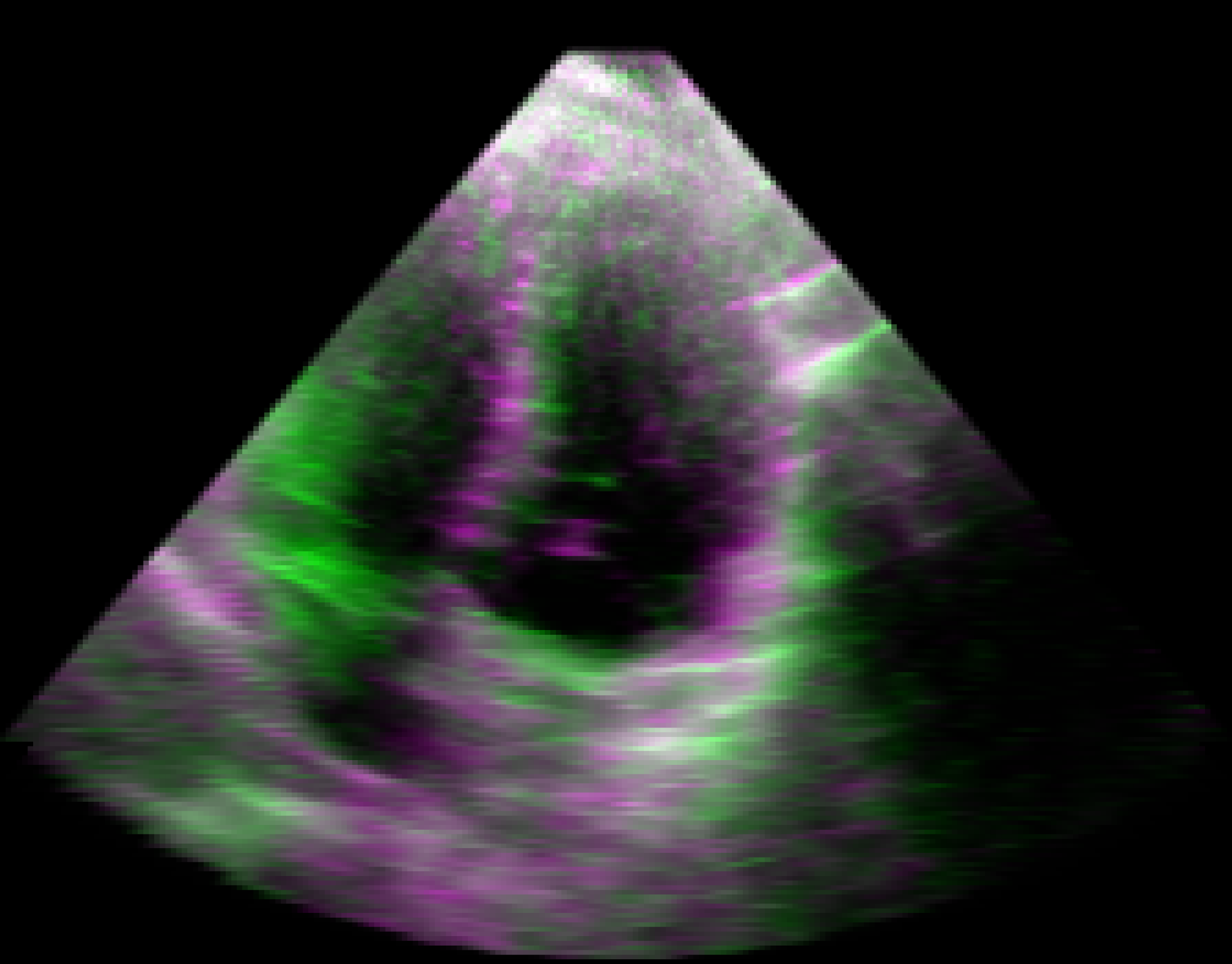}} &
        {\includegraphics[width=0.12\textwidth]{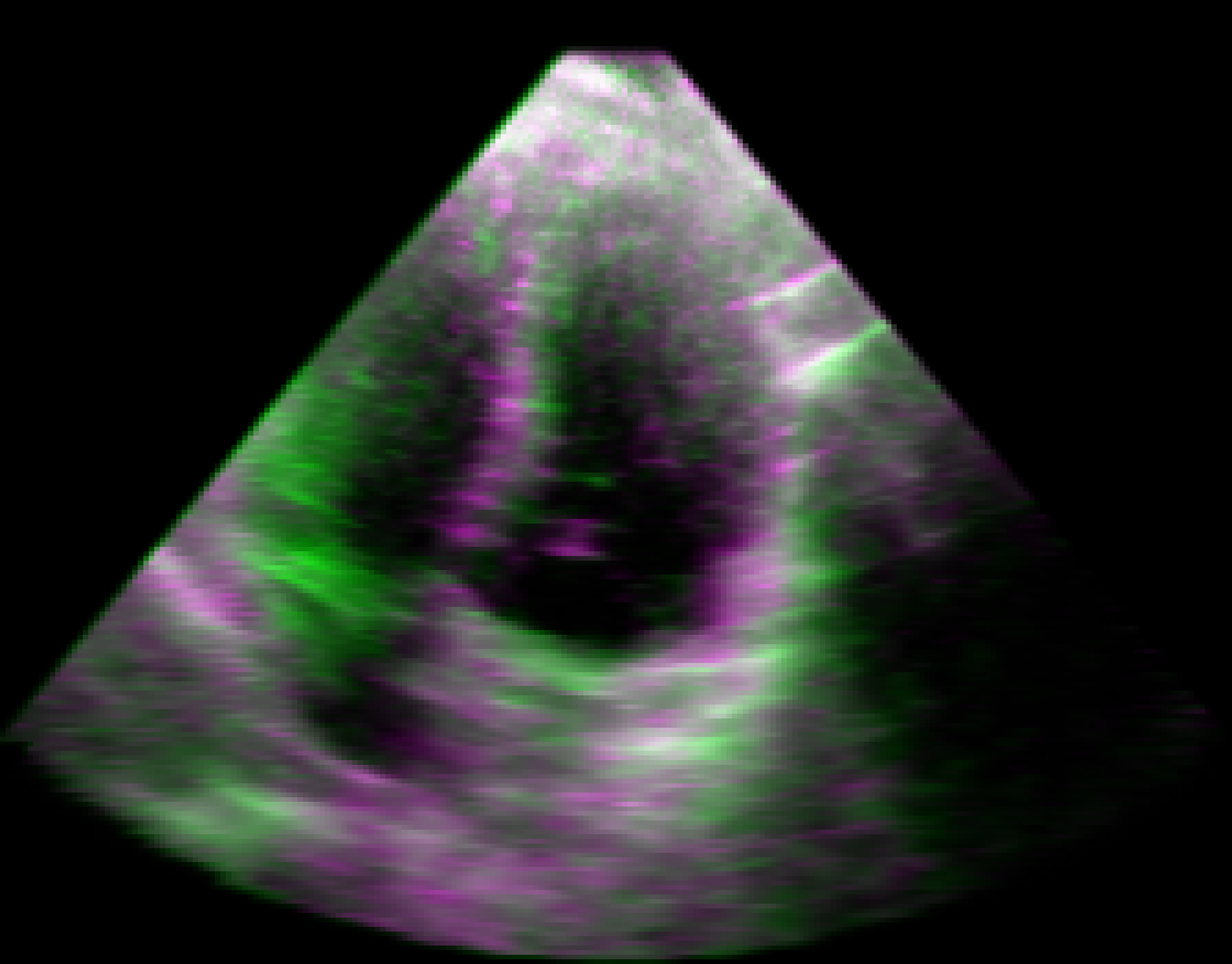}} &
        {\includegraphics[width=0.12\textwidth]{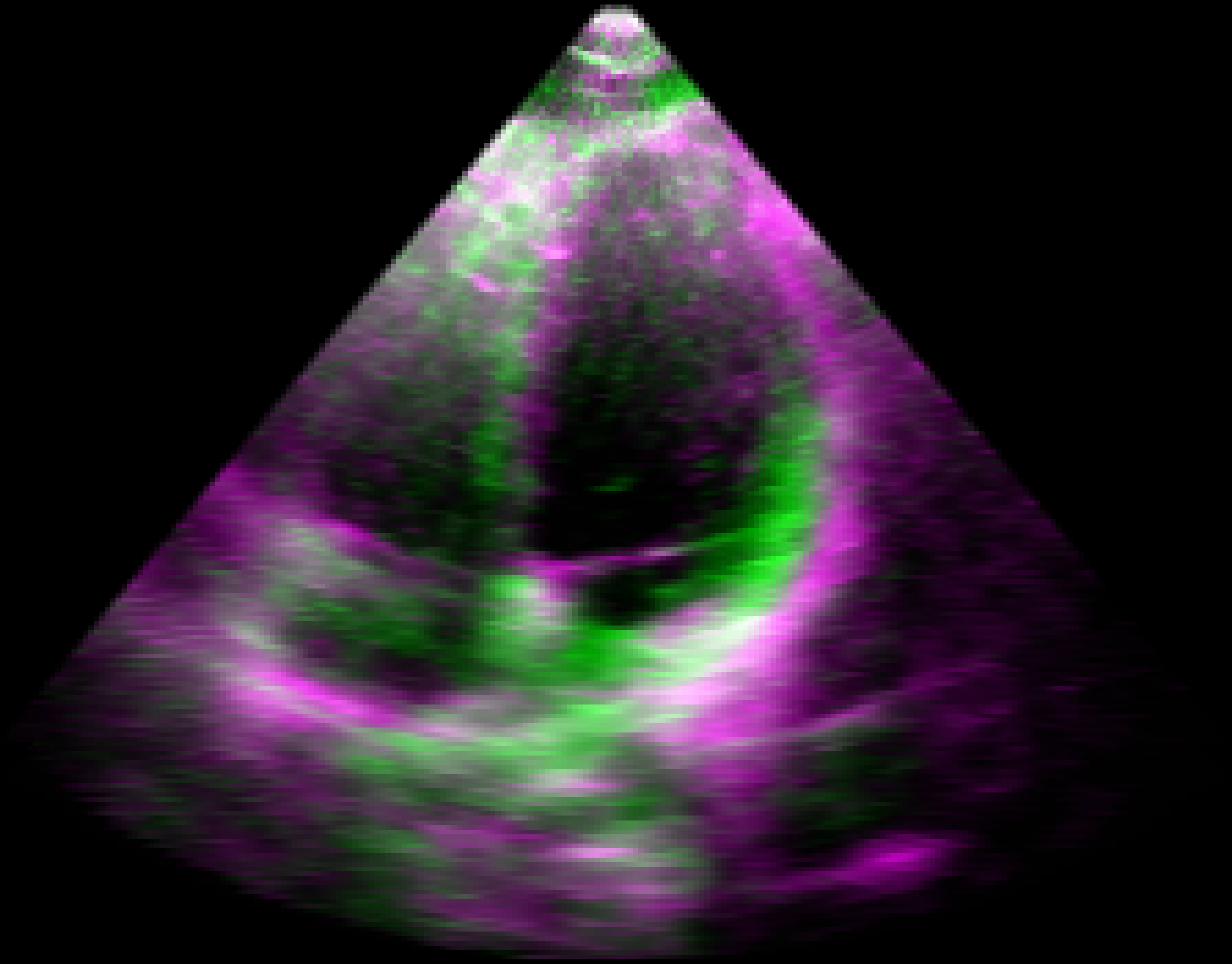}} &
        {\includegraphics[width=0.12\textwidth]{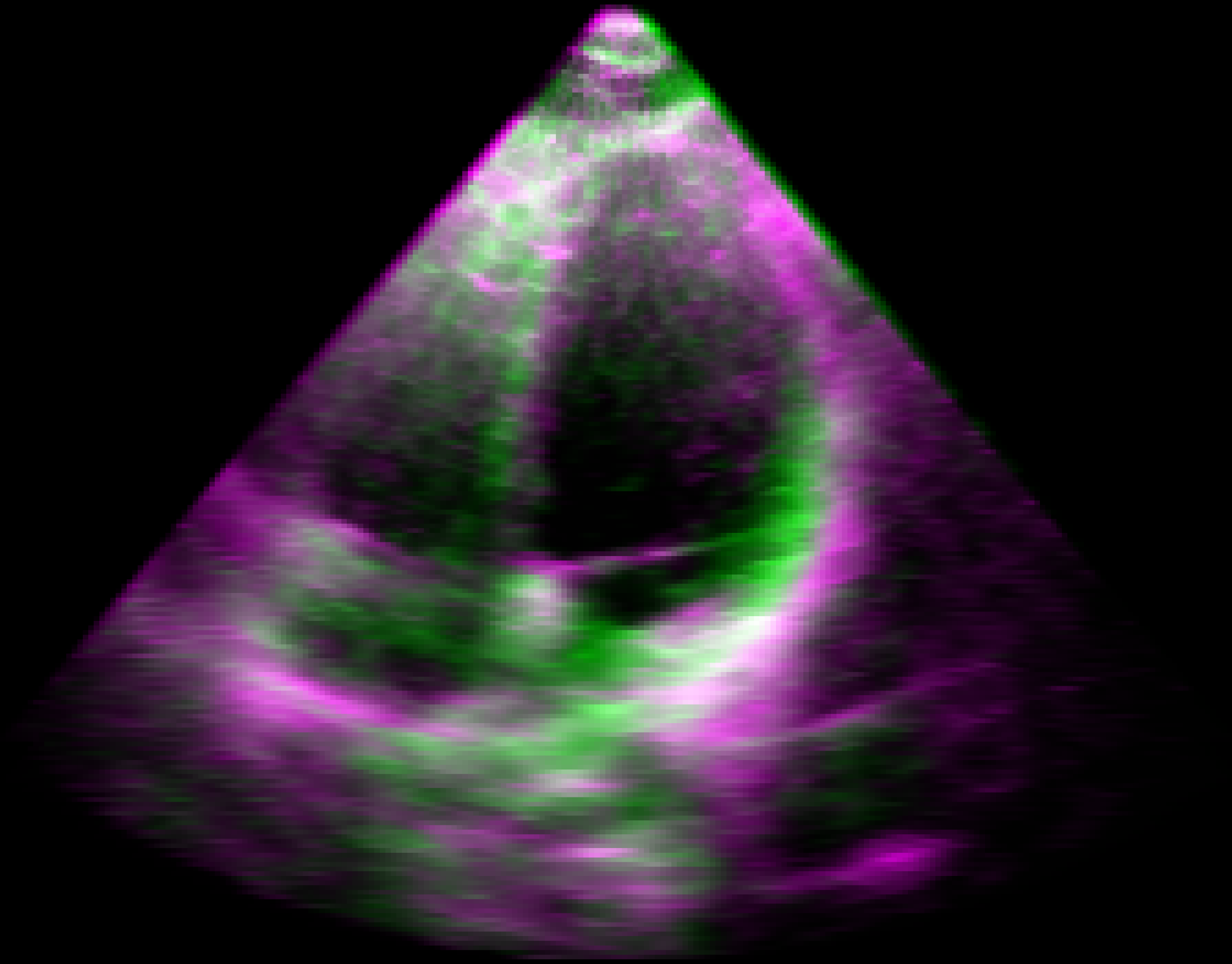}} \\ 
        \makecell[b]{Q2\vspace{15pt}} & 
        {\includegraphics[width=0.12\textwidth]{figures/fig_before_Im_019_20230817_104435_3D-Im_008_20230817_103907_3D_Coronal_0.pdf}} &
        {\includegraphics[width=0.12\textwidth]{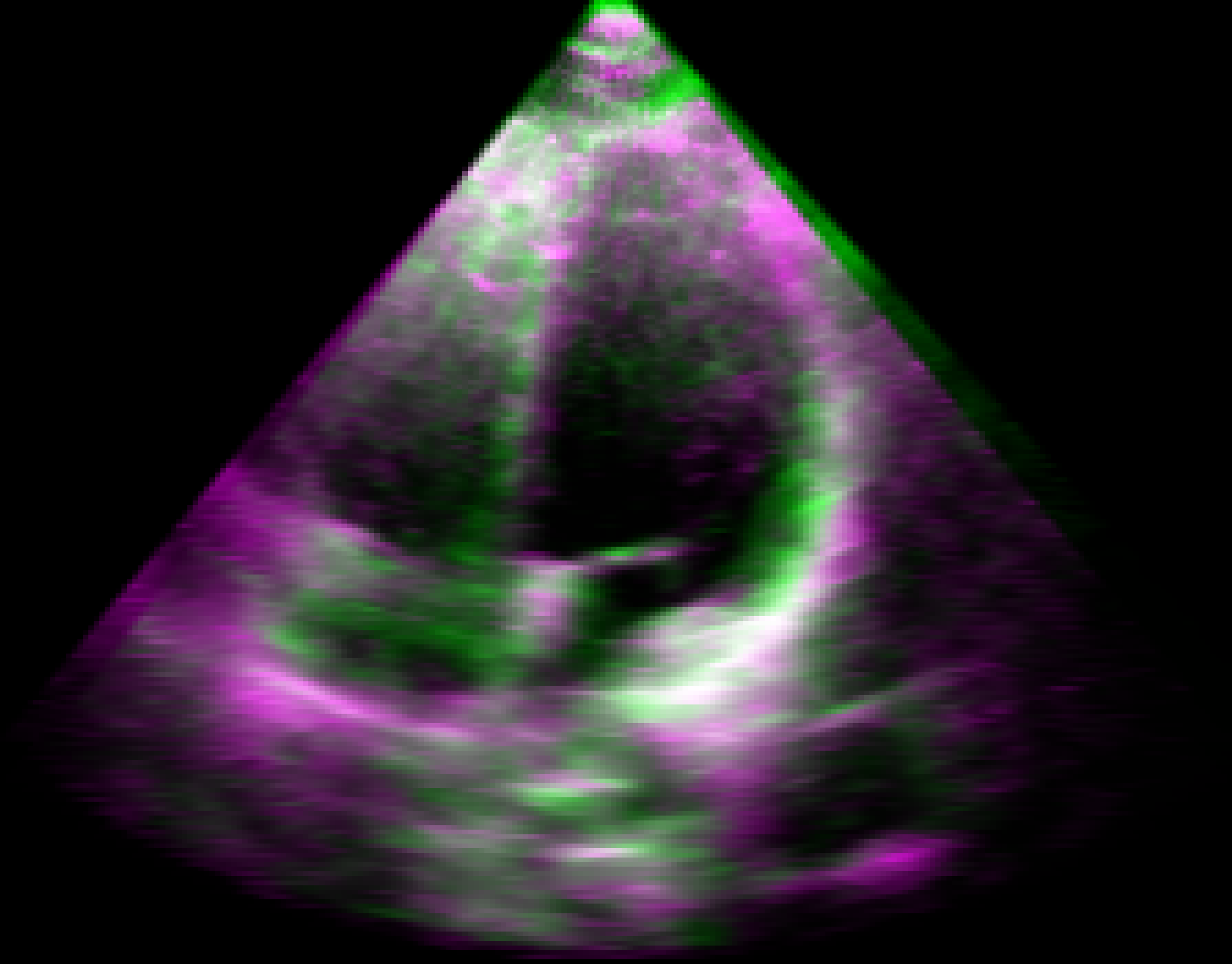}} &
        {\includegraphics[width=0.12\textwidth]{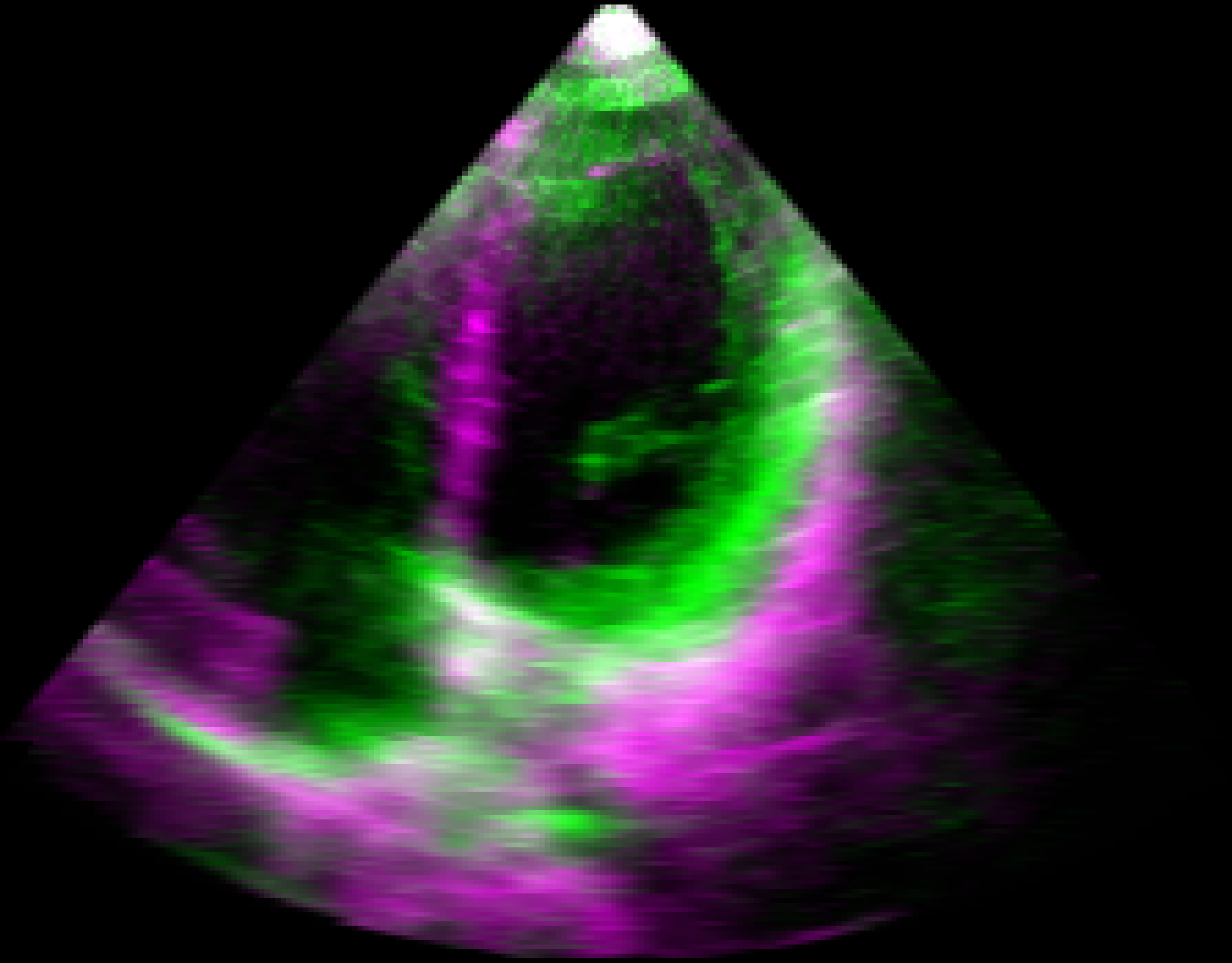}} &
        {\includegraphics[width=0.12\textwidth]{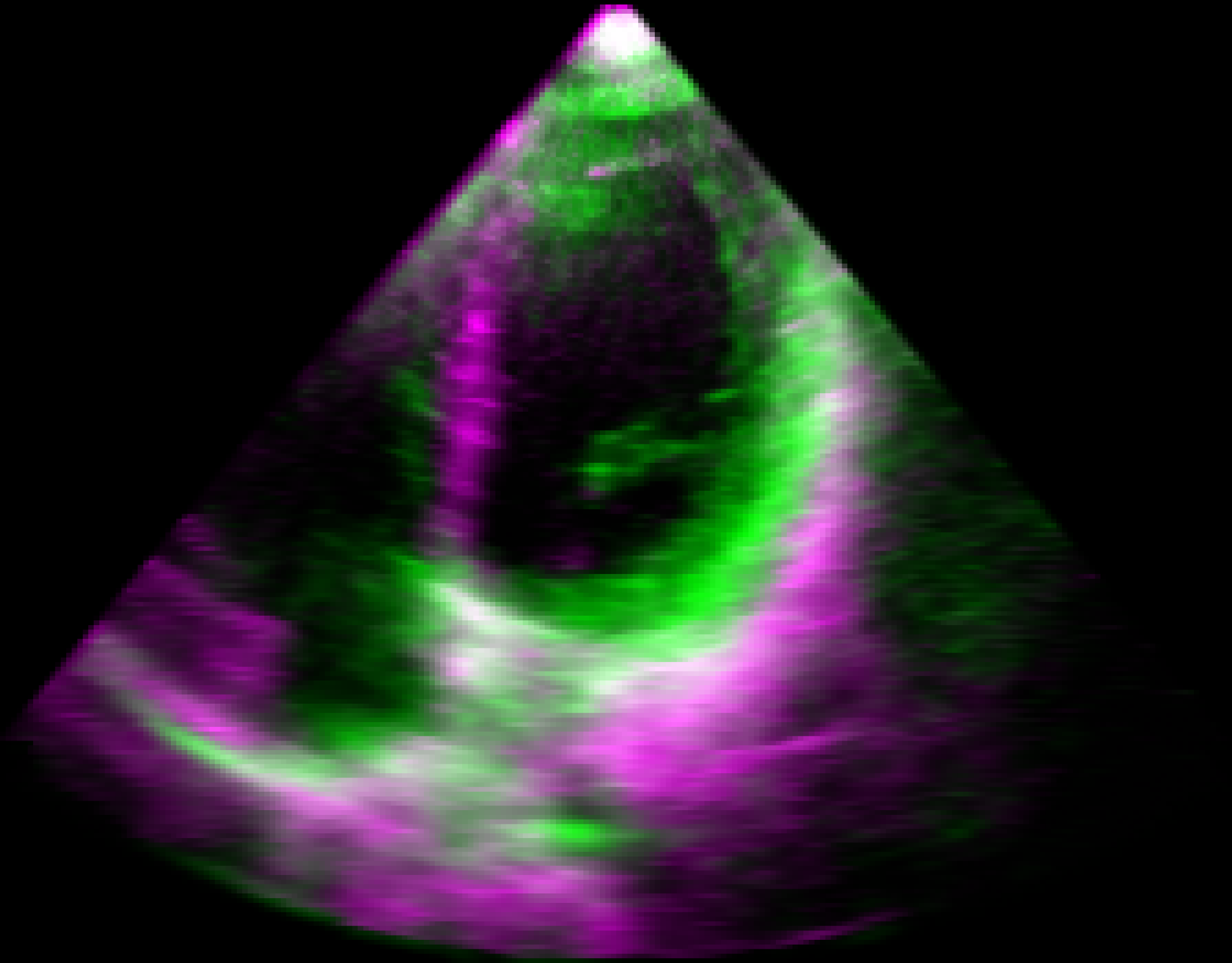}} &
        {\includegraphics[width=0.12\textwidth]{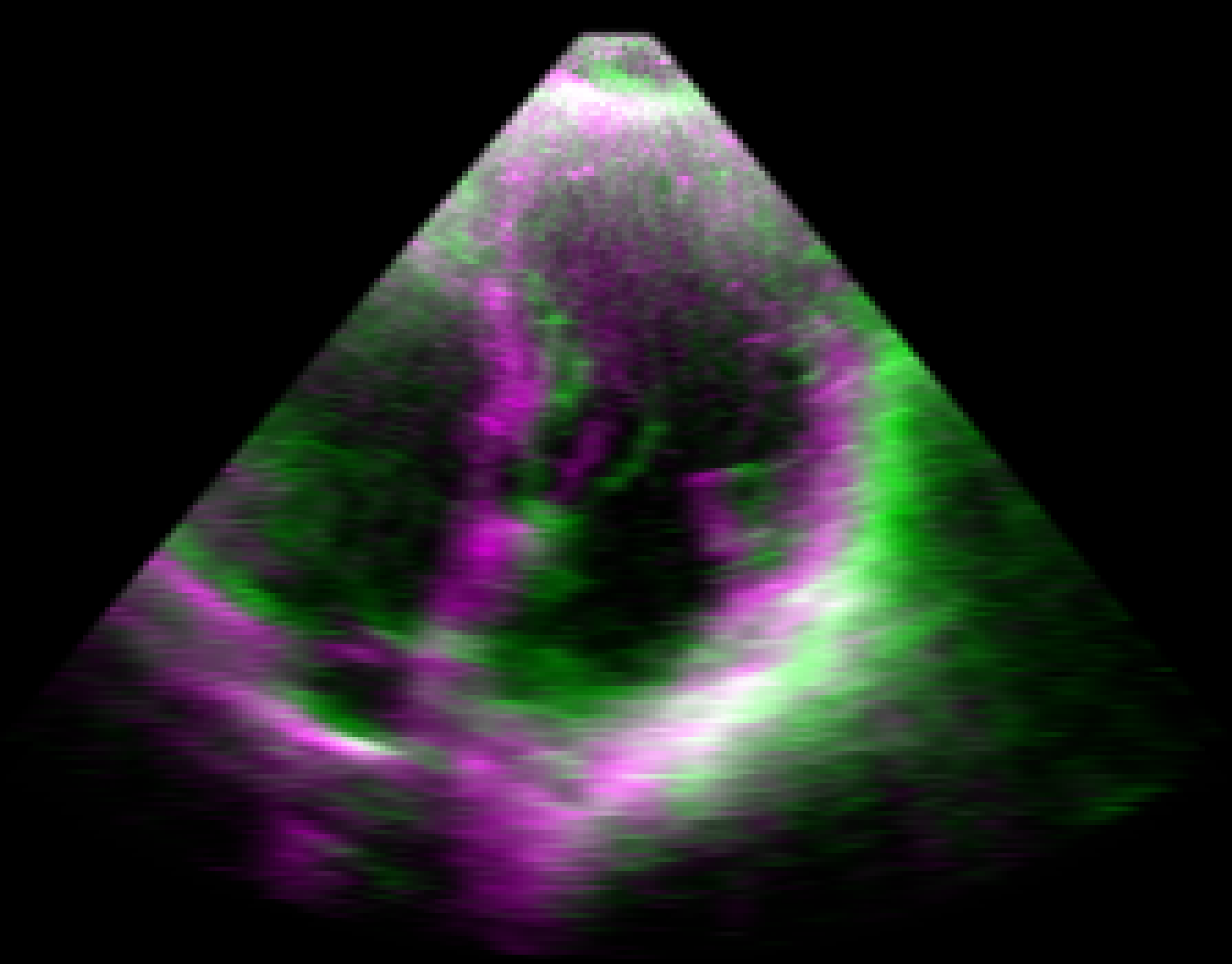}} &
        {\includegraphics[width=0.12\textwidth]{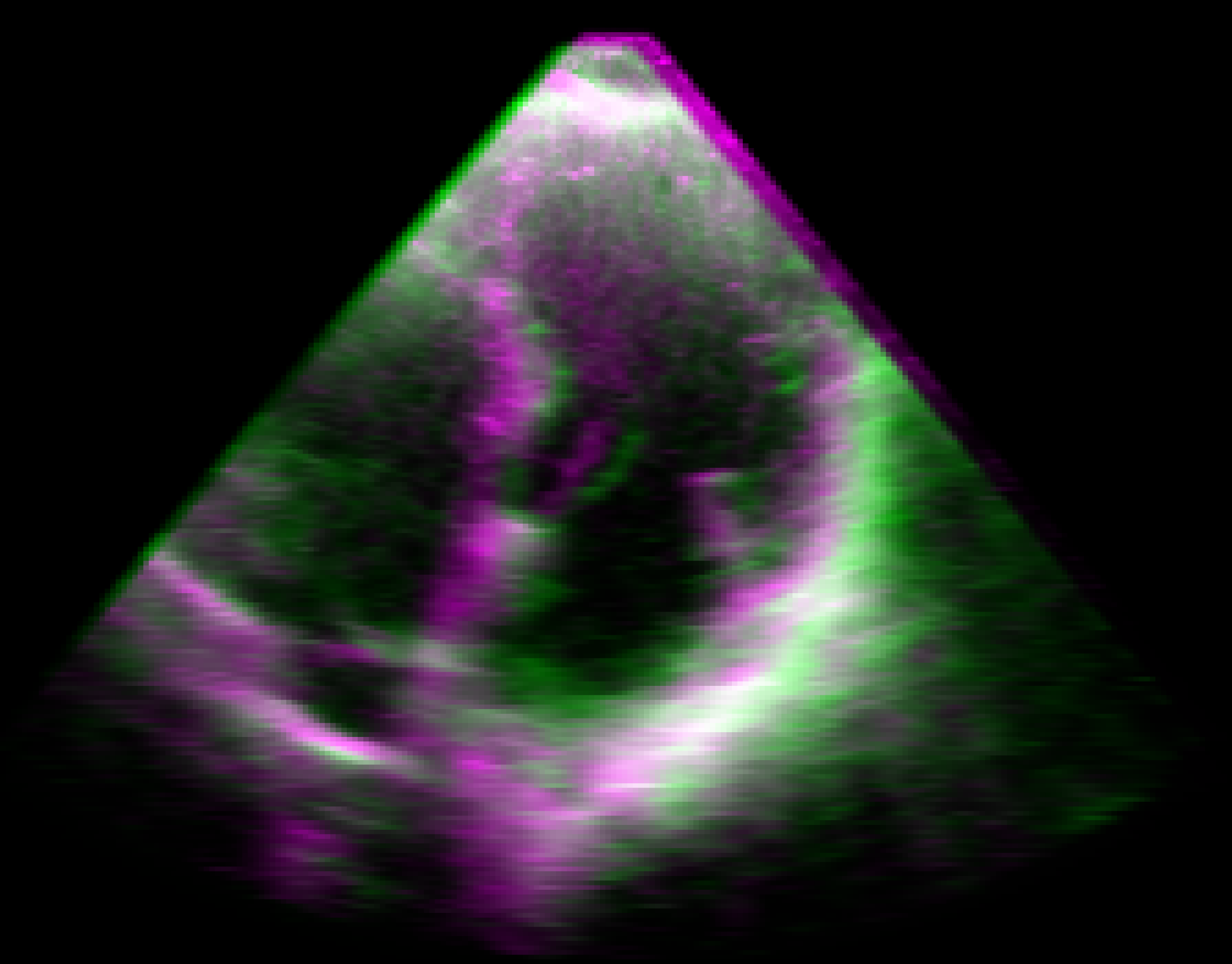}} \\
        \makecell[b]{Q3\vspace{15pt}} & 
        {\includegraphics[width=0.12\textwidth]{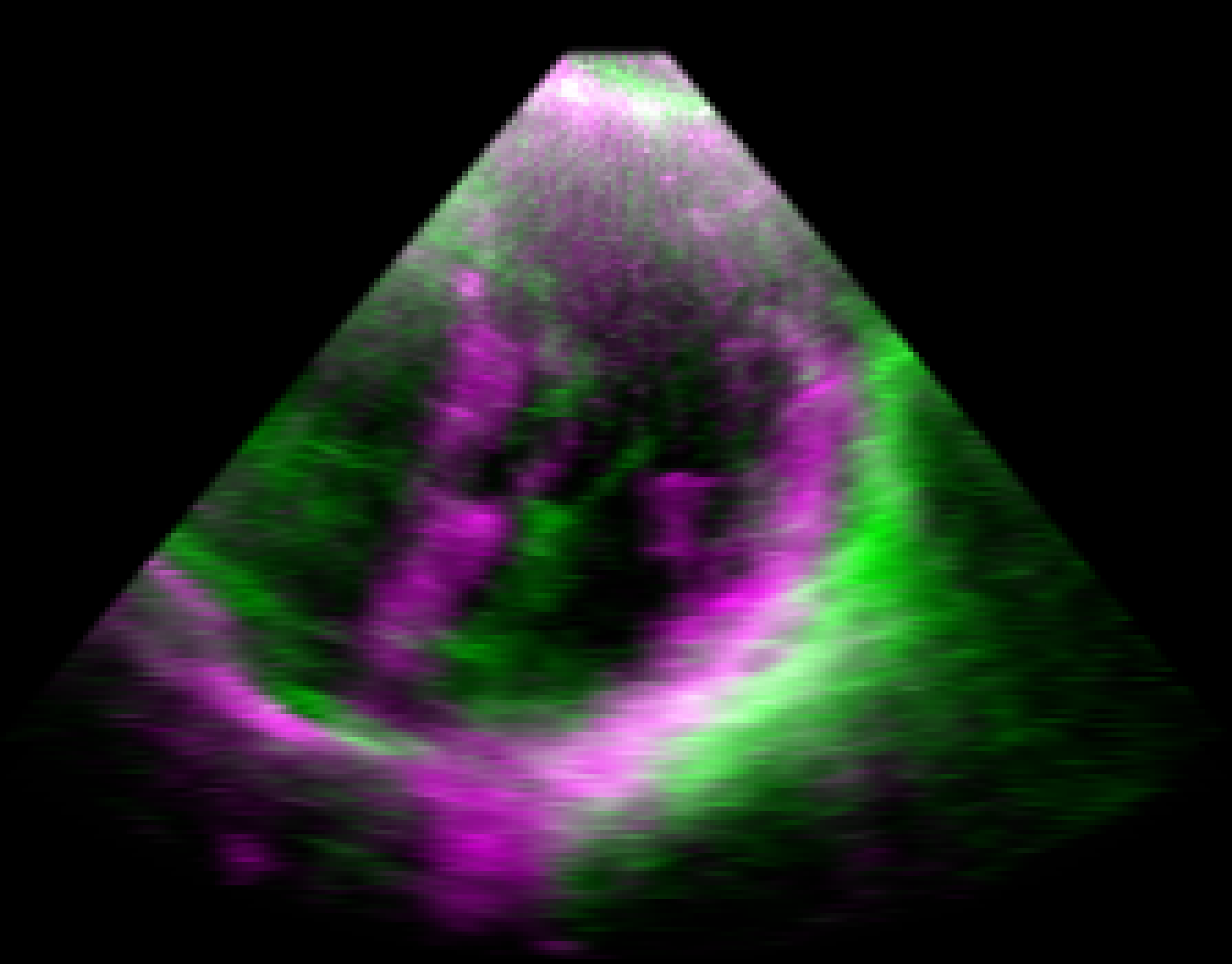}} &
        {\includegraphics[width=0.12\textwidth]{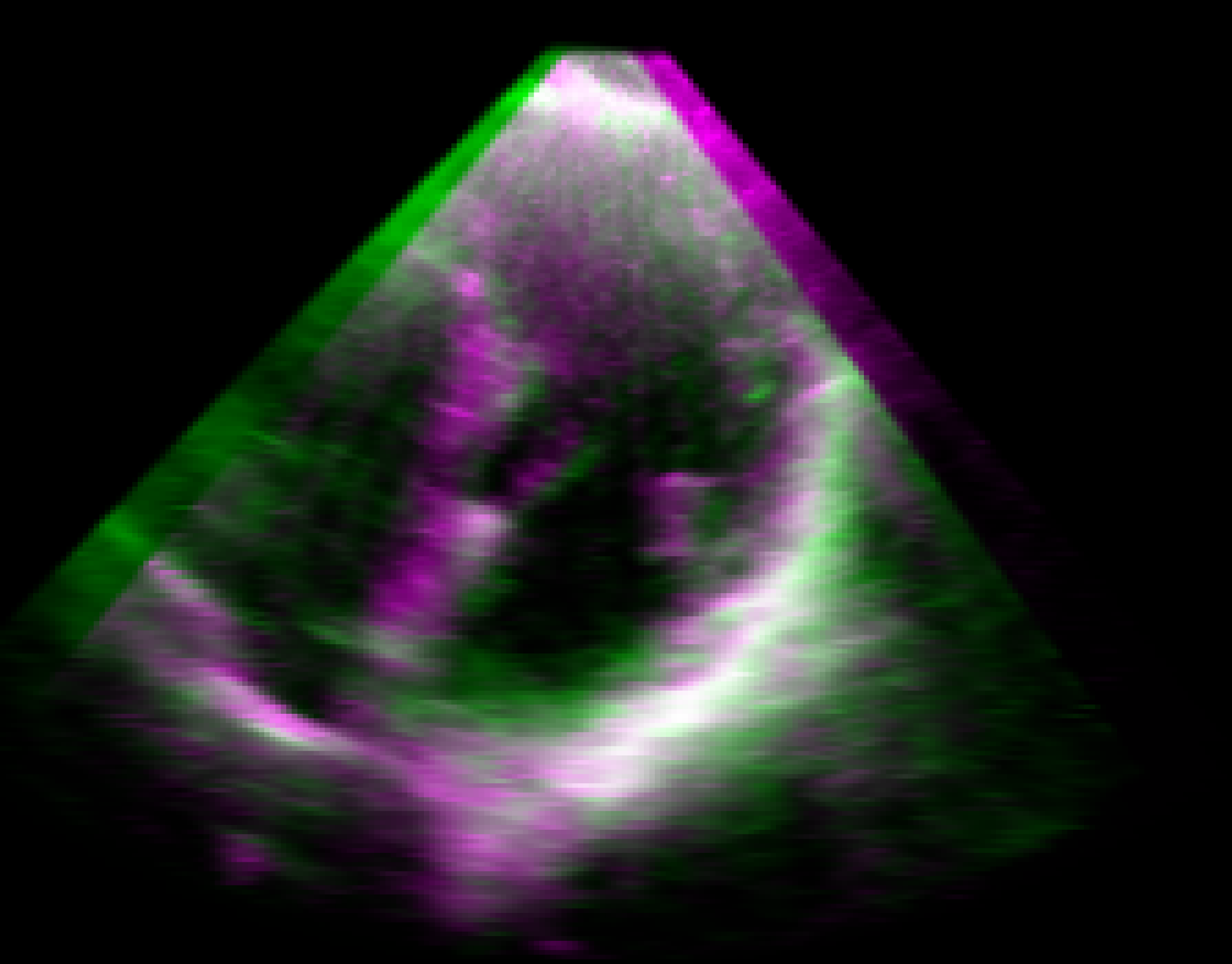}} &
        {\includegraphics[width=0.12\textwidth]{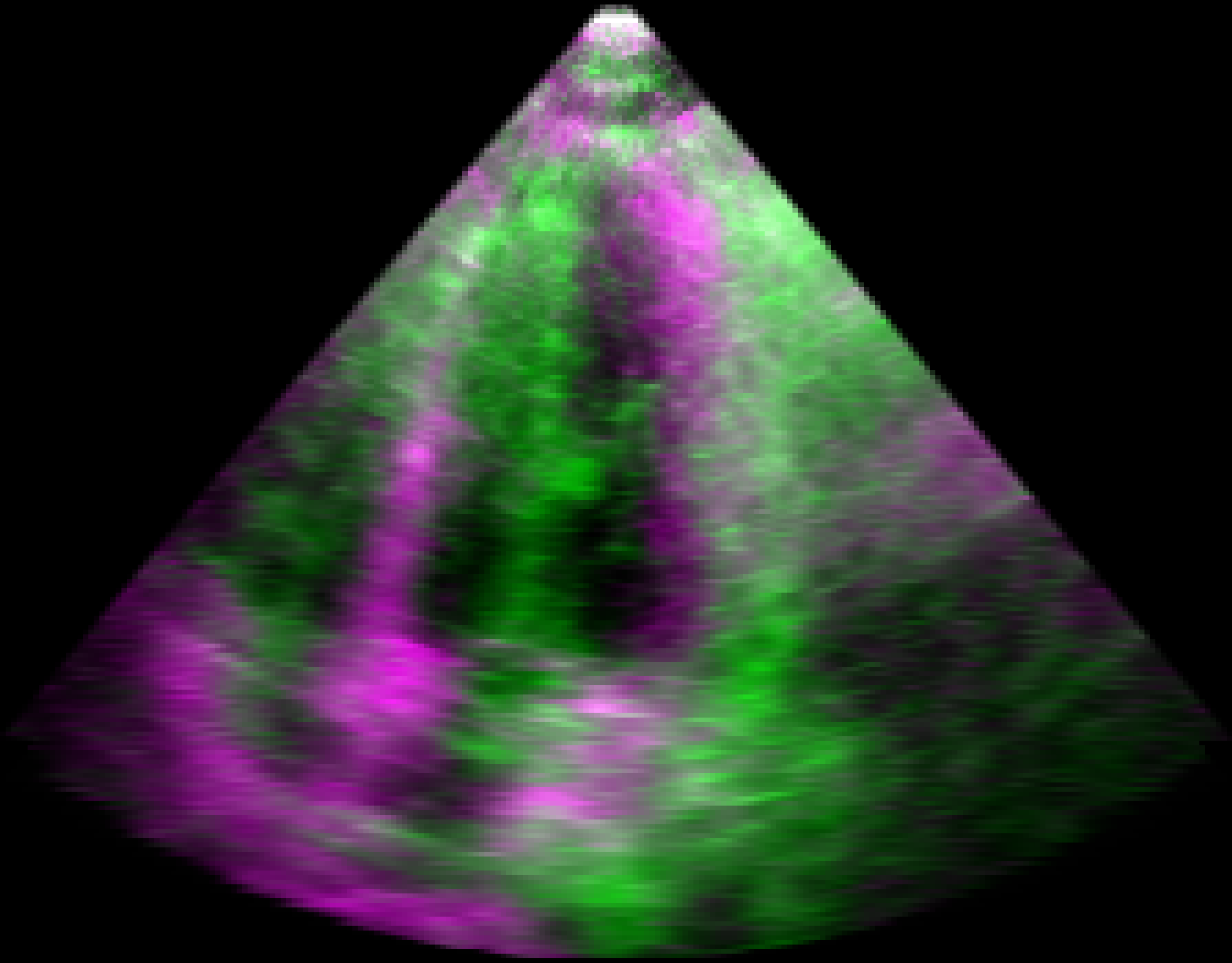}} &
        {\includegraphics[width=0.12\textwidth]{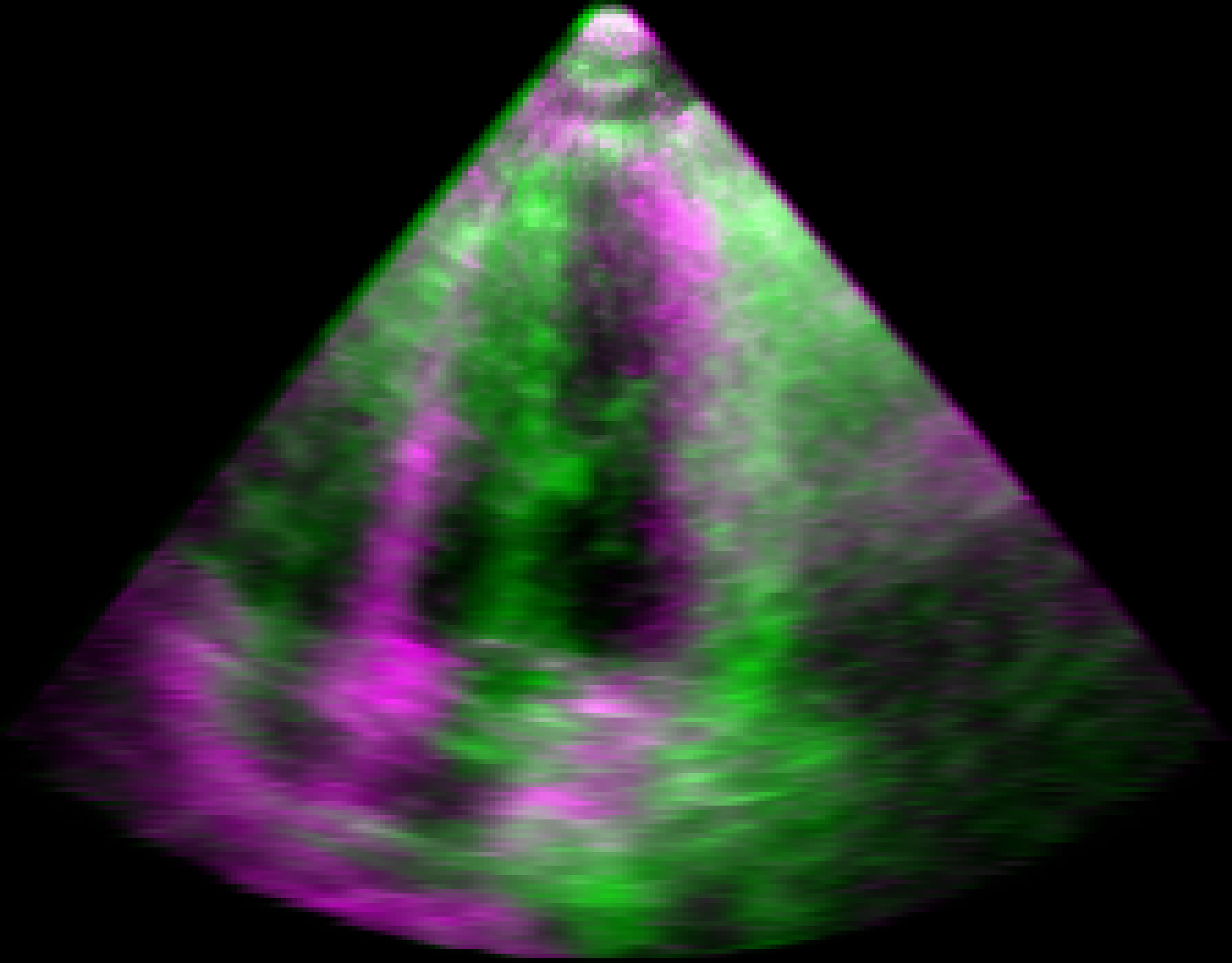}} &
        {\includegraphics[width=0.12\textwidth]{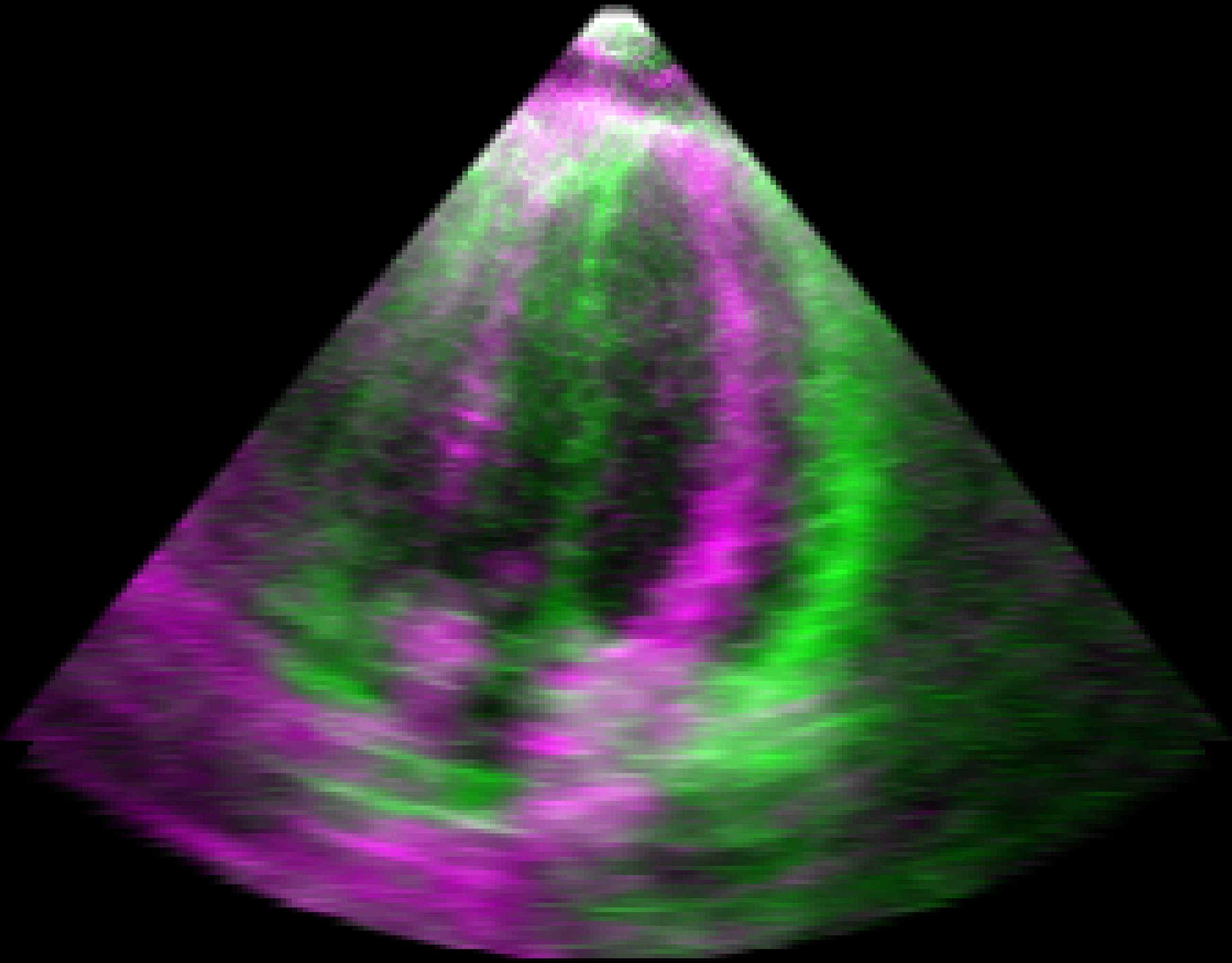}} &
        {\includegraphics[width=0.12\textwidth]{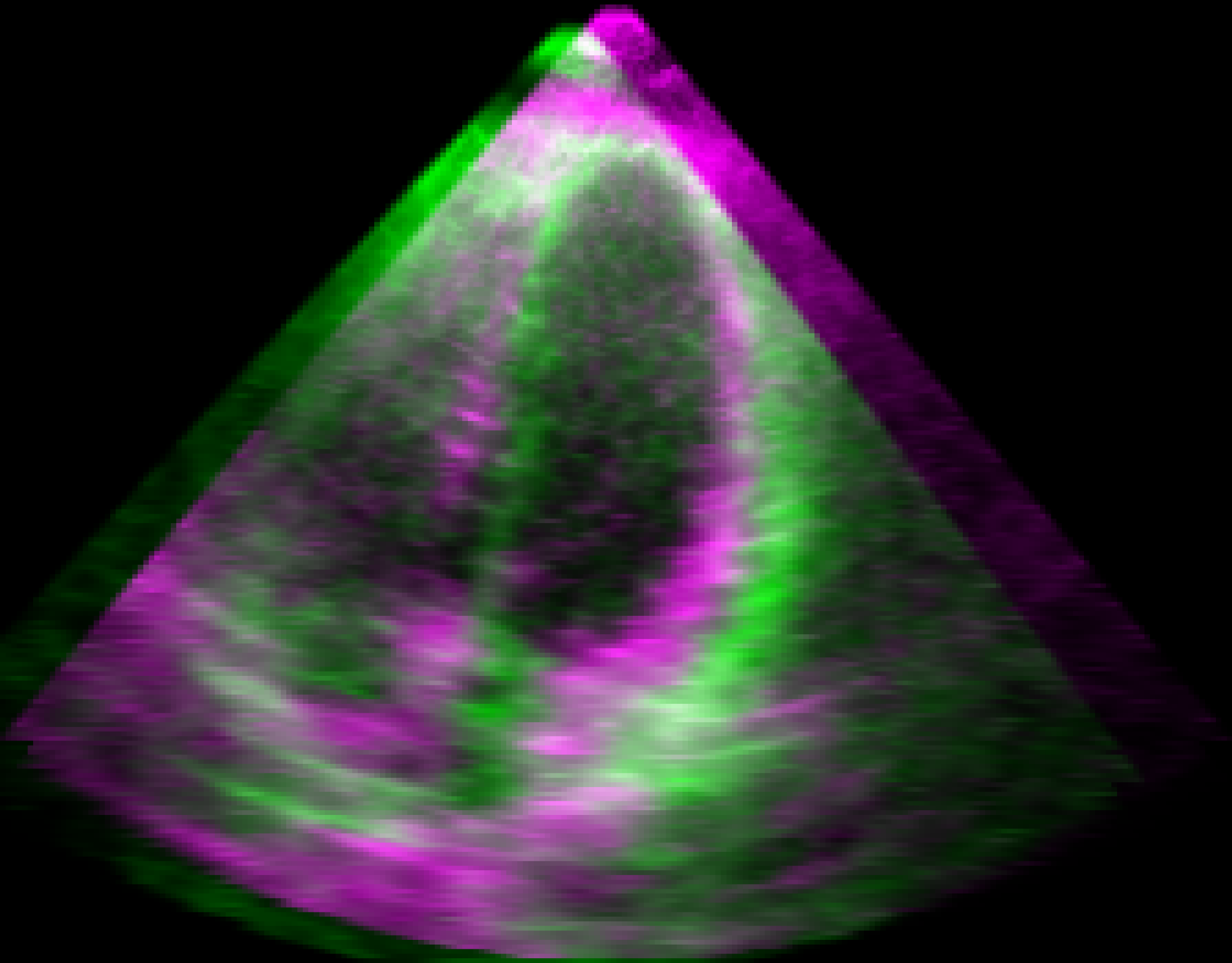}} \\
        \makecell[b]{Max\vspace{15pt}} & 
        {\includegraphics[width=0.12\textwidth]{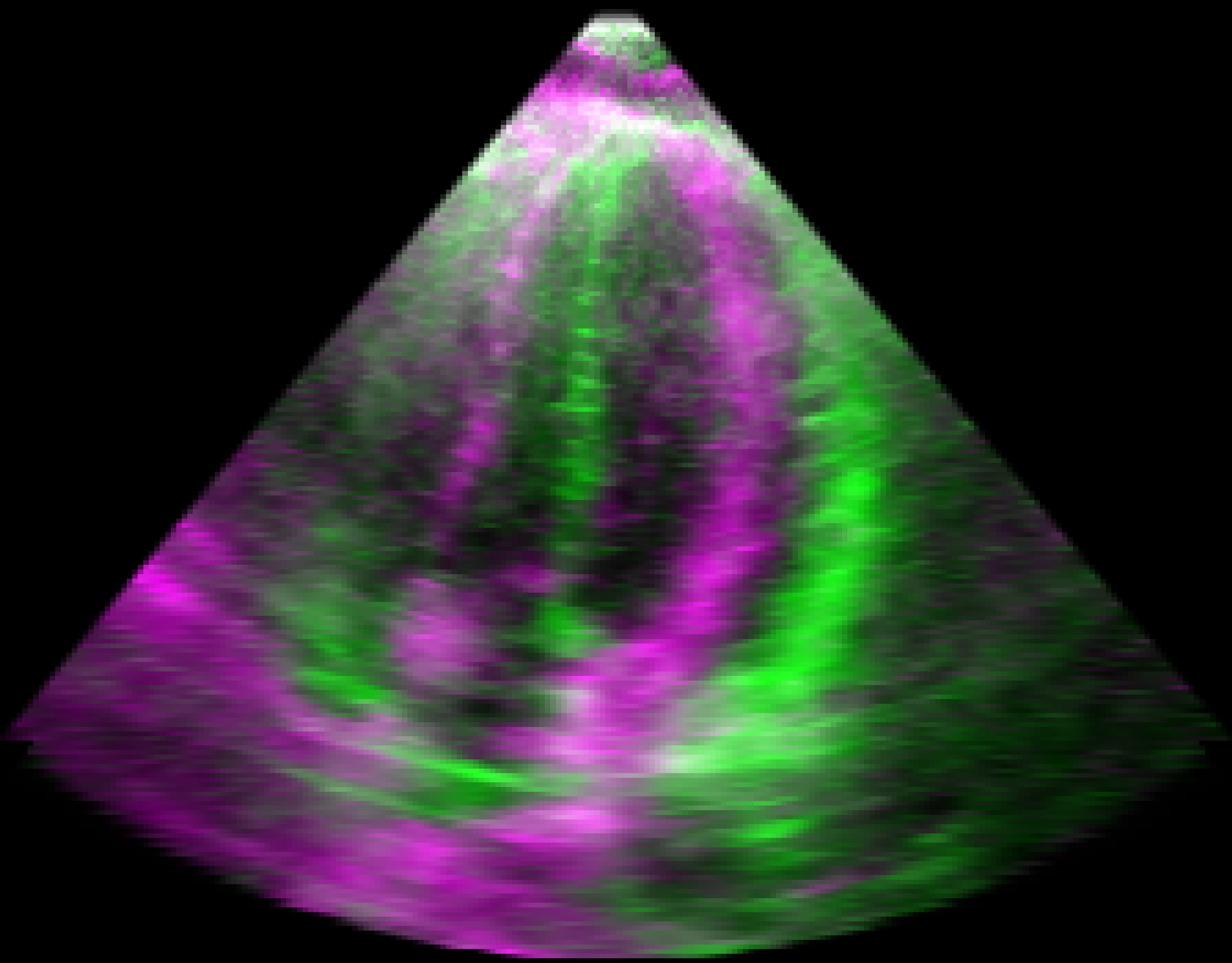}} &
        {\includegraphics[width=0.12\textwidth]{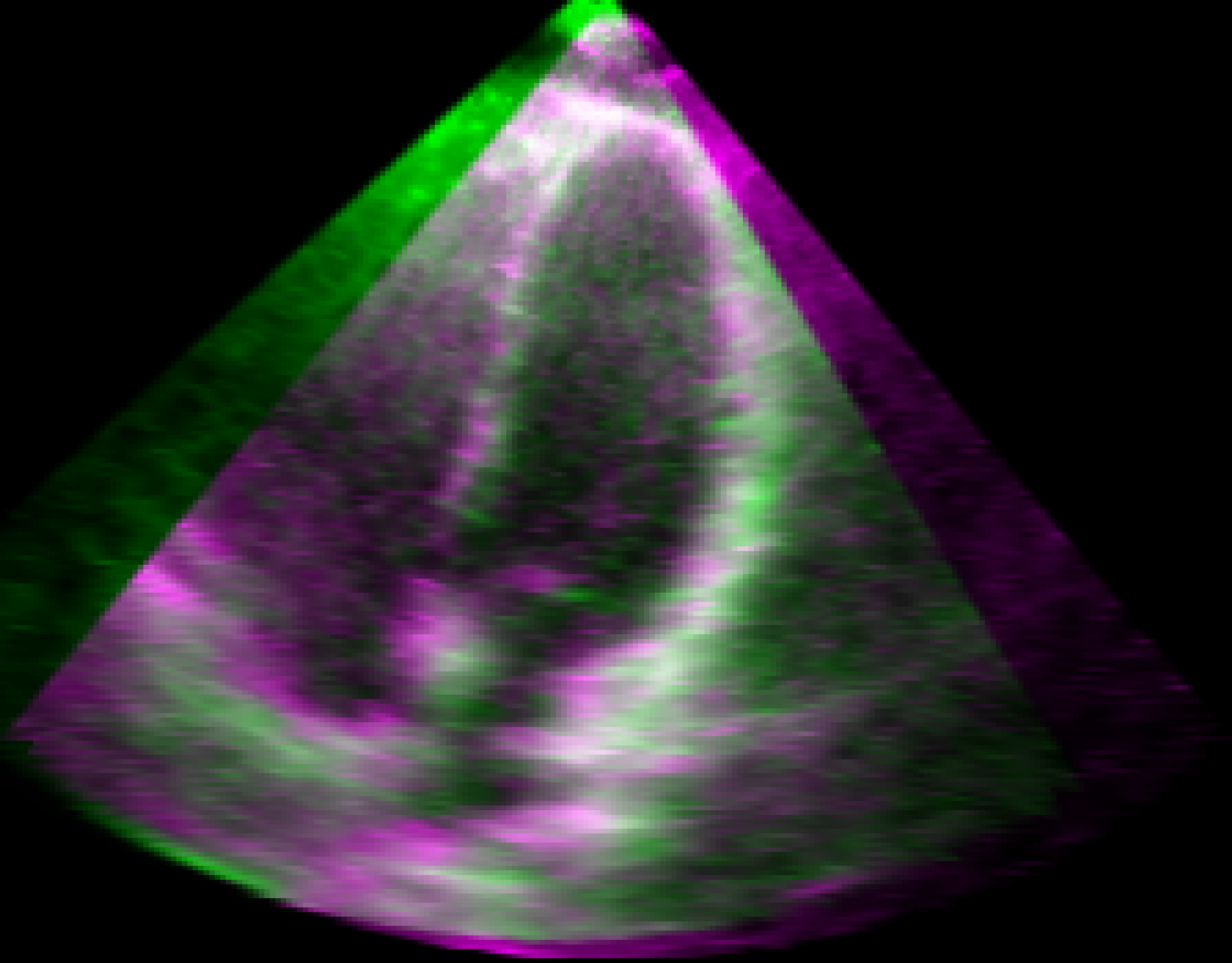}} &
        {\includegraphics[width=0.12\textwidth]{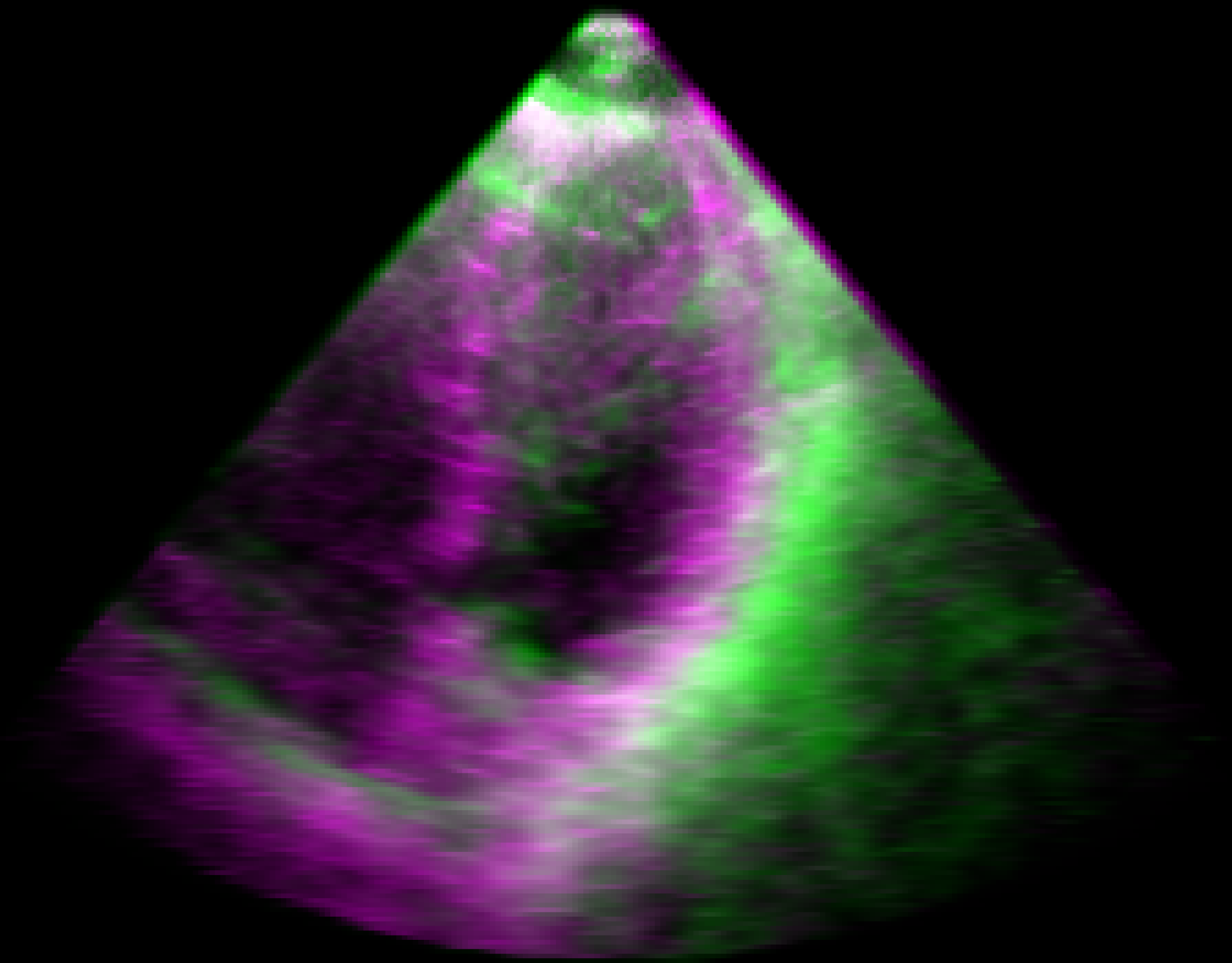}} &
        {\includegraphics[width=0.12\textwidth]{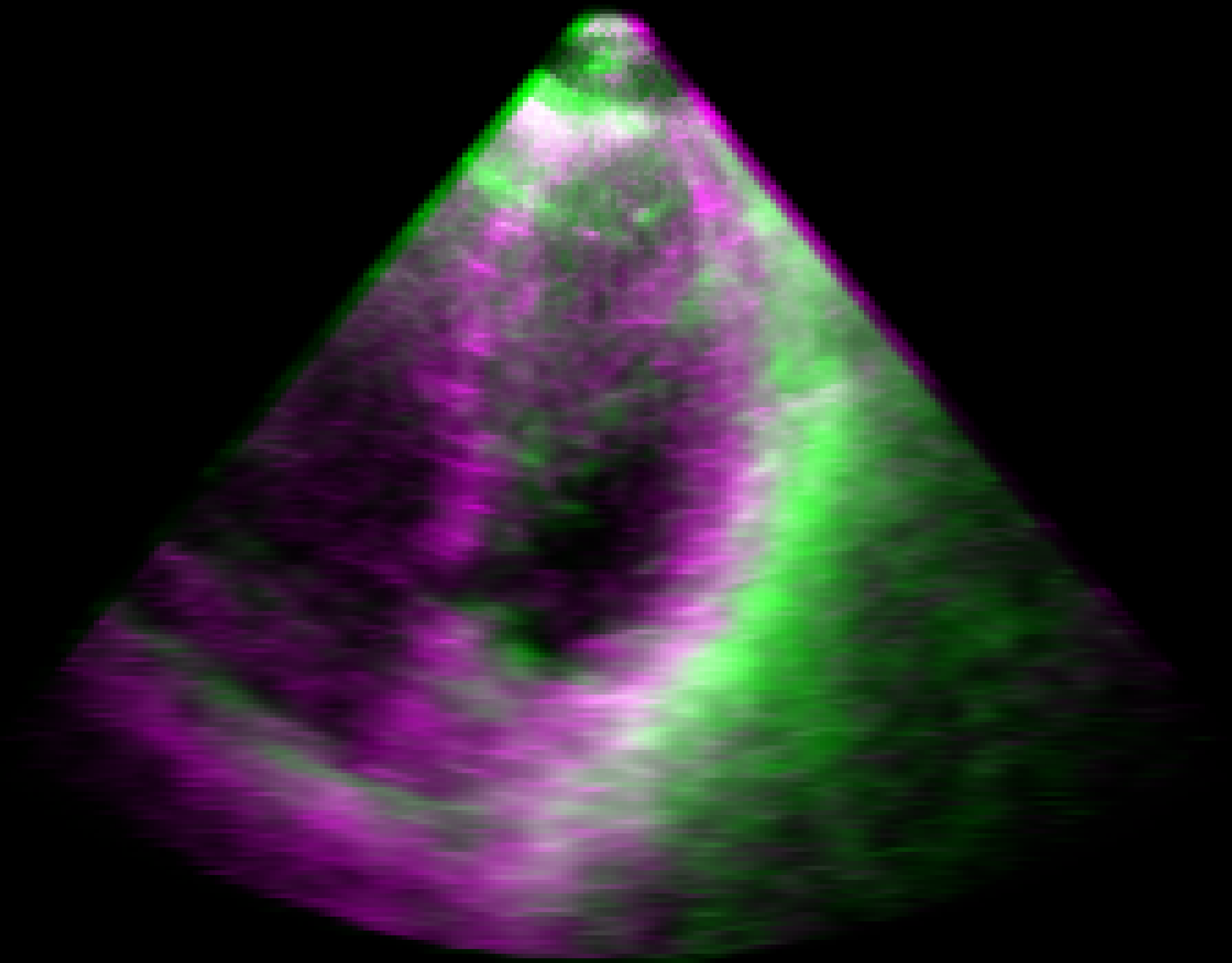}} &
        {\includegraphics[width=0.12\textwidth]{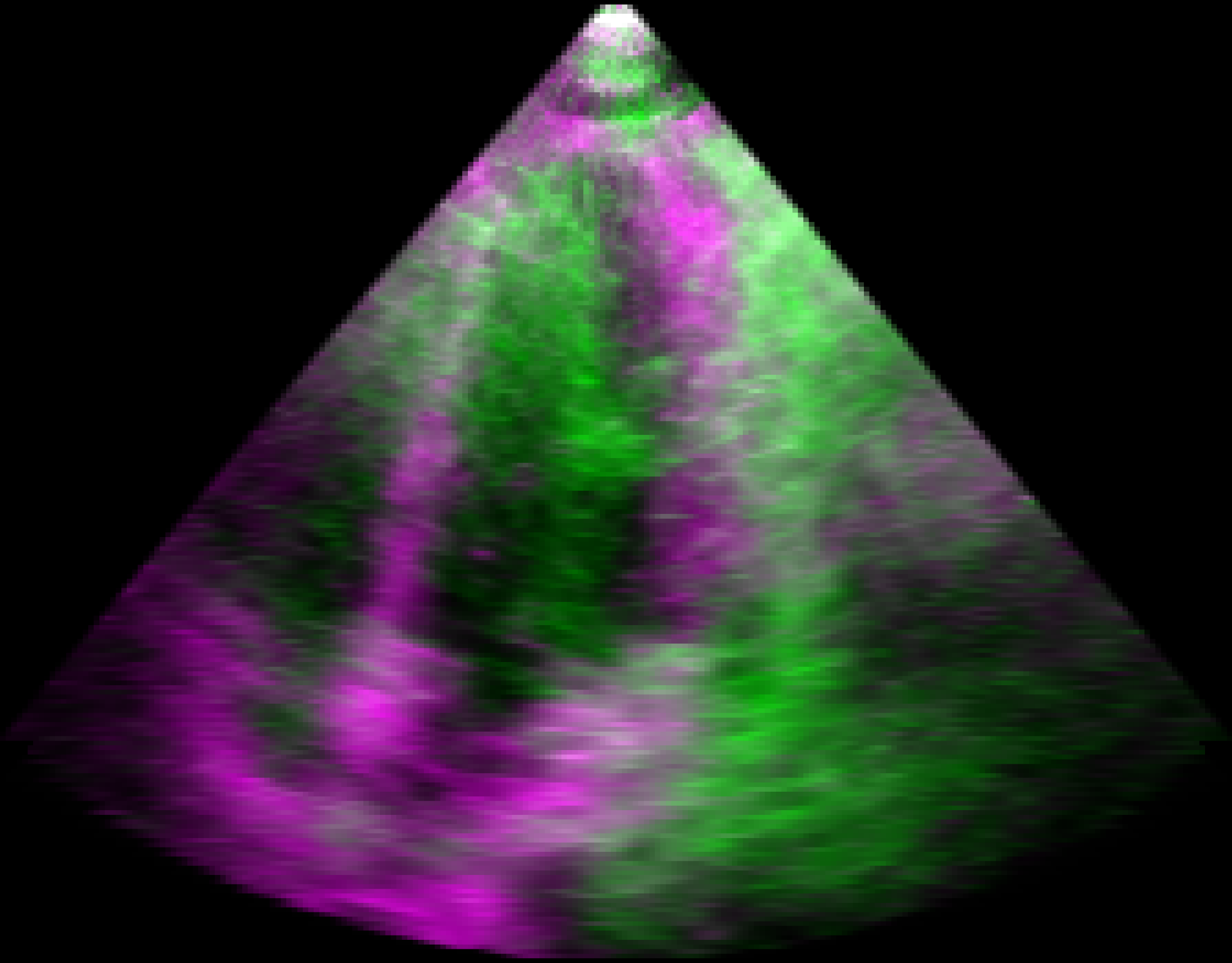}} &
        {\includegraphics[width=0.12\textwidth]{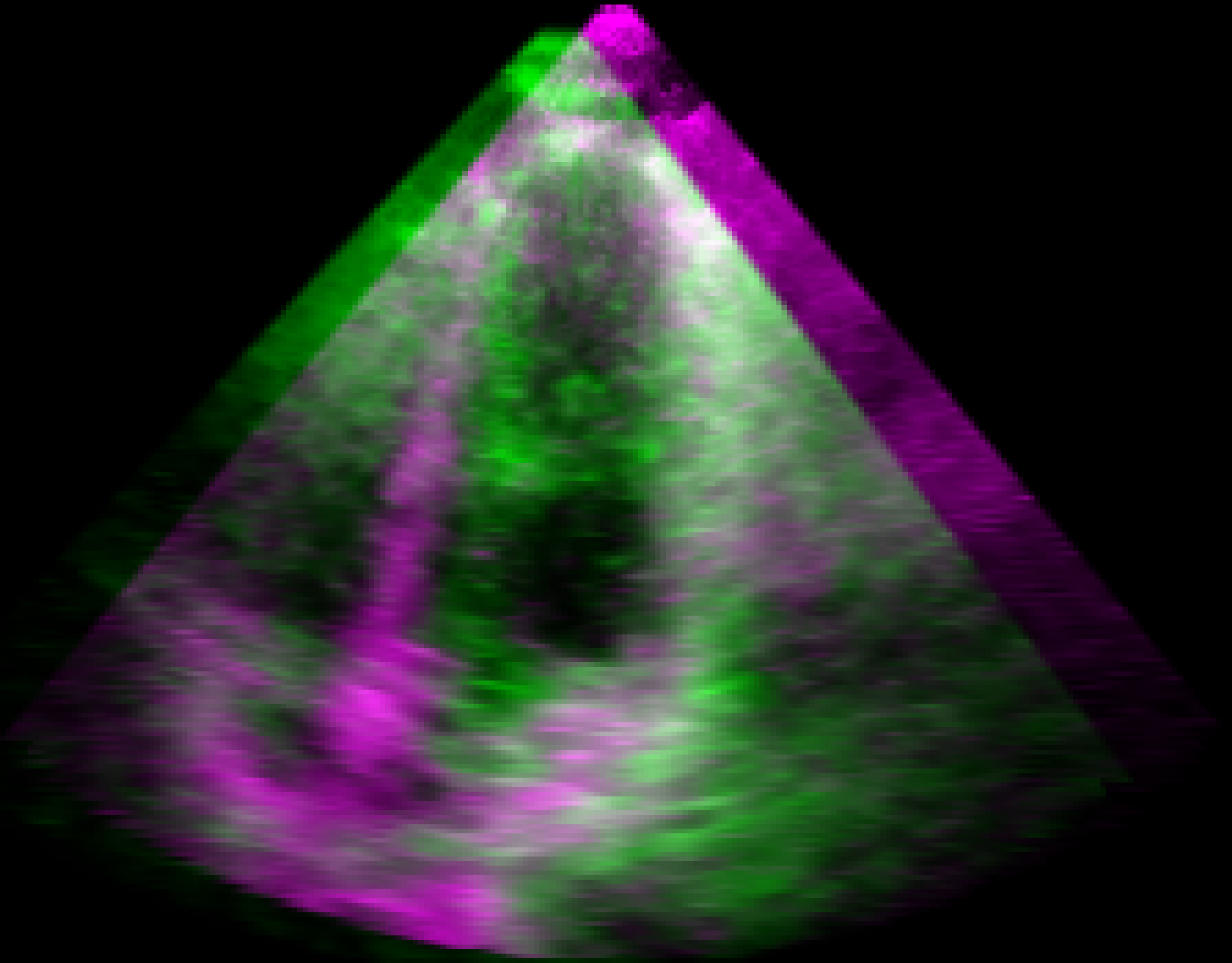}} \\
    \end{tabular}
    \caption{Image-based rigid (PF and EX) registration results of the ED frame of image pairs for percentile DSC difference values of the coronal view. Two consecutive columns show images before and after the registration for each category. The source and target images are shown in green and purple colors.}
    \label{fig:fig_perc_images_coronal_rigid_img}
\end{figure*}

\begin{figure*}[!ht]
    \centering
    \begin{tabular}{SSSSSSS}
         & \multicolumn{2}{c}{\scriptsize PF (CPU)} & \multicolumn{2}{c}{\scriptsize PF (GPU)} & \multicolumn{2}{c}{\scriptsize EX (CPU)} \\
         & Before & After & Before & After & Before & After \\
        \makecell[b]{Min\vspace{15pt}} & 
        {\includegraphics[width=0.12\textwidth]{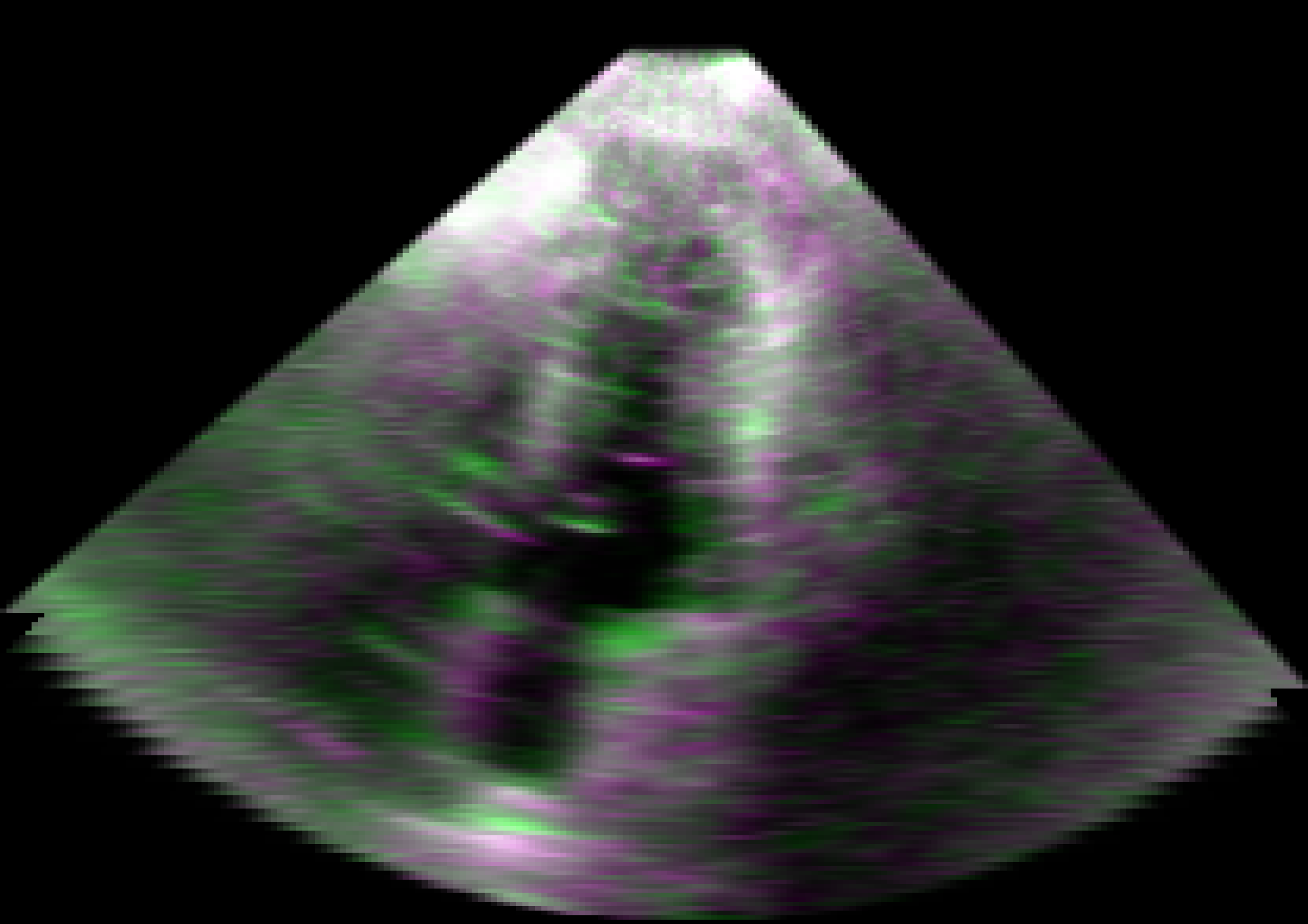}} &
        {\includegraphics[width=0.12\textwidth]{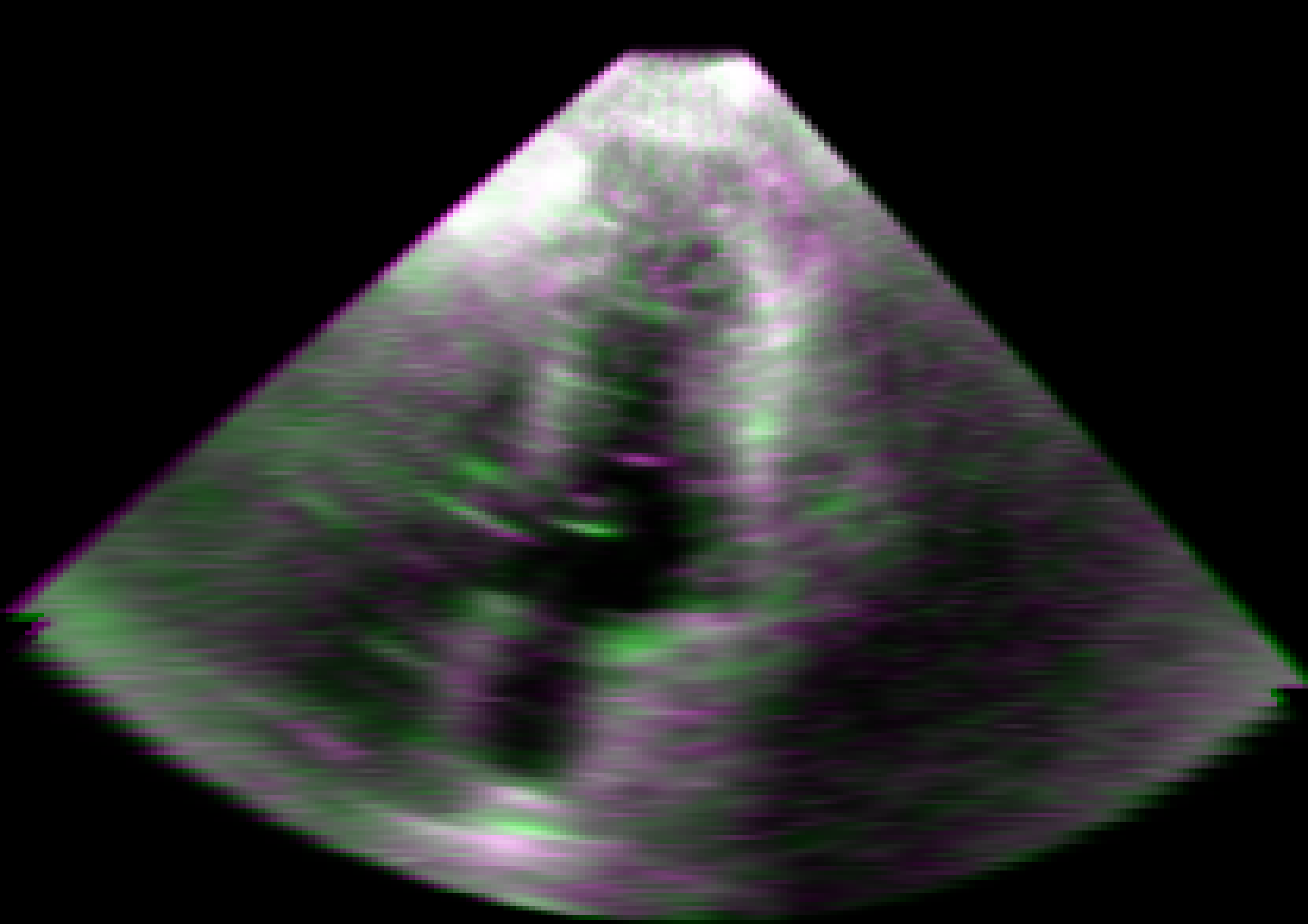}} &
        {\includegraphics[width=0.12\textwidth]{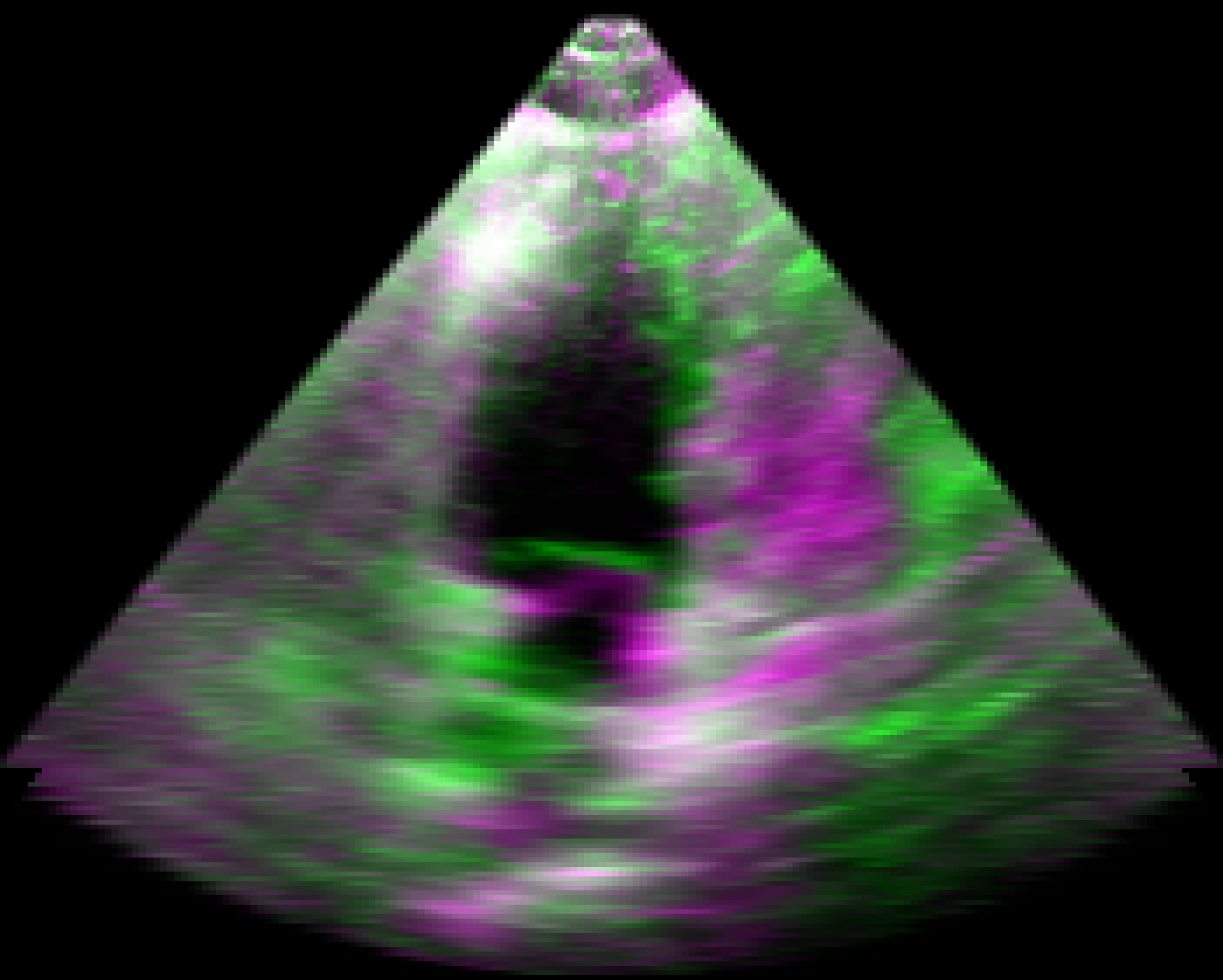}} &
        {\includegraphics[width=0.12\textwidth]{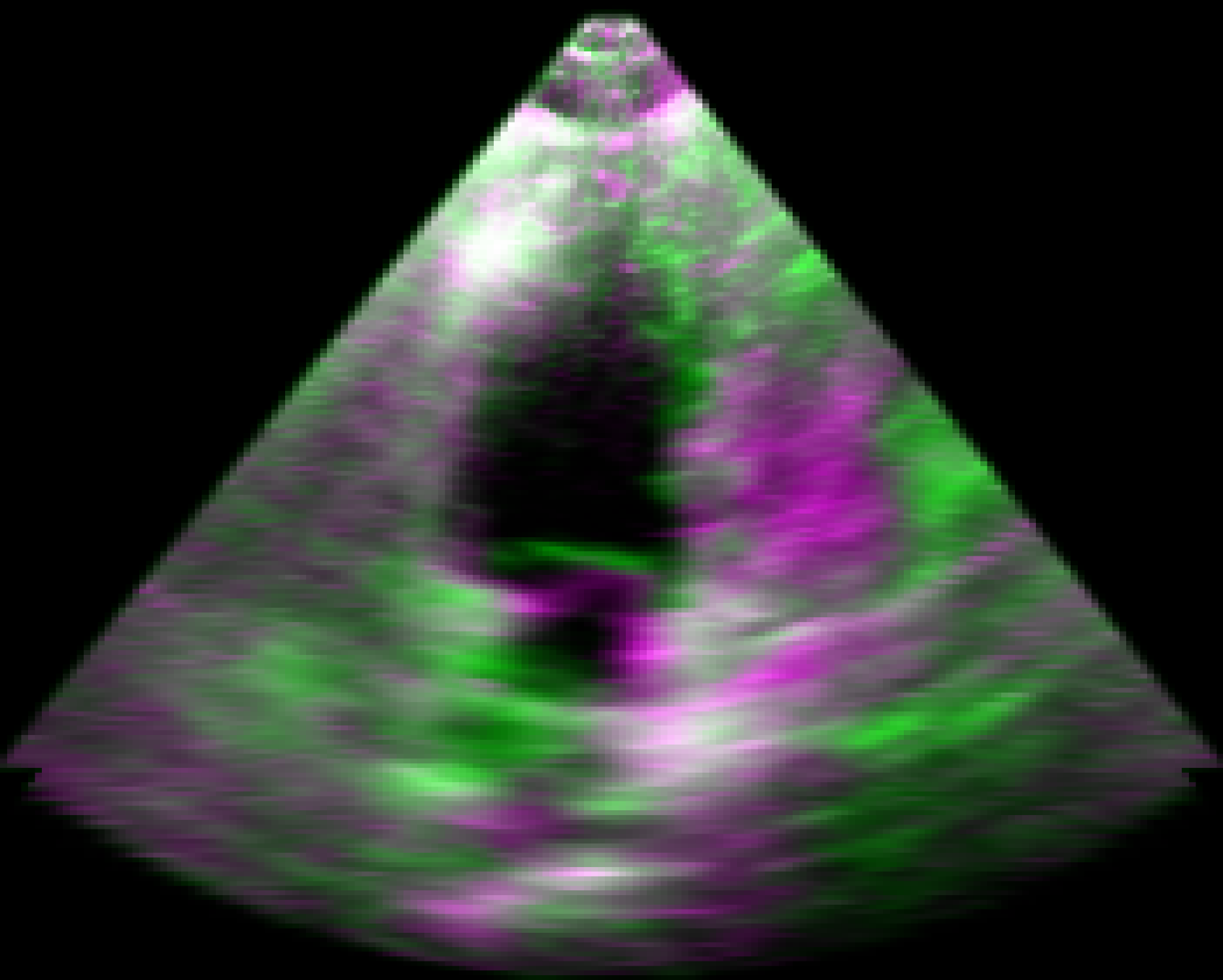}} &
        {\includegraphics[width=0.12\textwidth]{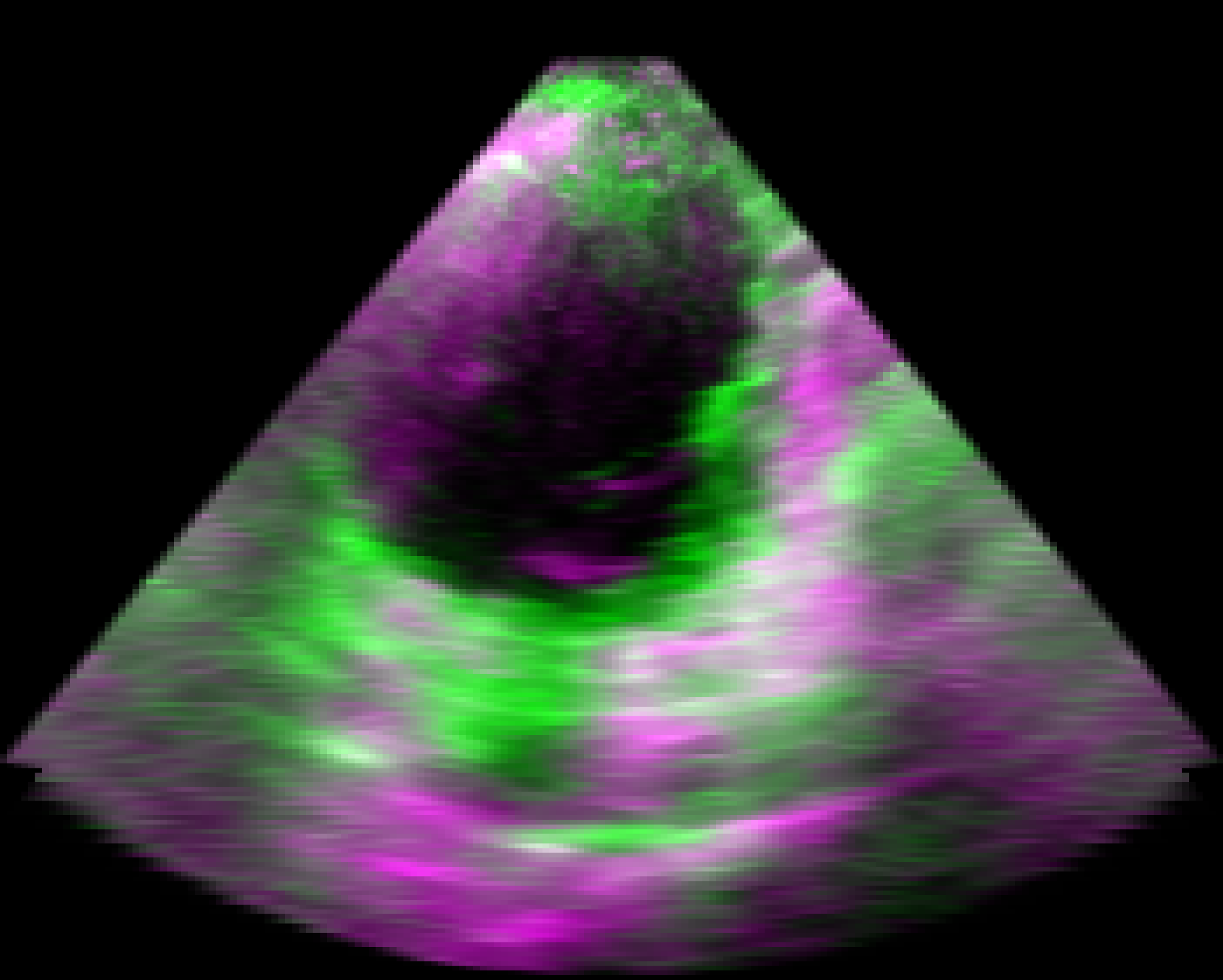}} &
        {\includegraphics[width=0.12\textwidth]{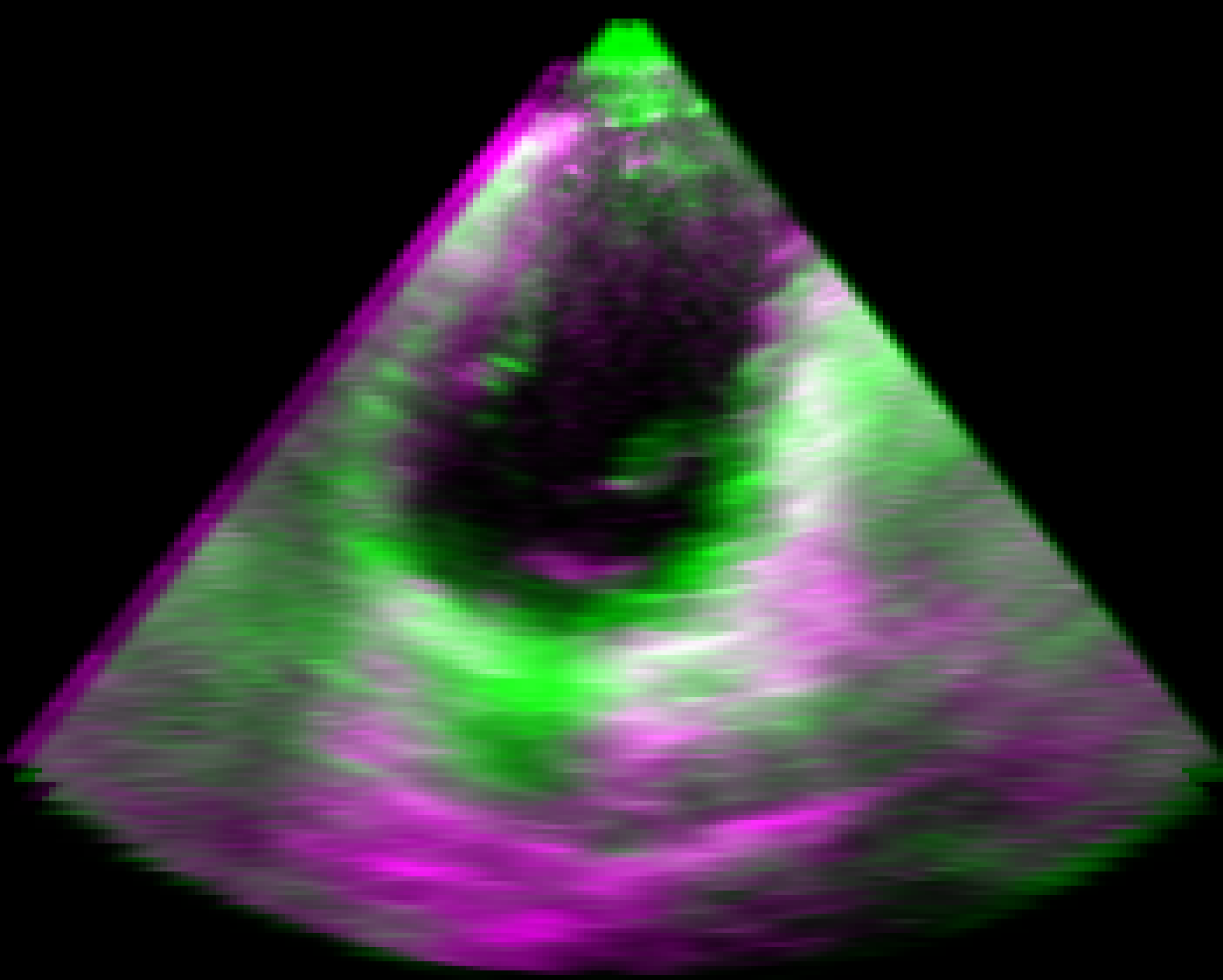}} \\
        \makecell[b]{Q1\vspace{15pt}} & 
        {\includegraphics[width=0.12\textwidth]{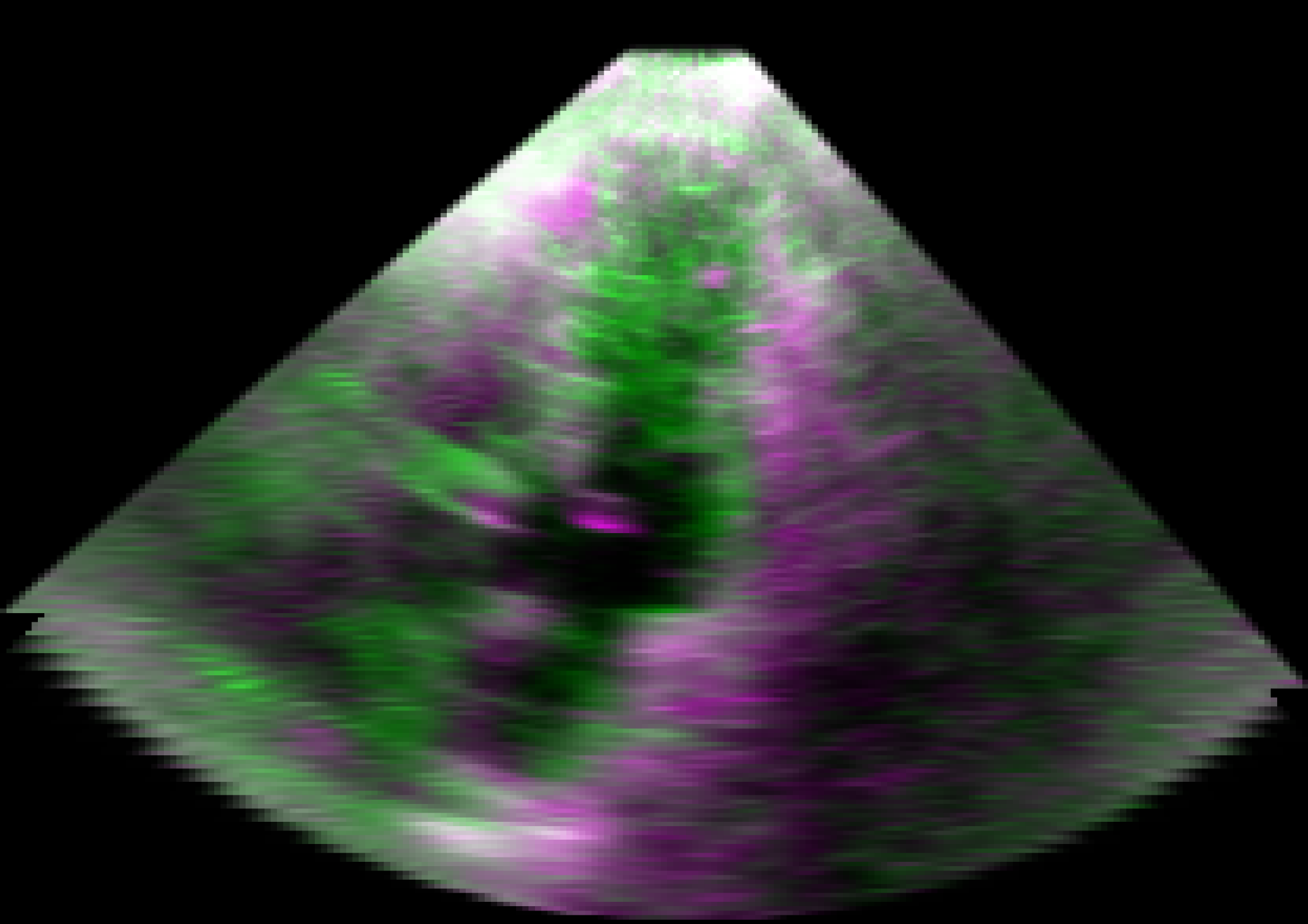}} &
        {\includegraphics[width=0.12\textwidth]{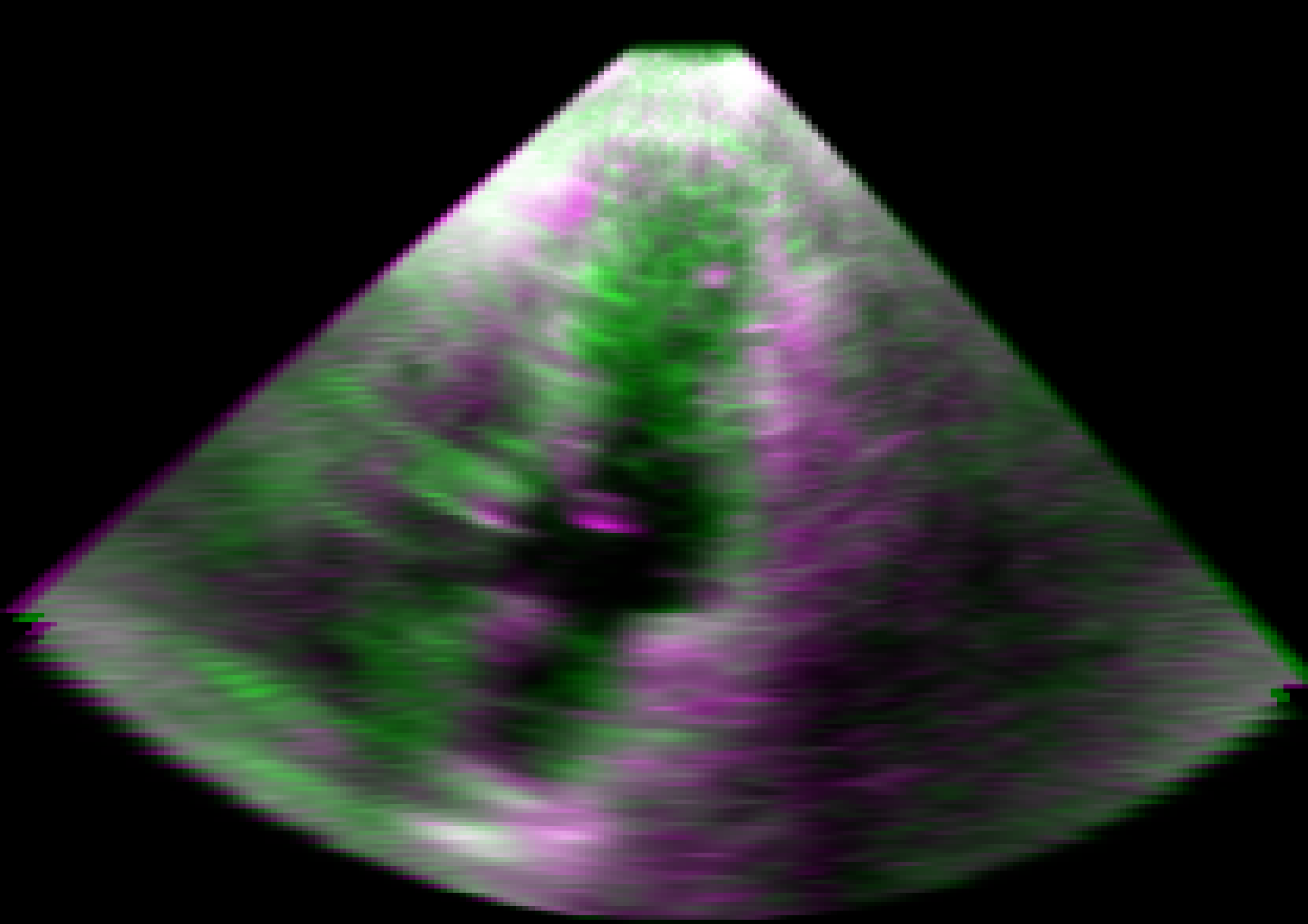}} &
        {\includegraphics[width=0.12\textwidth]{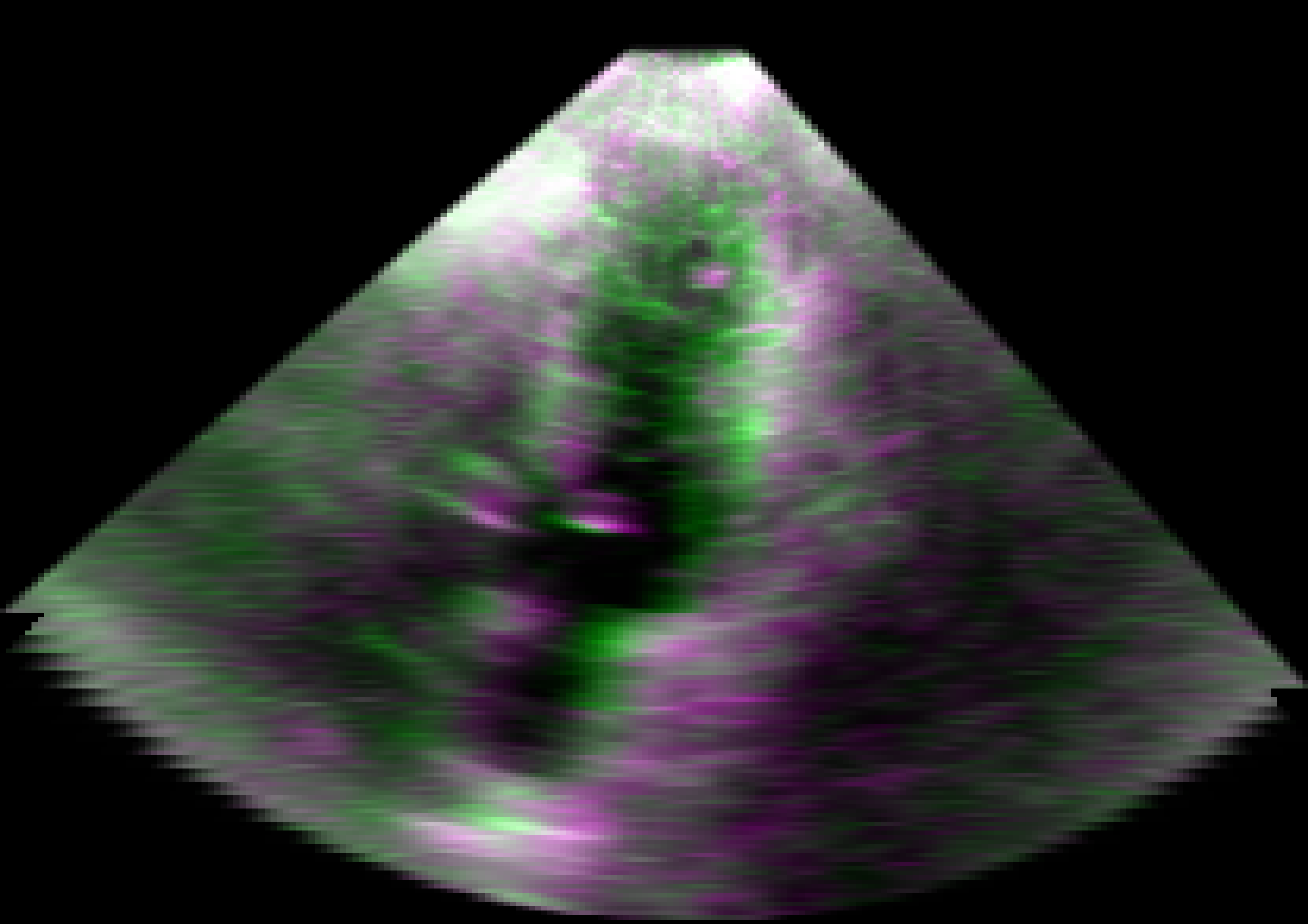}} &
        {\includegraphics[width=0.12\textwidth]{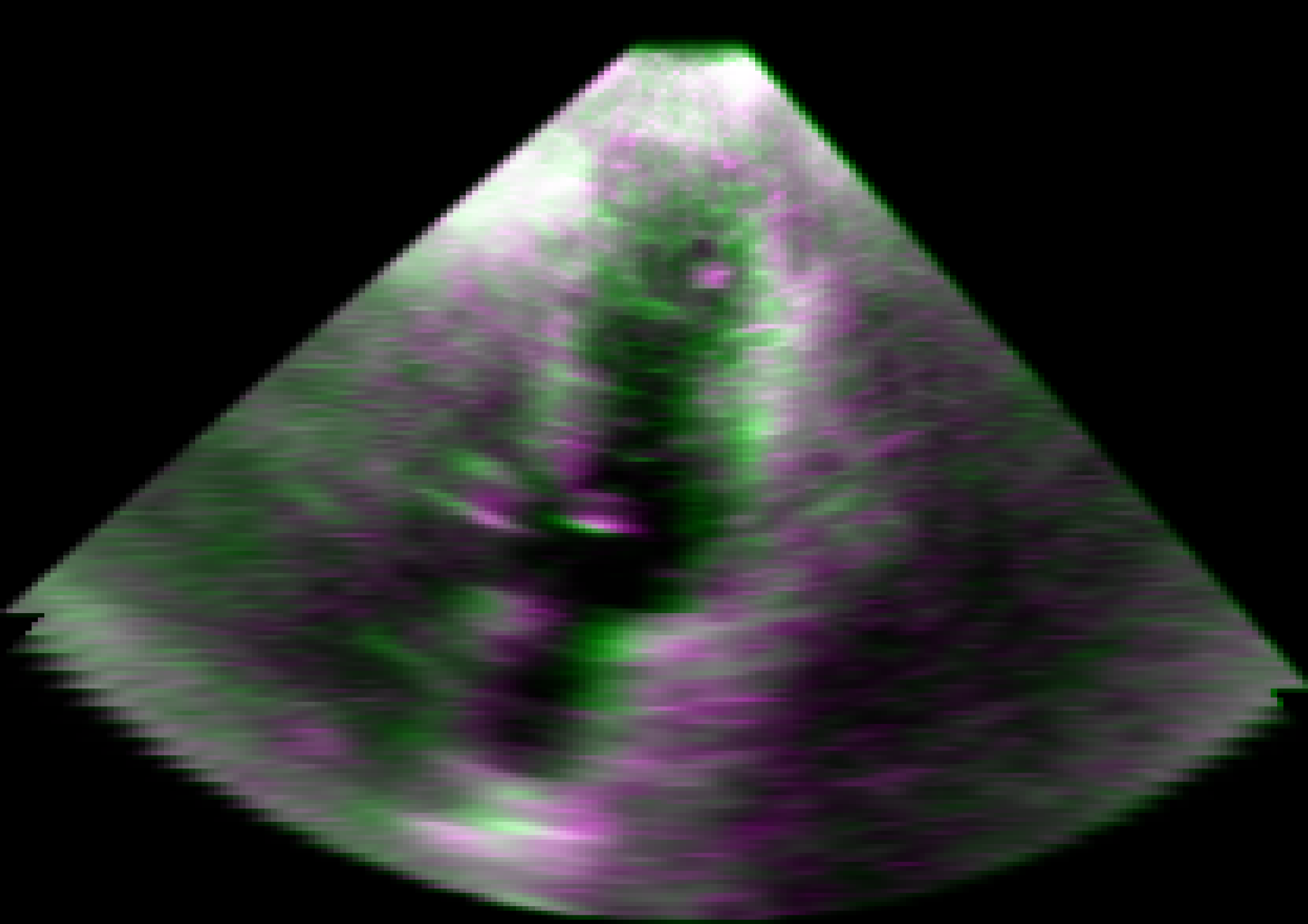}} &
        {\includegraphics[width=0.12\textwidth]{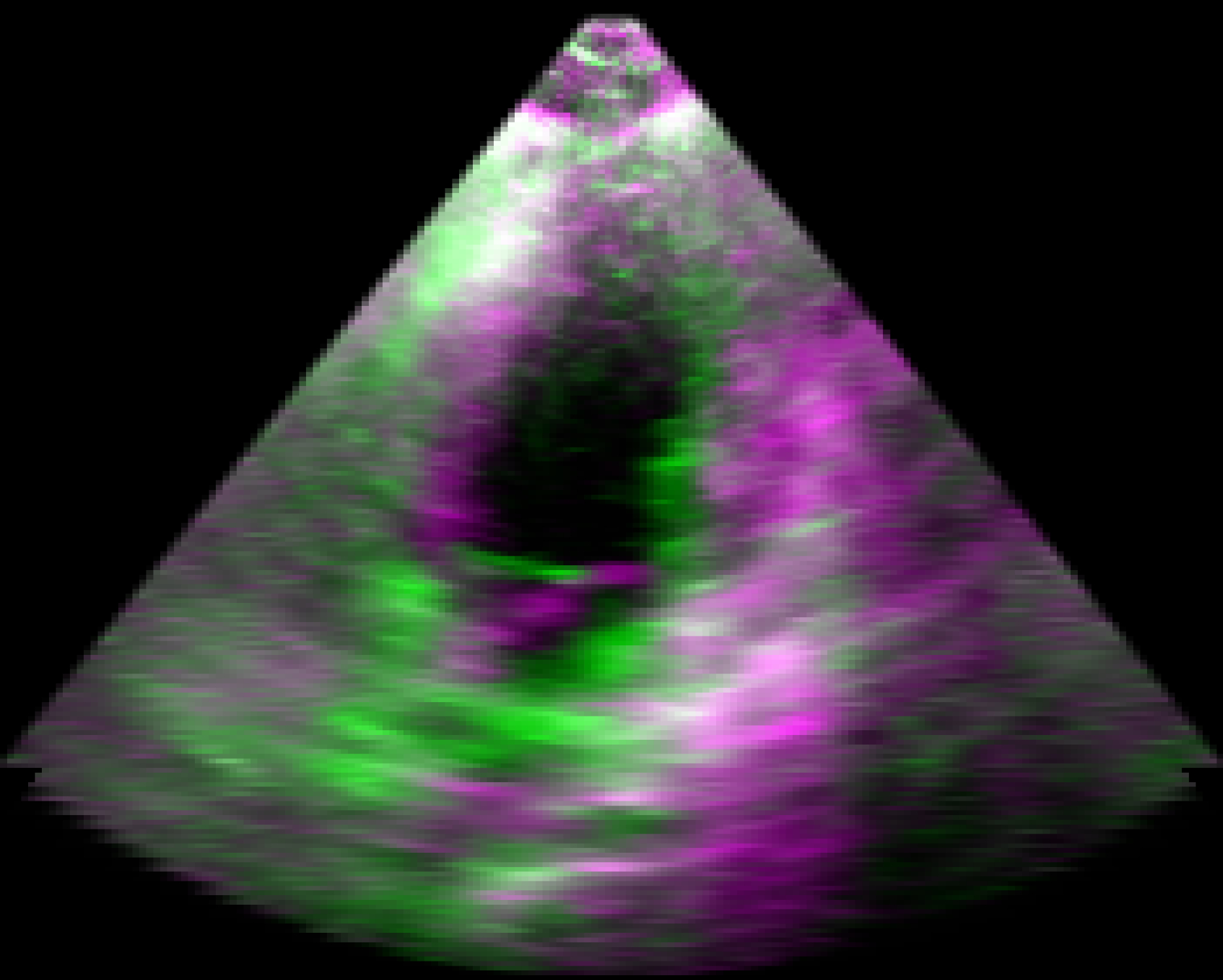}} &
        {\includegraphics[width=0.12\textwidth]{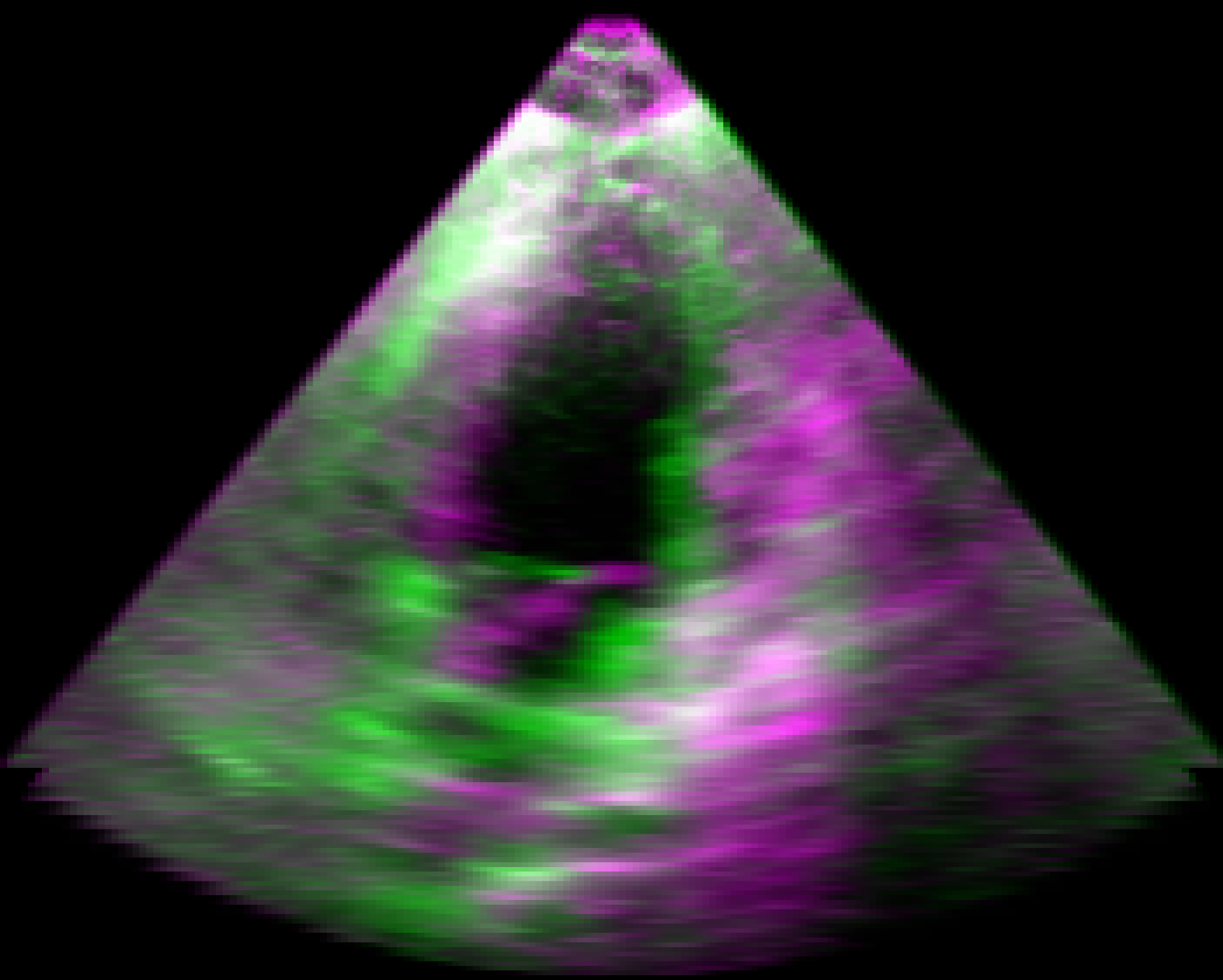}} \\ 
        \makecell[b]{Q2\vspace{15pt}} & 
        {\includegraphics[width=0.12\textwidth]{figures/fig_before_Im_019_20230817_104435_3D-Im_008_20230817_103907_3D_Sagittal_0.pdf}} &
        {\includegraphics[width=0.12\textwidth]{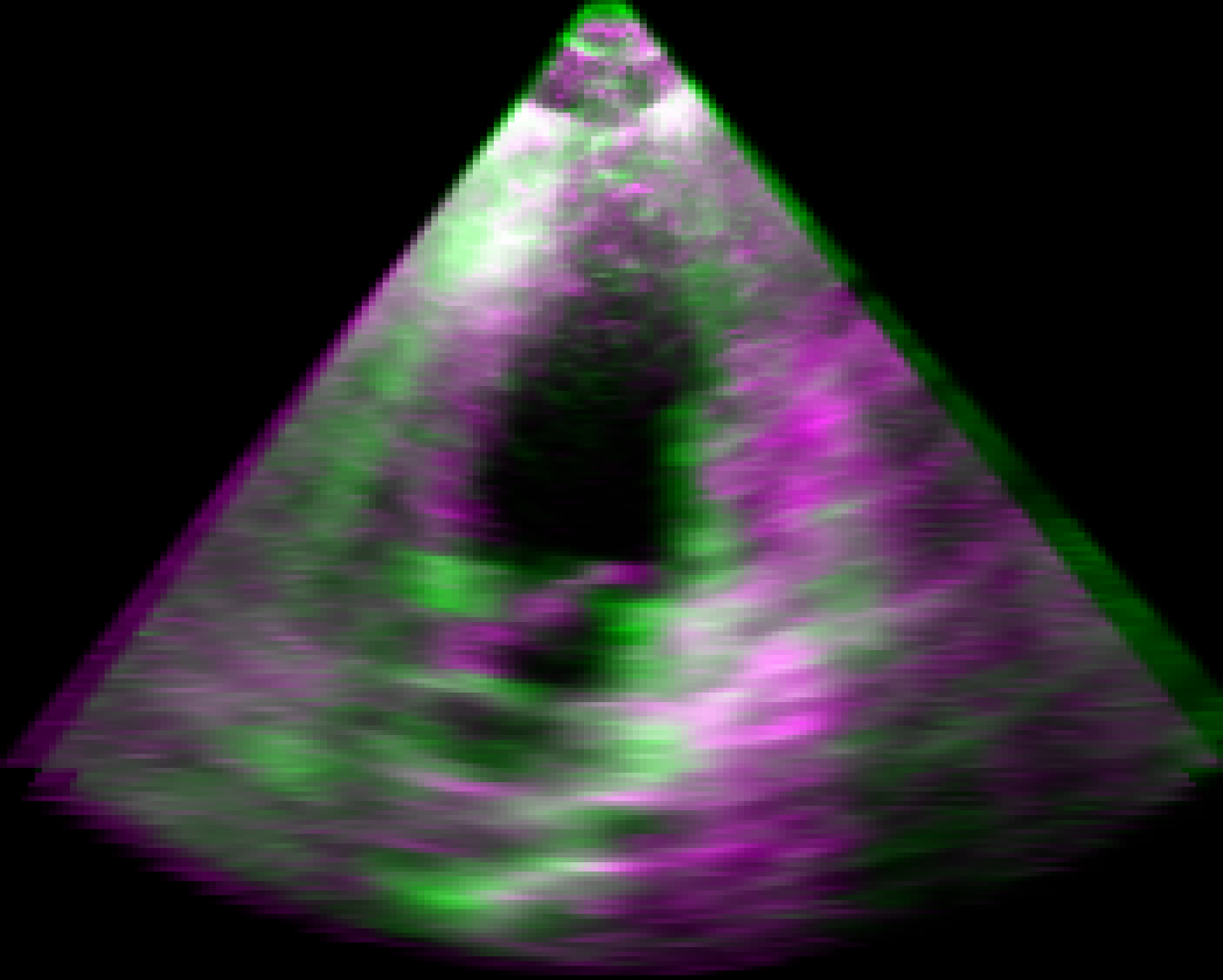}} &
        {\includegraphics[width=0.12\textwidth]{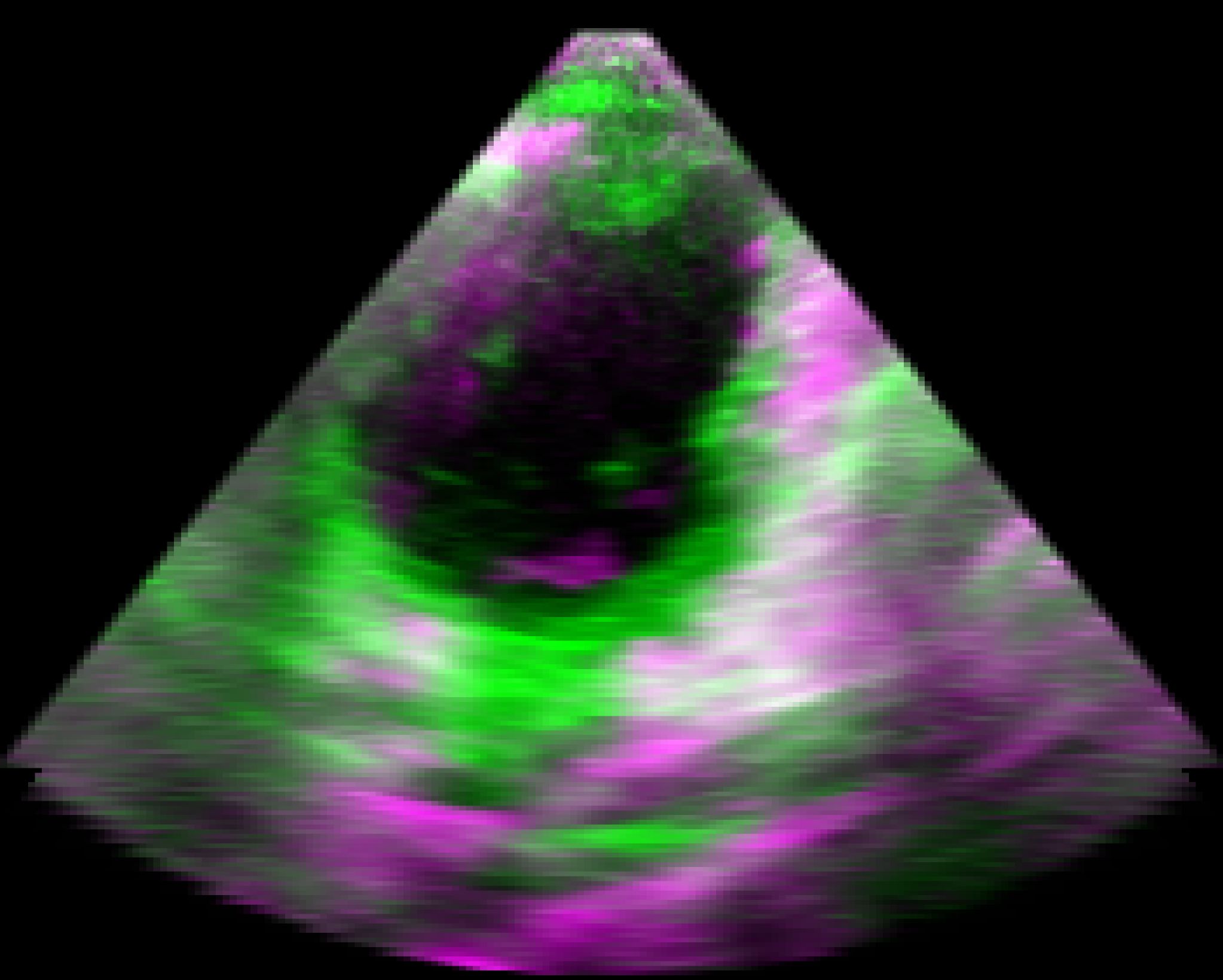}} &
        {\includegraphics[width=0.12\textwidth]{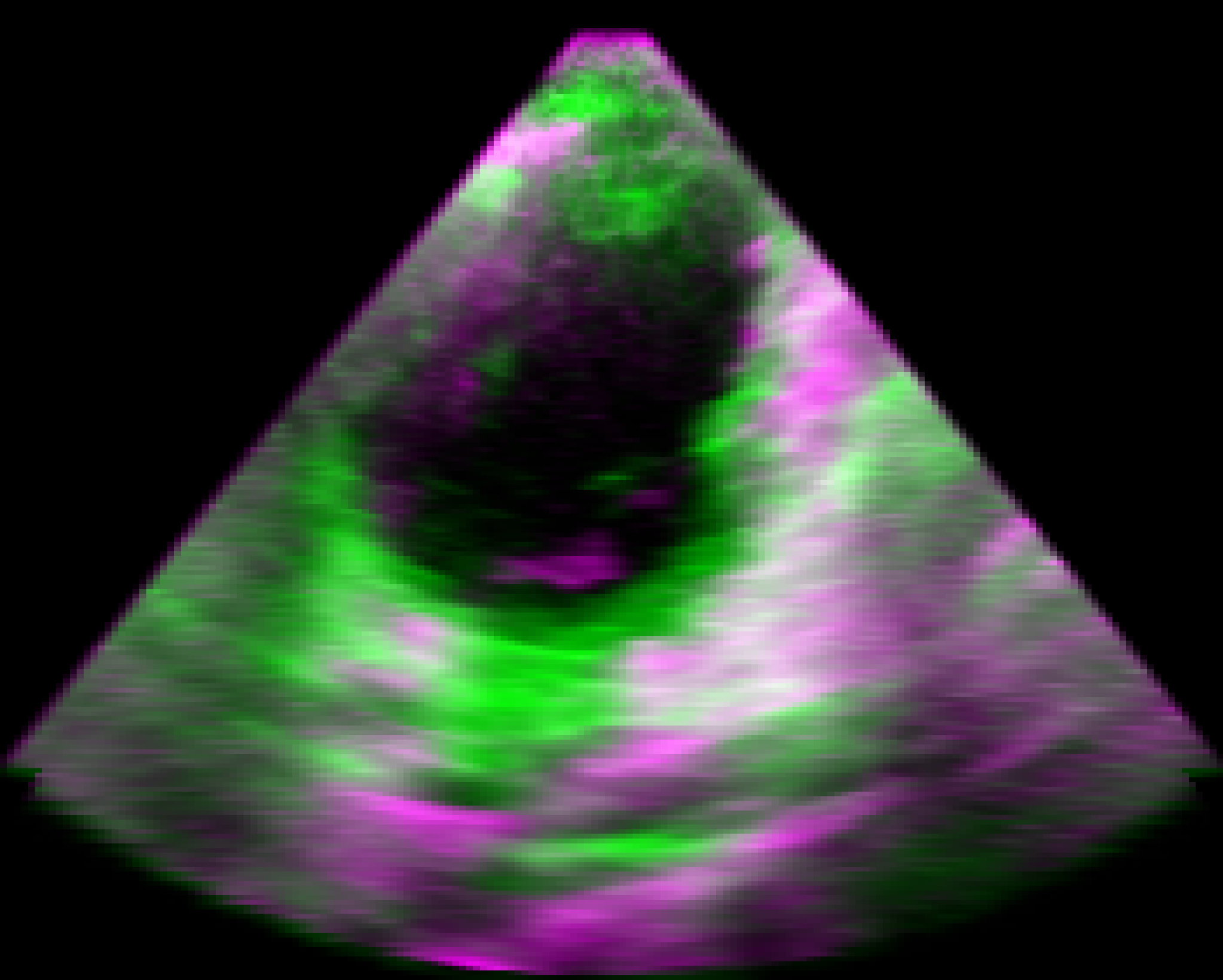}} &
        {\includegraphics[width=0.12\textwidth]{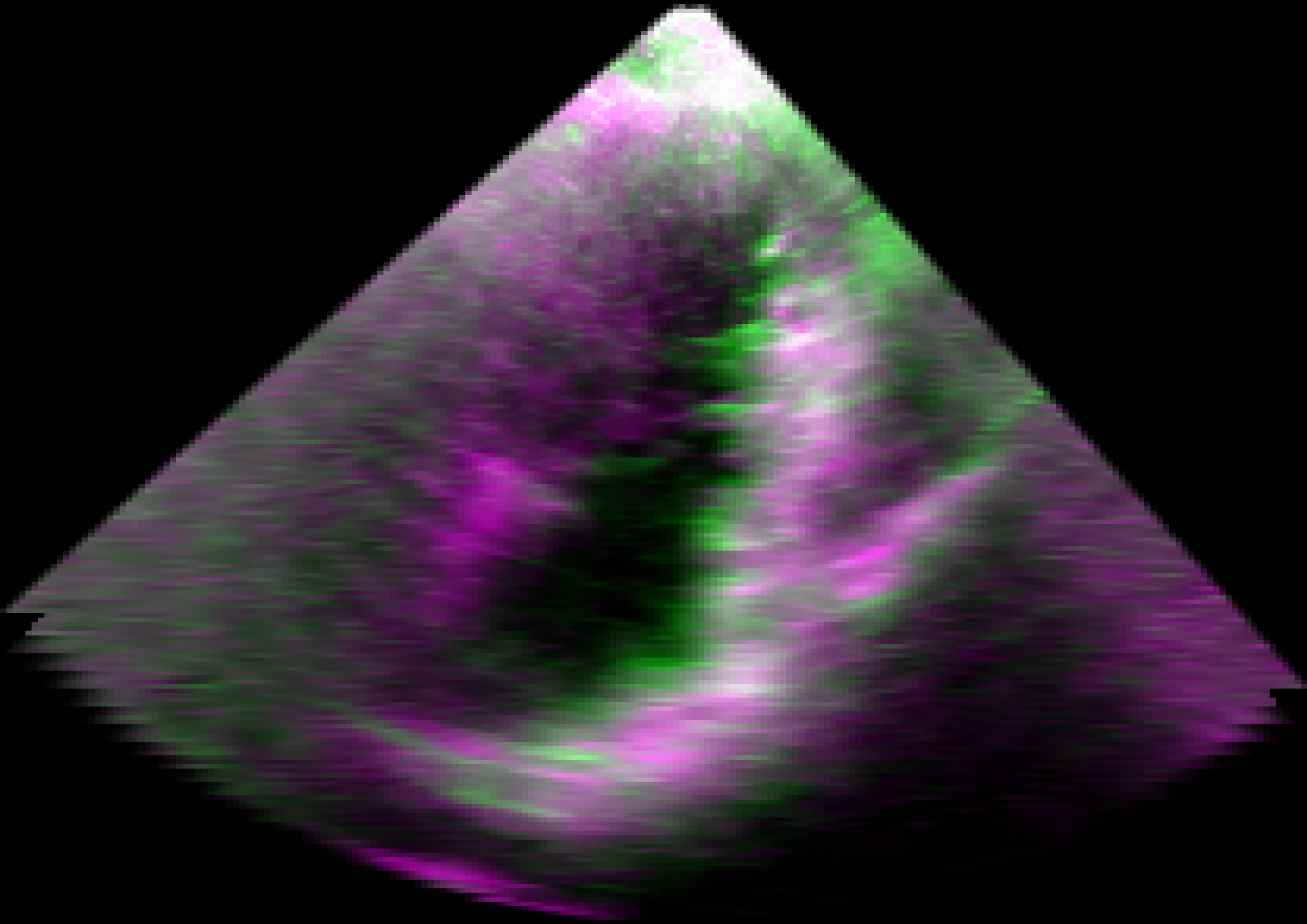}} &
        {\includegraphics[width=0.12\textwidth]{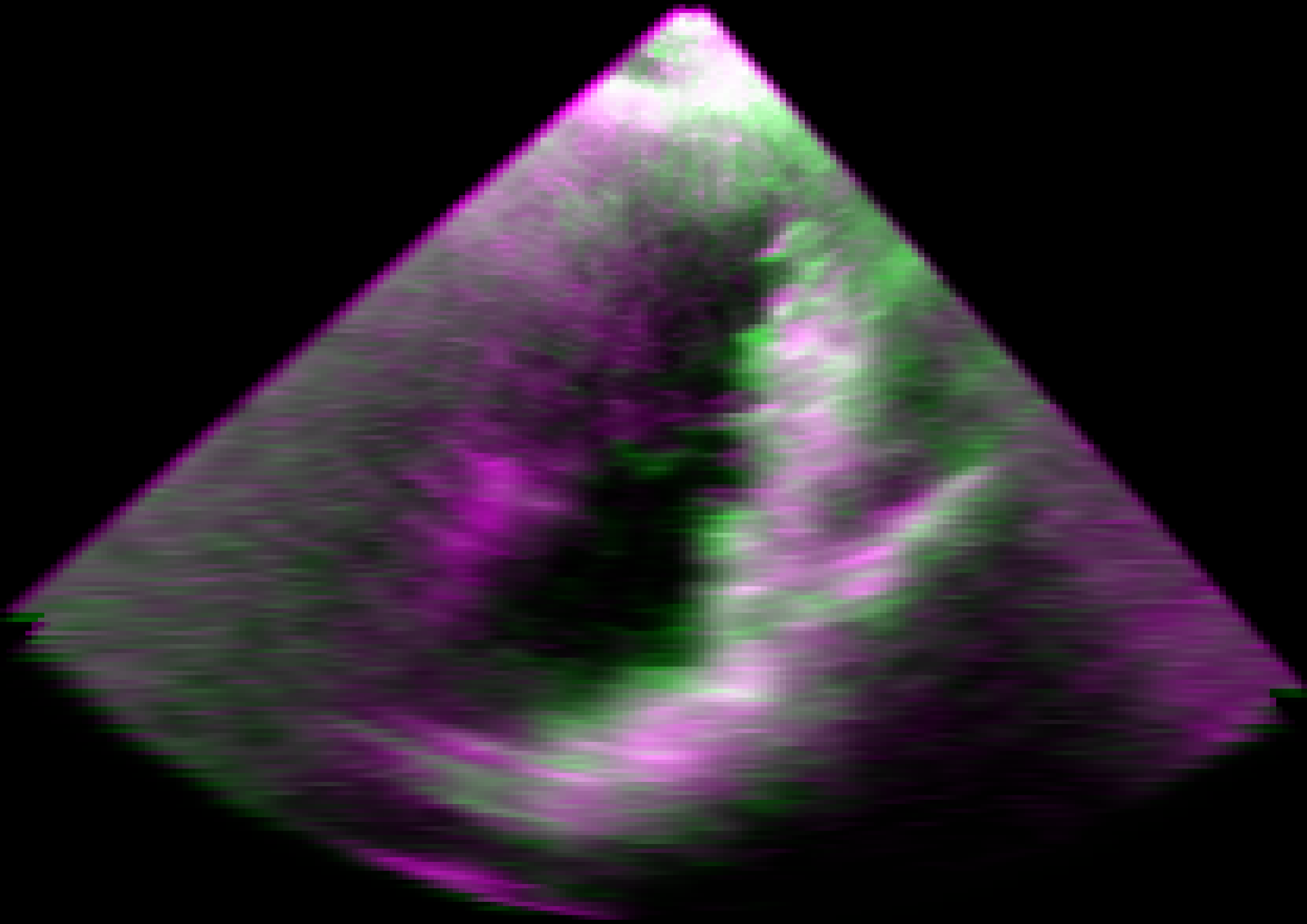}} \\
        \makecell[b]{Q3\vspace{15pt}} & 
        {\includegraphics[width=0.12\textwidth]{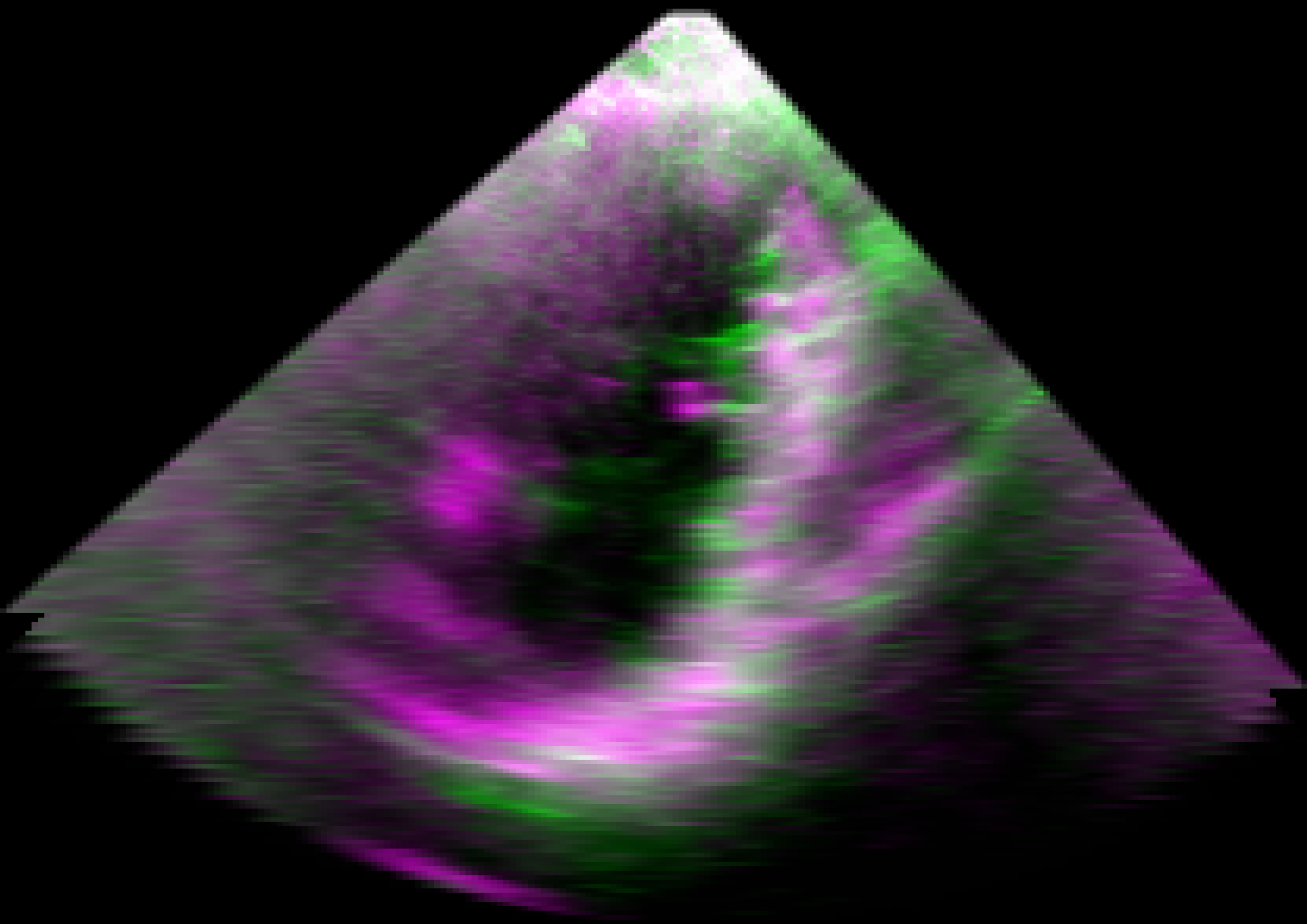}} &
        {\includegraphics[width=0.12\textwidth]{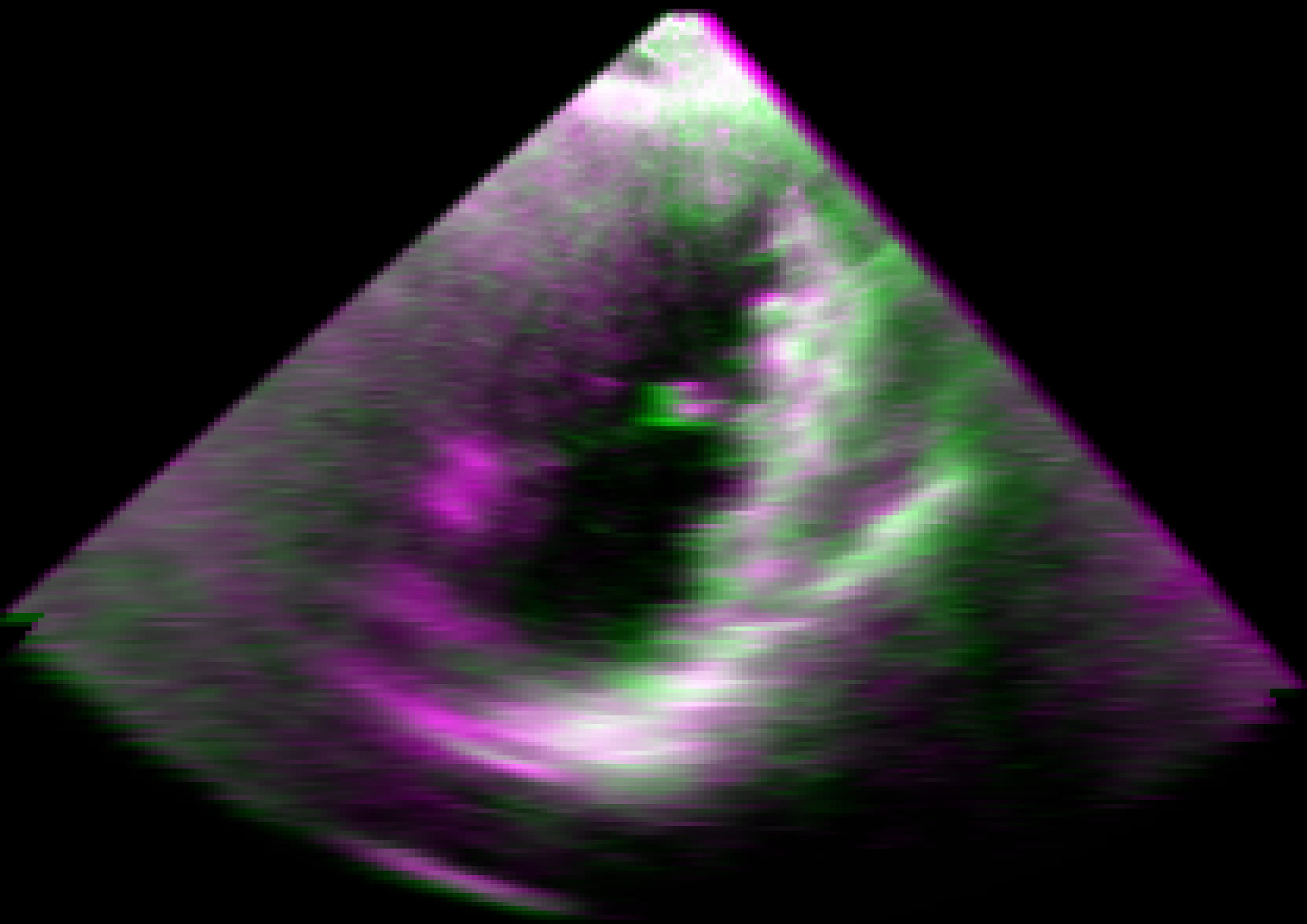}} &
        {\includegraphics[width=0.12\textwidth]{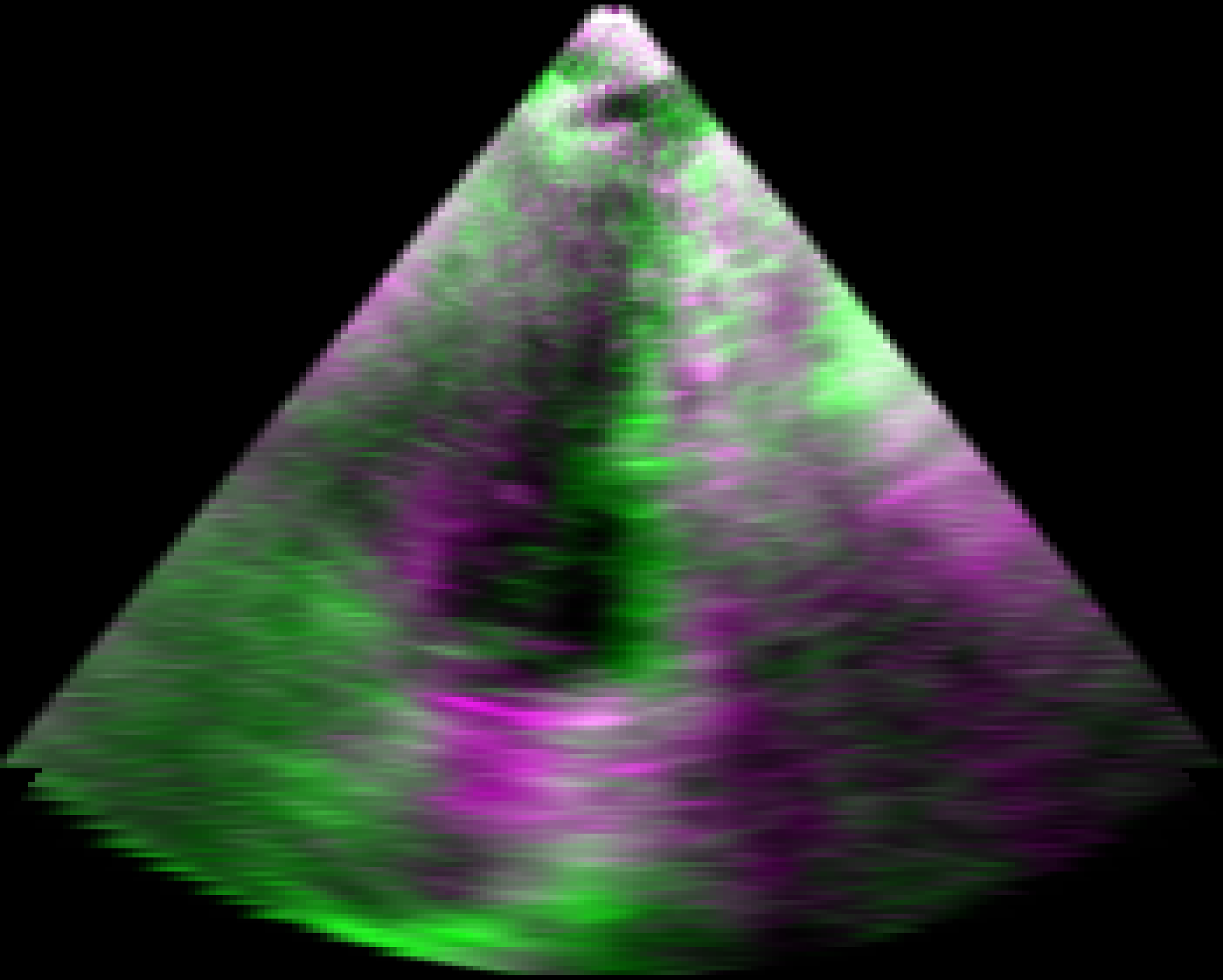}} &
        {\includegraphics[width=0.12\textwidth]{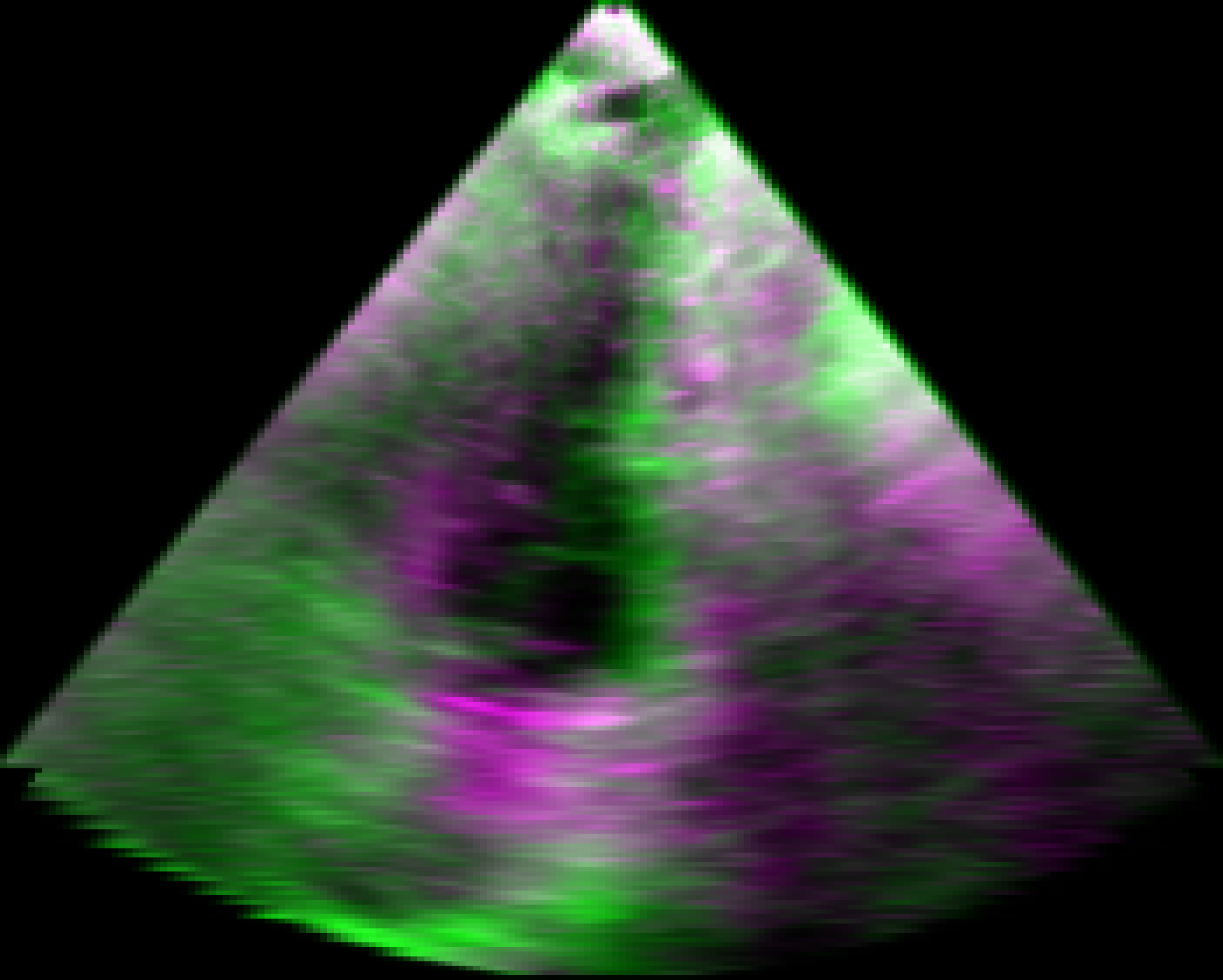}} &
        {\includegraphics[width=0.12\textwidth]{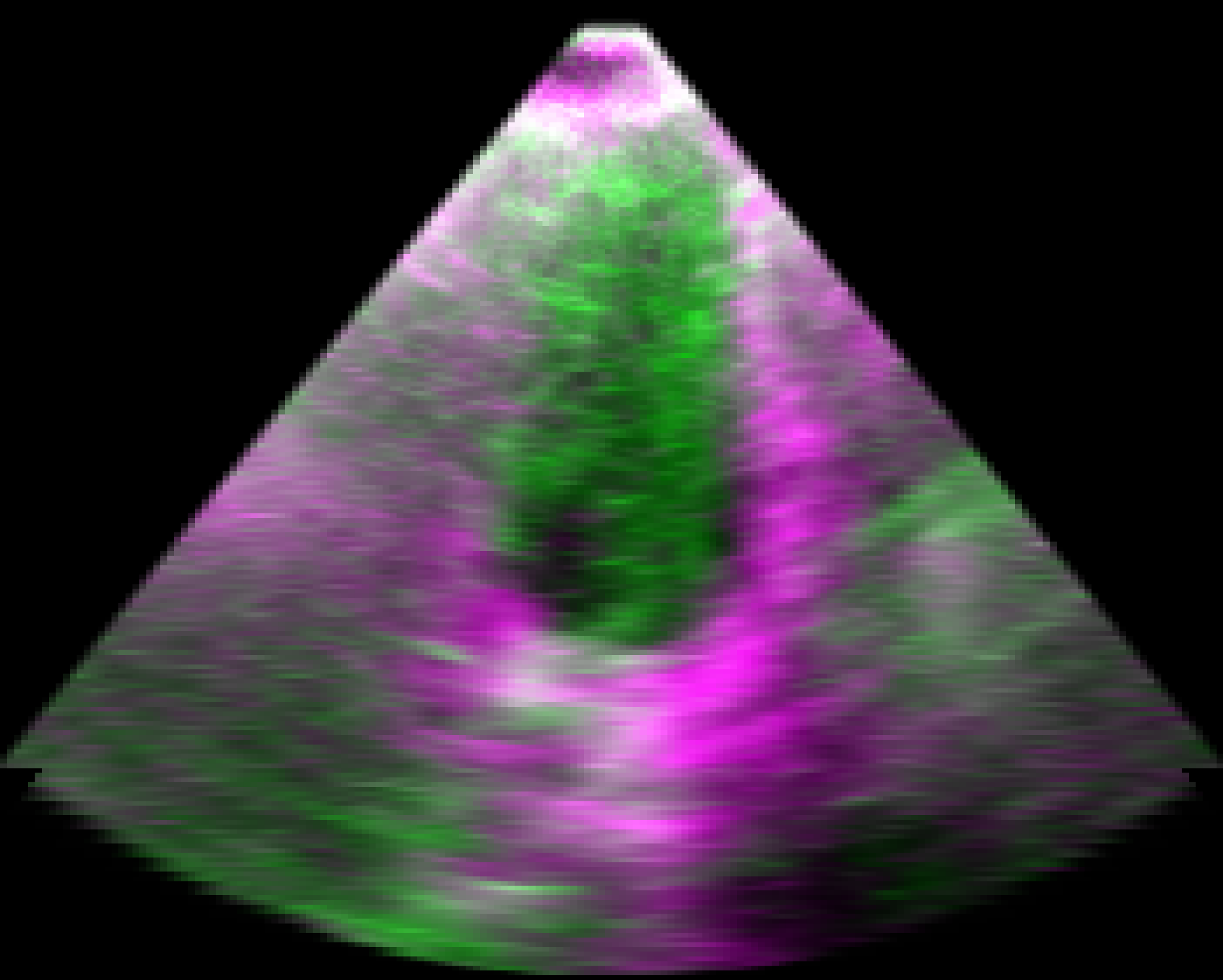}} &
        {\includegraphics[width=0.12\textwidth]{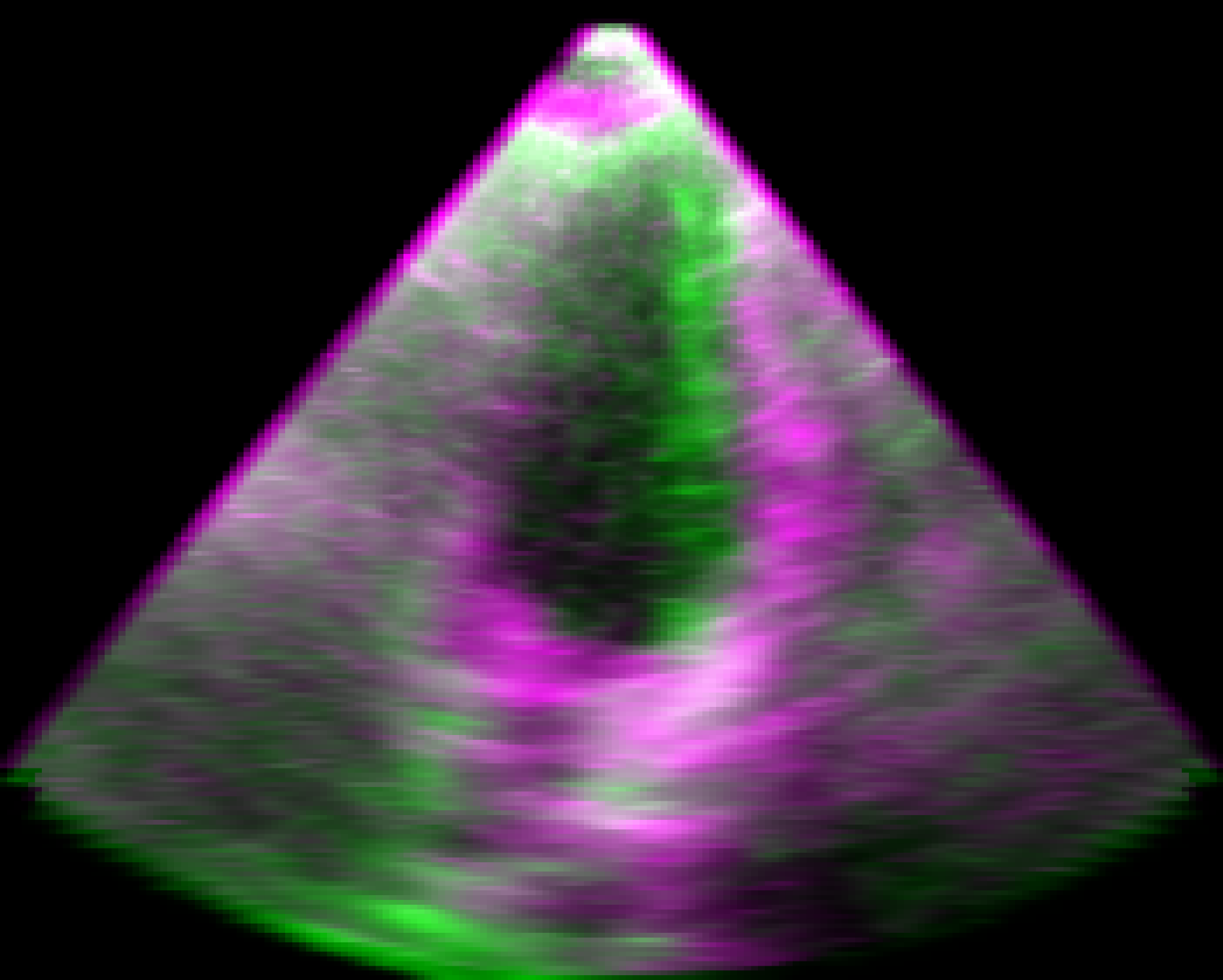}} \\
        \makecell[b]{Max\vspace{15pt}} &
        {\includegraphics[width=0.12\textwidth]{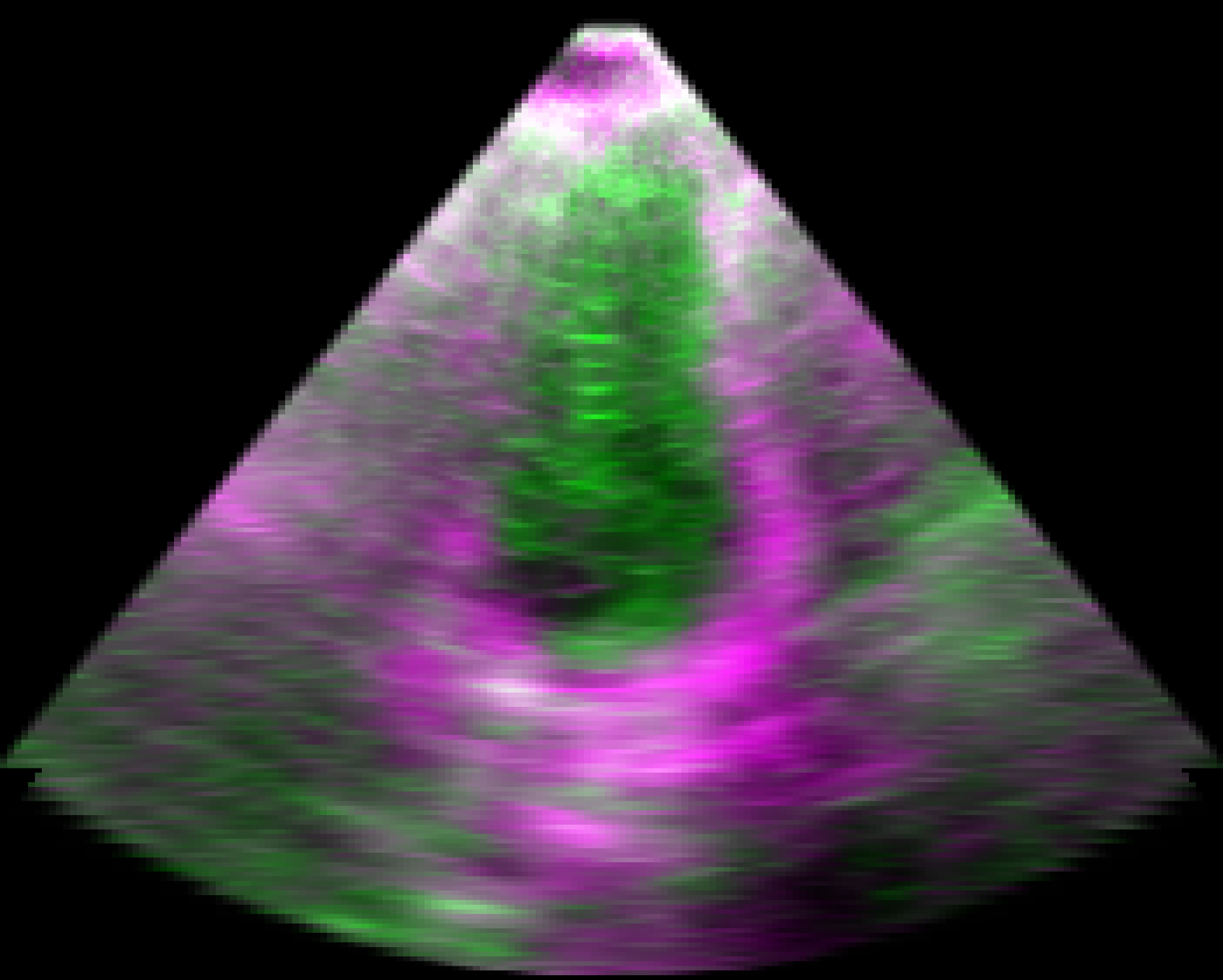}} &
        {\includegraphics[width=0.12\textwidth]{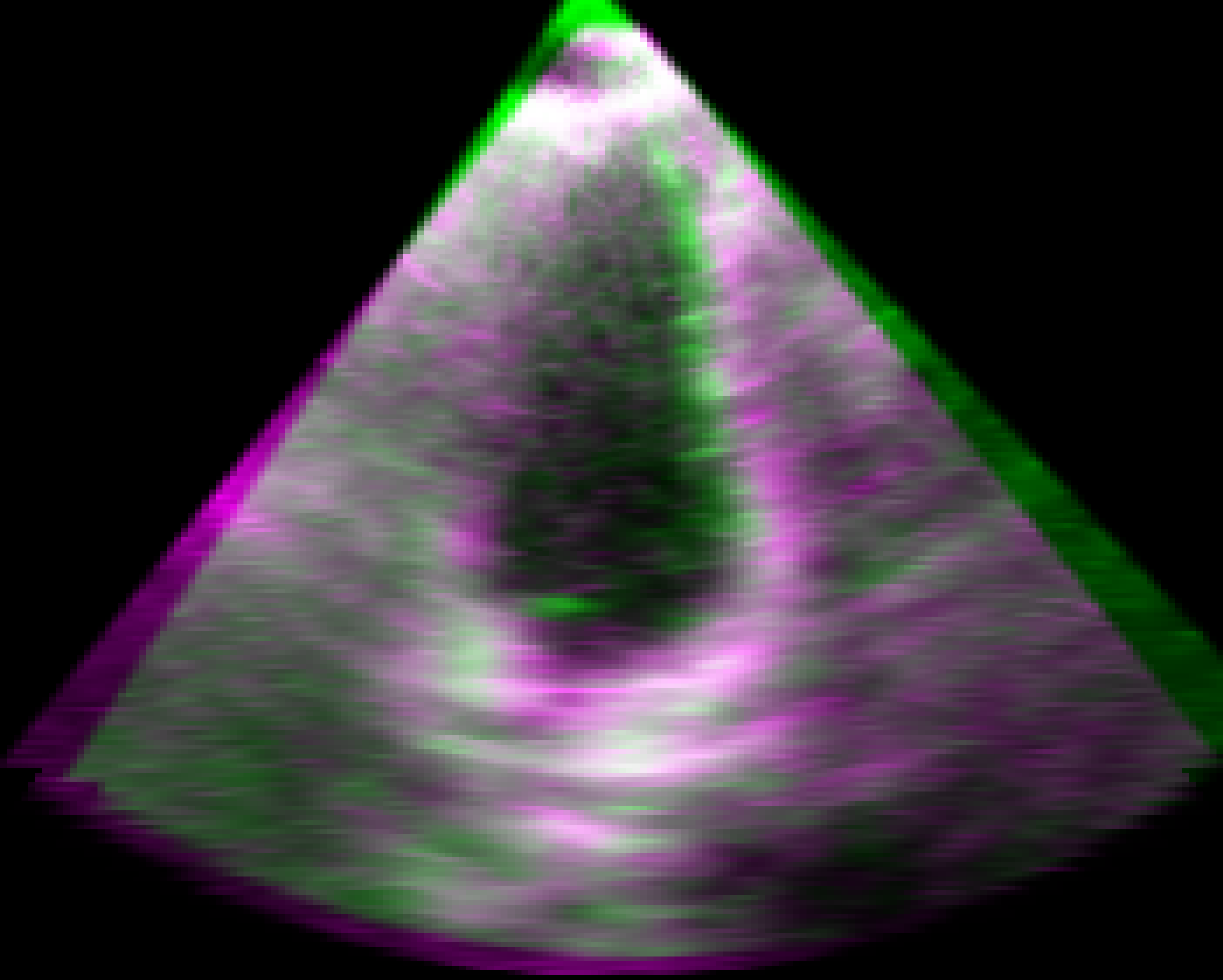}} &
        {\includegraphics[width=0.12\textwidth]{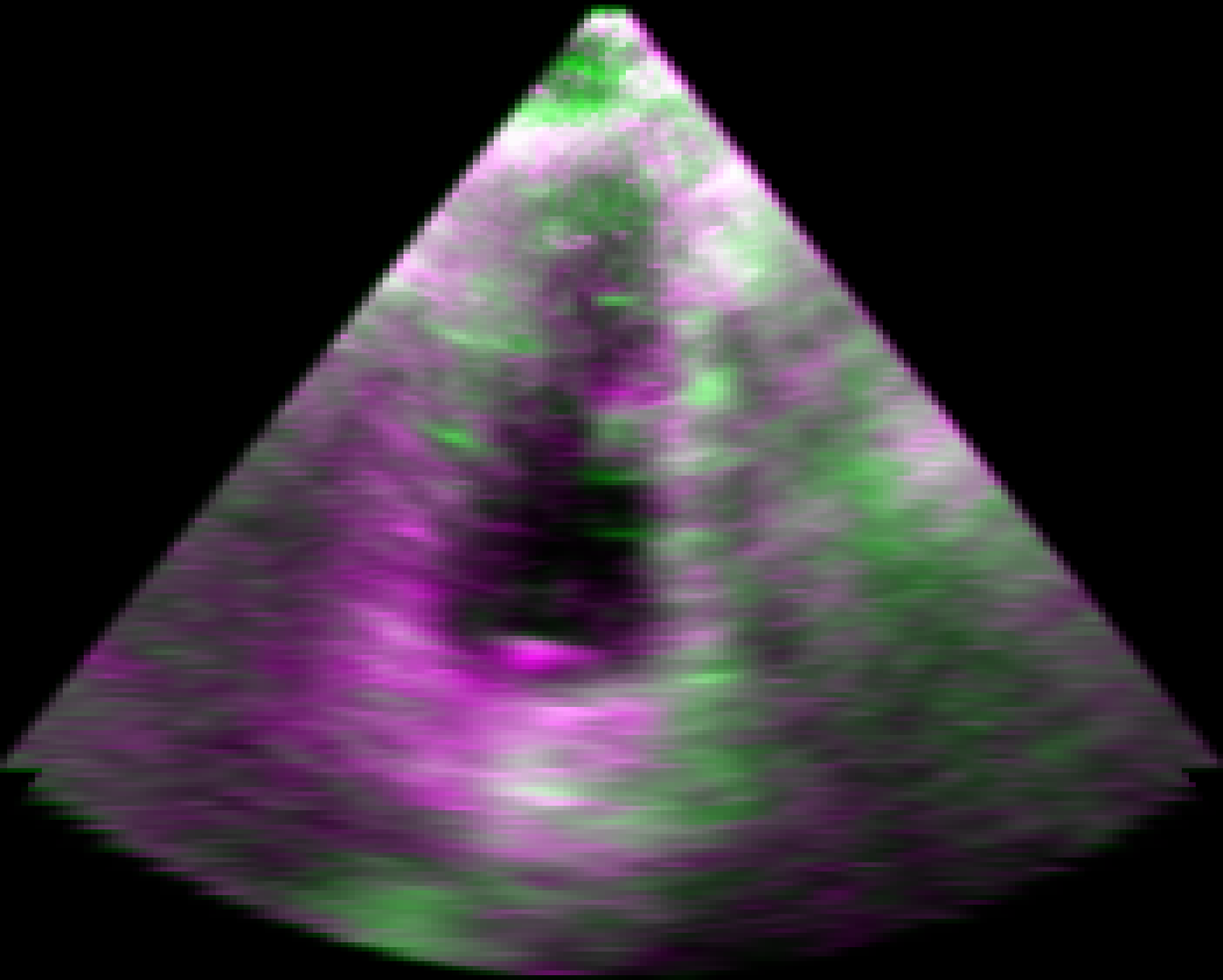}} &
        {\includegraphics[width=0.12\textwidth]{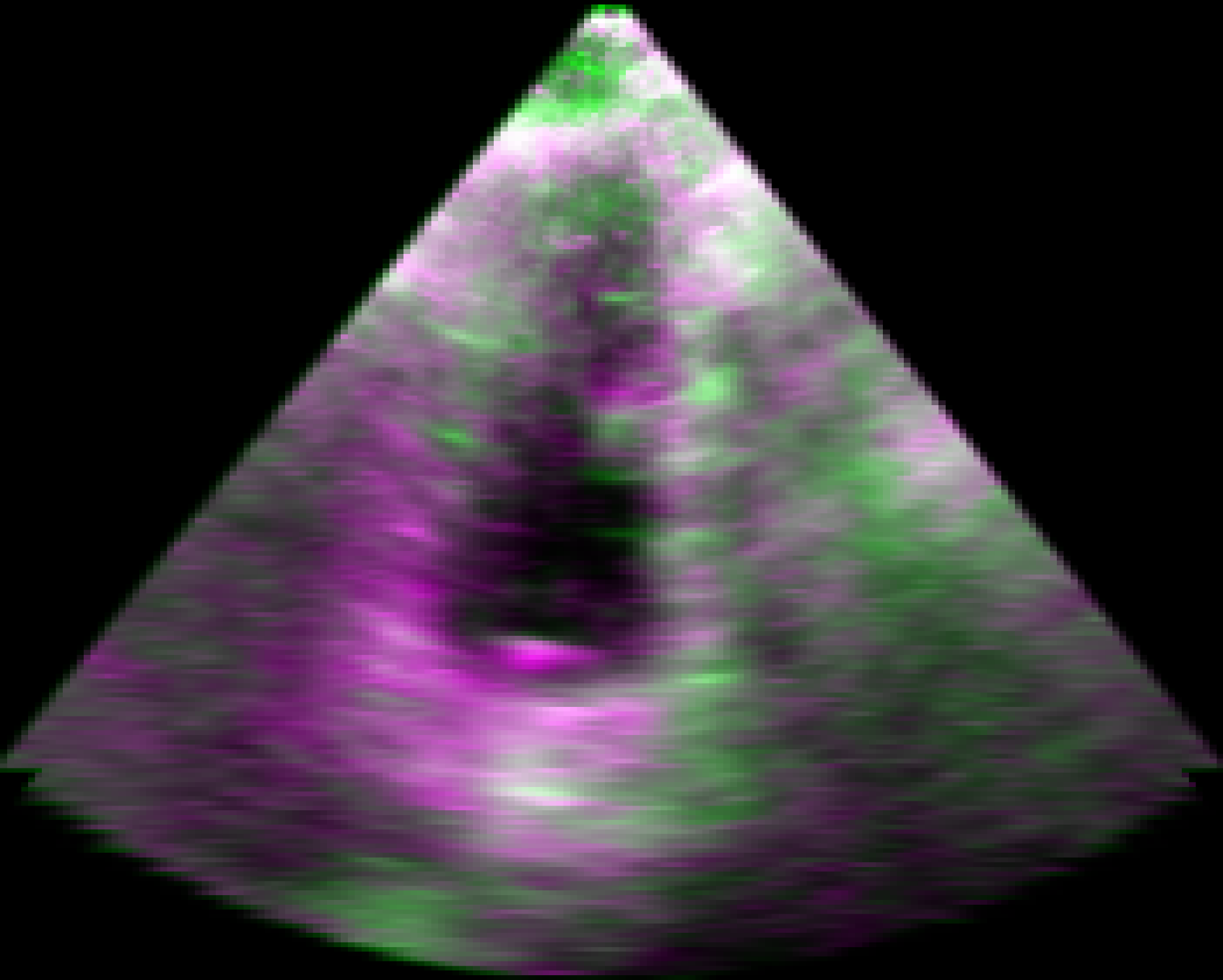}} &
        {\includegraphics[width=0.12\textwidth]{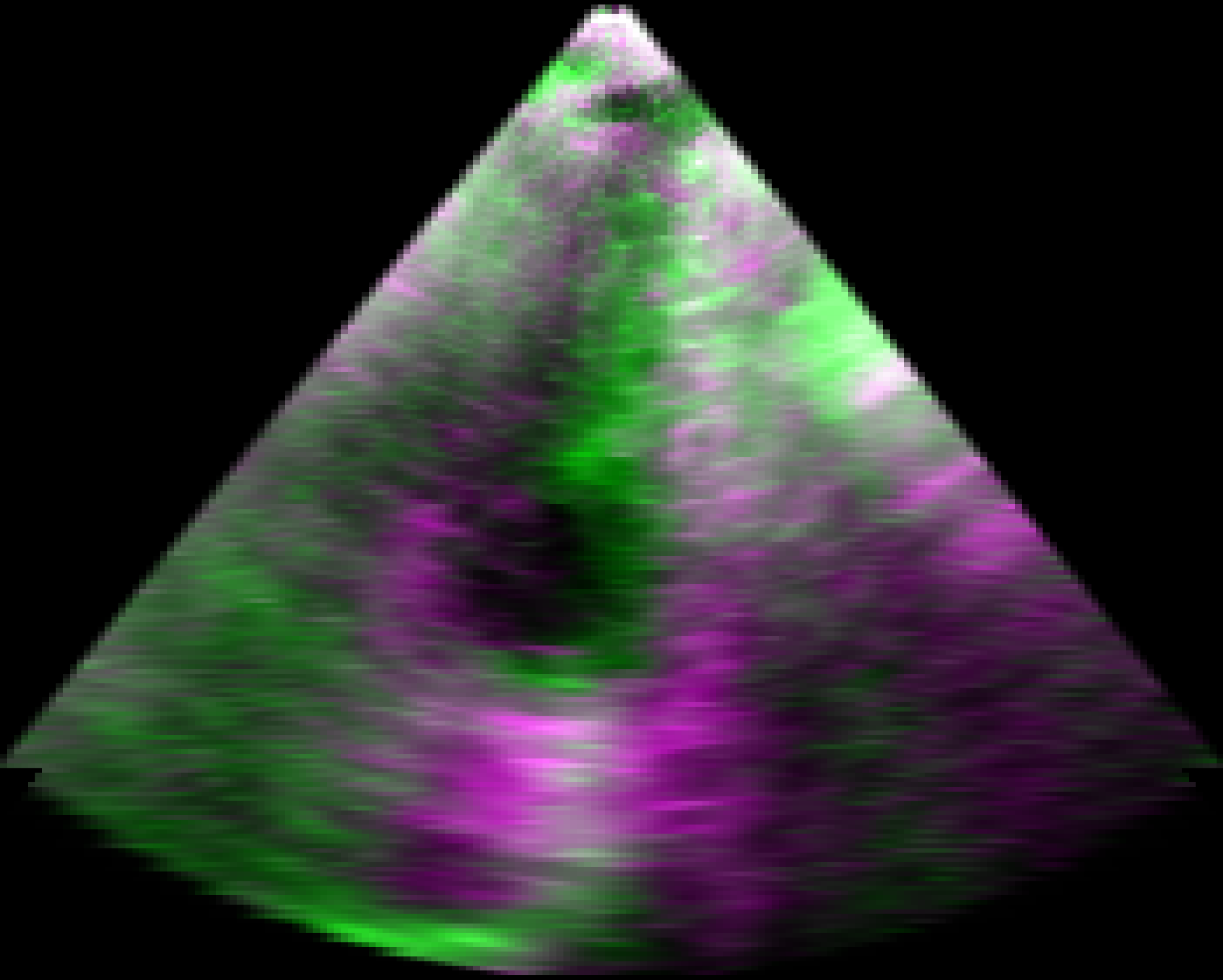}} &
        {\includegraphics[width=0.12\textwidth]{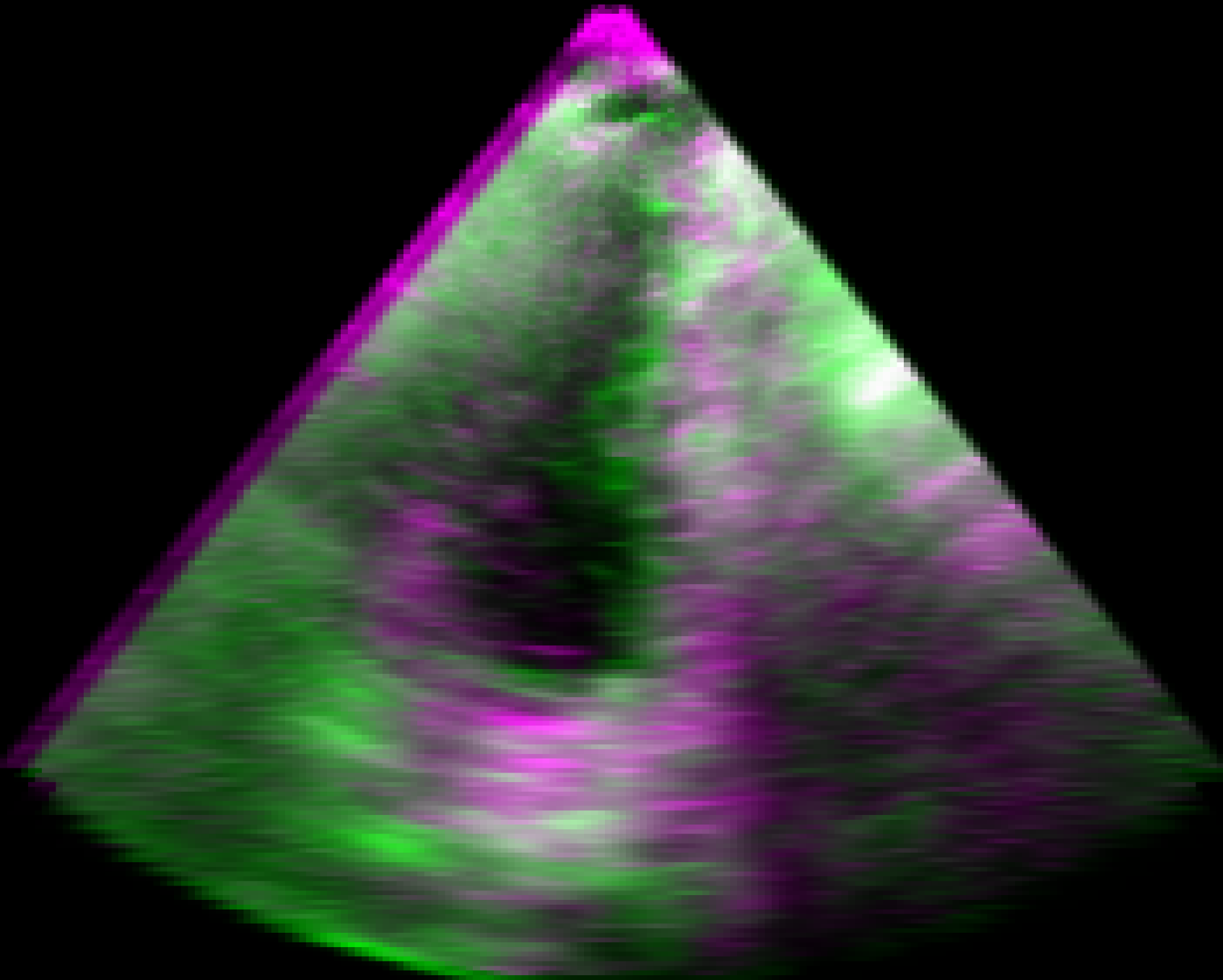}} \\
    \end{tabular}
    \caption{Image-based rigid (PF and EX) registration results of the ED frame of image pairs for percentile DSC difference values of the sagittal view. Two consecutive columns show images before and after the registration for each category. The source and target images are shown in green and purple colors.}
    \label{fig:fig_perc_images_sagittal_rigid_img}
\end{figure*}

Table \ref{tab:table_ncc_dsc} presents the average similarity metric (NCC) and the overlapping score (DSC) changes before and after the registration for the non-parallel (CPU) and parallel (GPU) versions of the PF and ES algorithms for 1073 frames of 66 image pairs of the volunteer dataset. The rigid registration was performed independently using the intensities of images and LV binary masks to evaluate the appropriate approach for 3D TTE images. Figure \ref{fig:fig_pf_es_se_ncc} shows the DSC difference values for the minimum, 25\%, 50\%, 75\%, and maximum values that are summarized in Tables \ref{tab:table_percentile} and \ref{tab:table_percentile_se} for rigid and nonrigid registration. Corresponding images of the pairs listed in Table \ref{tab:table_percentile} for rigid registration using image intensities (PF(CPU), PF(GPU), and ES(CPU)) are shown in Figures \ref{fig:fig_perc_images_axial_rigid_img}, \ref{fig:fig_perc_images_coronal_rigid_img}, and \ref{fig:fig_perc_images_sagittal_rigid_img} for axial, coronal, and sagittal views, respectively. It can be seen that the maximum DSC improvement is achieved for the CPU version of PF compared to the GPU version of PF and ES methods. The image pairs corresponding to the minimum DSC difference of the PF show that they are already aligned (DSC before registration of PF(CPU) and PF(GPU) are 0.846 and 0.811). The maximum DSC difference of PF(CPU) and ES indicates that there is a limited overlap between the images (DSC before registration of PF(CPU) and ES are 0.211 and 0.104, respectively). Since the ES performs a predefined grid search to align the source image to the target image, the registration fails for more images than the CPU and GPU versions of PF.

\renewcommand{\arraystretch}{1.5}
\begin{table}[H]
\caption{The change of similarity metric (NCC) and overlapping score (DSC) before and after rigid (PF and EX) registration. IC and IG refer to image-based registration performed on CPU and GPU hardware, respectively. The MC and MG refer to mask-based registration in the same way. The best values of the NCC and DSC after registration are highlighted.}
\label{tab:table_ncc_dsc}
\setlength{\tabcolsep}{18pt}
    \begin{tabular}{llc}
        \toprule
        Metric & Method  & Value\\
        \midrule
        NCC & Before reg	& 0.715 $\pm$ 0.008	\\
         & PF$_{IC}$ & 0.666 $\pm$ 0.011 \\
         & \textbf{PF}\pmb{$_{IG}$} & \textbf{0.718} \pmb{$\pm$ 0.008} \\
         & EX$_{IC}$ & 0.645 $\pm$ 0.014 \\
         & PF$_{MC}$ & 0.556 $\pm$ 0.014 \\
         & PF$_{MG}$ & 0.576 $\pm$ 0.048 \\
         & EX$_{MC}$ & 0.514 $\pm$ 0.009 \\
        DSC & Before reg	& 0.497 $\pm$ 0.051	\\
         & PF$_{IC}$ & 0.680 $\pm$ 0.011\\
         & PF$_{IG}$ & 0.550 $\pm$ 0.047\\
         & EX$_{IC}$ & 0.641 $\pm$ 0.013\\
         & PF$_{MC}$ & 0.811 $\pm$ 0.006\\
         & \textbf{PF}\pmb{$_{MG}$} & \textbf{0.819} \pmb{$\pm$ 0.045}\\
         & EX$_{MC}$ & 0.743 $\pm$ 0.003\\
        \bottomrule
    \end{tabular}
\end{table}
\renewcommand{\arraystretch}{1.0}

\begin{figure*}[!ht]
    \centering
    \begin{tabular}{SSSSSSS}
         & \multicolumn{2}{c}{\scriptsize PF (CPU)} & \multicolumn{2}{c}{\scriptsize PF (GPU)} & \multicolumn{2}{c}{\scriptsize EX (CPU)} \\
         & Before & After & Before & After & Before & After \\
        \makecell[b]{Min\vspace{20pt}} & 
        {\includegraphics[width=0.12\textwidth]{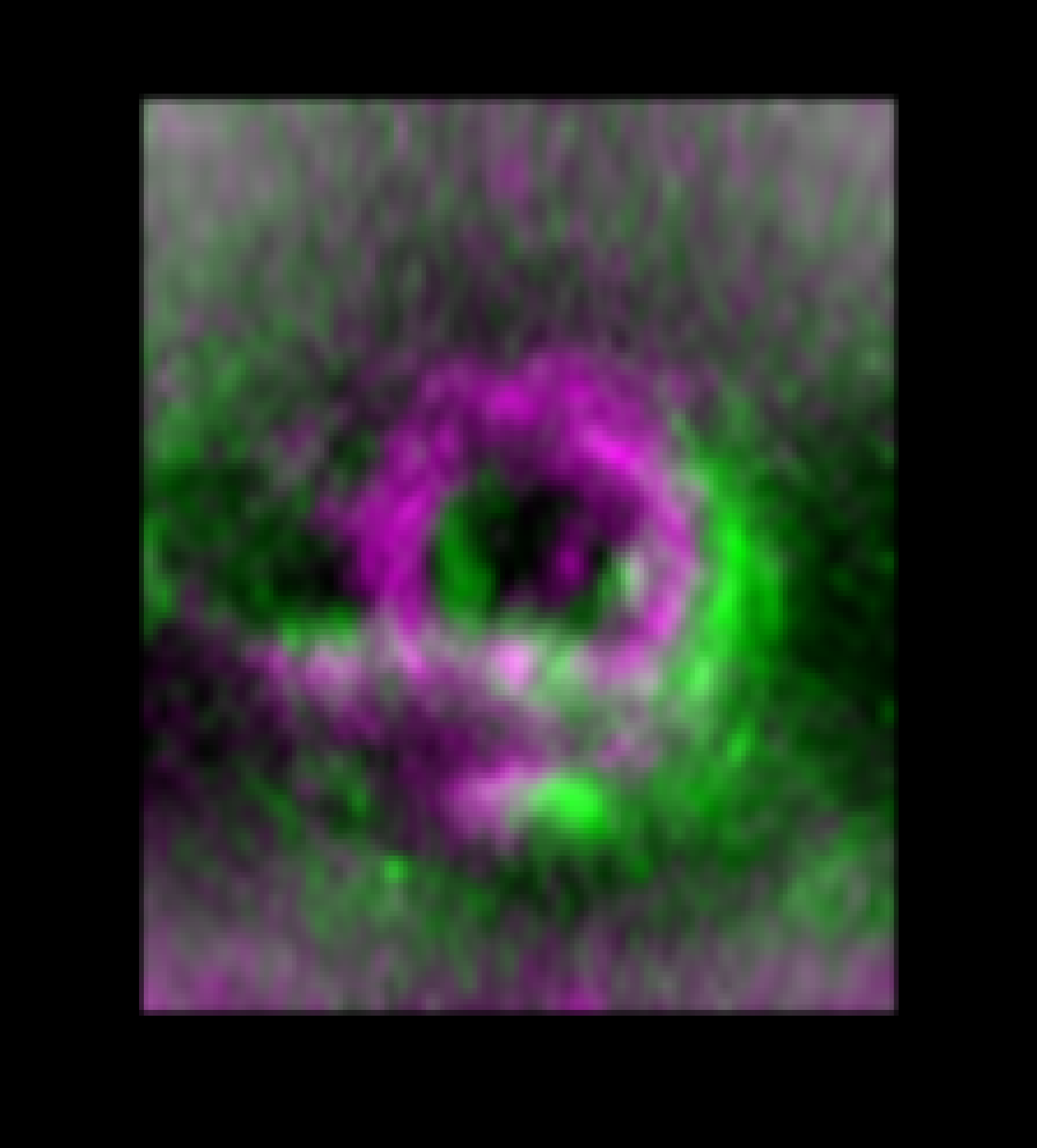}} &
        {\includegraphics[width=0.12\textwidth]{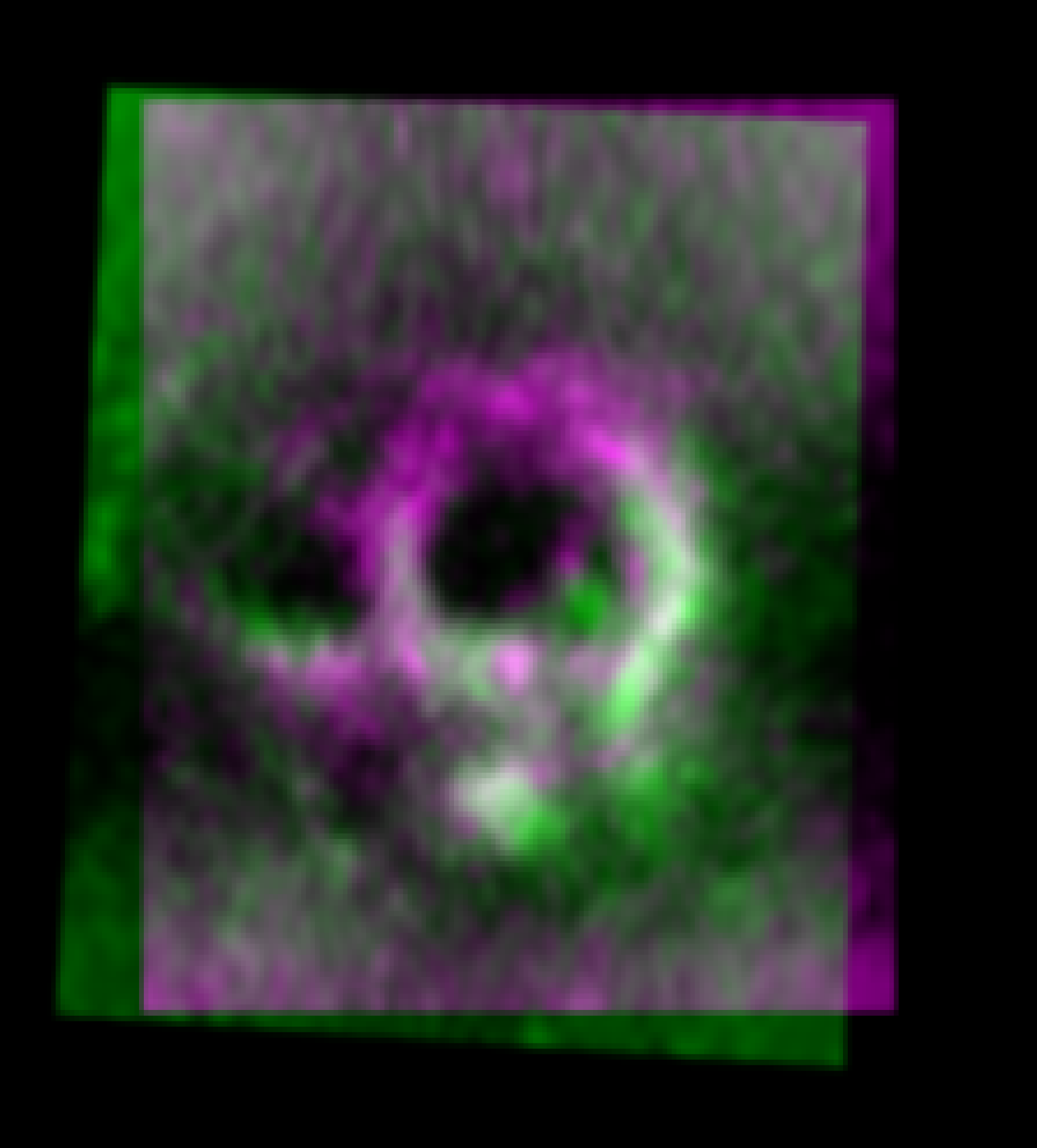}} &
        {\includegraphics[width=0.12\textwidth]{figures/fig_before_Im_016_20230817_104344_3D-Im_012_20230817_104241_3D_Axial_0.pdf}} &
        {\includegraphics[width=0.12\textwidth]{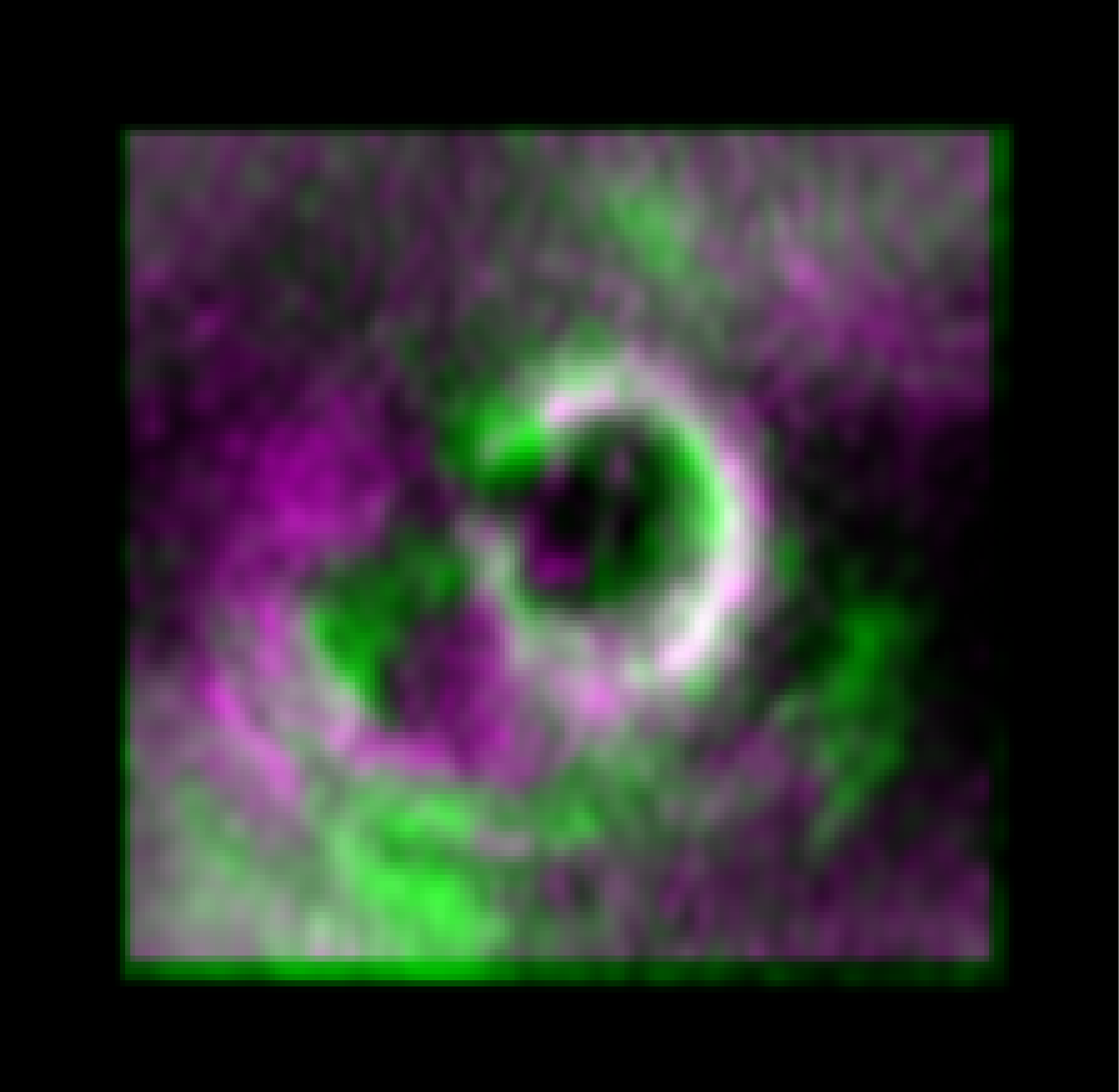}} &
        {\includegraphics[width=0.12\textwidth]{figures/fig_before_Im_001_20240215_151507_3D-Im_007_20240215_151552_3D_Axial_0.pdf}} &
        {\includegraphics[width=0.12\textwidth]{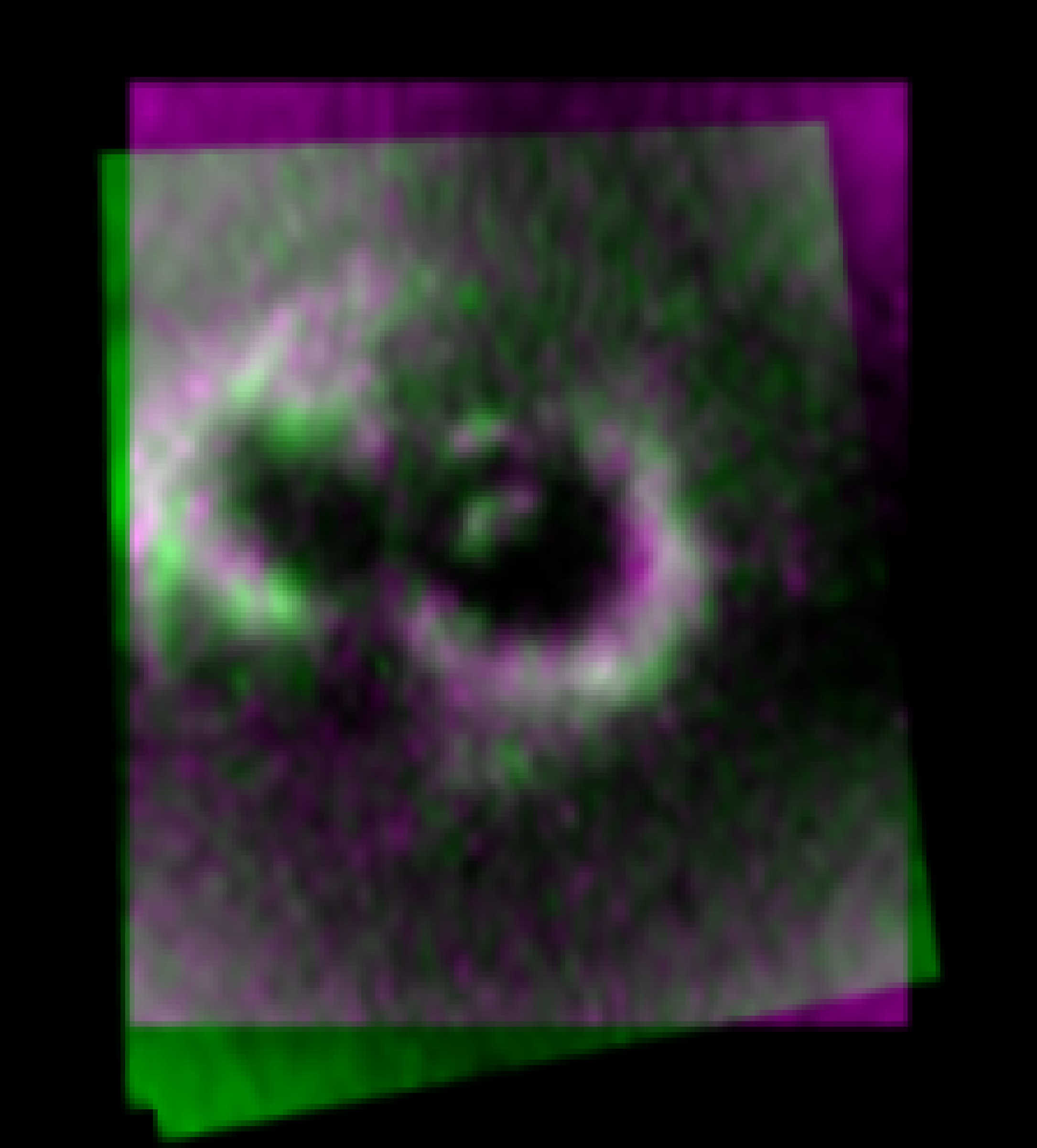}} \\
        \makecell[b]{Q1\vspace{20pt}} & 
        {\includegraphics[width=0.12\textwidth]{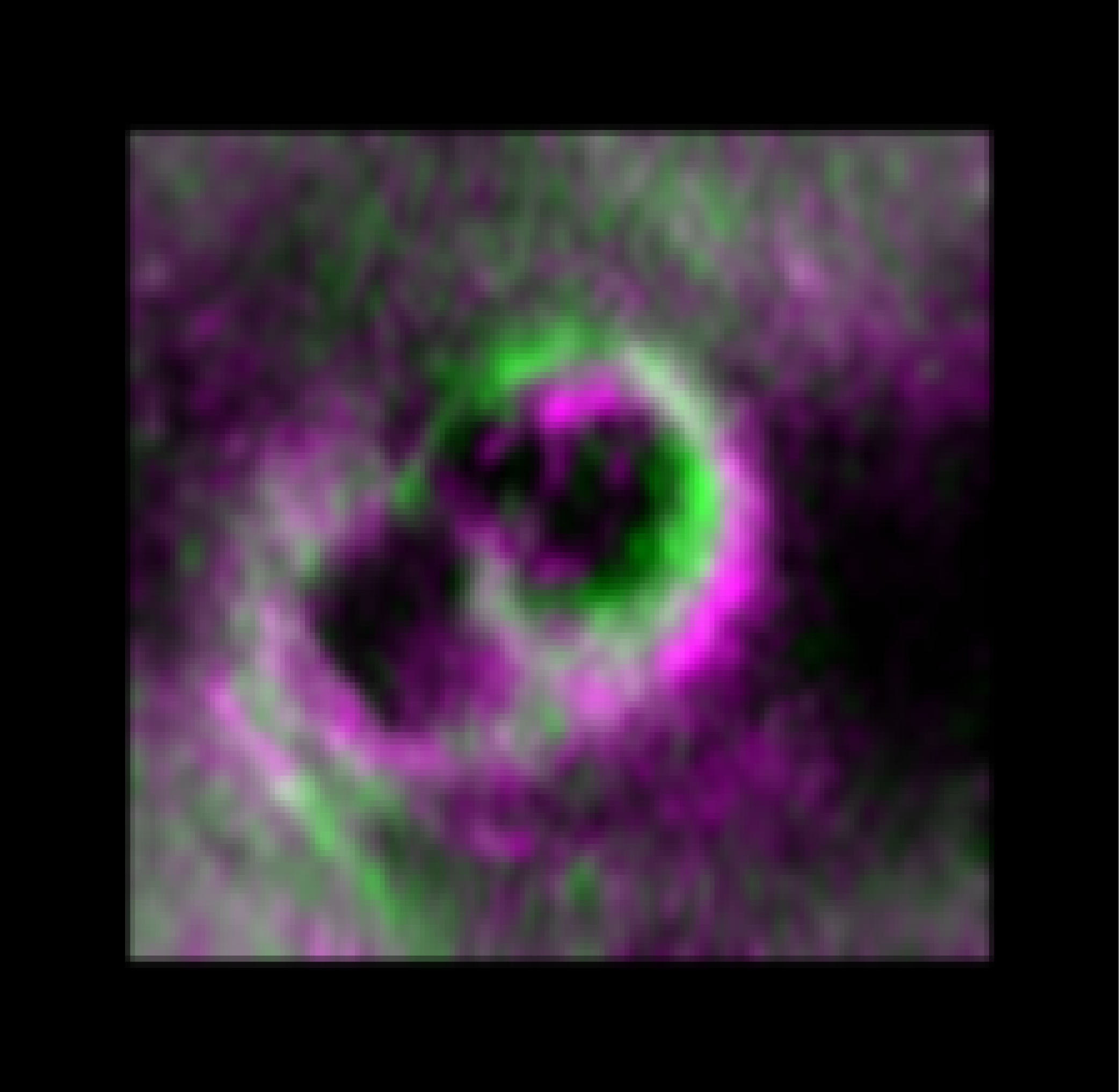}} &
        {\includegraphics[width=0.12\textwidth]{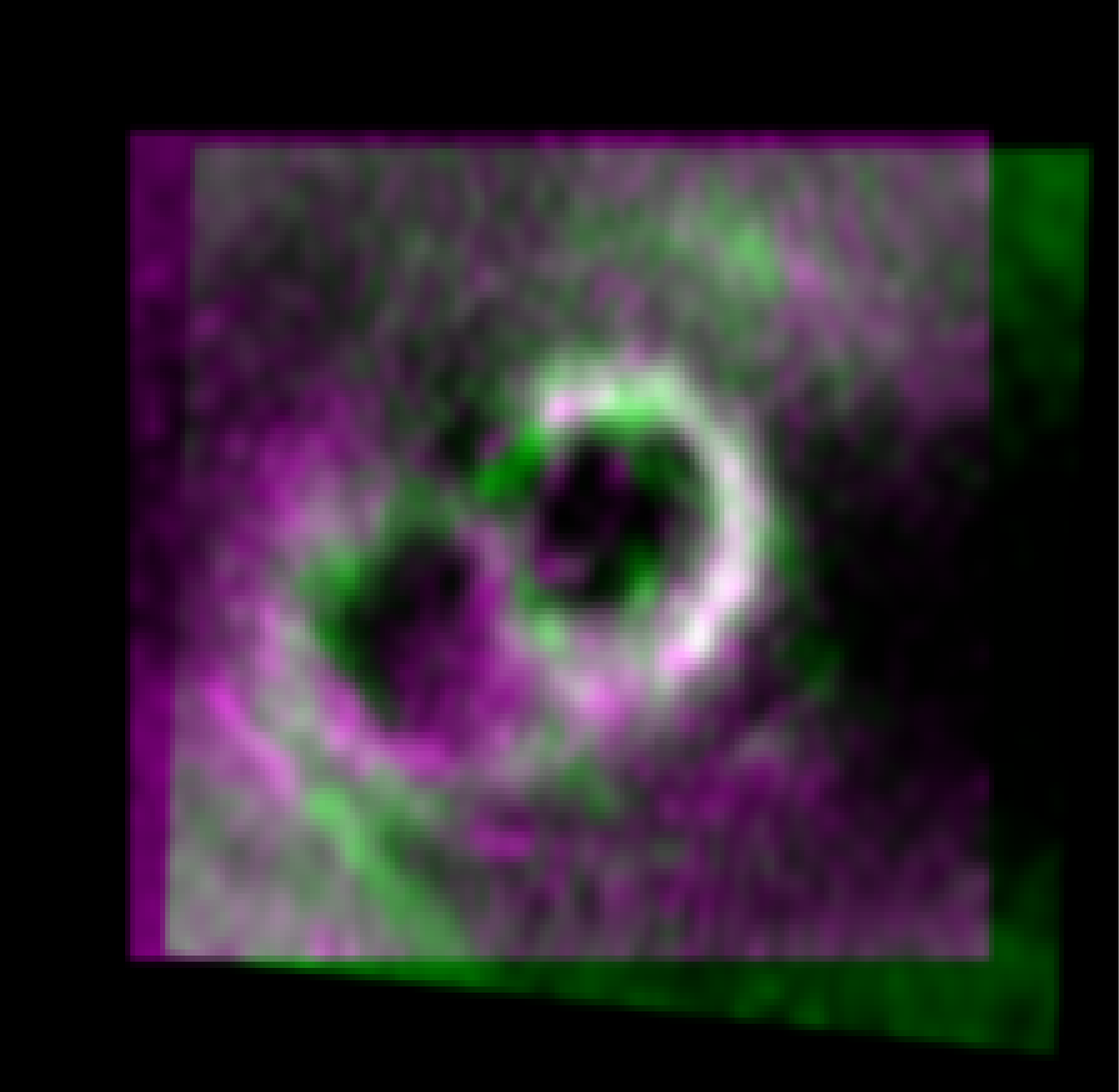}} &
        {\includegraphics[width=0.12\textwidth]{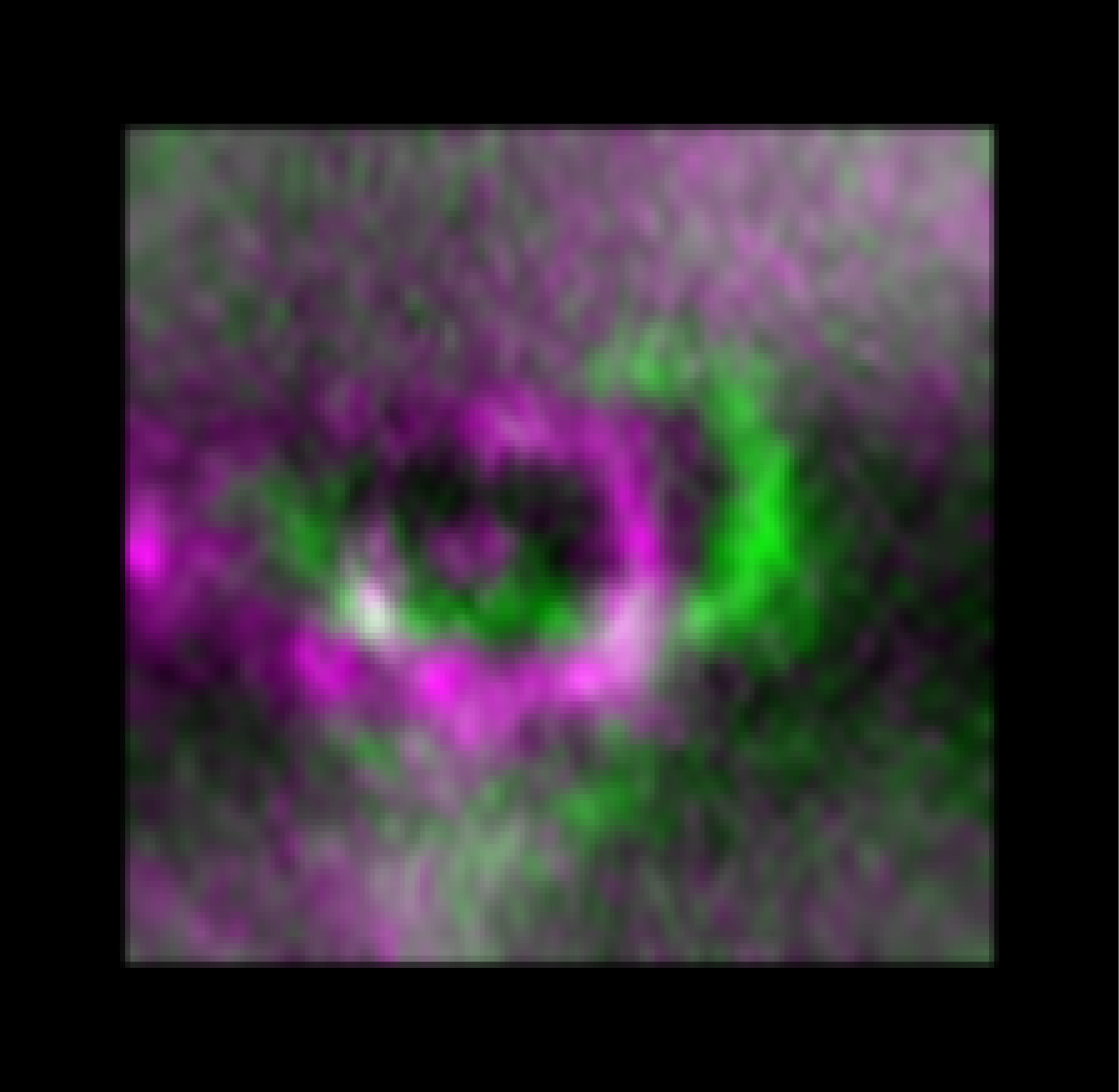}} &
        {\includegraphics[width=0.12\textwidth]{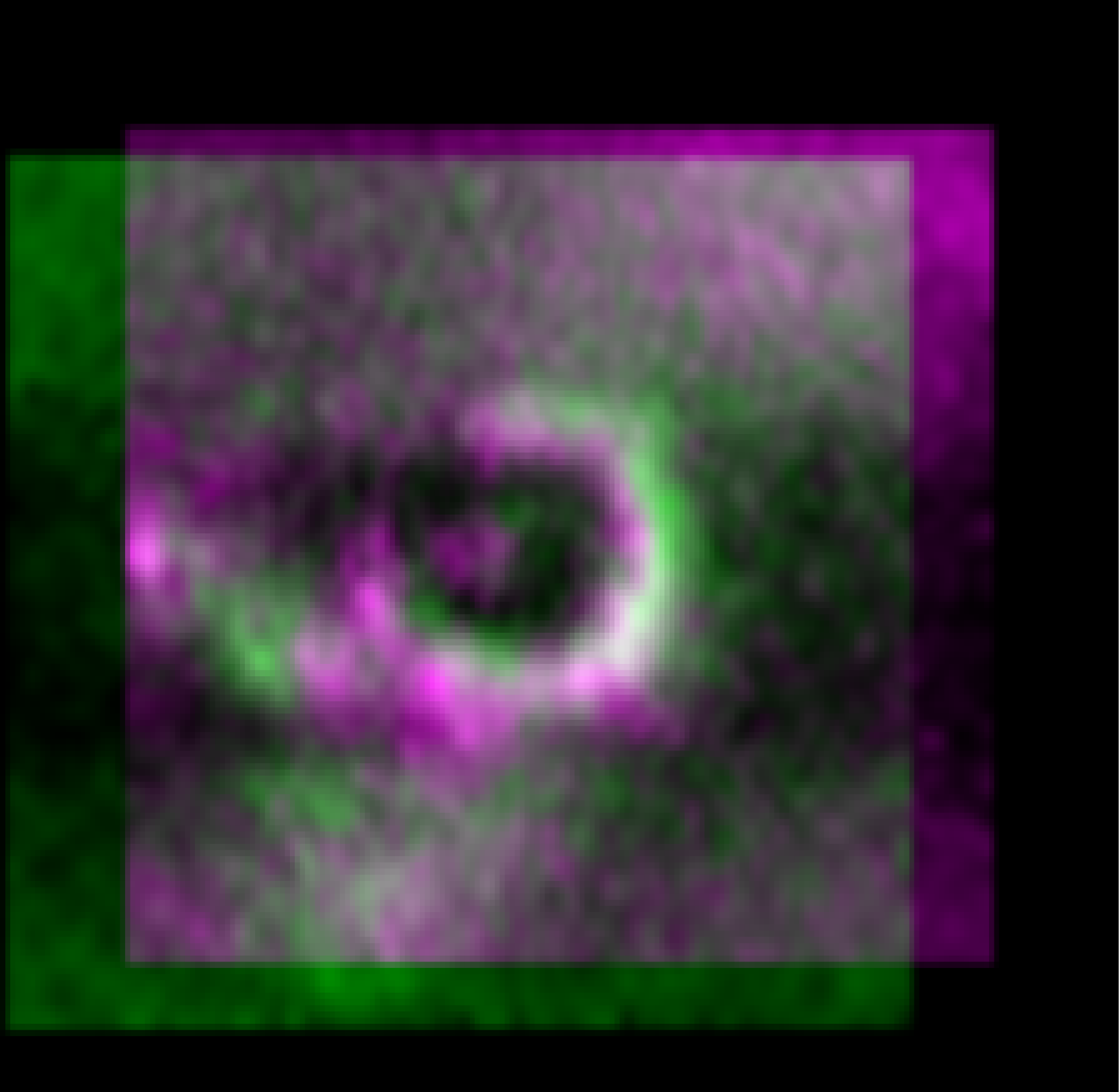}} & 
        {\includegraphics[width=0.12\textwidth]{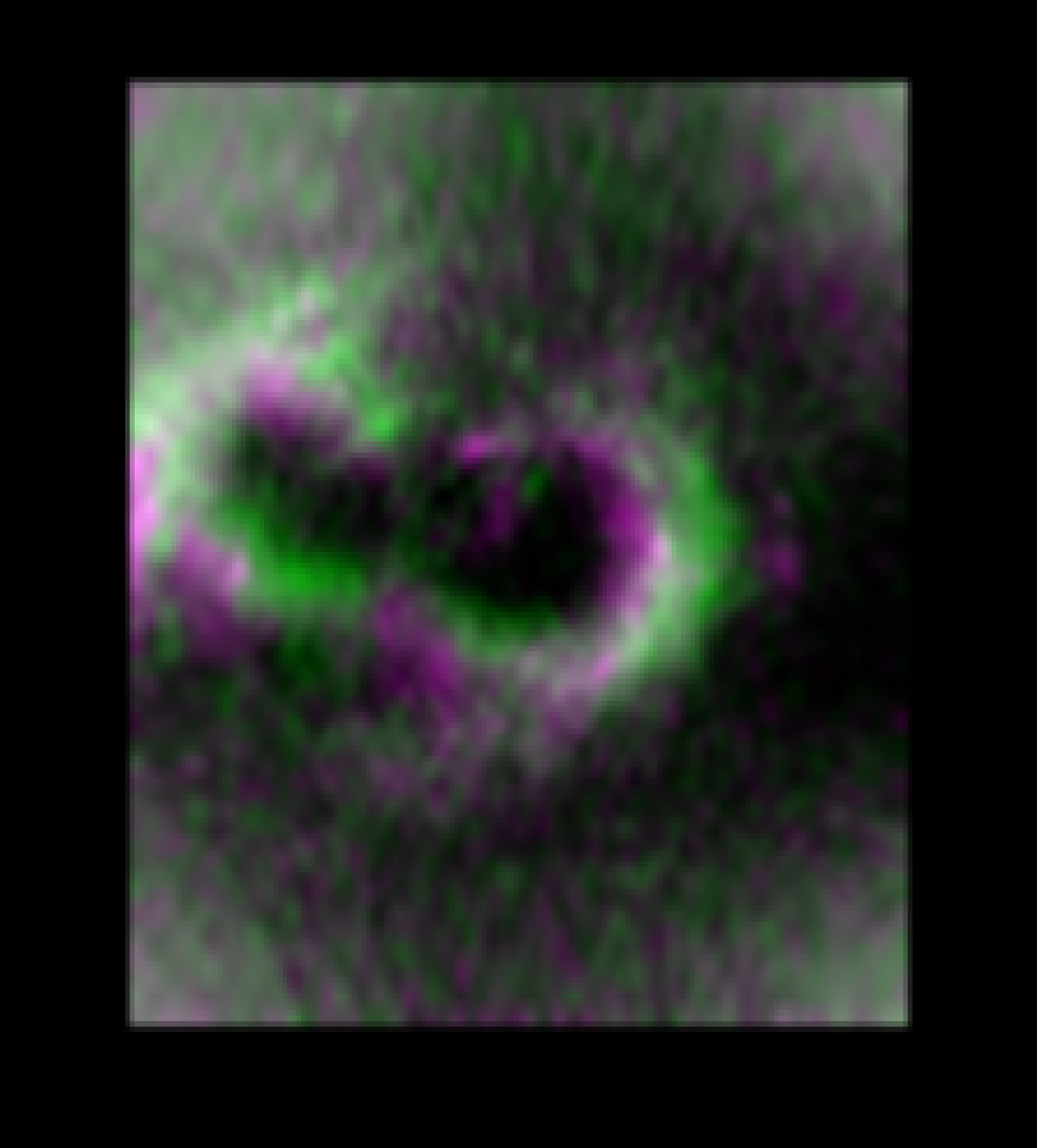}} &
        {\includegraphics[width=0.12\textwidth]{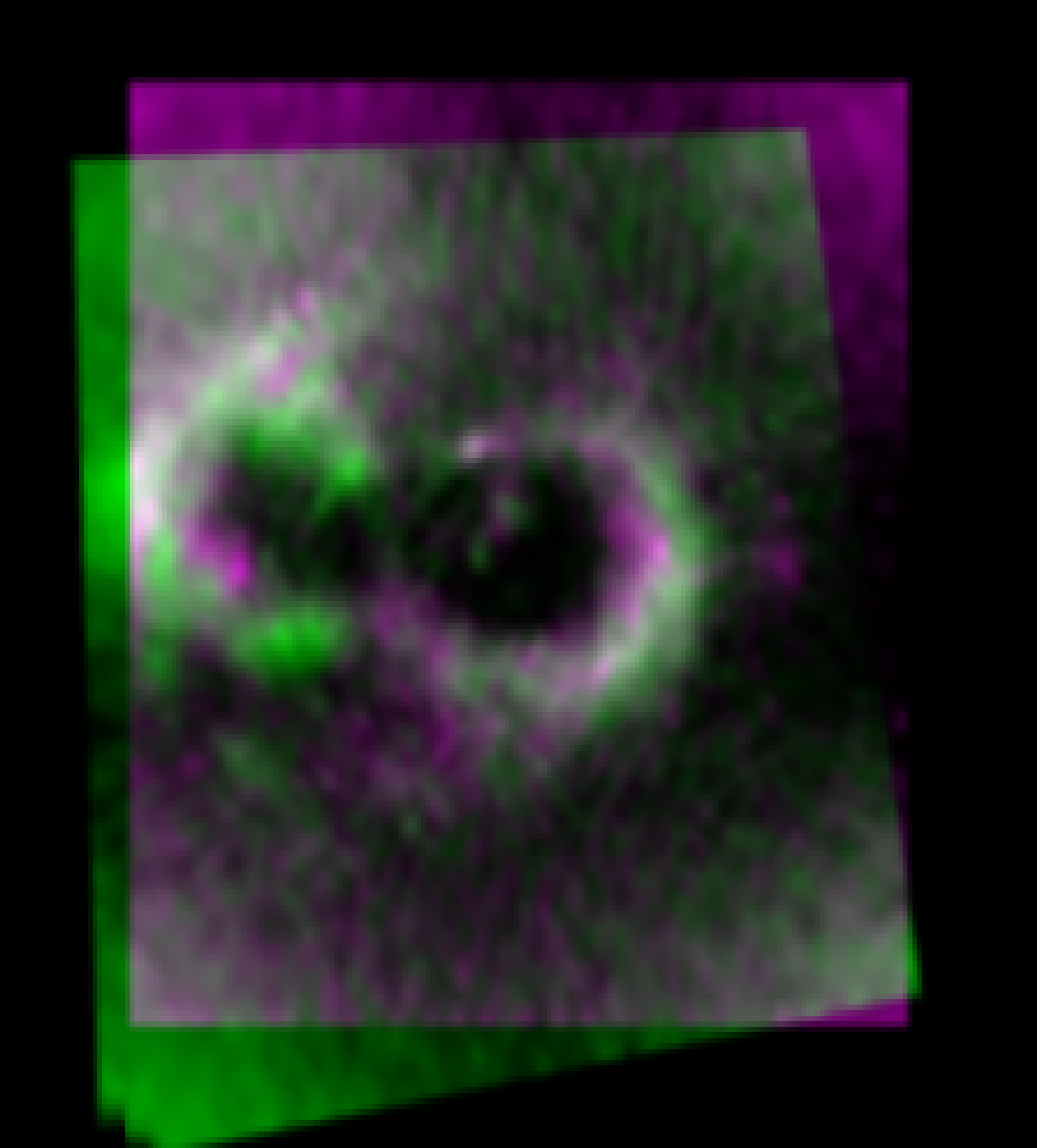}} \\ 
        \makecell[b]{Q2\vspace{20pt}} &
        {\includegraphics[width=0.12\textwidth]{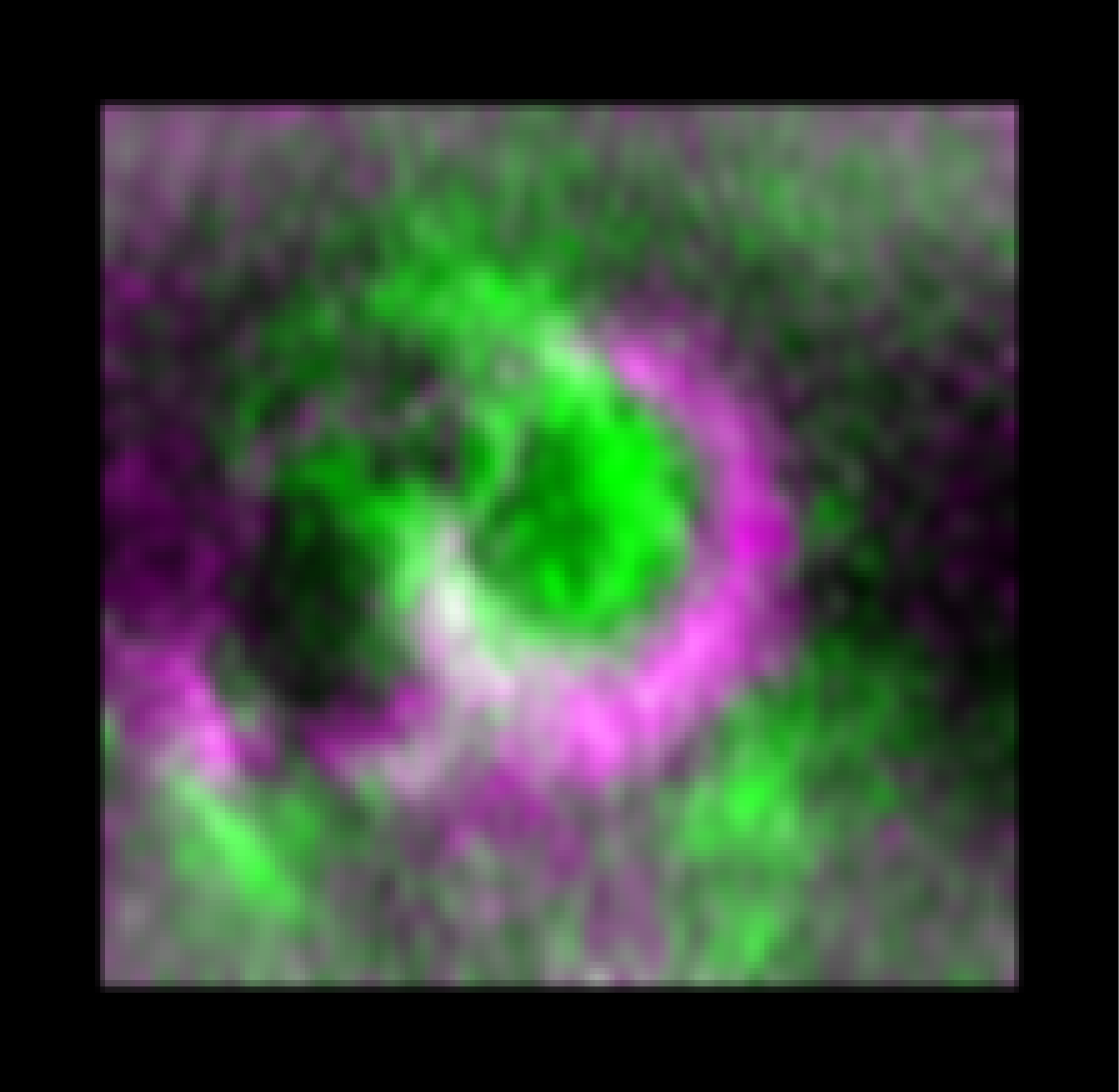}} &
        {\includegraphics[width=0.12\textwidth]{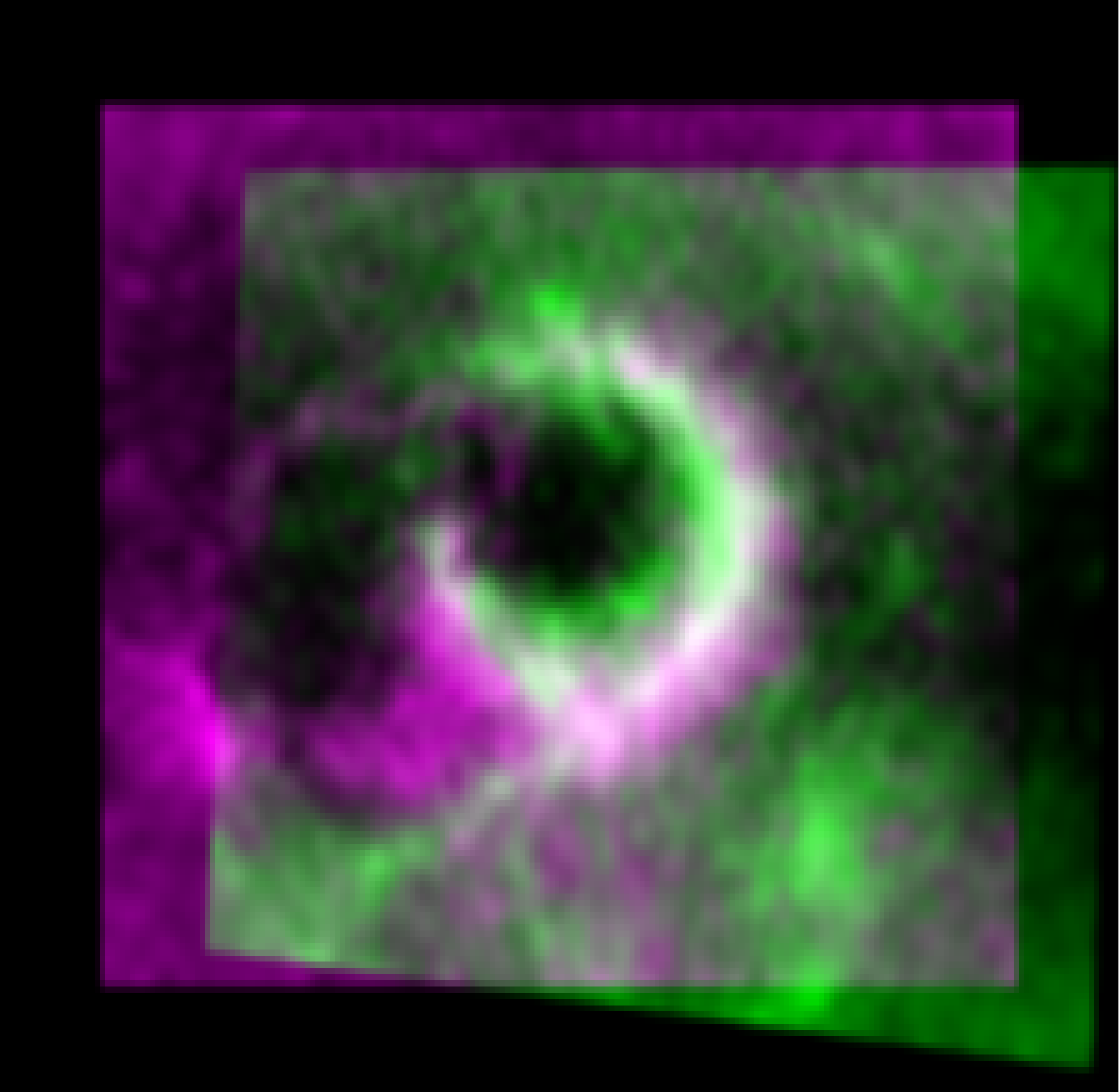}} &
        {\includegraphics[width=0.12\textwidth]{figures/fig_before_Im_016_20240119_144948_3D-Im_019_20240119_145037_3D_Axial_0.pdf}} &
        {\includegraphics[width=0.12\textwidth]{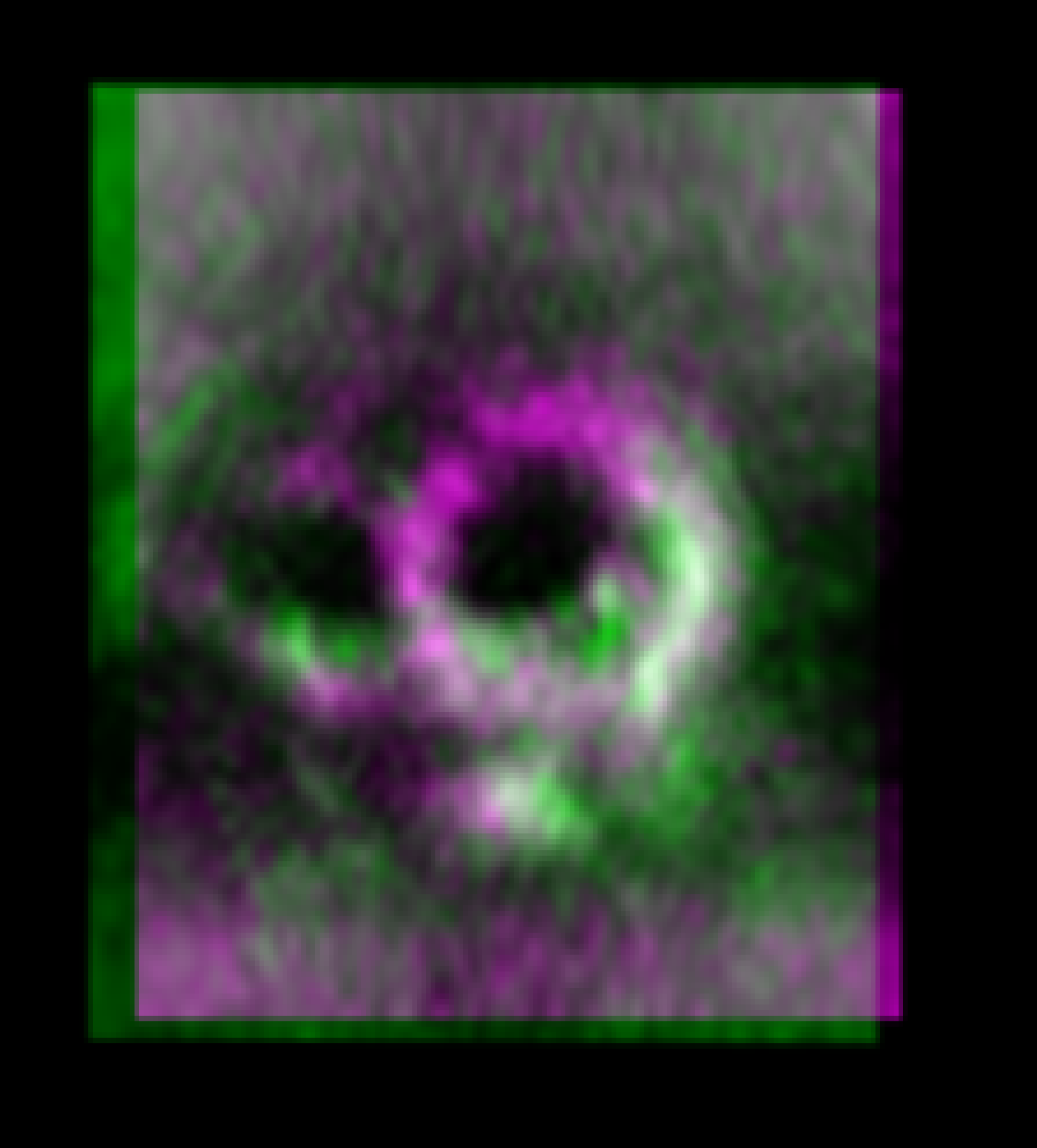}} &
        {\includegraphics[width=0.12\textwidth]{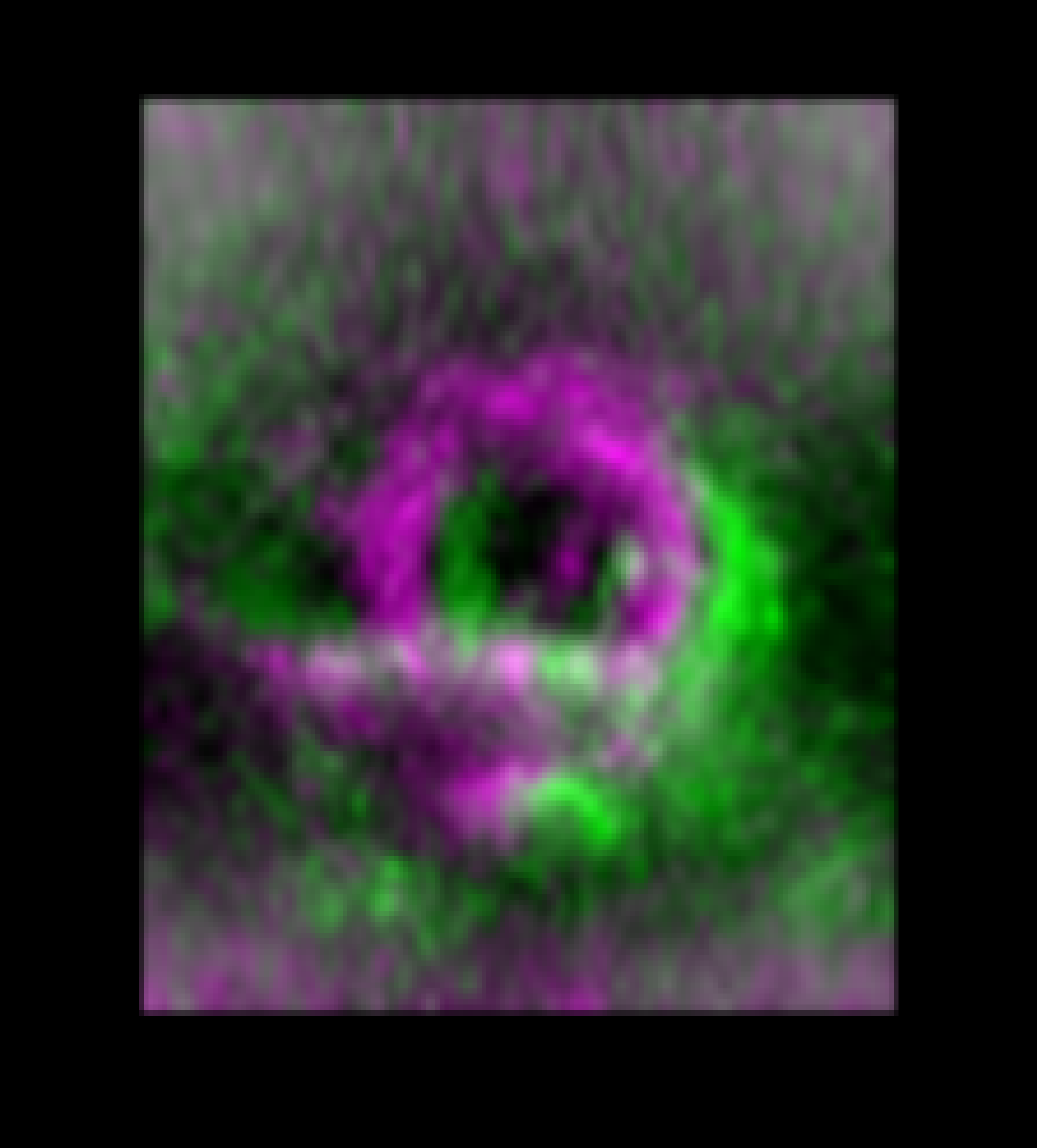}} &
        {\includegraphics[width=0.12\textwidth]{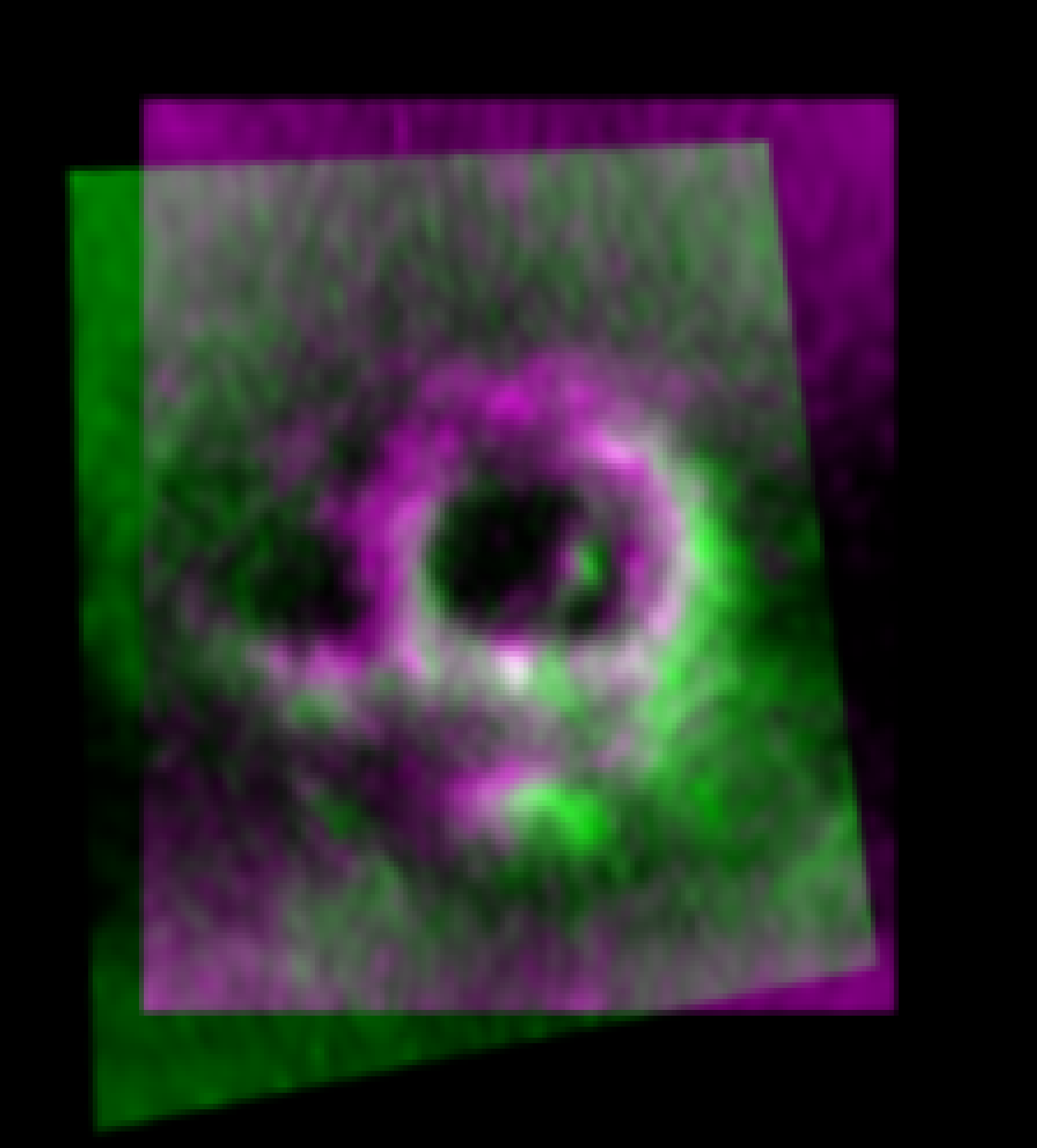}} \\
        \makecell[b]{Q3\vspace{20pt}} &
        {\includegraphics[width=0.12\textwidth]{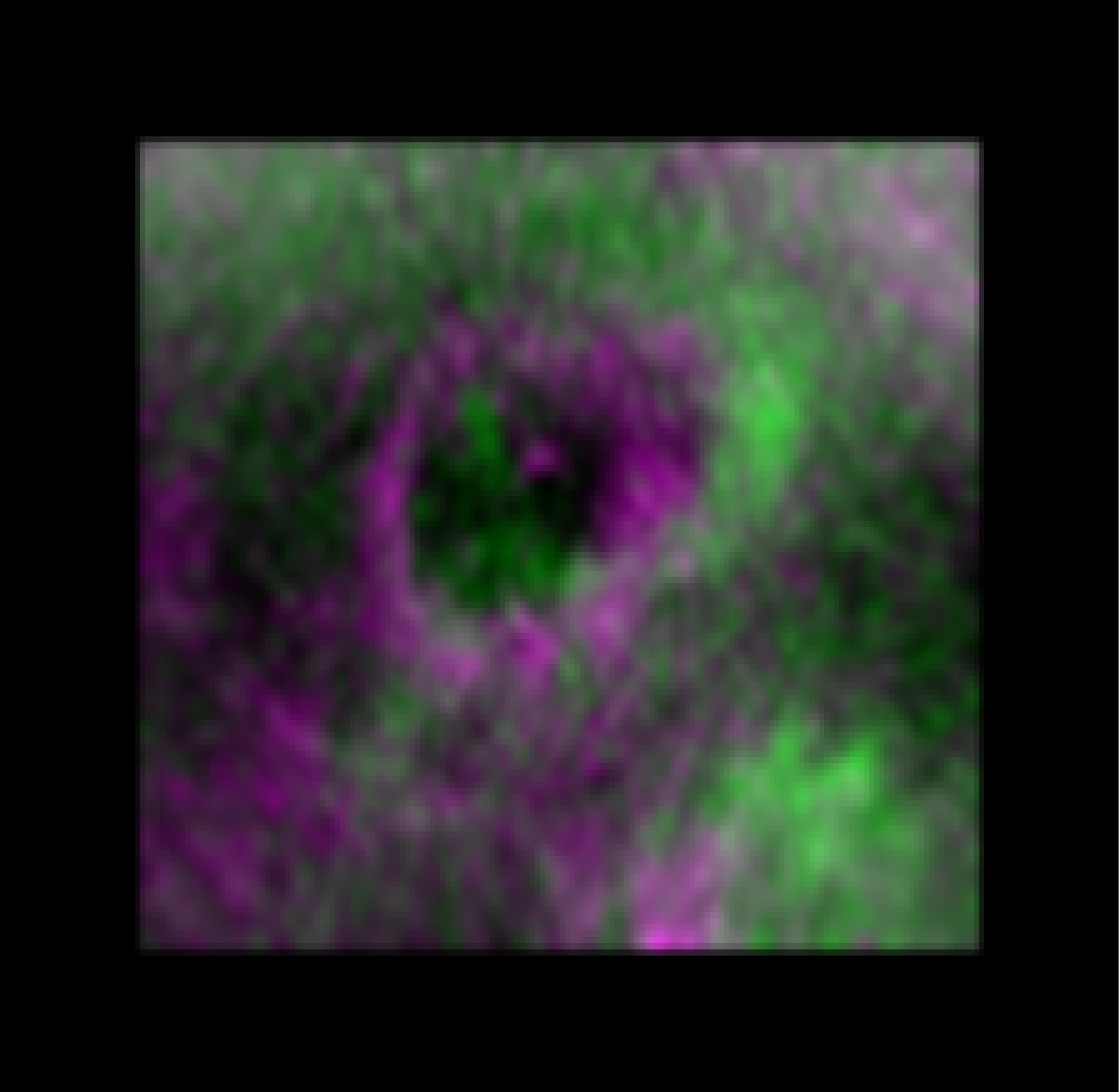}} &
        {\includegraphics[width=0.12\textwidth]{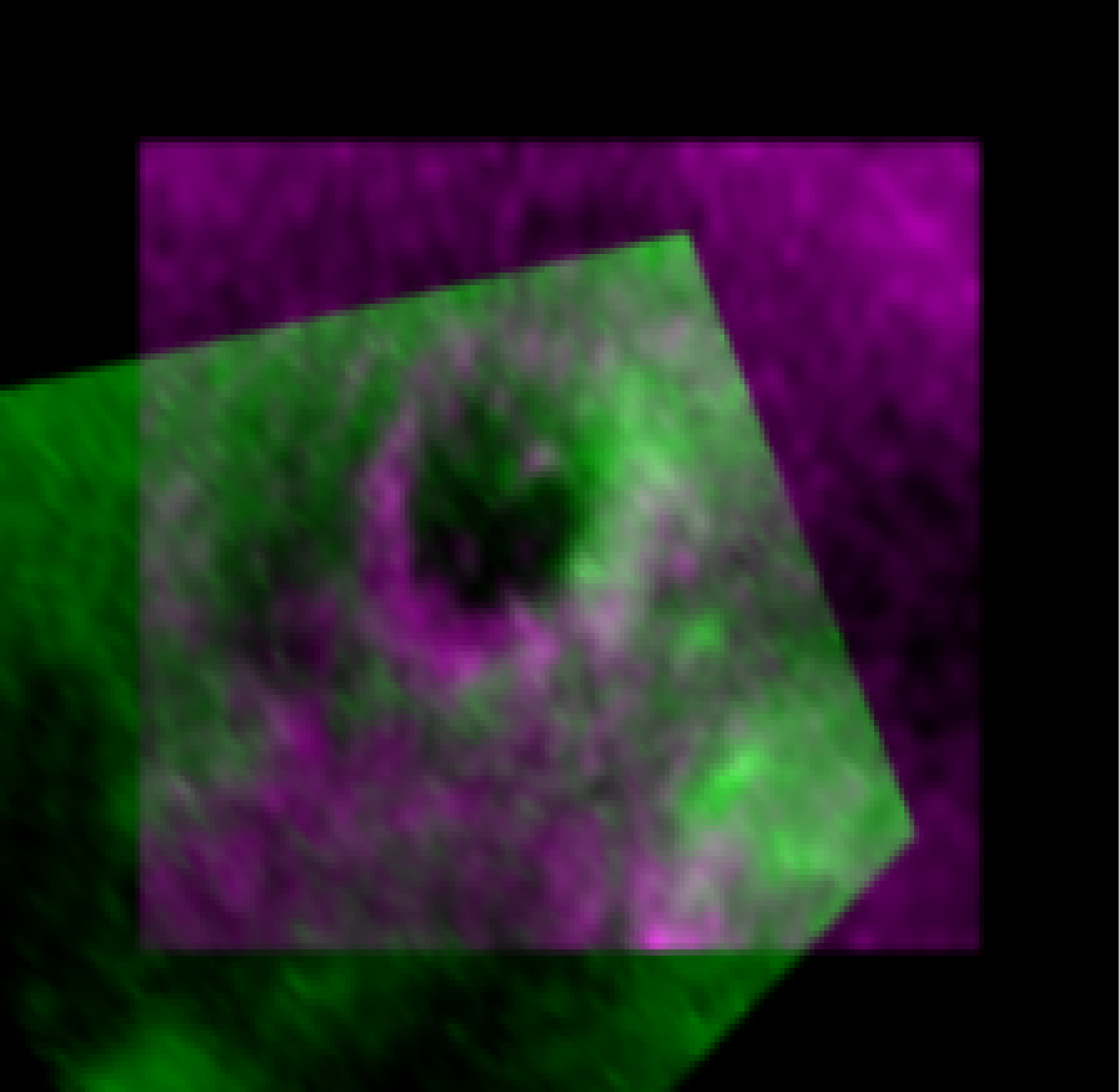}} &
        {\includegraphics[width=0.12\textwidth]{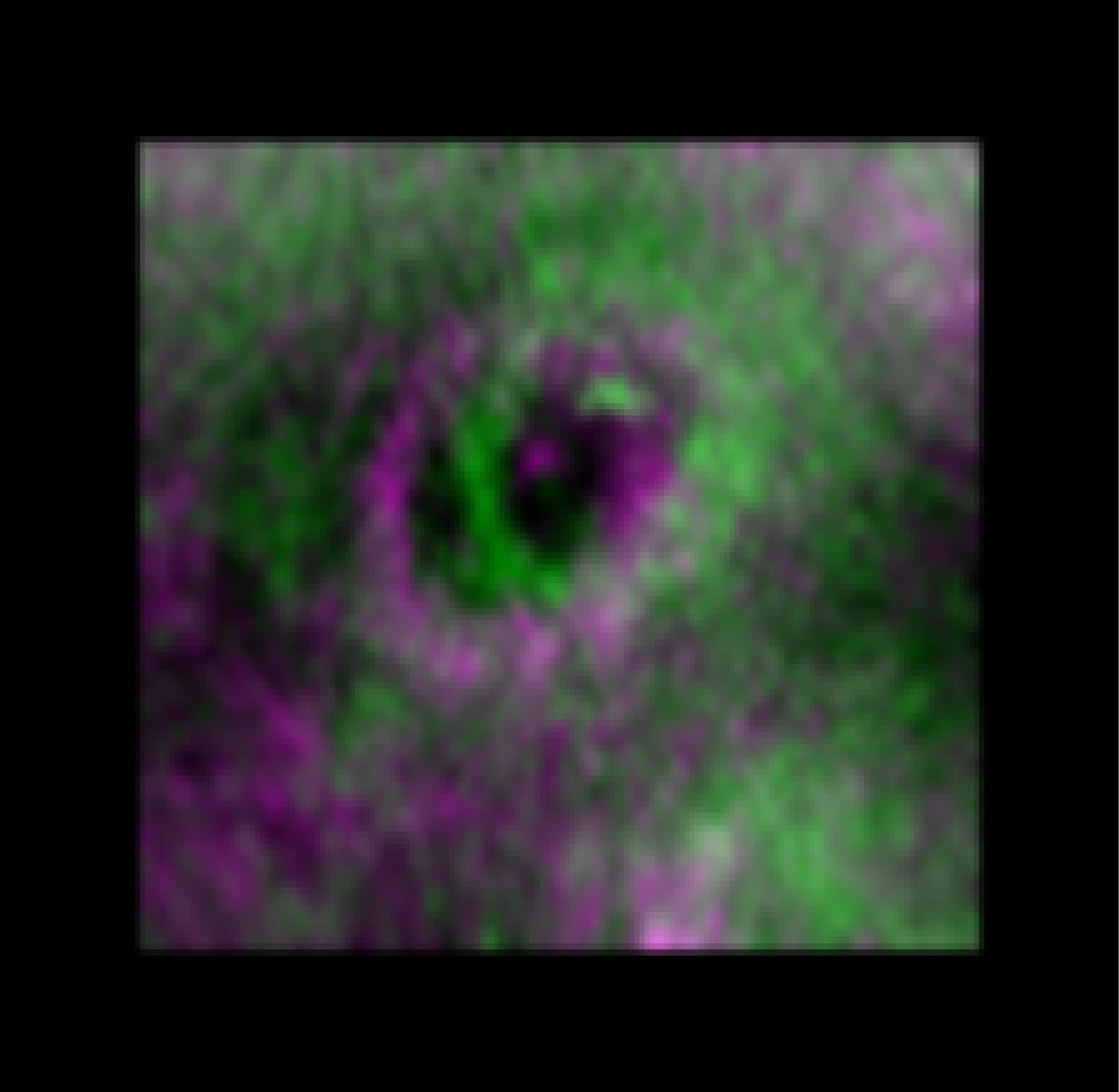}} &
        {\includegraphics[width=0.12\textwidth]{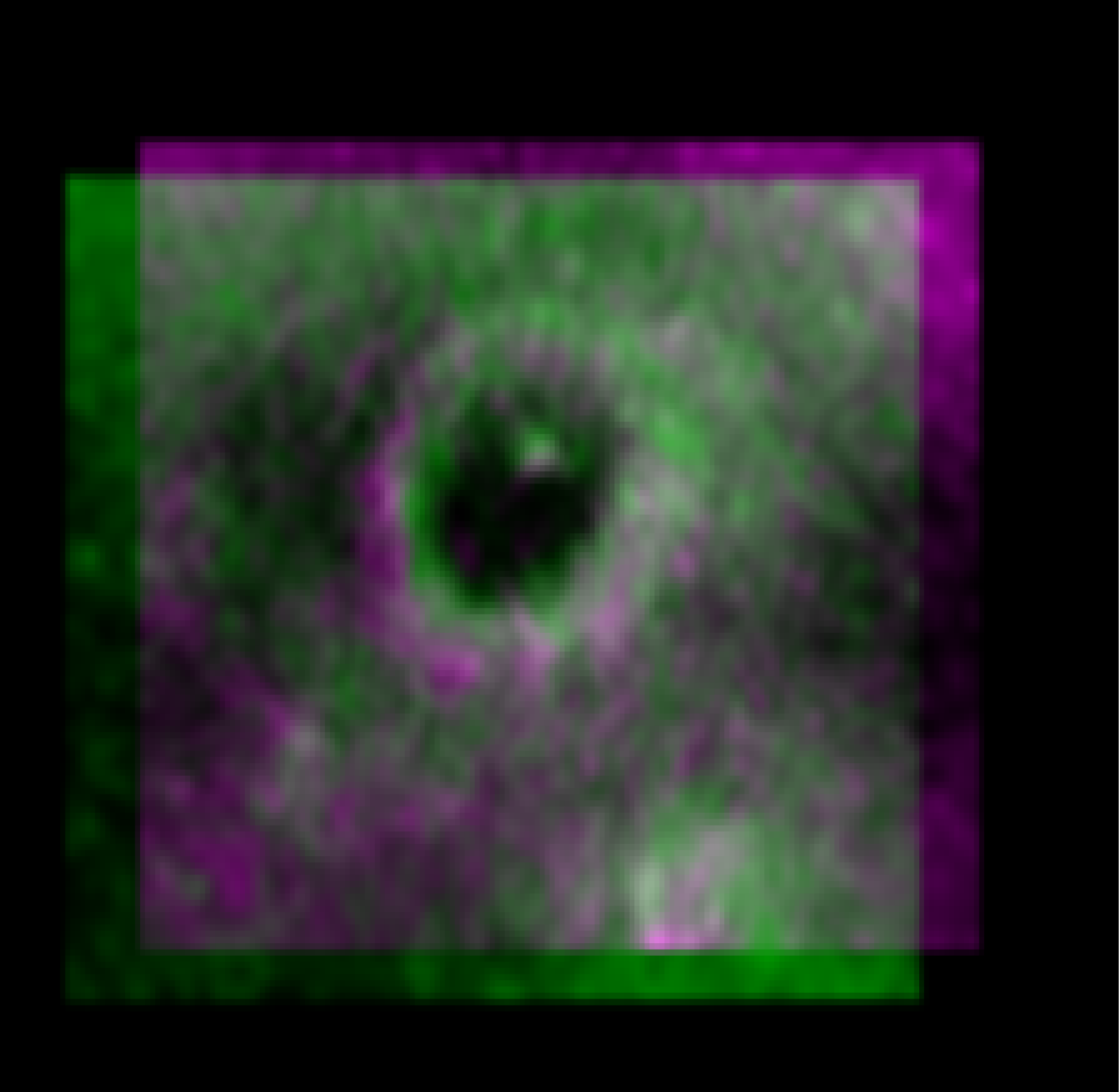}} &
        {\includegraphics[width=0.12\textwidth]{figures/fig_before_Im_013_20240221_143621_3D-Im_004_20240221_143057_3D_Axial_0.pdf}} &
        {\includegraphics[width=0.12\textwidth]{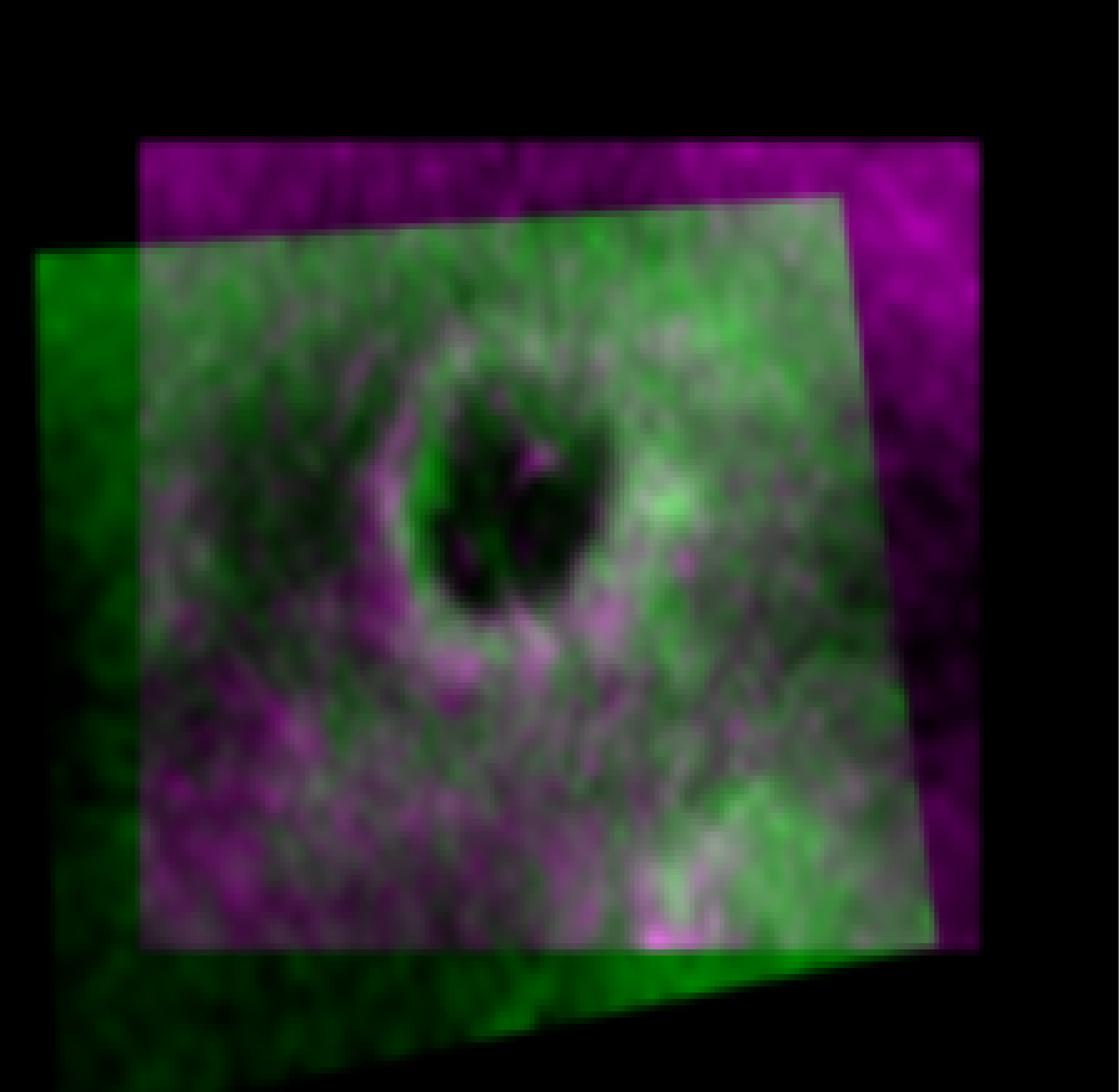}} \\
        \makecell[b]{Max\vspace{20pt}} & 
        {\includegraphics[width=0.12\textwidth]{figures/fig_before_Im_003_20240215_151511_3D-Im_010_20240215_151637_3D_Axial_0.pdf}} &
        {\includegraphics[width=0.12\textwidth]{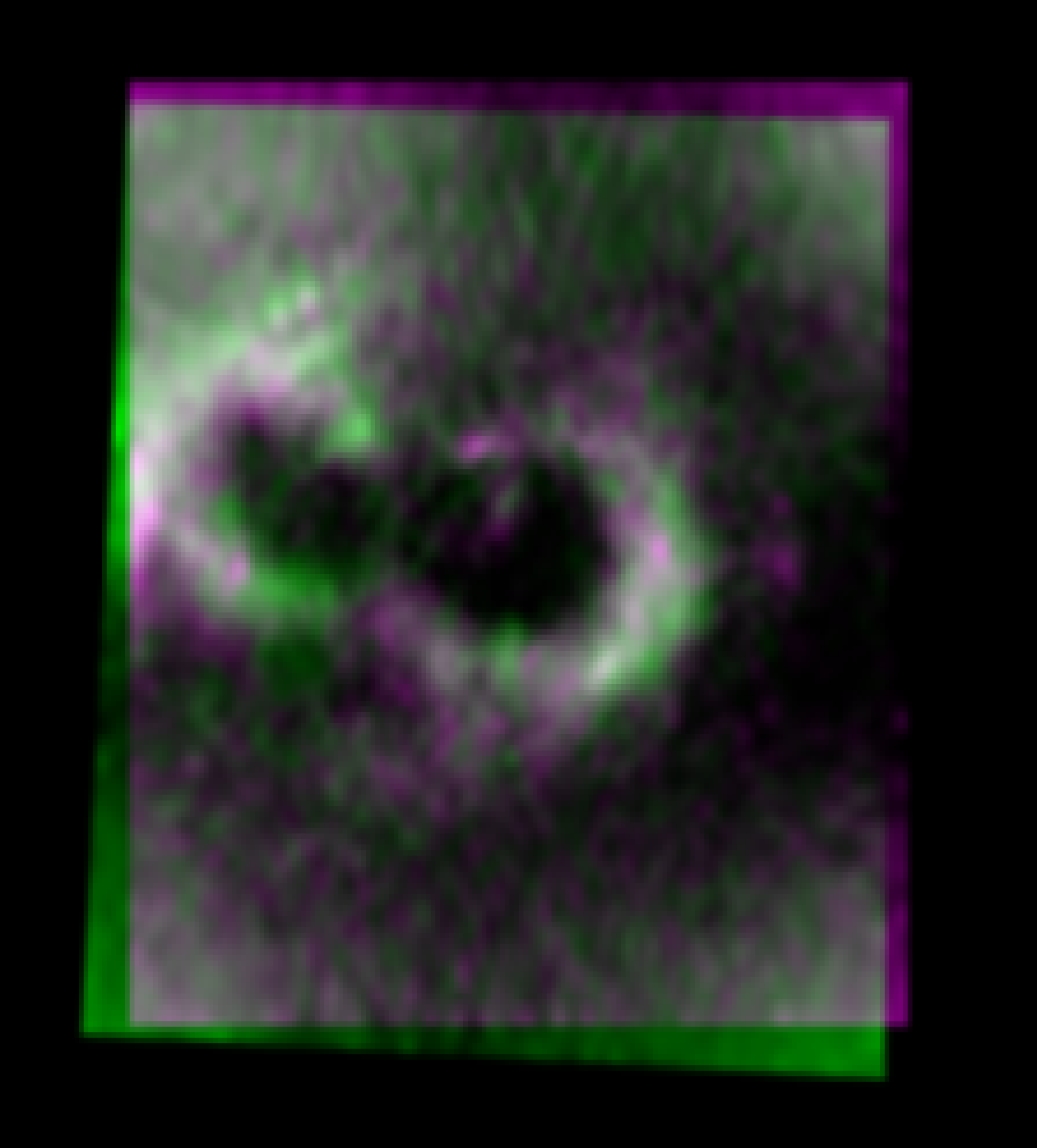}} &
        {\includegraphics[width=0.12\textwidth]{figures/fig_before_Im_013_20240221_143621_3D-Im_002_20240221_143029_3D_Axial_0.pdf}} &
        {\includegraphics[width=0.12\textwidth]{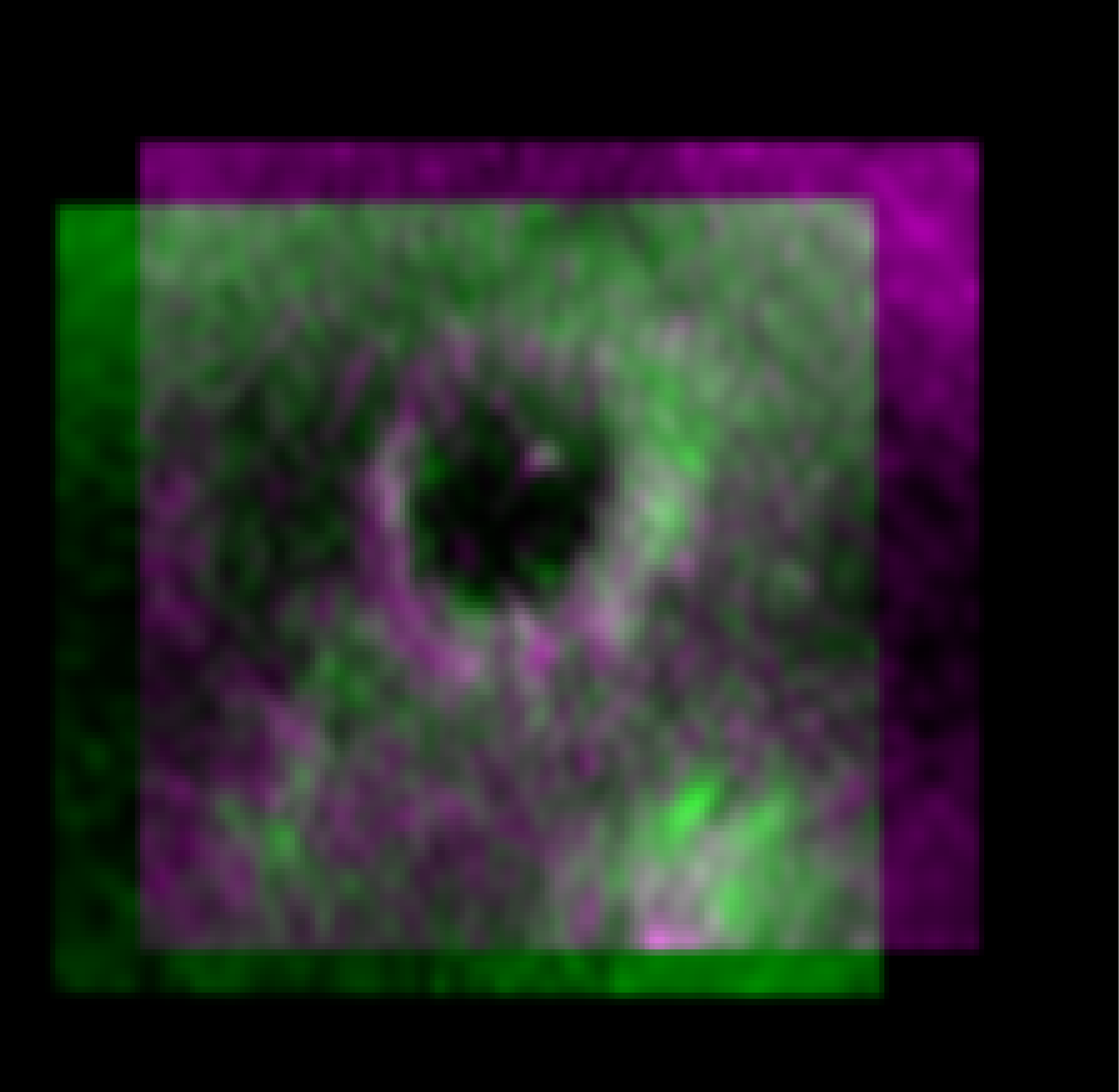}} &
        {\includegraphics[width=0.12\textwidth]{figures/fig_before_Im_004_20240123_153322_3D-Im_010_20240123_153651_3D_Axial_0.pdf}} &
        {\includegraphics[width=0.12\textwidth]{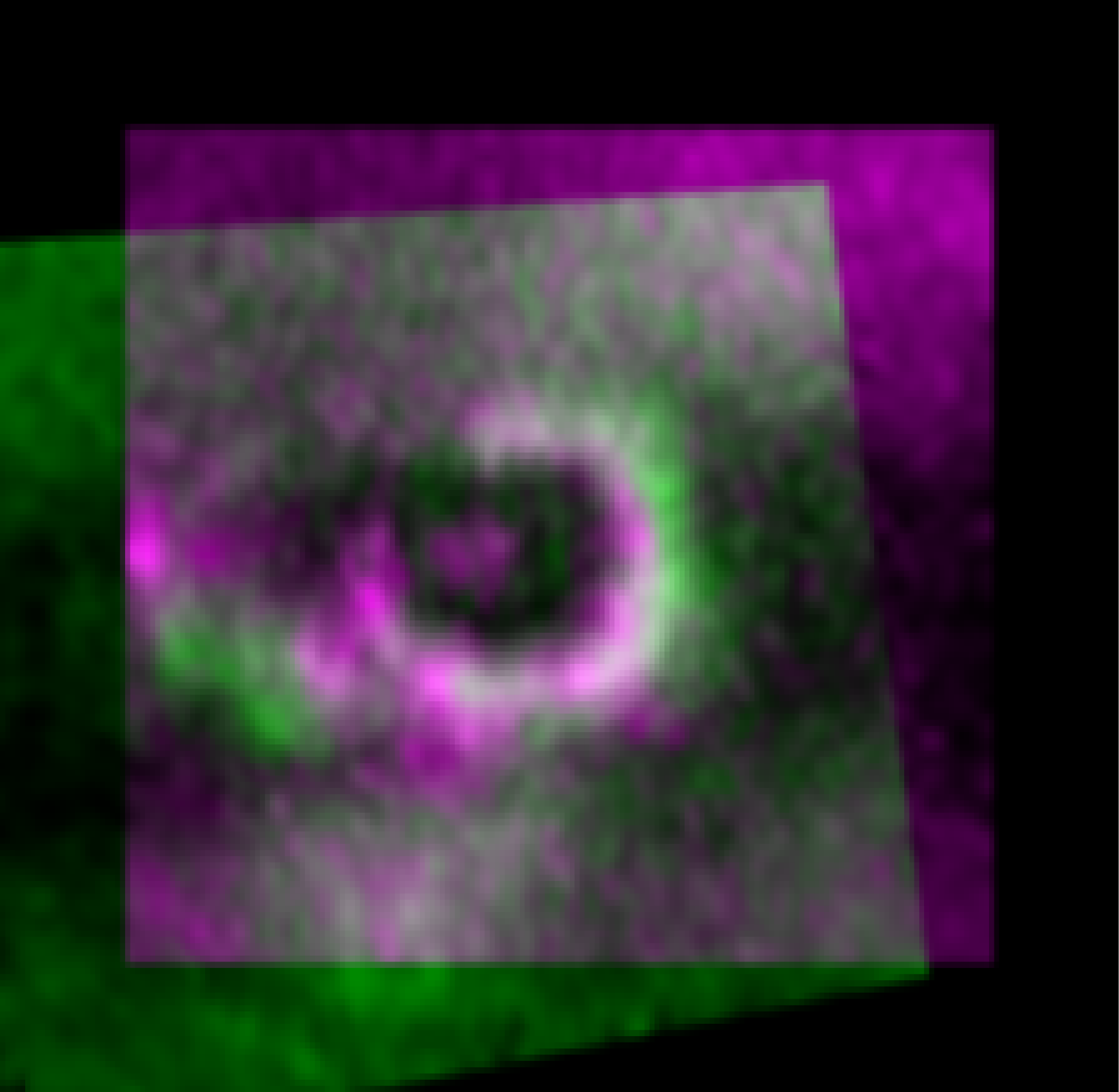}} \\
    \end{tabular}
    \caption{Mask-based rigid (PF and EX) registration results of the ED frame of image pairs for percentile DSC difference values of the axial view. Two consecutive columns show images before and after the registration for each category. The source and target images are shown in green and purple colors.}
    \label{fig:fig_perc_images_axial_rigid_mask}
\end{figure*}

\begin{figure*}[!ht]
    \centering
    \begin{tabular}{SSSSSSS}
         & \multicolumn{2}{c}{\scriptsize PF (CPU)} & \multicolumn{2}{c}{\scriptsize PF (GPU)} & \multicolumn{2}{c}{\scriptsize EX (CPU)} \\
         & Before & After & Before & After & Before & After \\
         \makecell[b]{Min\vspace{15pt}} & 
        {\includegraphics[width=0.12\textwidth]{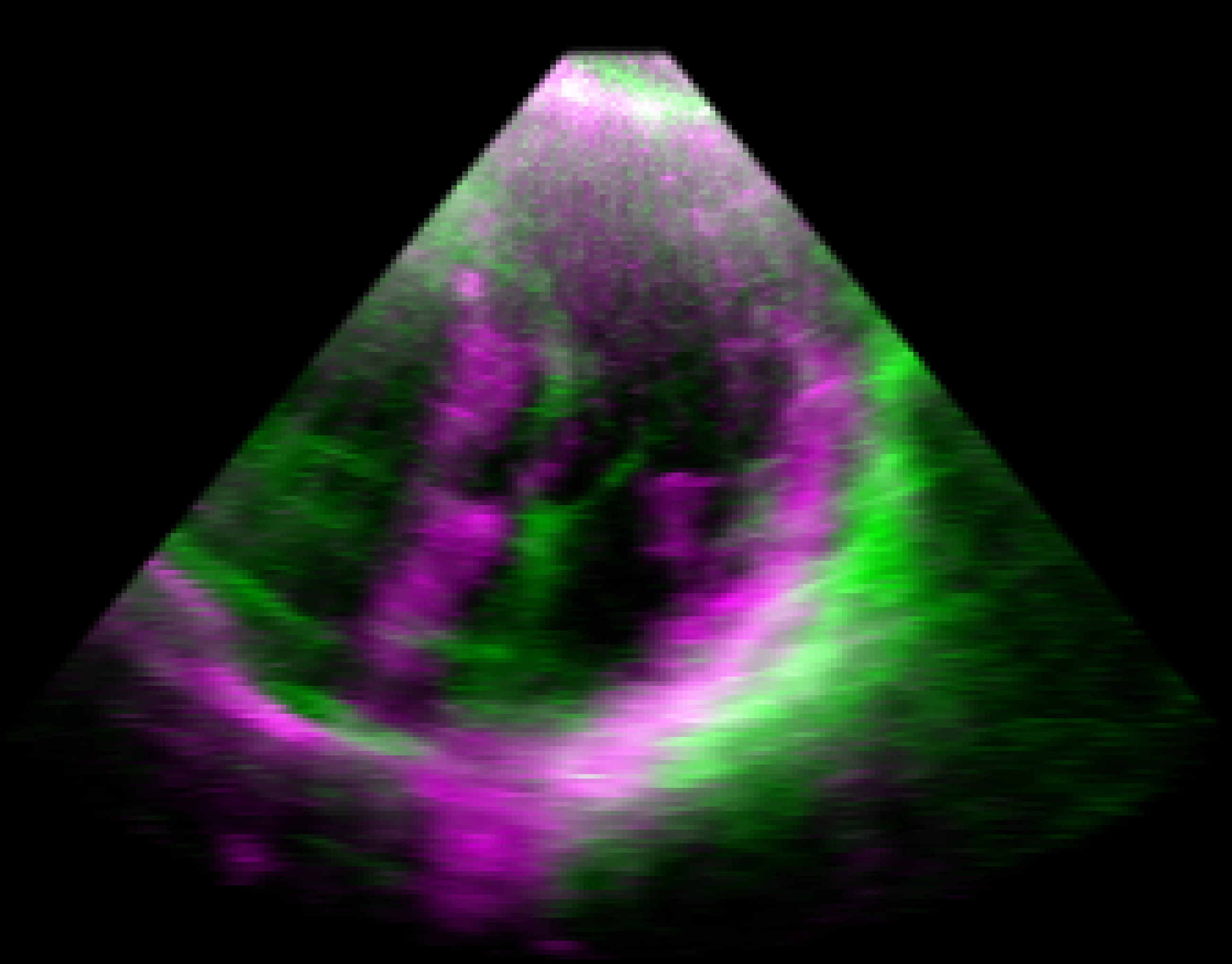}} &
        {\includegraphics[width=0.12\textwidth]{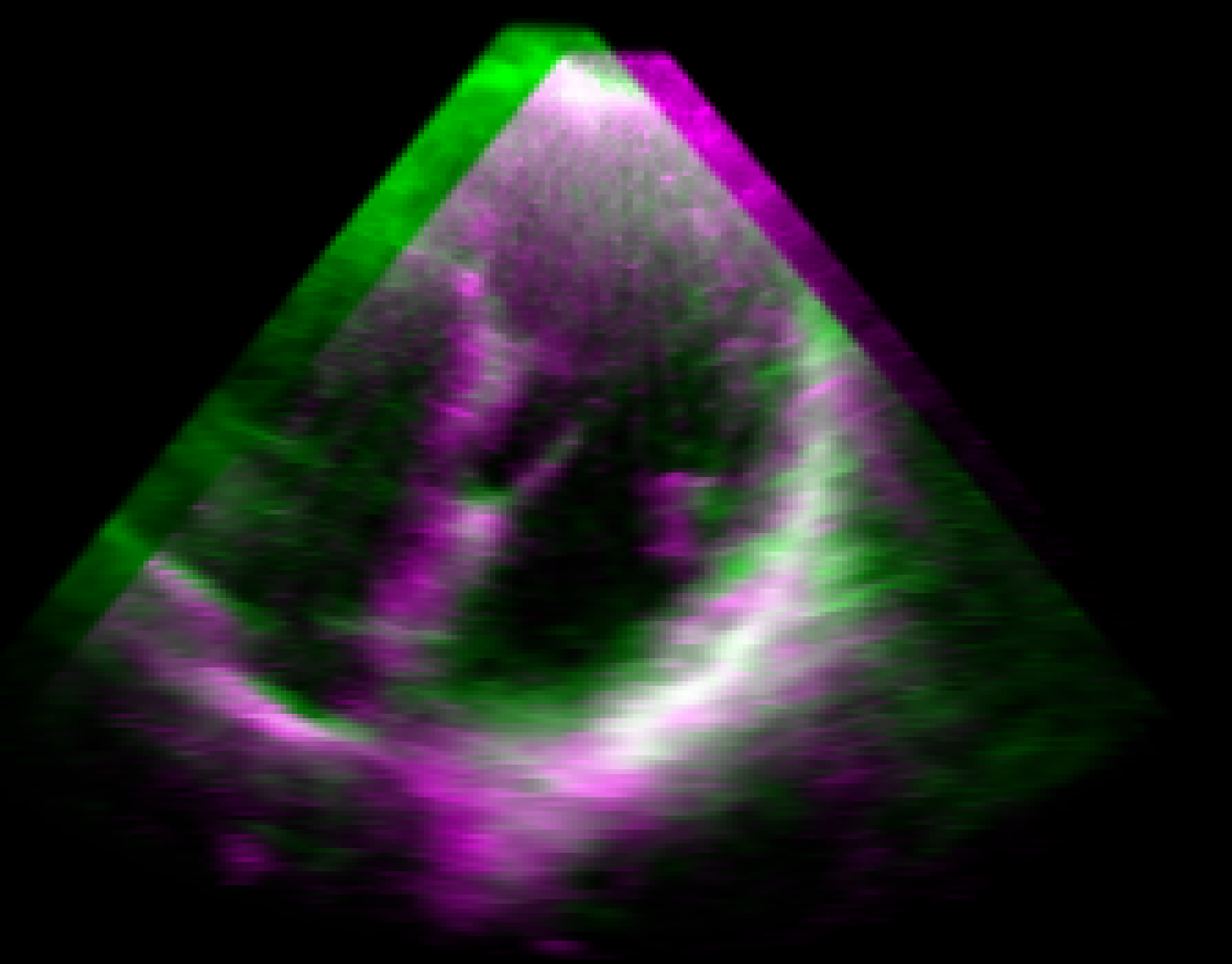}} &
        {\includegraphics[width=0.12\textwidth]{figures/fig_before_Im_016_20230817_104344_3D-Im_012_20230817_104241_3D_Coronal_0.pdf}} &
        {\includegraphics[width=0.12\textwidth]{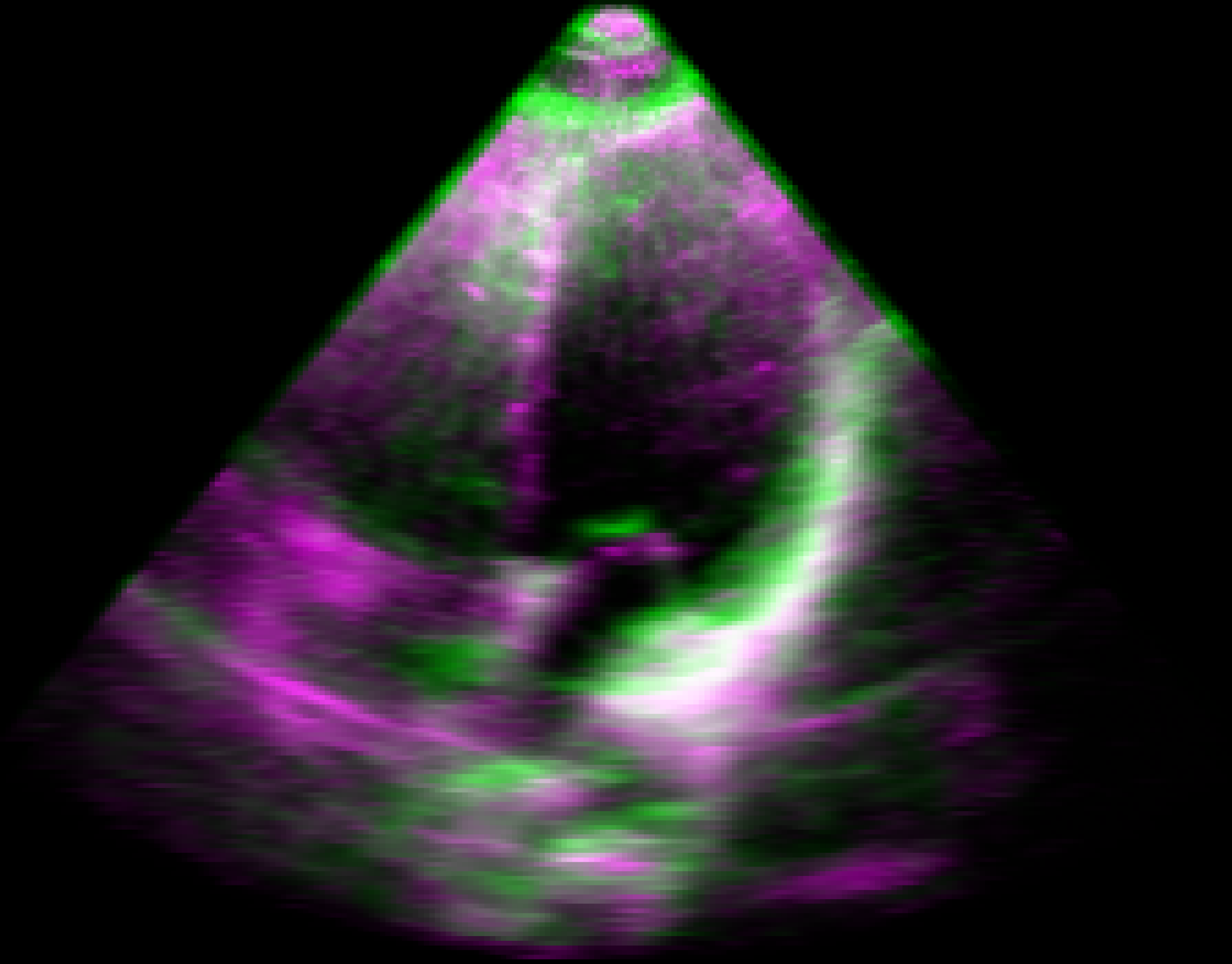}} &
        {\includegraphics[width=0.12\textwidth]{figures/fig_before_Im_001_20240215_151507_3D-Im_007_20240215_151552_3D_Coronal_0.pdf}} &
        {\includegraphics[width=0.12\textwidth]{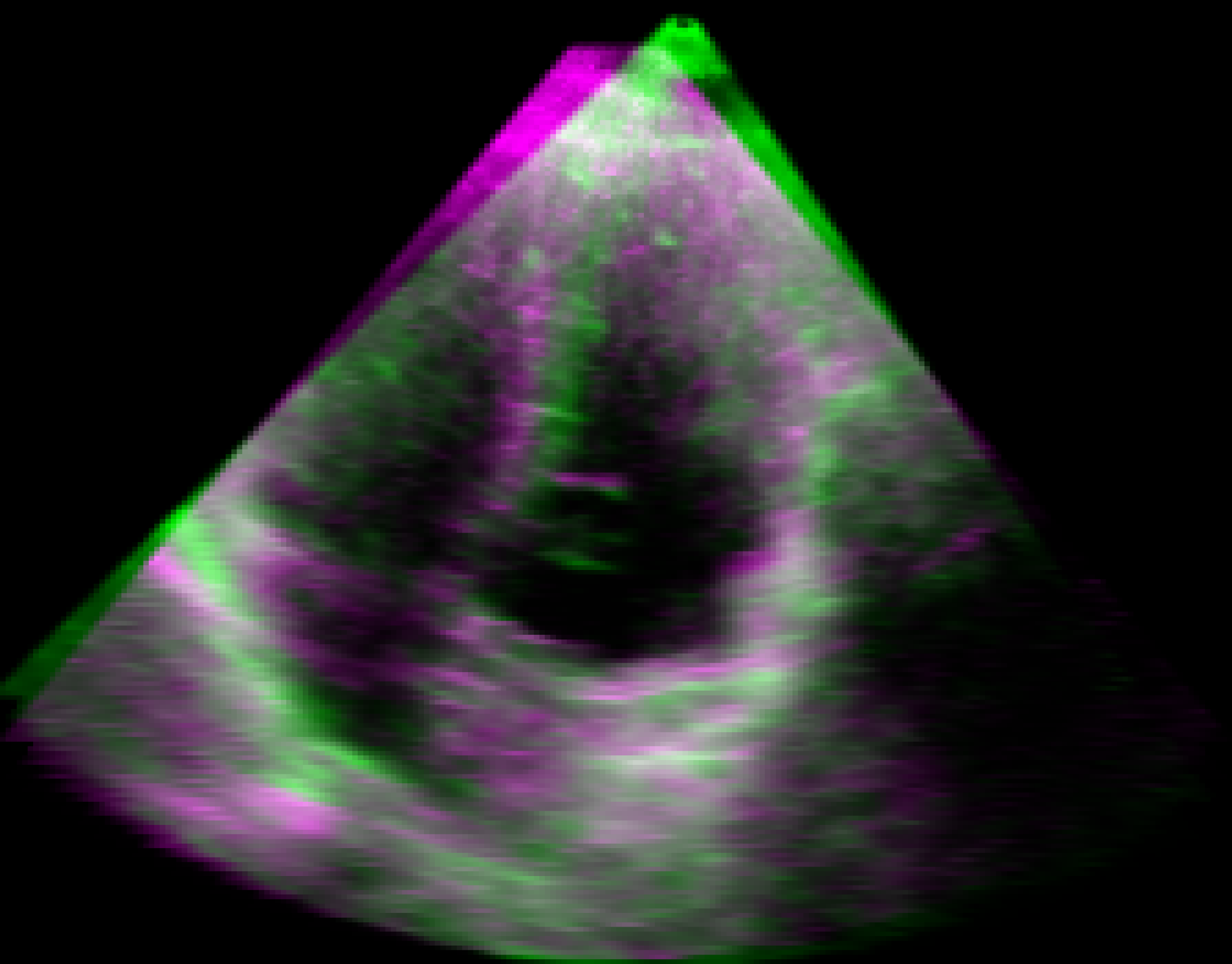}} \\
        \makecell[b]{Q1\vspace{15pt}} & 
        {\includegraphics[width=0.12\textwidth]{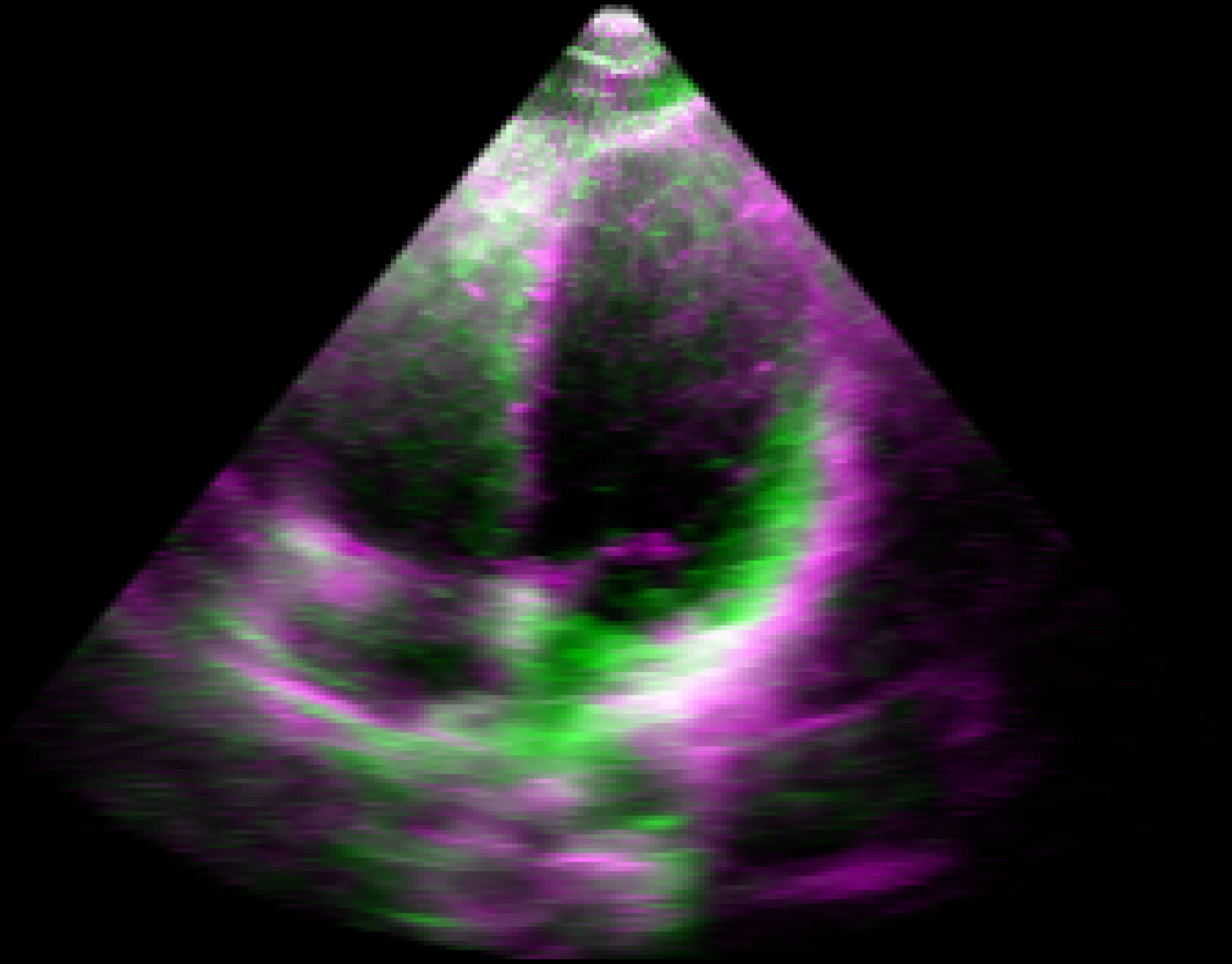}} &
        {\includegraphics[width=0.12\textwidth]{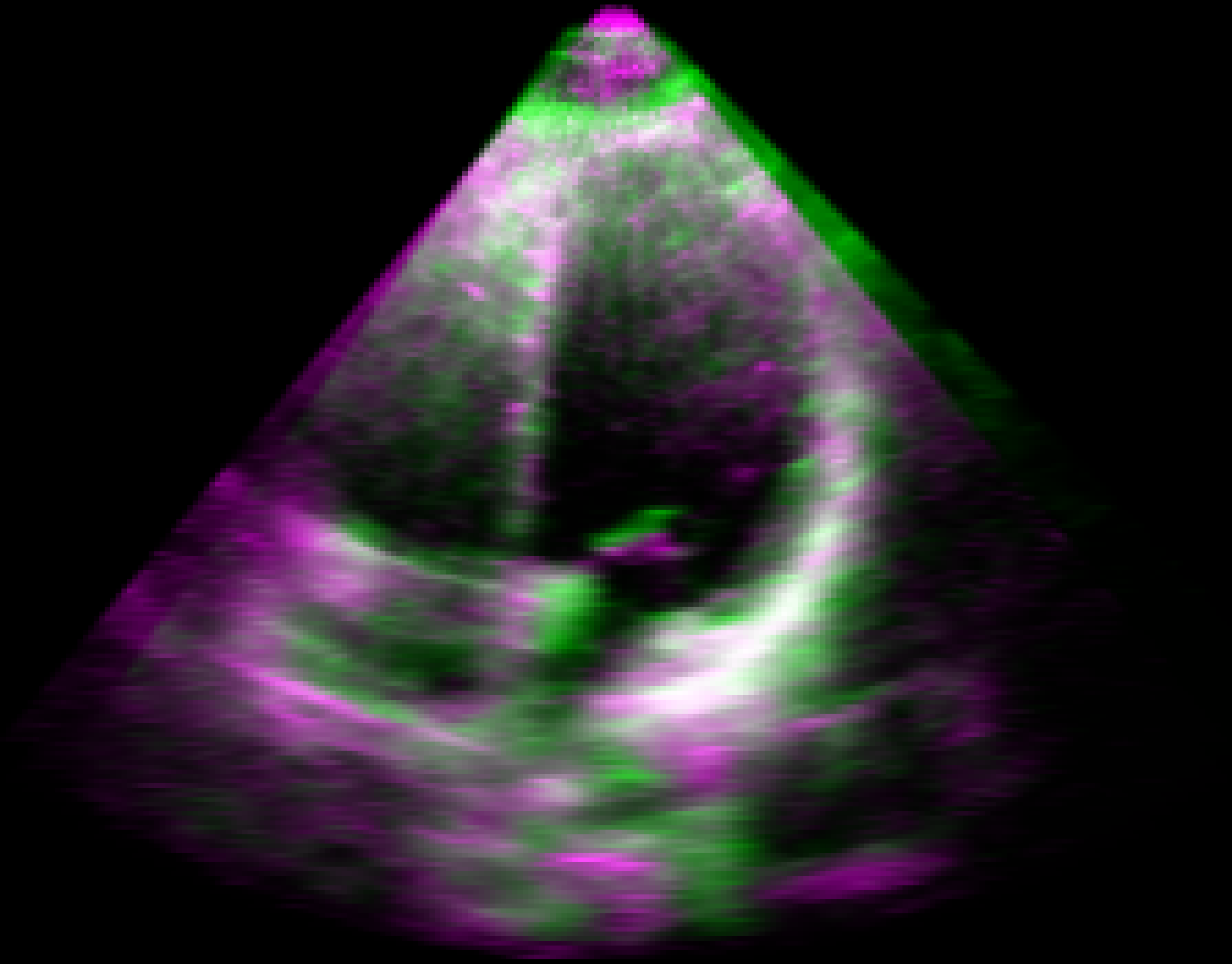}} &
        {\includegraphics[width=0.12\textwidth]{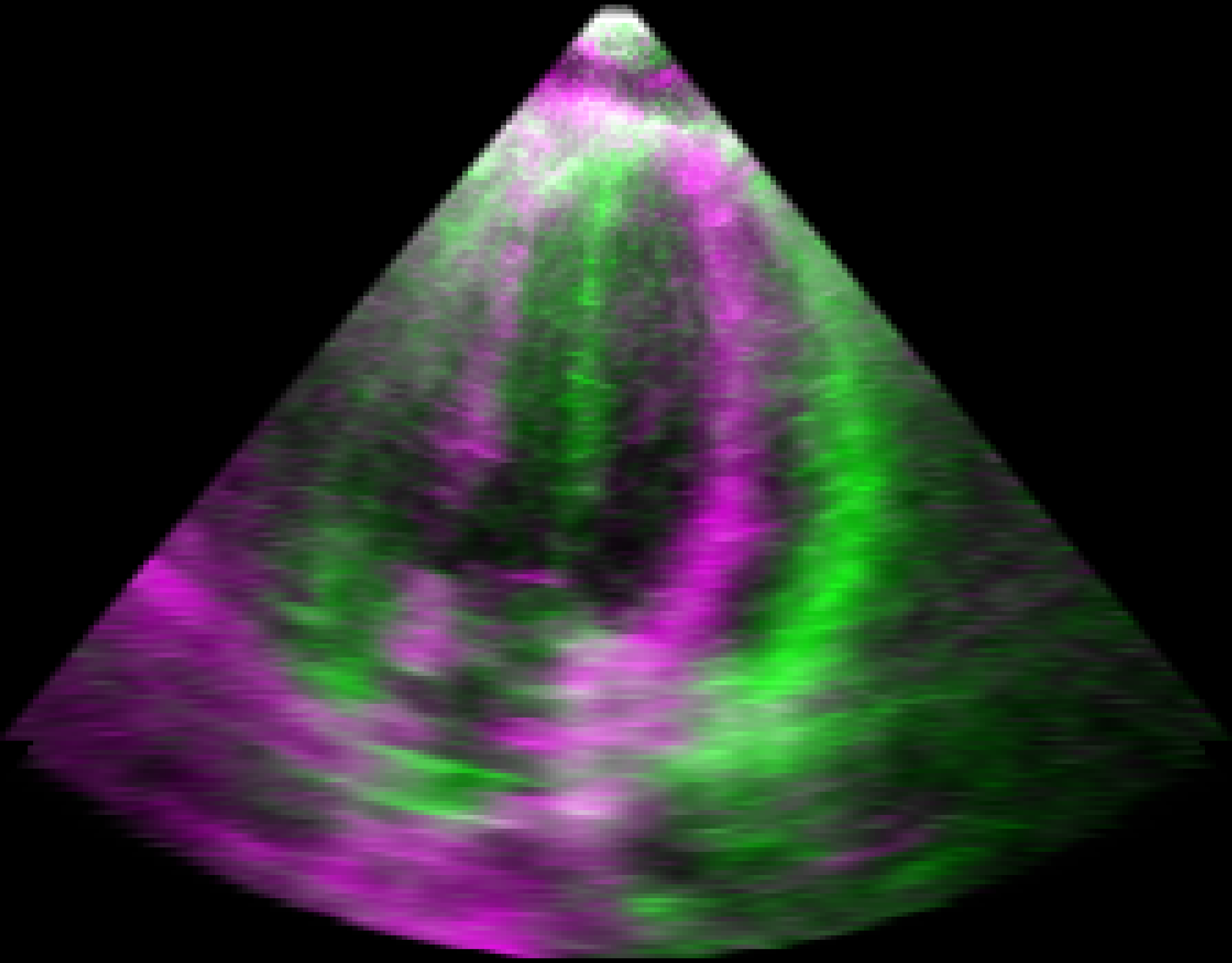}} &
        {\includegraphics[width=0.12\textwidth]{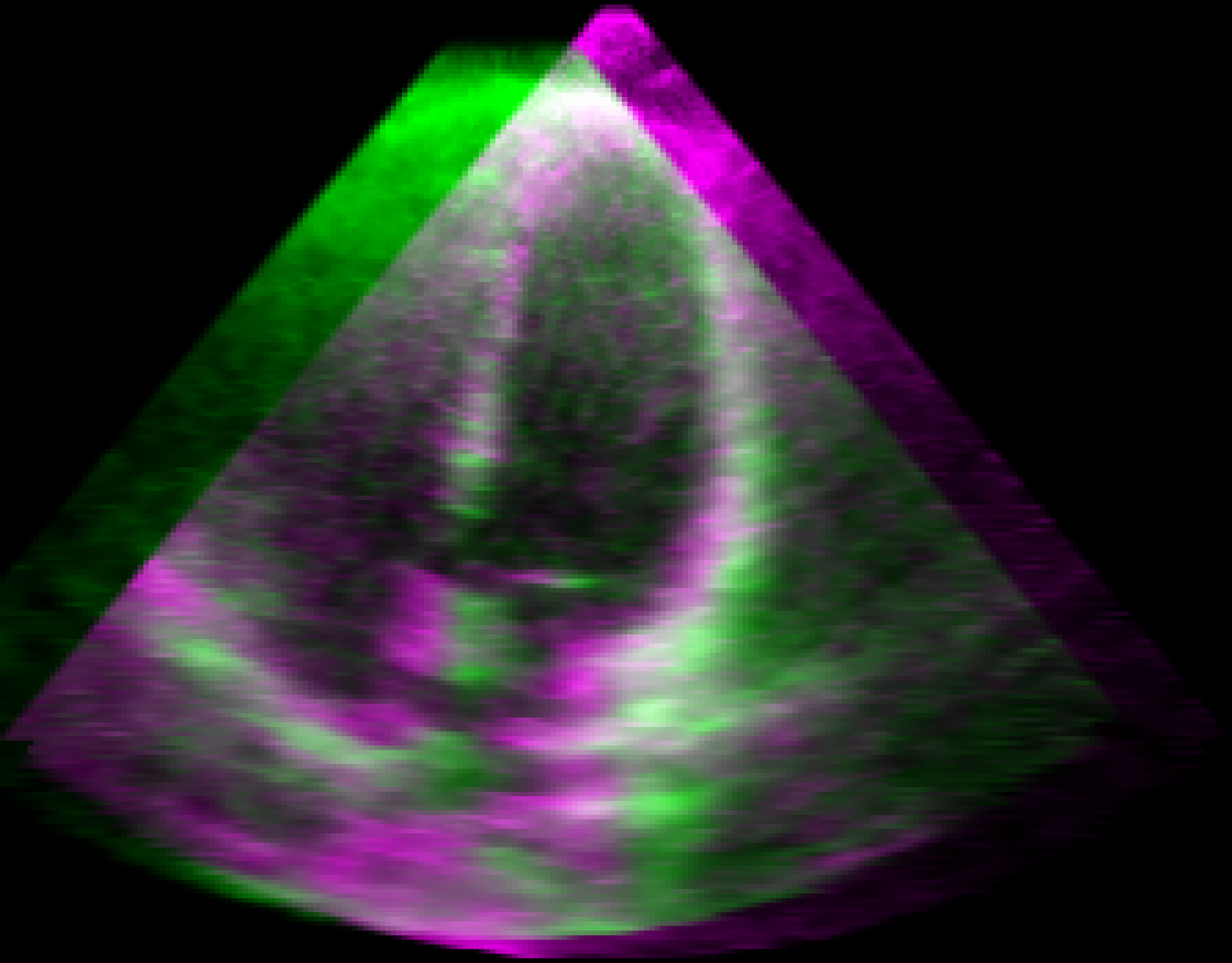}} &
        {\includegraphics[width=0.12\textwidth]{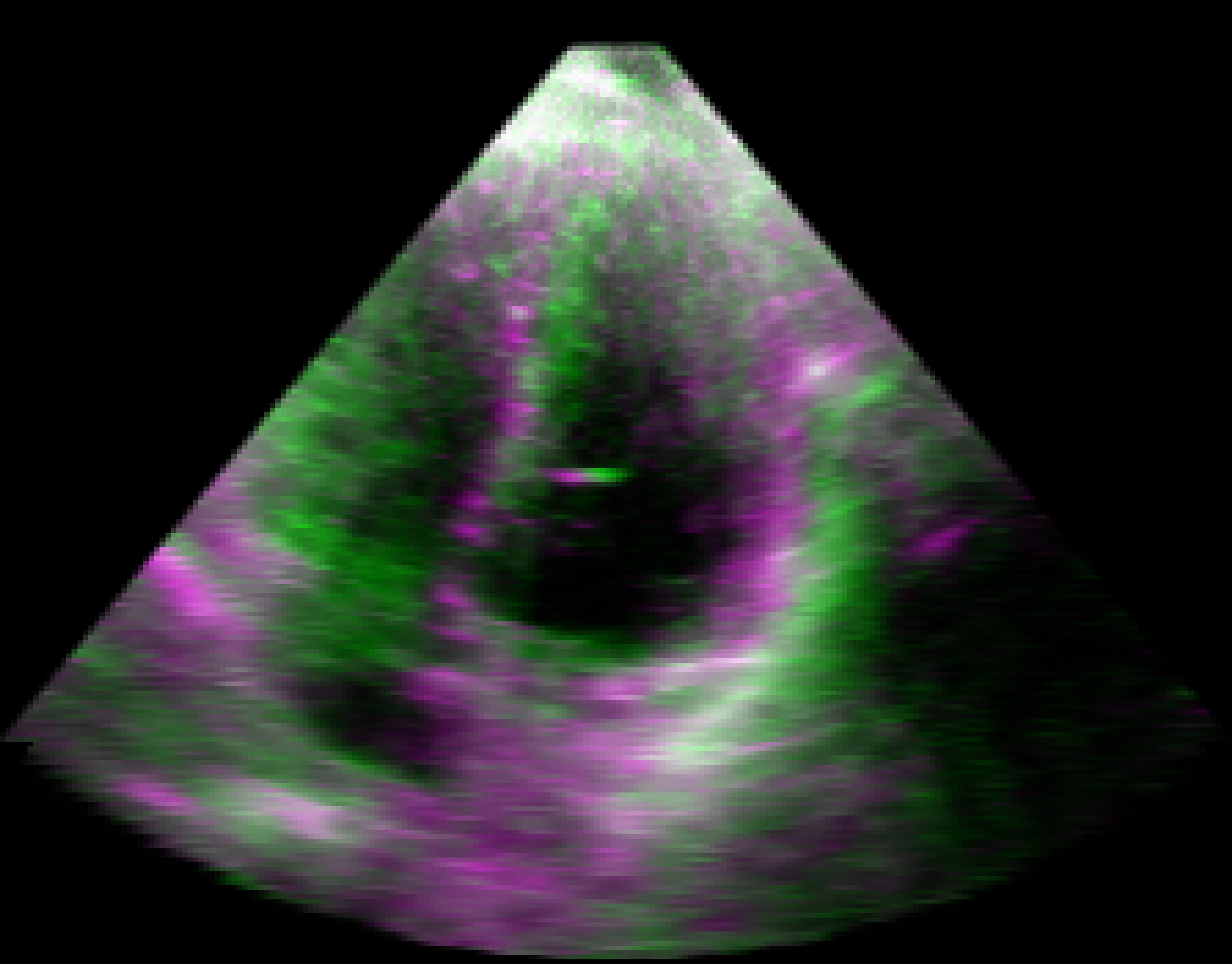}} &
        {\includegraphics[width=0.12\textwidth]{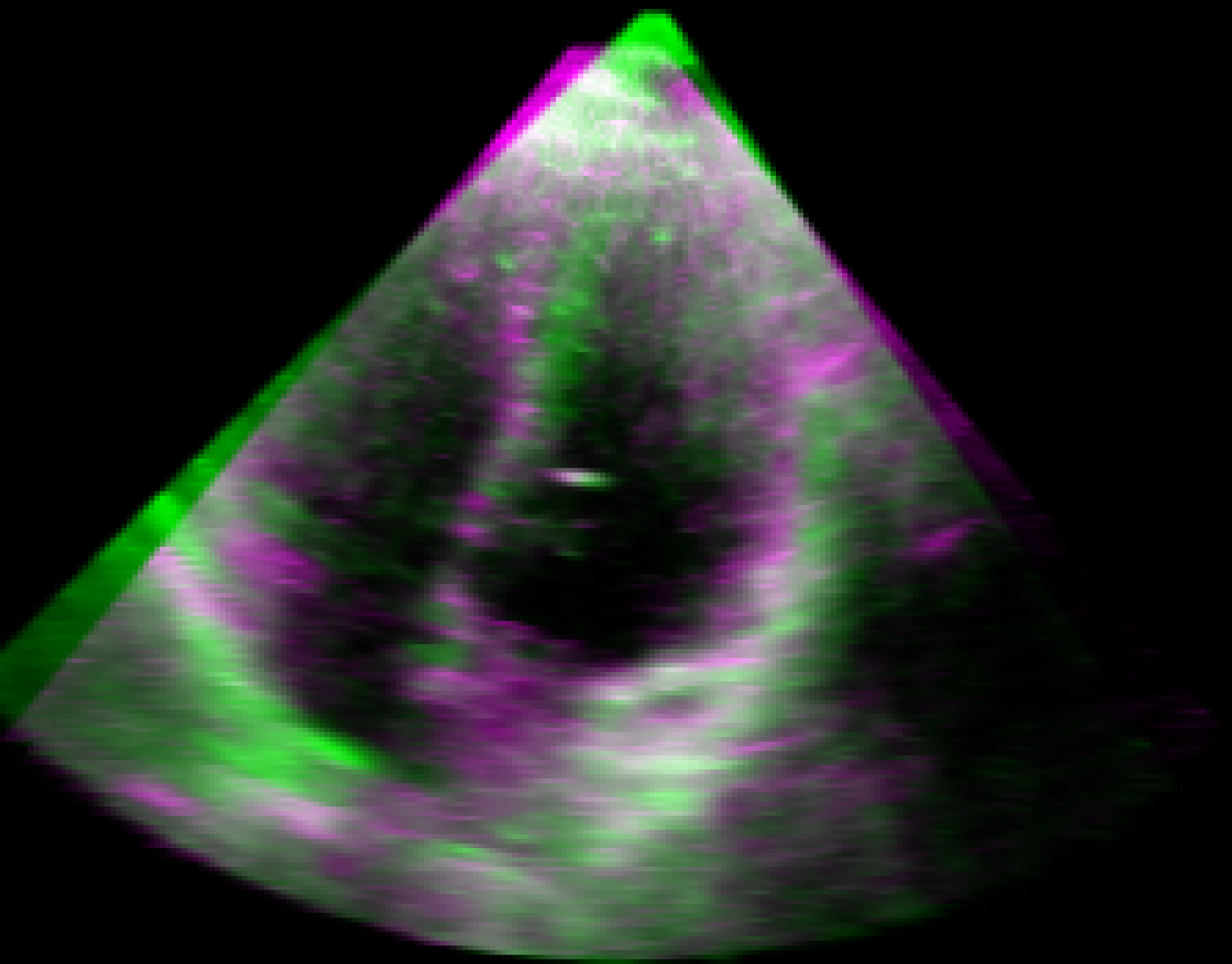}} \\
        \makecell[b]{Q2\vspace{15pt}} & 
        {\includegraphics[width=0.12\textwidth]{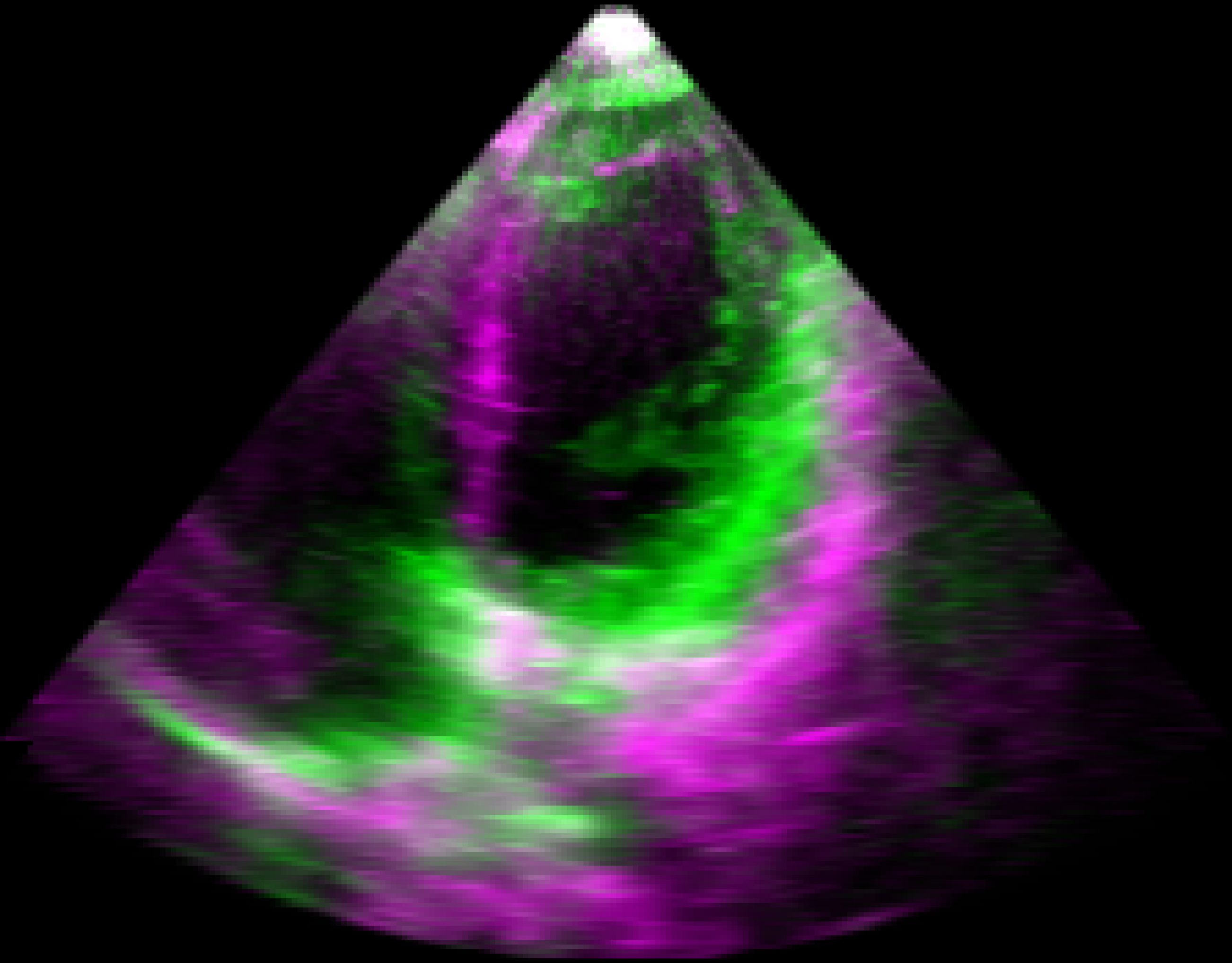}} &
        {\includegraphics[width=0.12\textwidth]{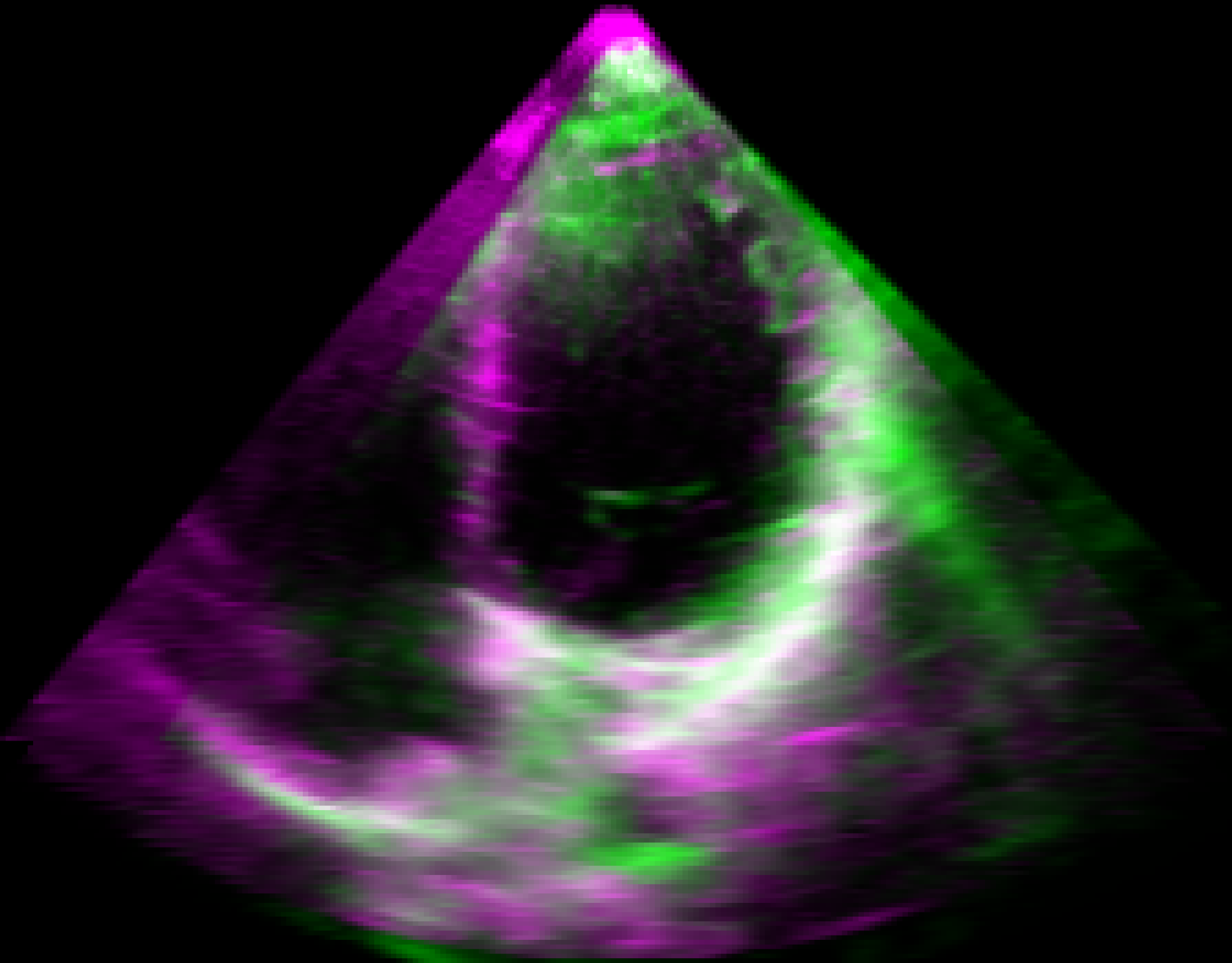}} &
        {\includegraphics[width=0.12\textwidth]{figures/fig_before_Im_016_20240119_144948_3D-Im_019_20240119_145037_3D_Coronal_0.pdf}} &
        {\includegraphics[width=0.12\textwidth]{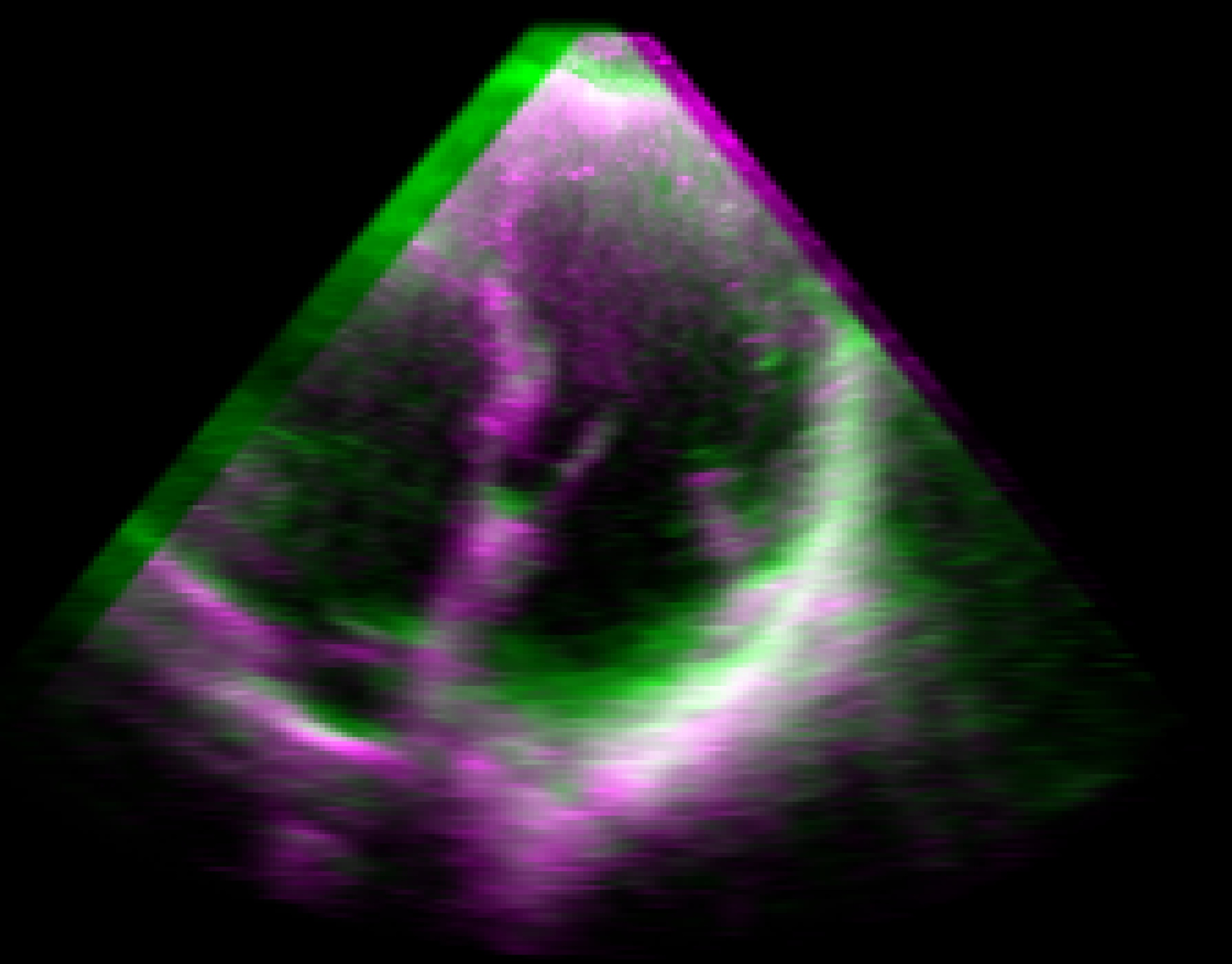}} &
        {\includegraphics[width=0.12\textwidth]{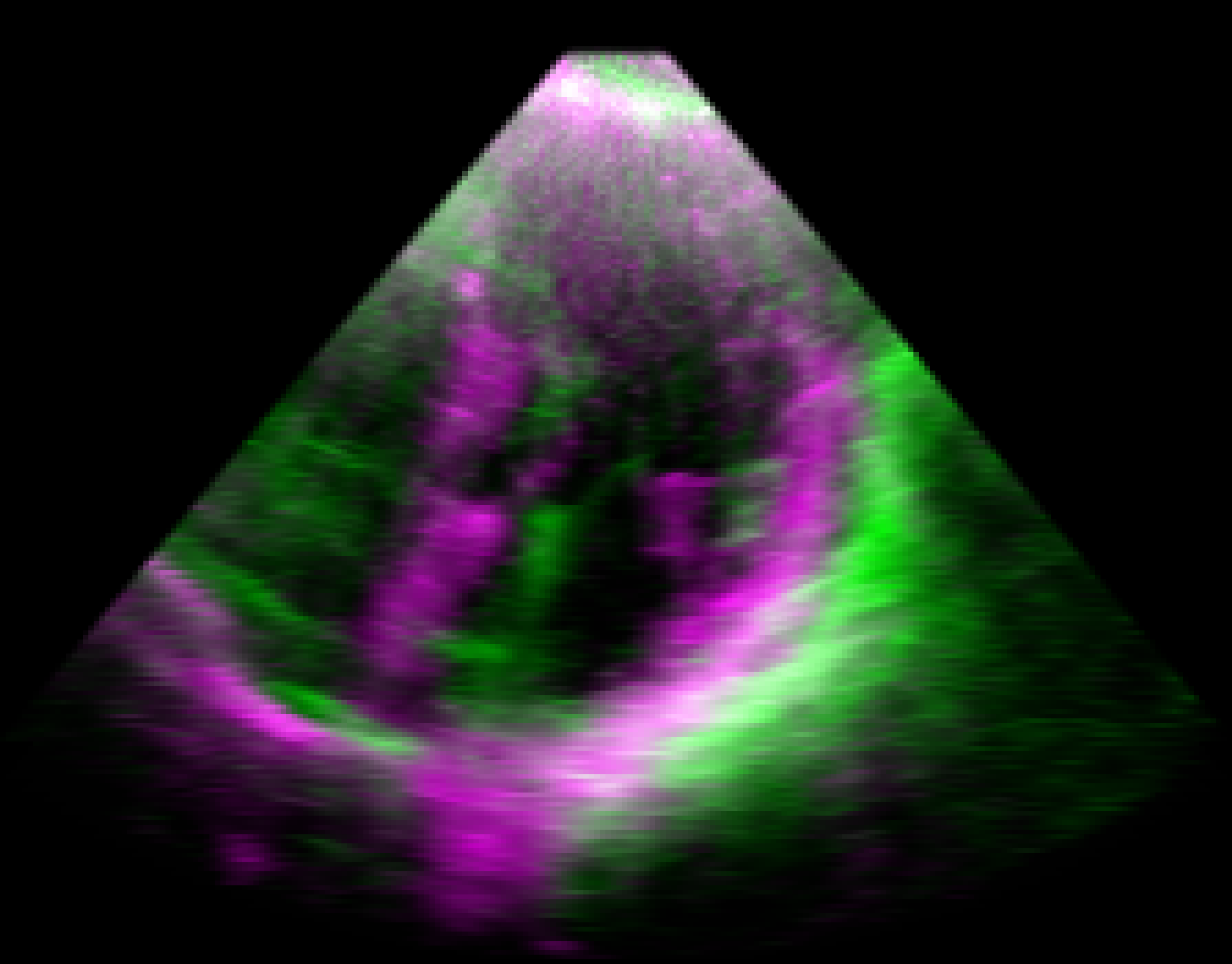}} &
        {\includegraphics[width=0.12\textwidth]{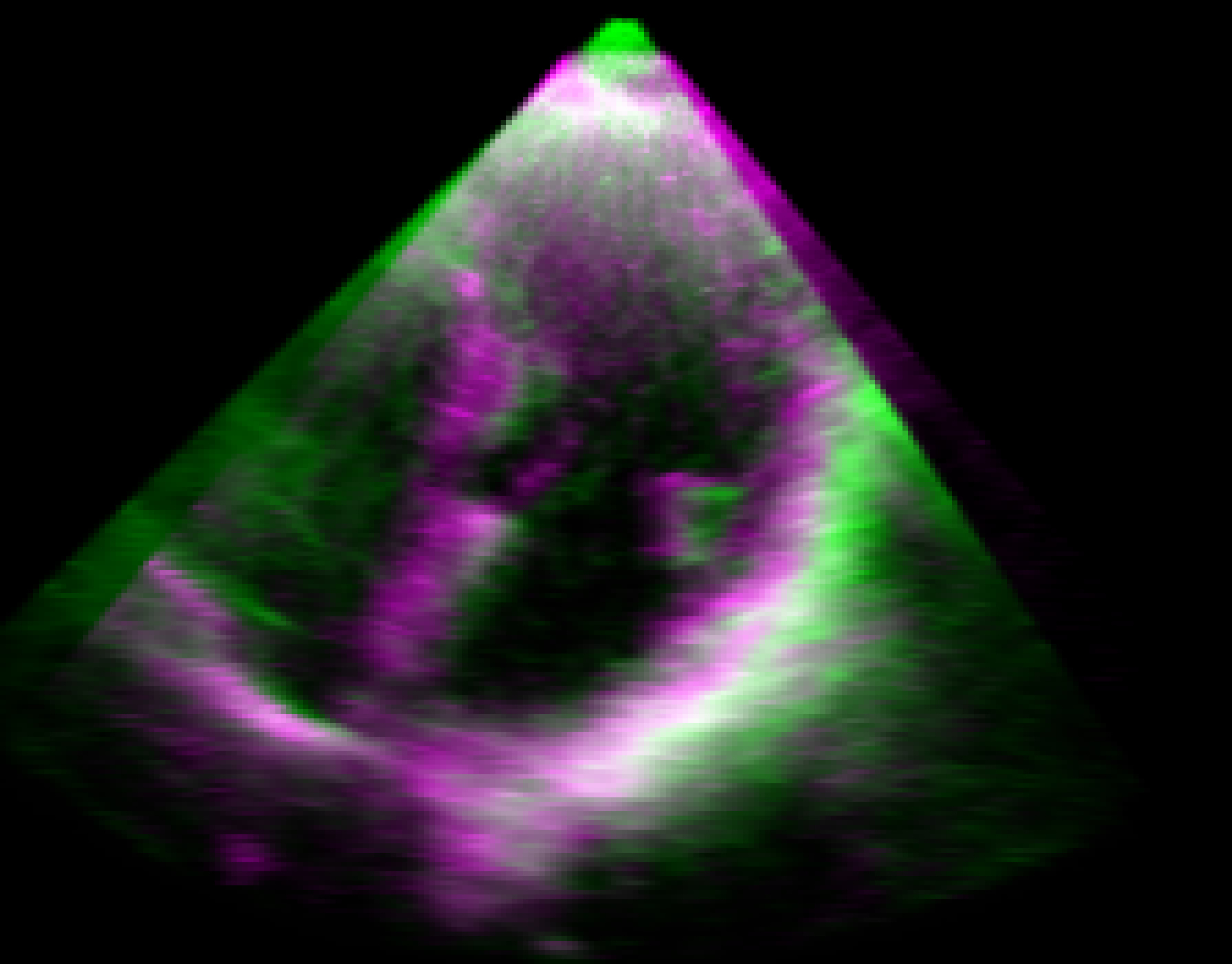}} \\ 
        \makecell[b]{Q3\vspace{15pt}} & 
        {\includegraphics[width=0.12\textwidth]{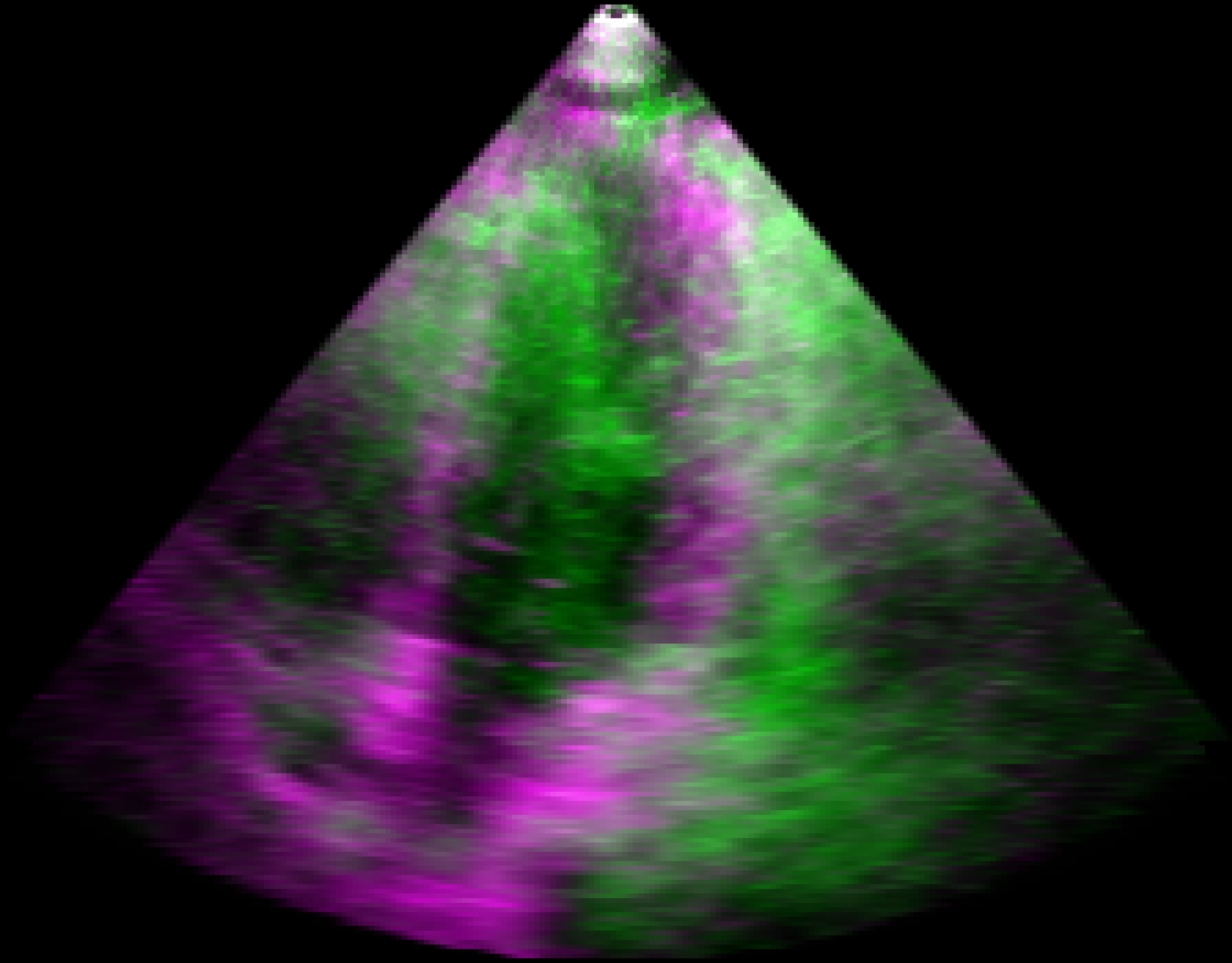}} &
        {\includegraphics[width=0.12\textwidth]{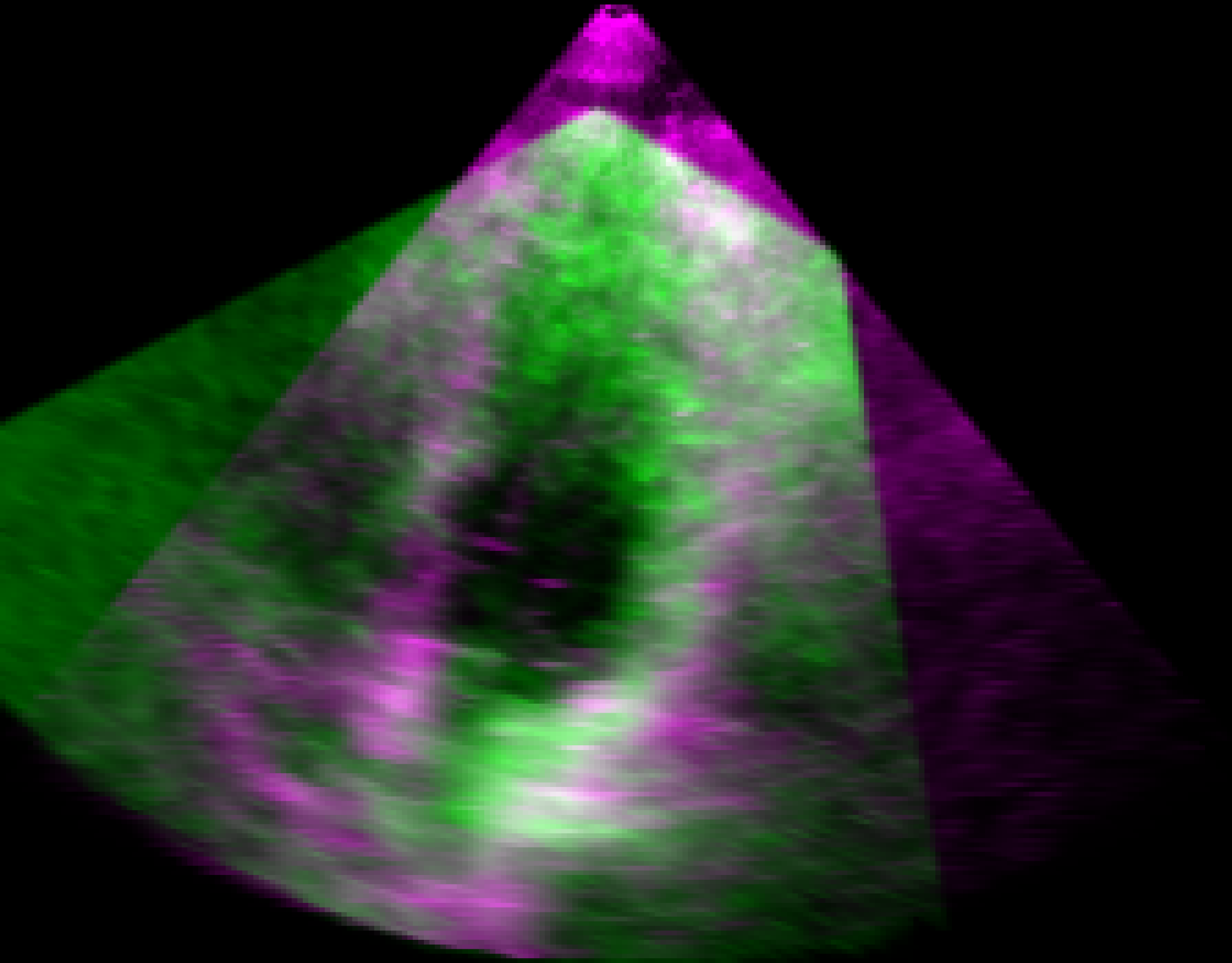}} &
        {\includegraphics[width=0.12\textwidth]{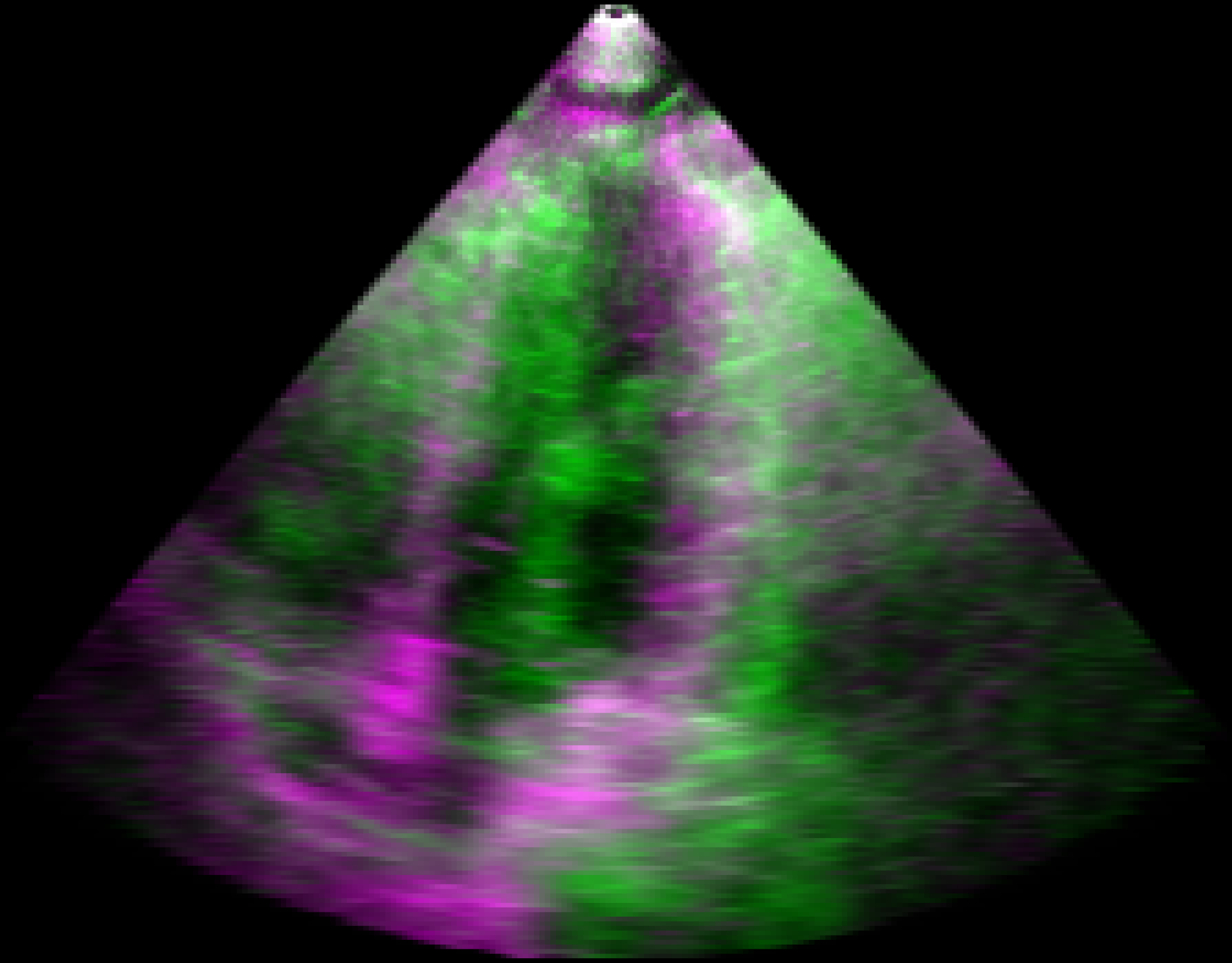}} &
        {\includegraphics[width=0.12\textwidth]{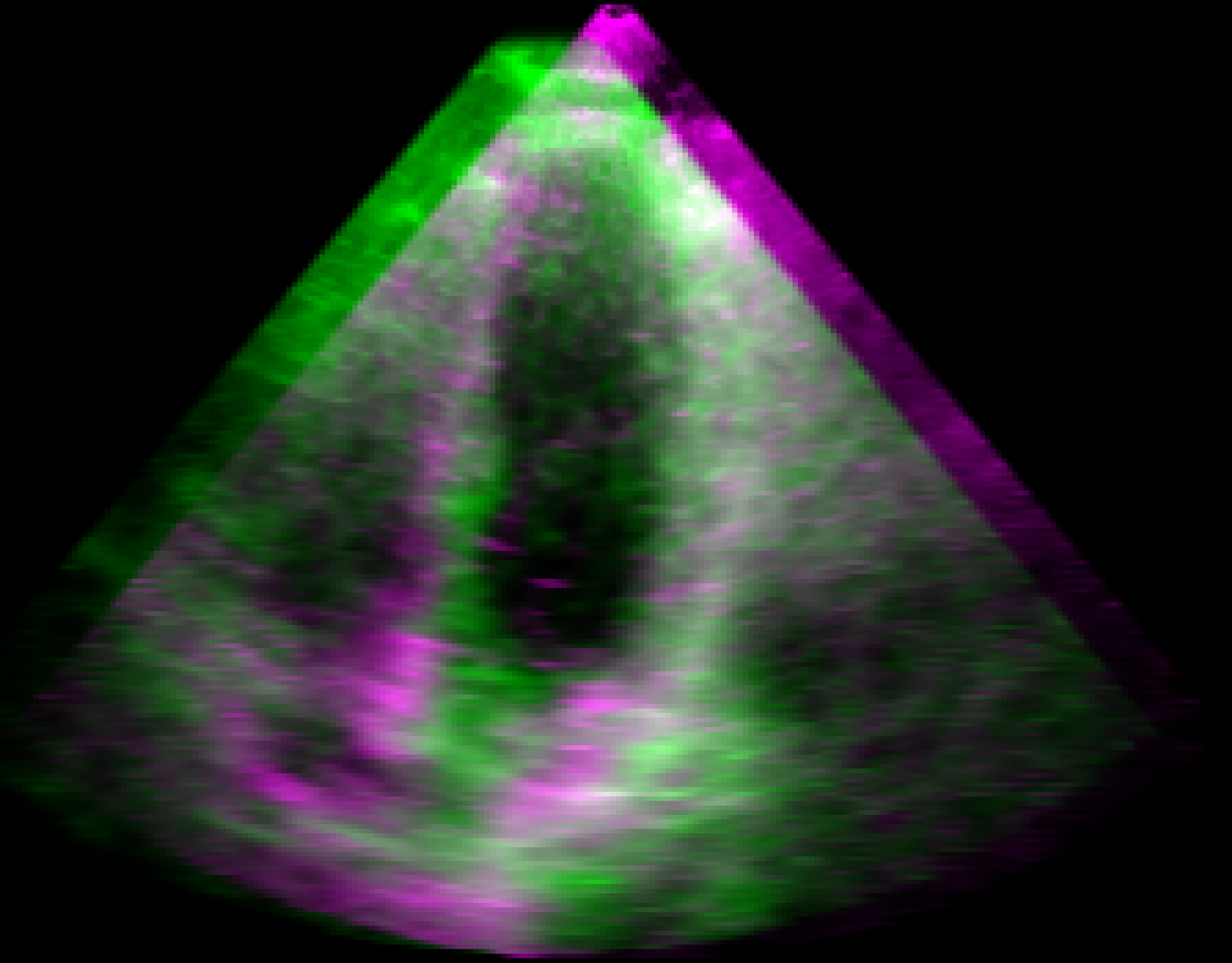}} &
        {\includegraphics[width=0.12\textwidth]{figures/fig_before_Im_013_20240221_143621_3D-Im_004_20240221_143057_3D_Coronal_0.pdf}} &
        {\includegraphics[width=0.12\textwidth]{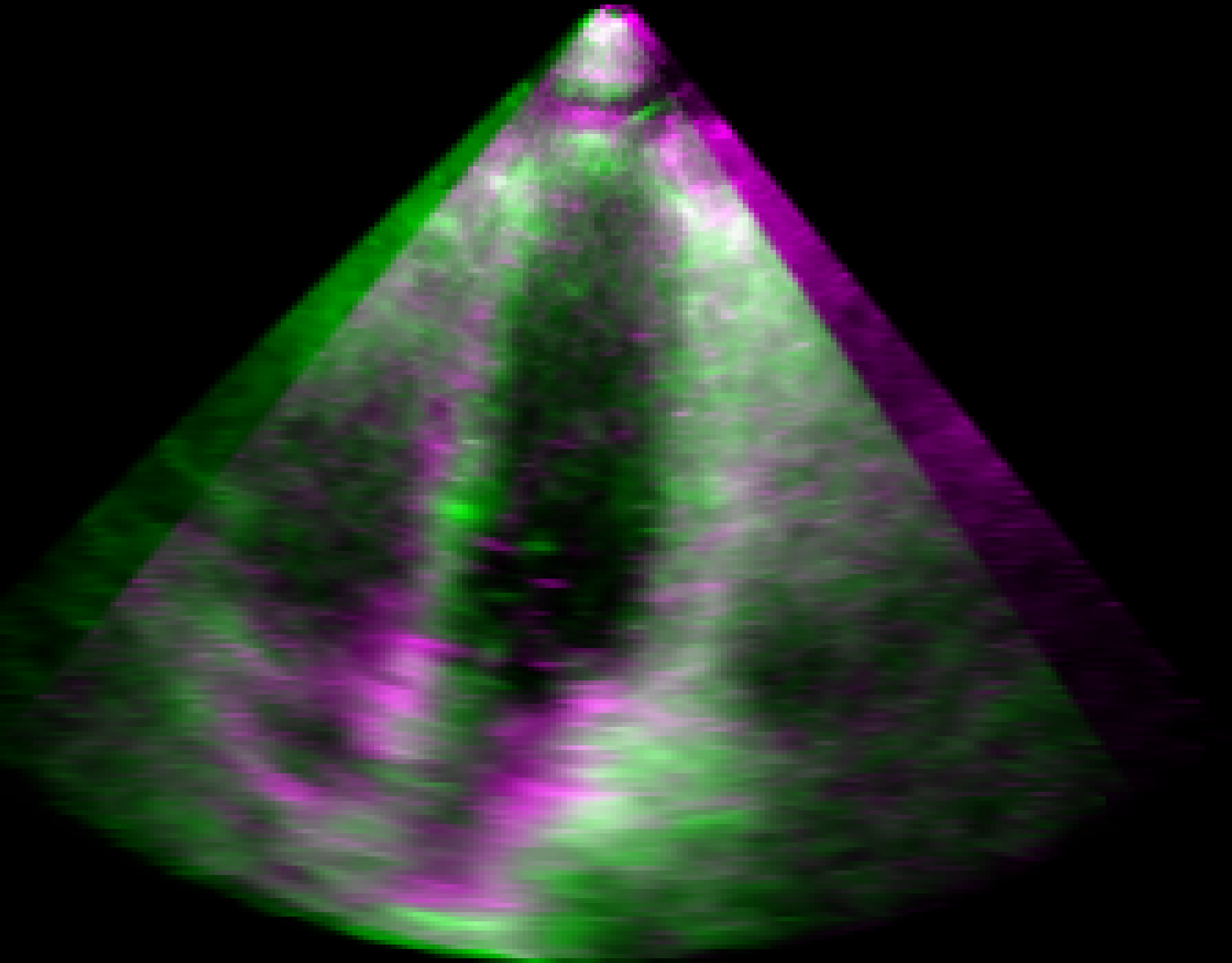}} \\
        \makecell[b]{Max\vspace{15pt}} & 
        {\includegraphics[width=0.12\textwidth]{figures/fig_before_Im_003_20240215_151511_3D-Im_010_20240215_151637_3D_Coronal_0.pdf}} &
        {\includegraphics[width=0.12\textwidth]{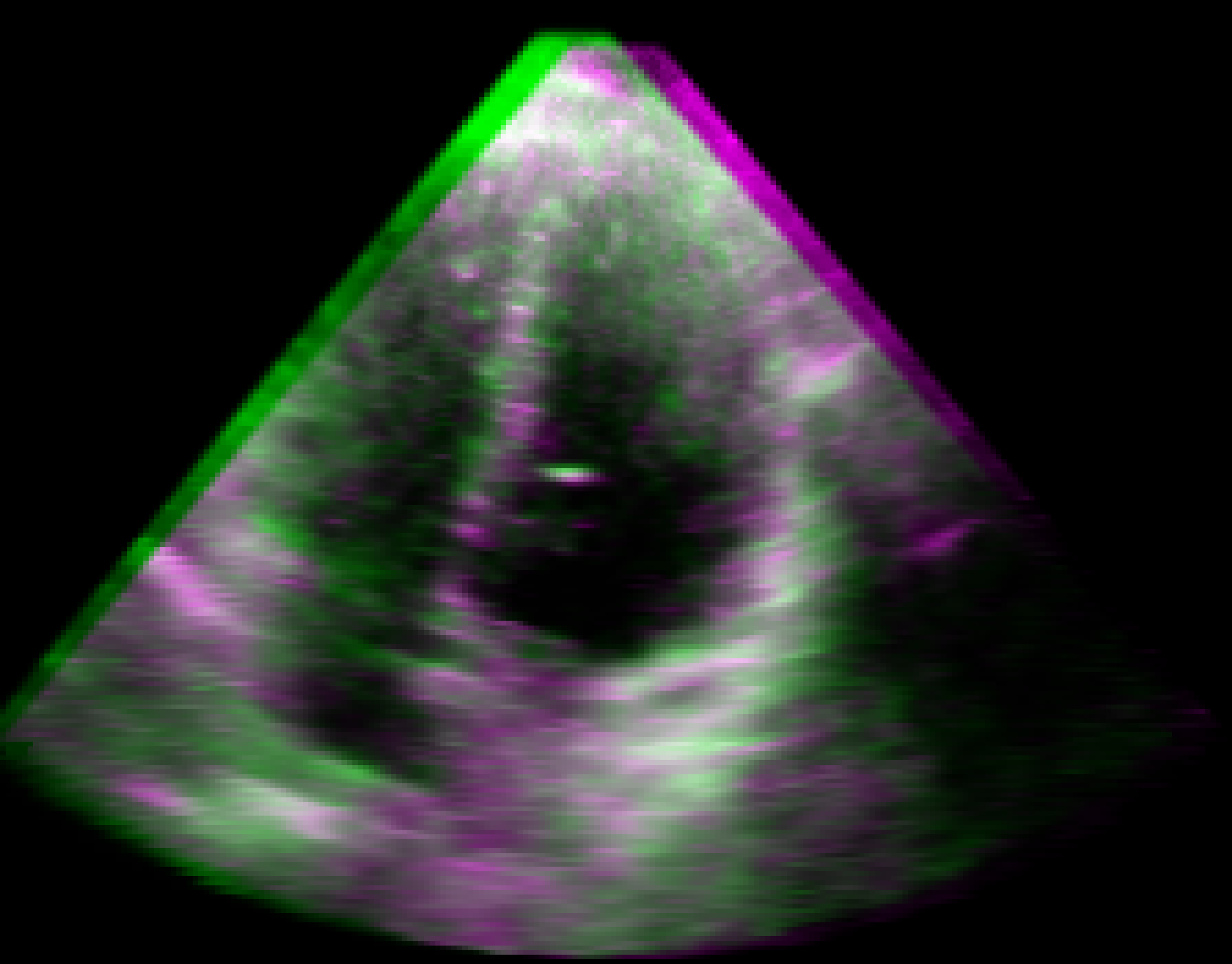}} &
        {\includegraphics[width=0.12\textwidth]{figures/fig_before_Im_013_20240221_143621_3D-Im_002_20240221_143029_3D_Coronal_0.pdf}} &
        {\includegraphics[width=0.12\textwidth]{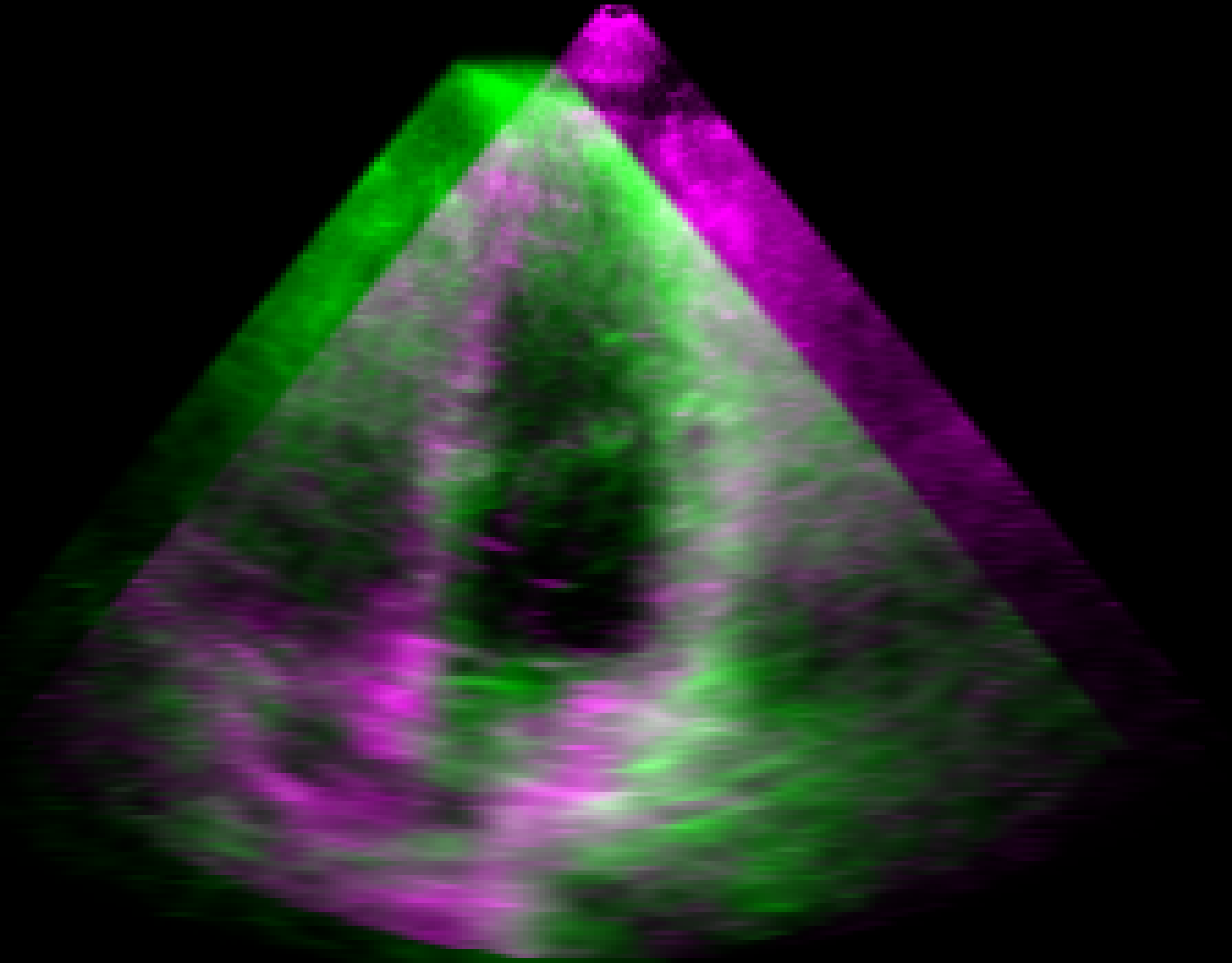}} &
        {\includegraphics[width=0.12\textwidth]{figures/fig_before_Im_004_20240123_153322_3D-Im_010_20240123_153651_3D_Coronal_0.pdf}} &
        {\includegraphics[width=0.12\textwidth]{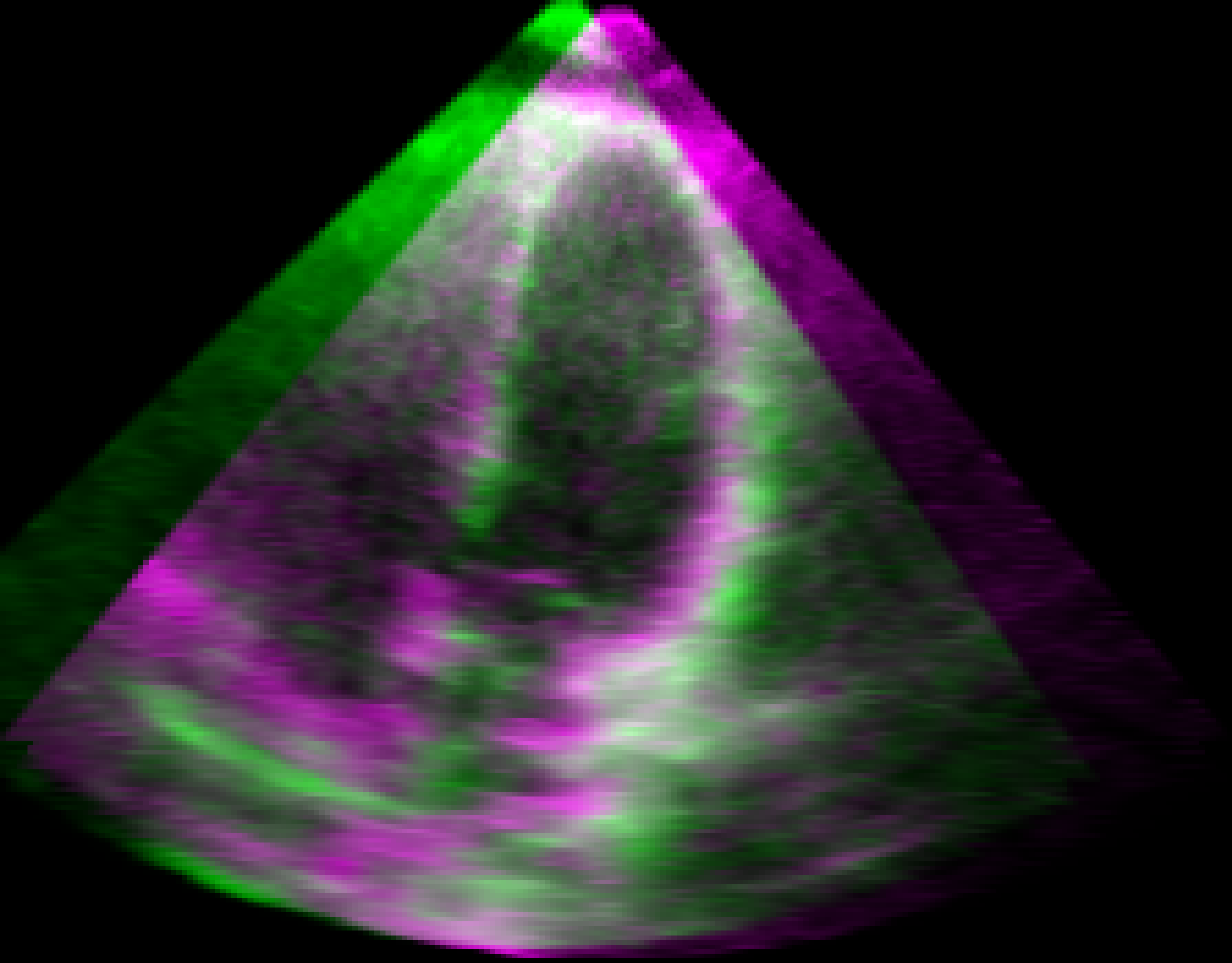}} \\
    \end{tabular}
    \caption{Mask-based rigid (PF and EX) registration results of the ED frame of image pairs for percentile DSC difference values of the coronal view. Two consecutive columns show images before and after the registration for each category. The source and target images are shown in green and purple colors.}
    \label{fig:fig_perc_images_coronal_rigid_mask}
\end{figure*}

\begin{figure*}[!ht]
    \centering
    \begin{tabular}{SSSSSSS}
         & \multicolumn{2}{c}{\scriptsize PF (CPU)} & \multicolumn{2}{c}{\scriptsize PF (GPU)} & \multicolumn{2}{c}{\scriptsize EX (CPU)} \\
         & Before & After & Before & After & Before & After \\
        \makecell[b]{Min\vspace{15pt}} & 
        {\includegraphics[width=0.12\textwidth]{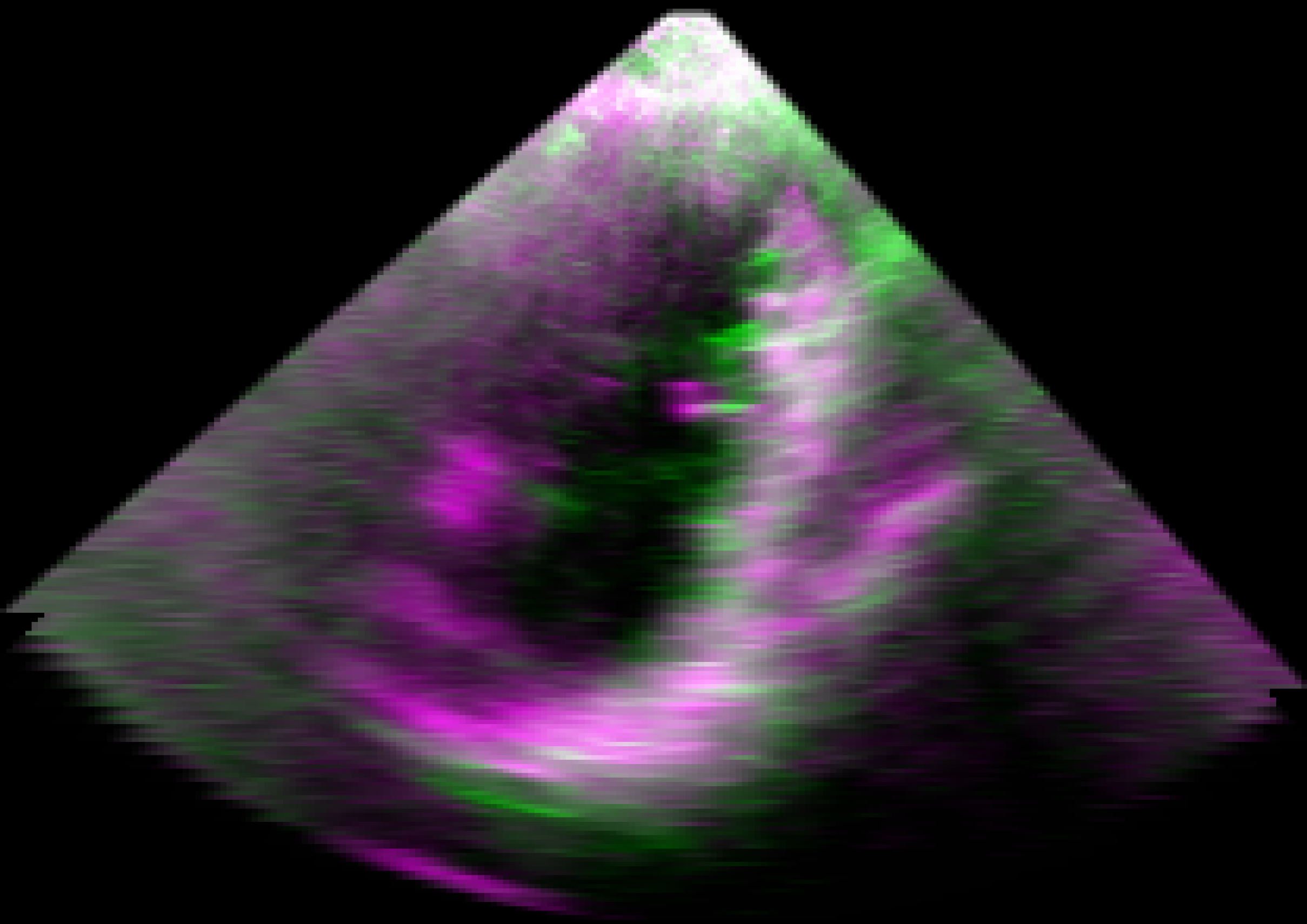}} &
        {\includegraphics[width=0.12\textwidth]{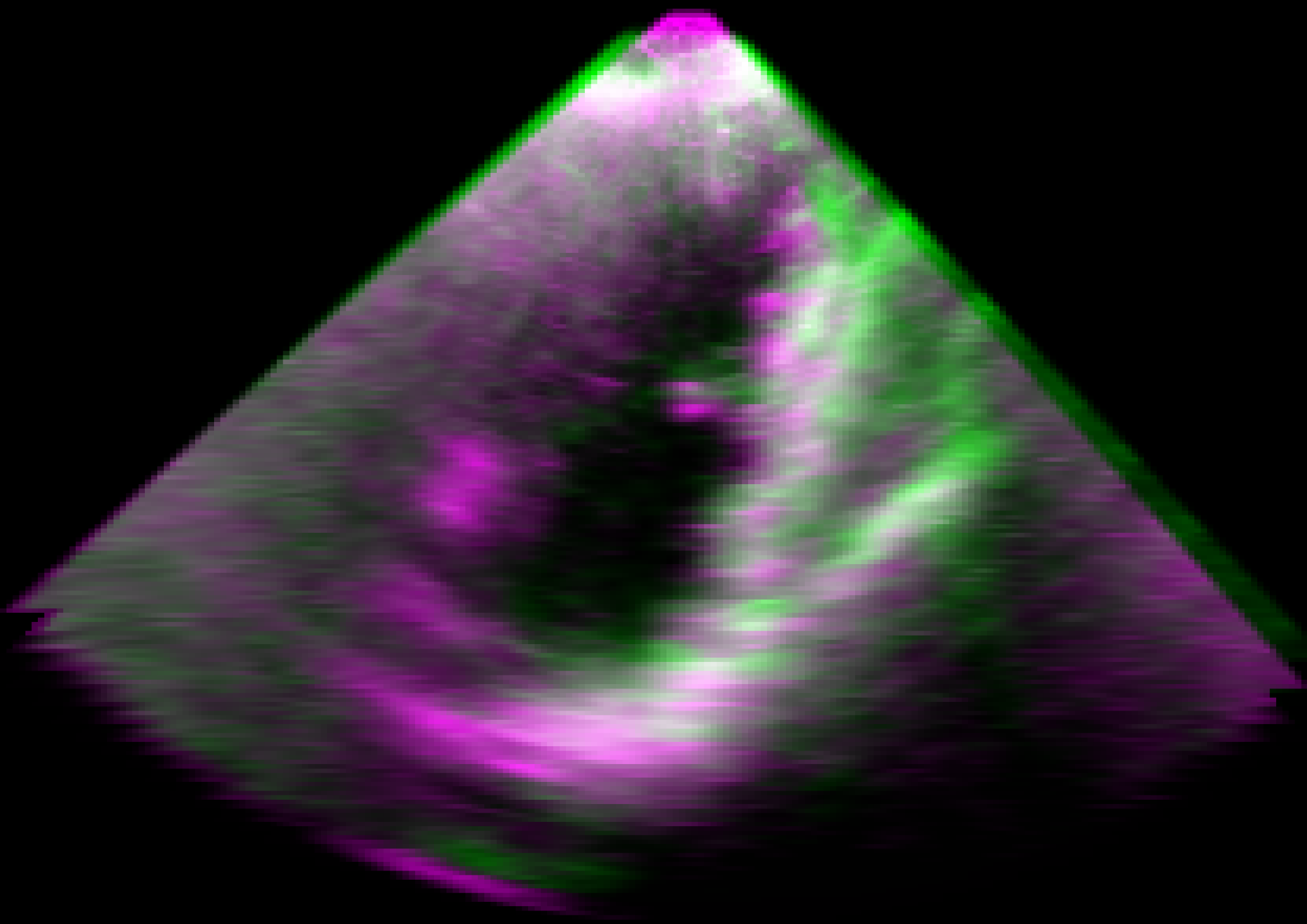}} &
        {\includegraphics[width=0.12\textwidth]{figures/fig_before_Im_016_20230817_104344_3D-Im_012_20230817_104241_3D_Sagittal_0.pdf}} &
        {\includegraphics[width=0.12\textwidth]{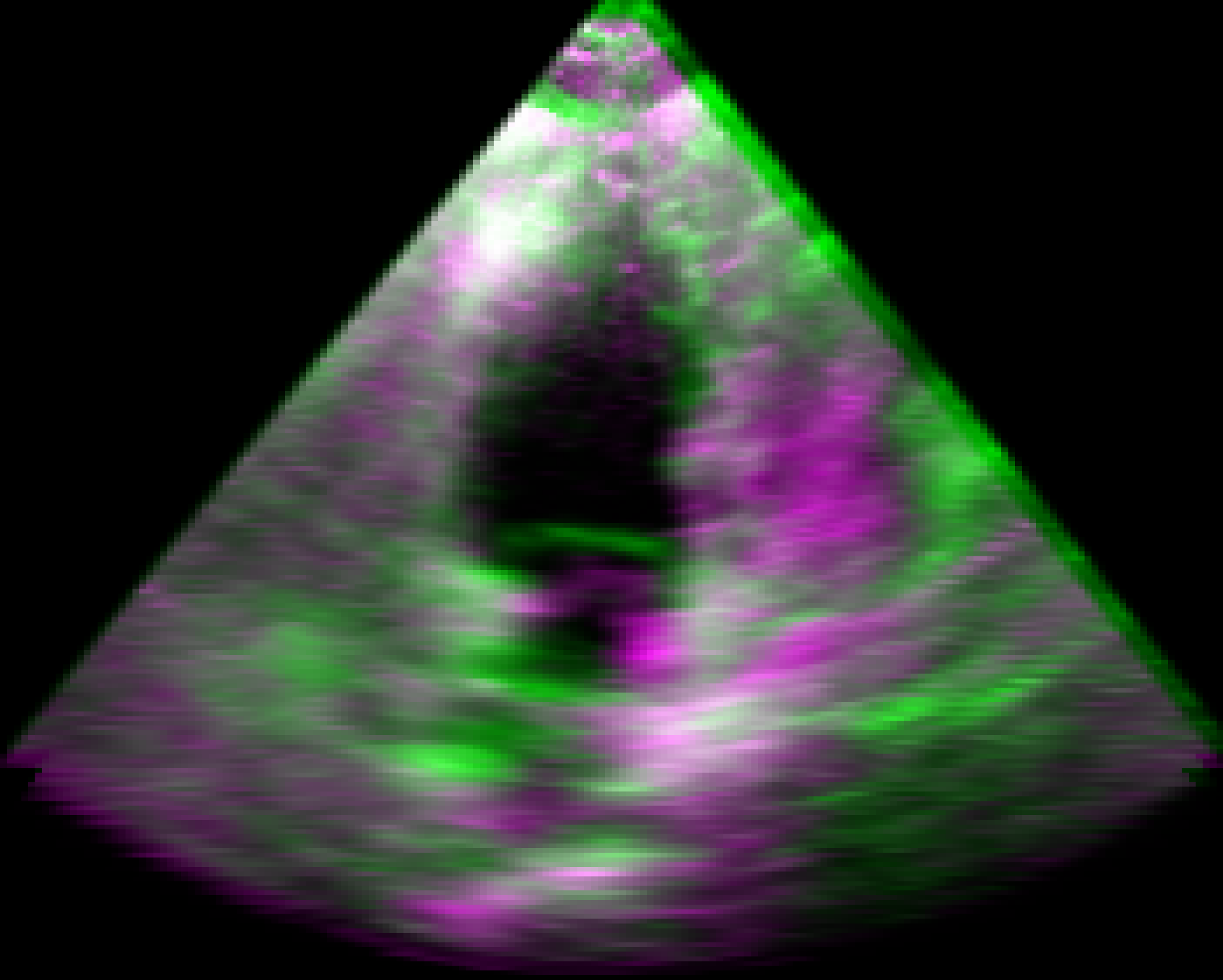}} &
        {\includegraphics[width=0.12\textwidth]{figures/fig_before_Im_001_20240215_151507_3D-Im_007_20240215_151552_3D_Sagittal_0.pdf}} &
        {\includegraphics[width=0.12\textwidth]{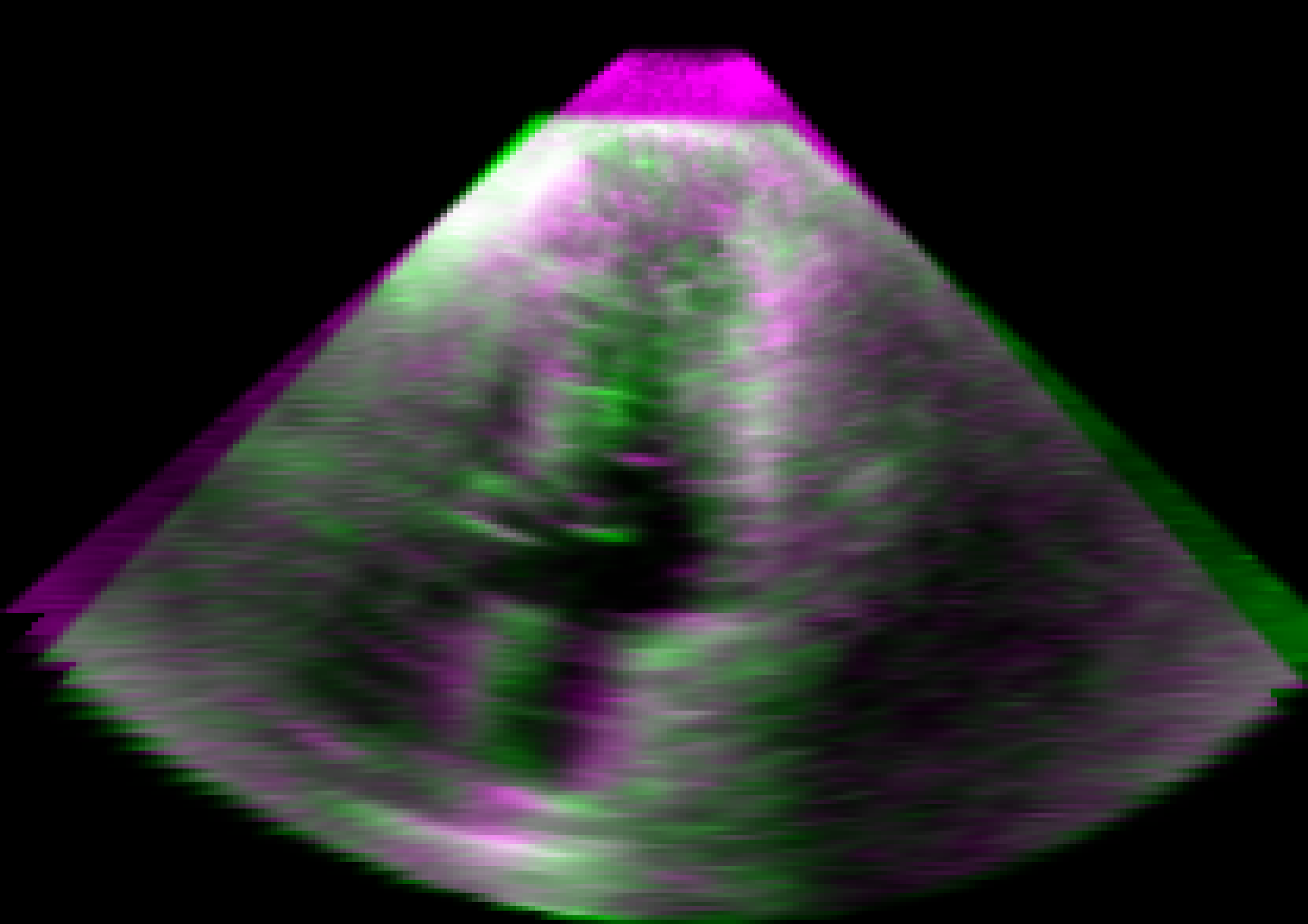}} \\
        \makecell[b]{Q1\vspace{15pt}} & 
        {\includegraphics[width=0.12\textwidth]{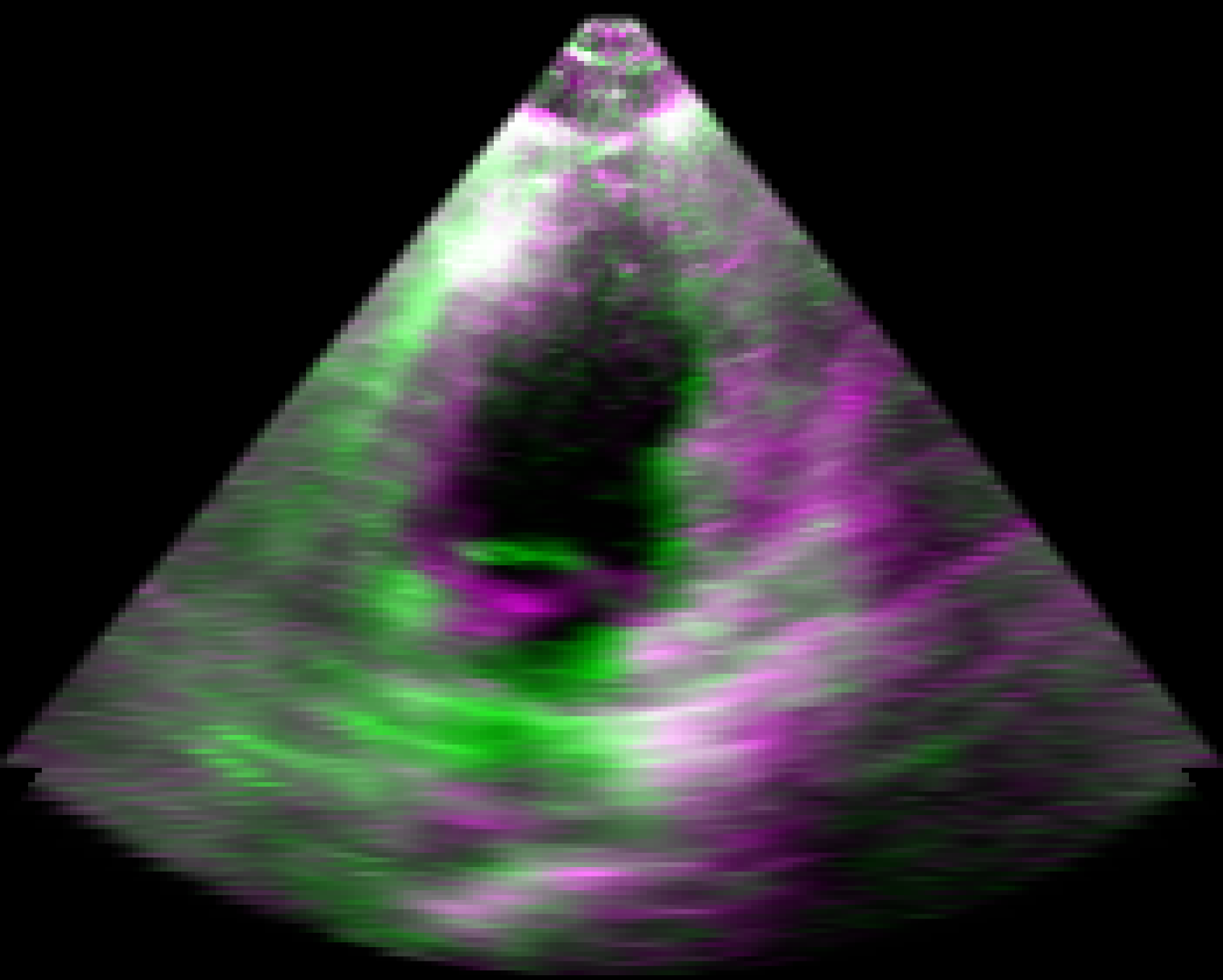}} &
        {\includegraphics[width=0.12\textwidth]{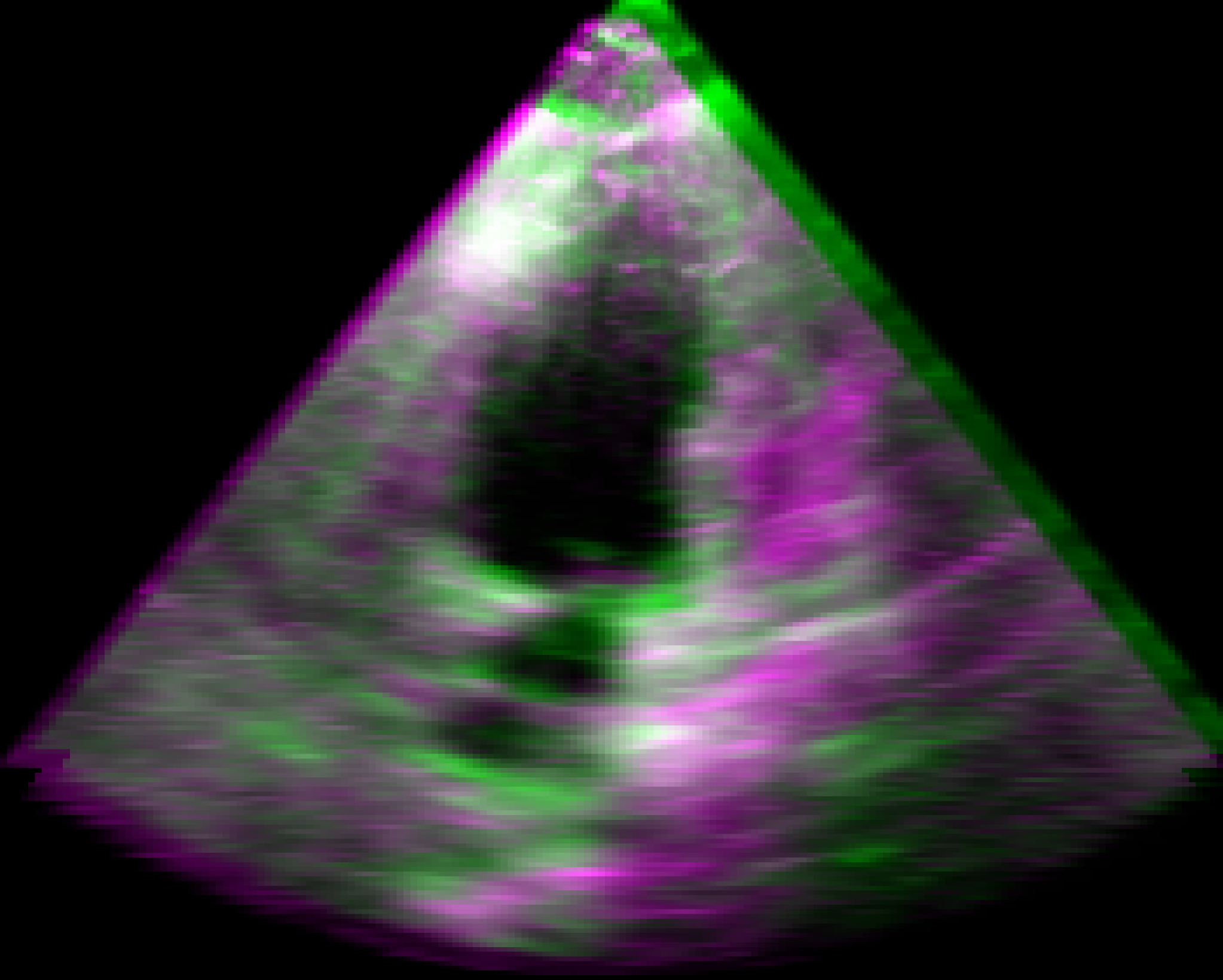}} &
        {\includegraphics[width=0.12\textwidth]{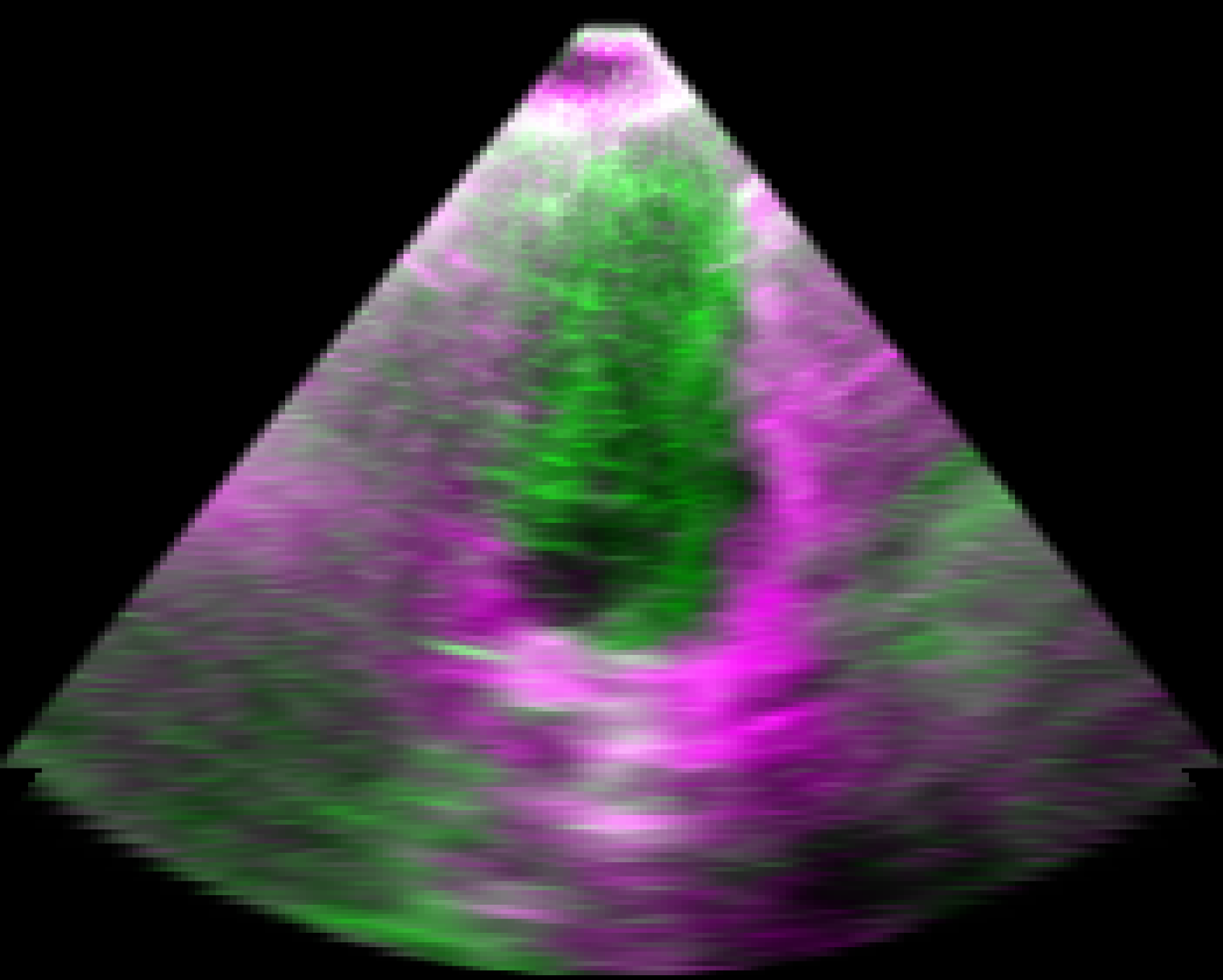}} &
        {\includegraphics[width=0.12\textwidth]{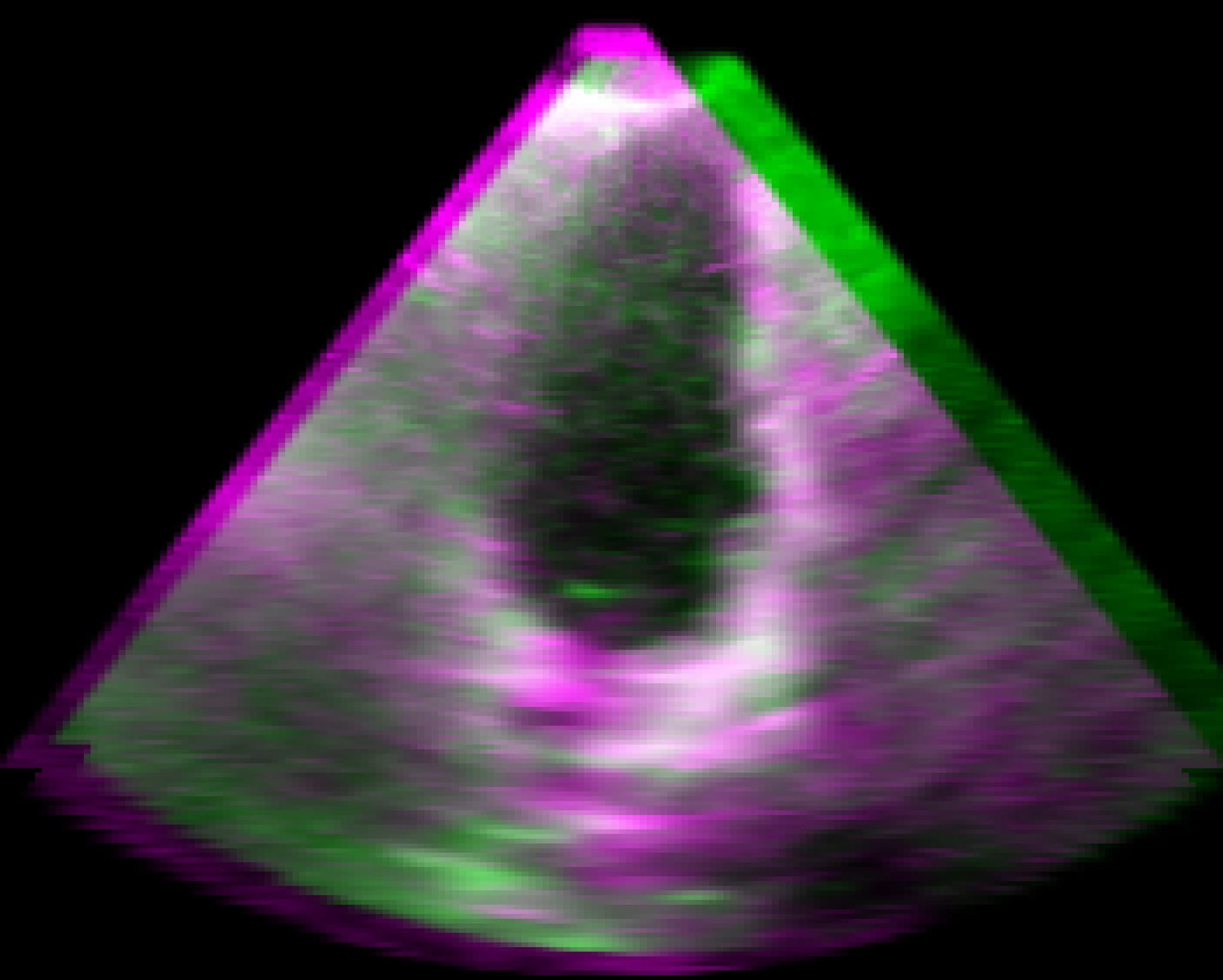}} &
        {\includegraphics[width=0.12\textwidth]{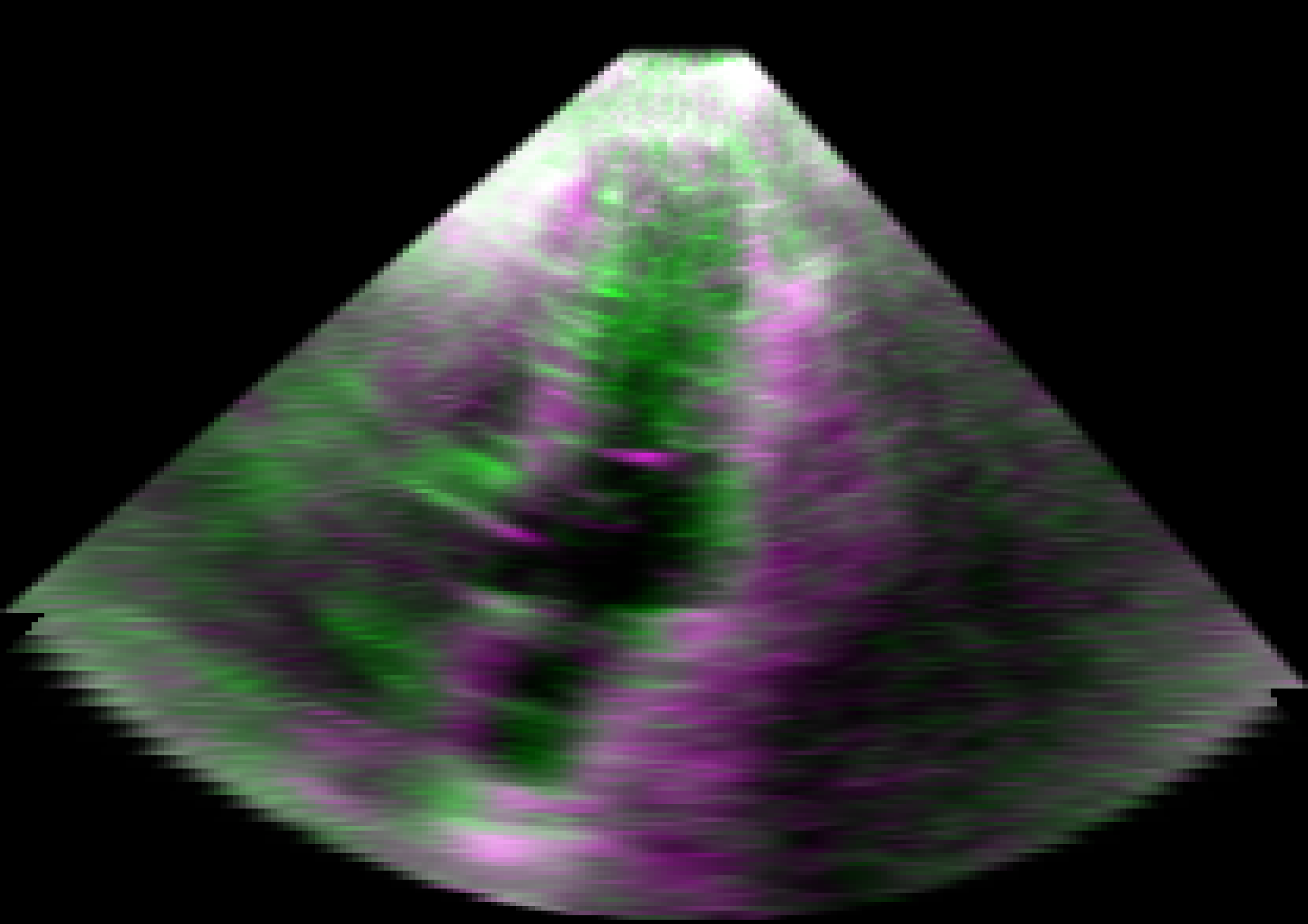}} &
        {\includegraphics[width=0.12\textwidth]{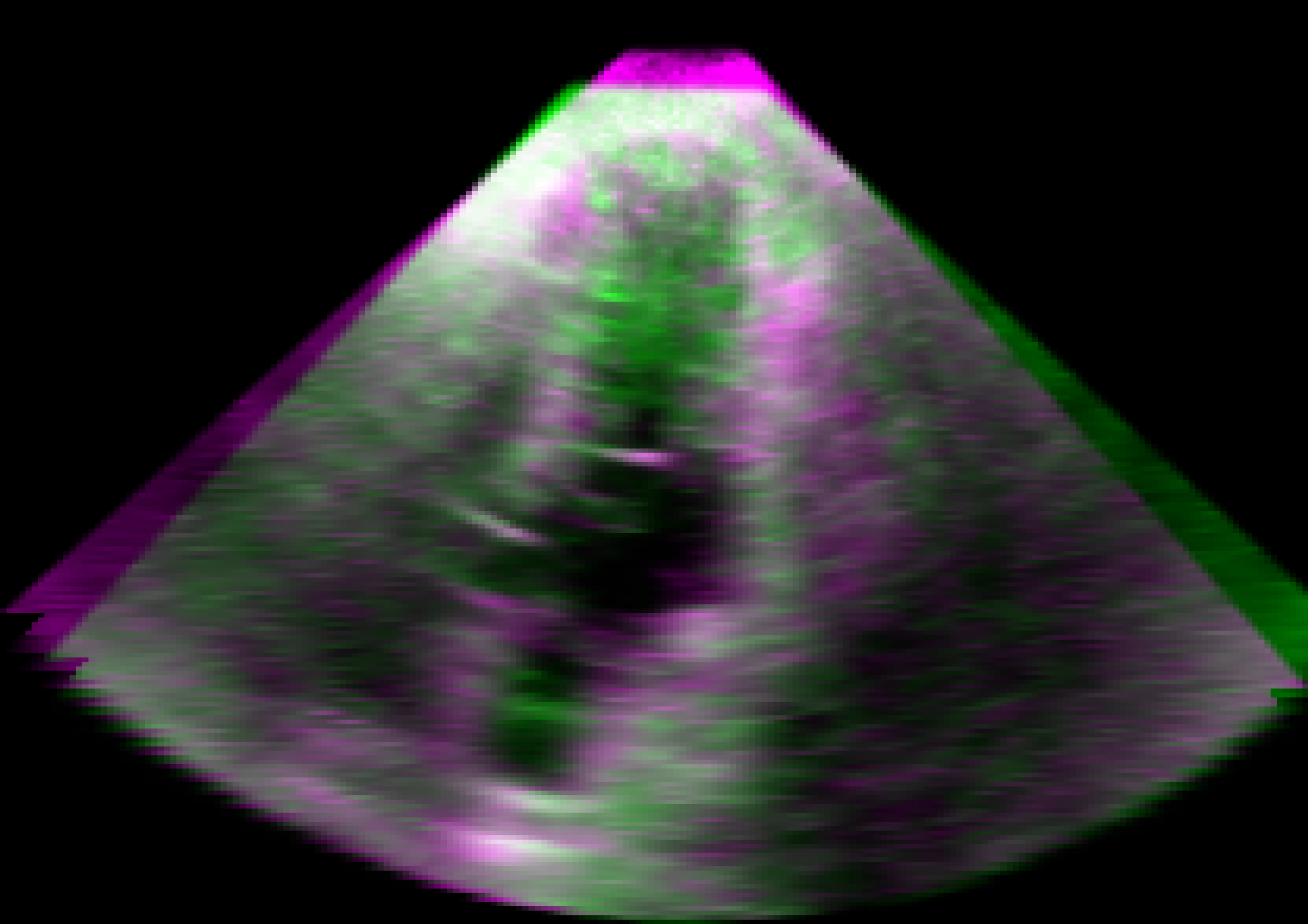}} \\ 
        \makecell[b]{Q2\vspace{15pt}} & 
        {\includegraphics[width=0.12\textwidth]{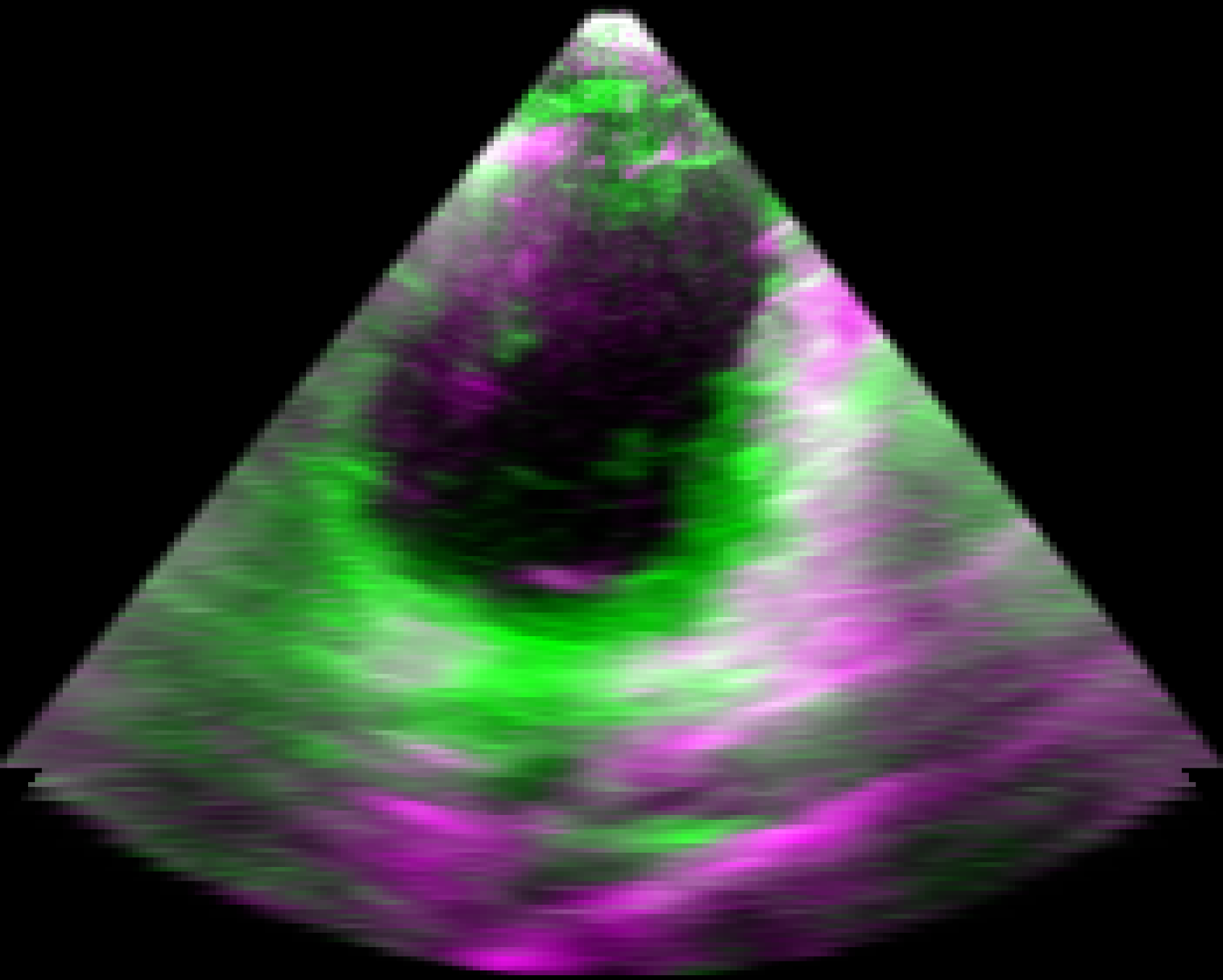}} &
        {\includegraphics[width=0.12\textwidth]{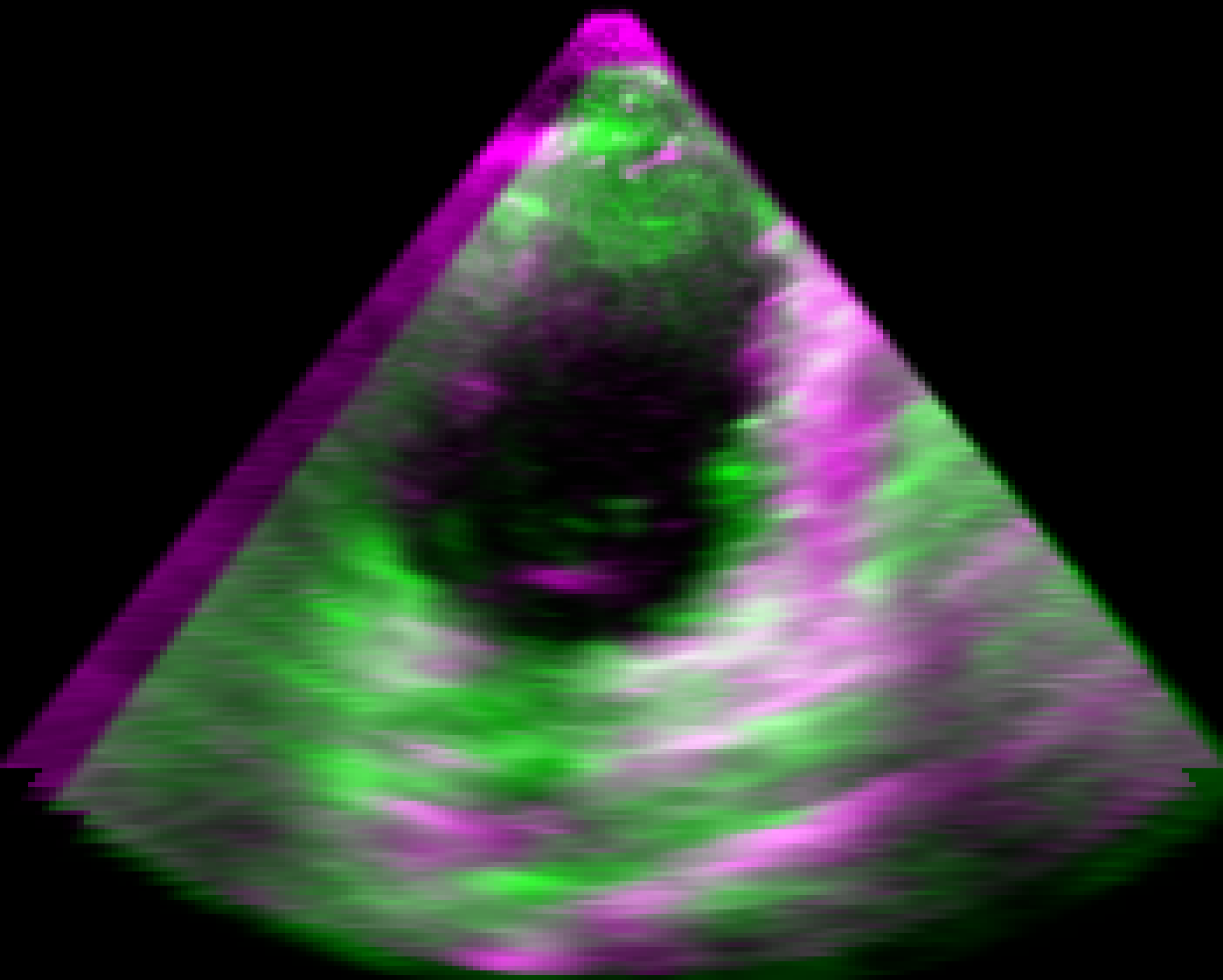}} &
        {\includegraphics[width=0.12\textwidth]{figures/fig_before_Im_016_20240119_144948_3D-Im_019_20240119_145037_3D_Sagittal_0.pdf}} &
        {\includegraphics[width=0.12\textwidth]{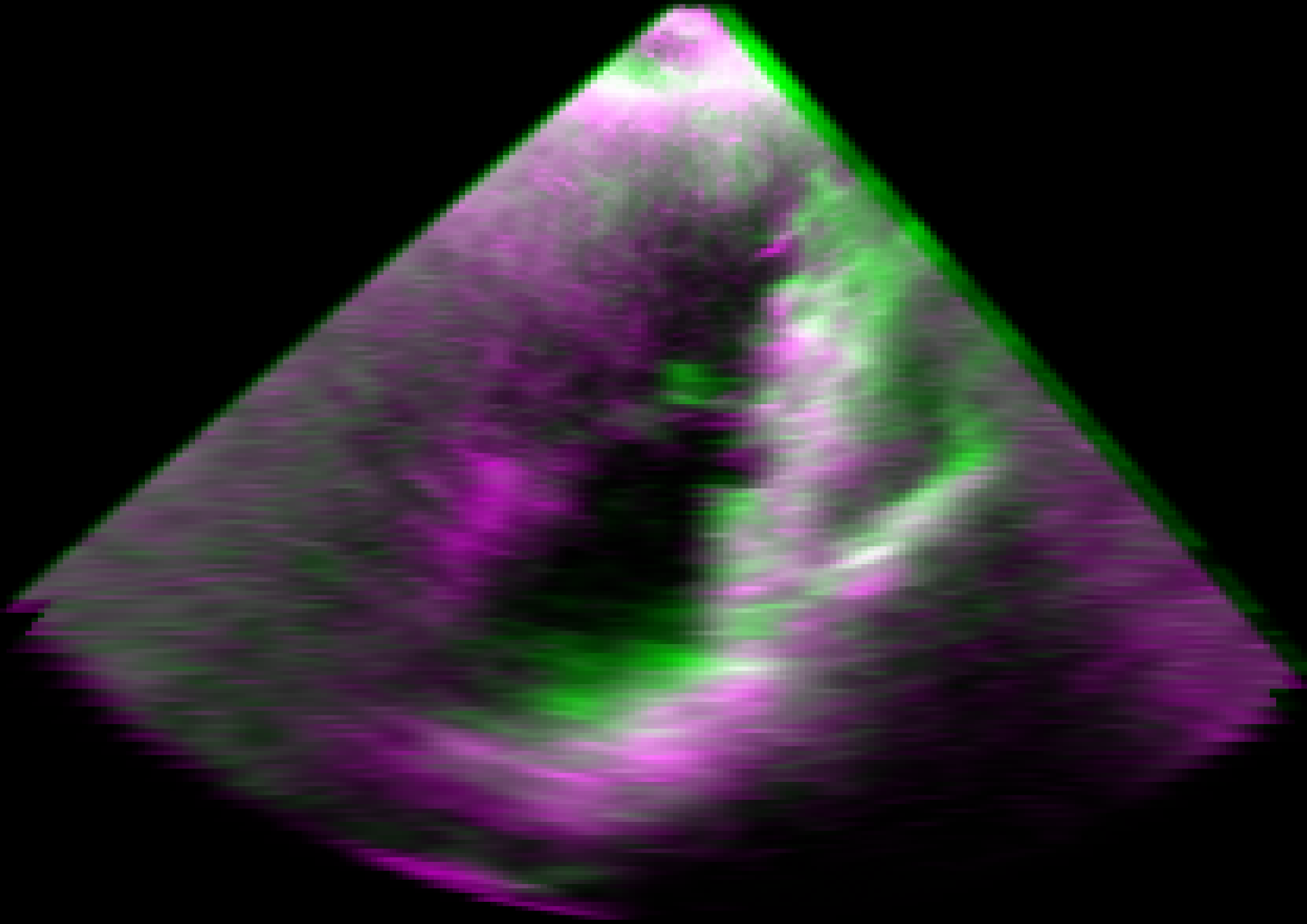}} &
        {\includegraphics[width=0.12\textwidth]{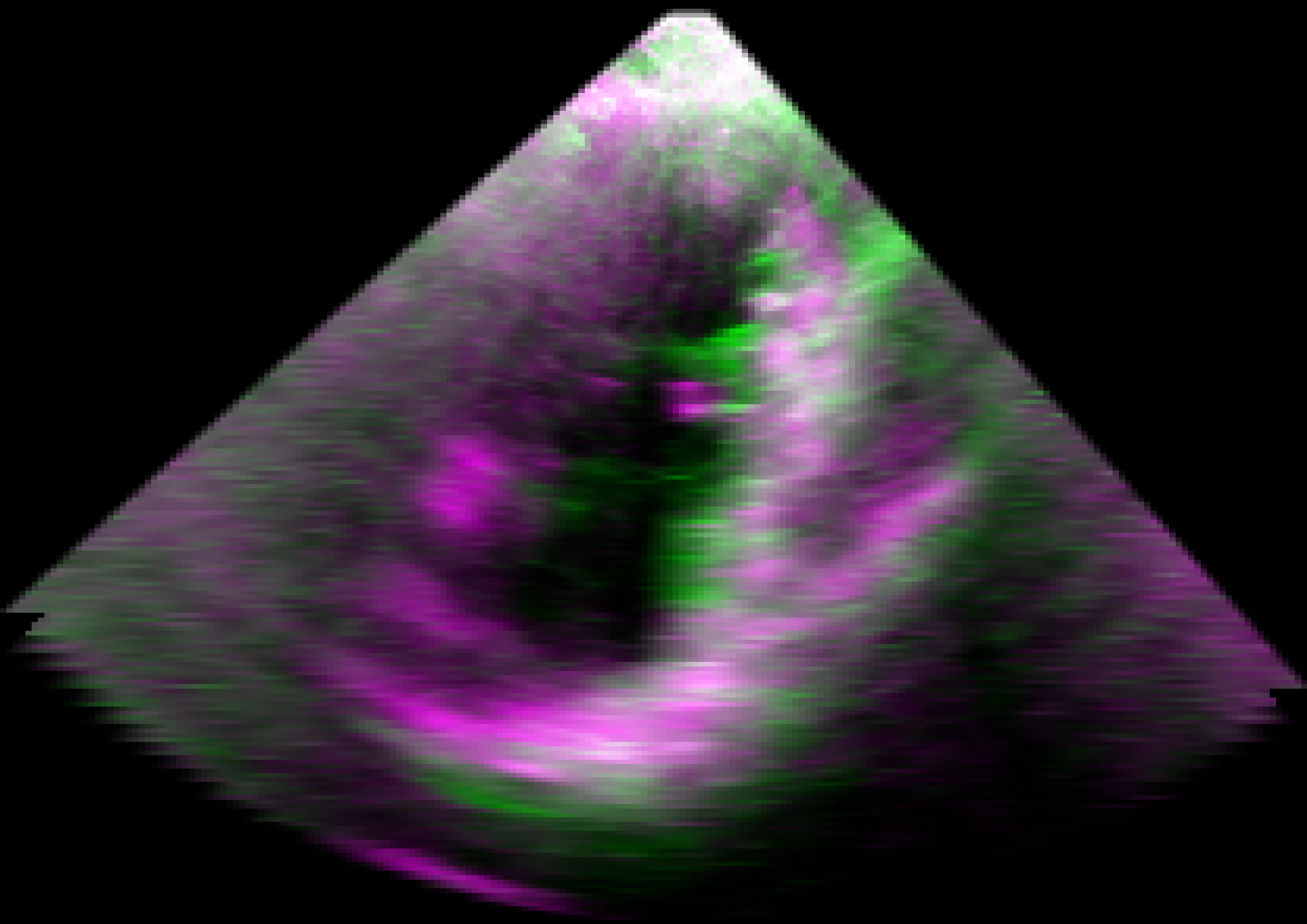}} &
        {\includegraphics[width=0.12\textwidth]{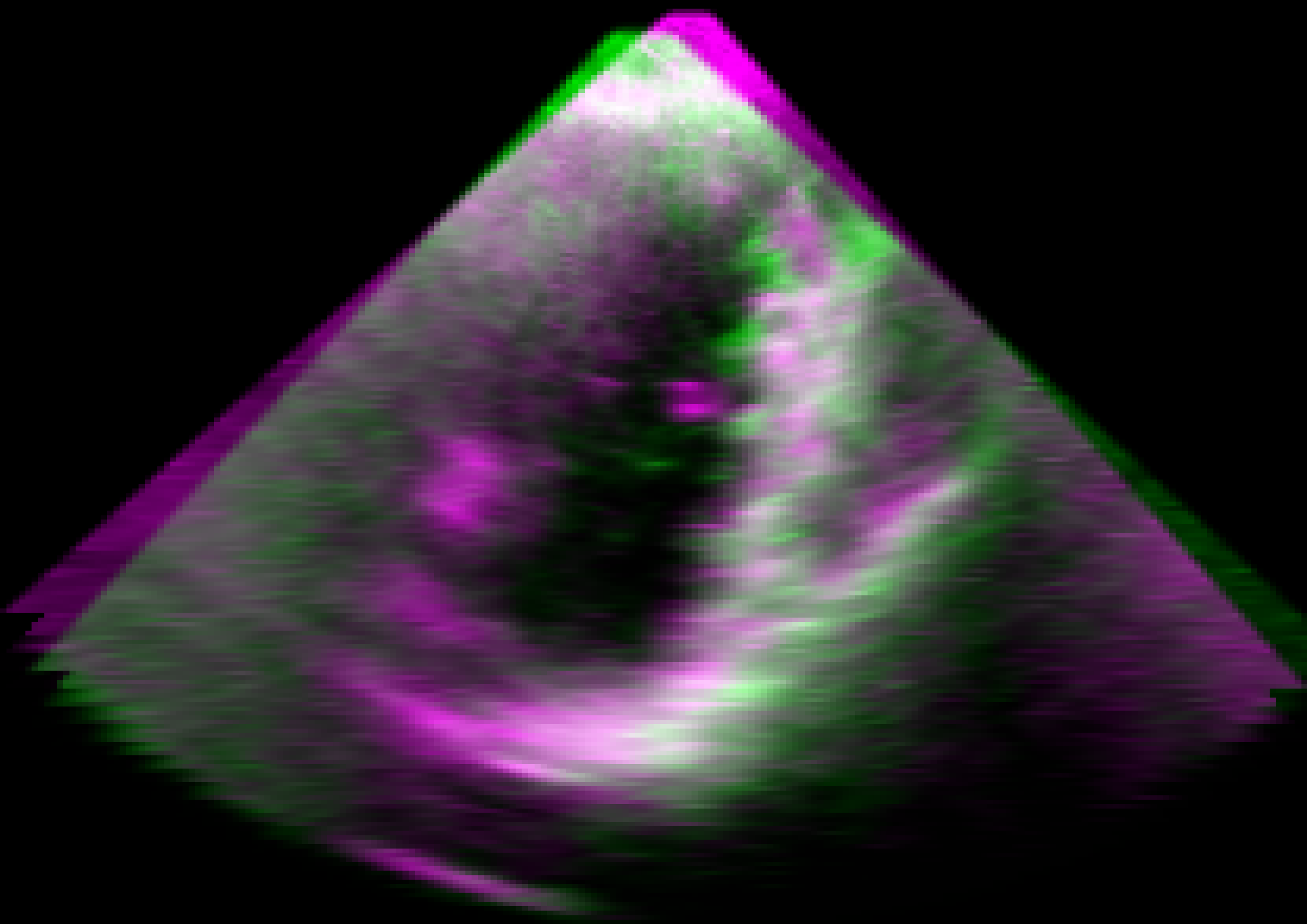}} \\
        \makecell[b]{Q3\vspace{15pt}} & 
        {\includegraphics[width=0.12\textwidth]{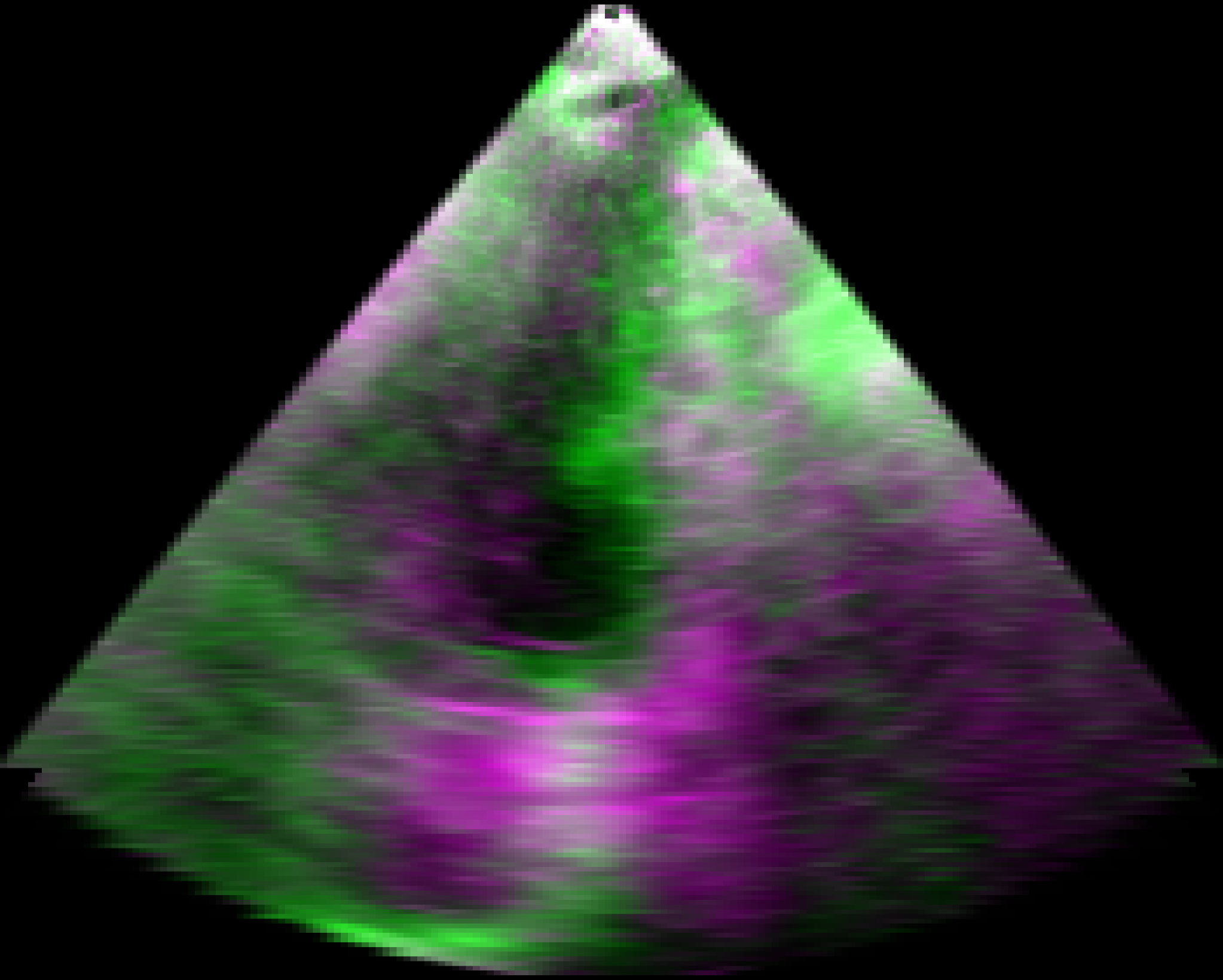}} &
        {\includegraphics[width=0.12\textwidth]{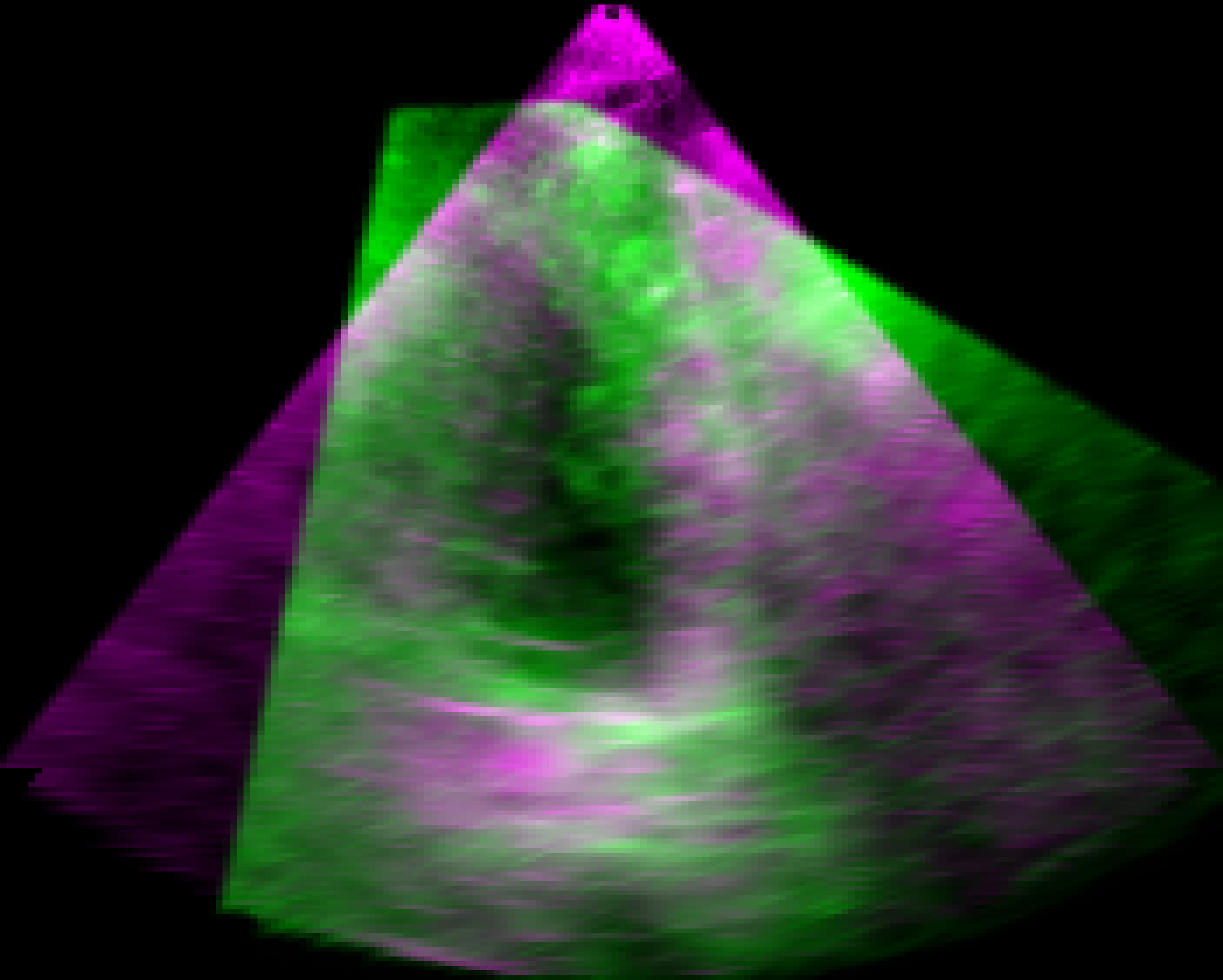}} &
        {\includegraphics[width=0.12\textwidth]{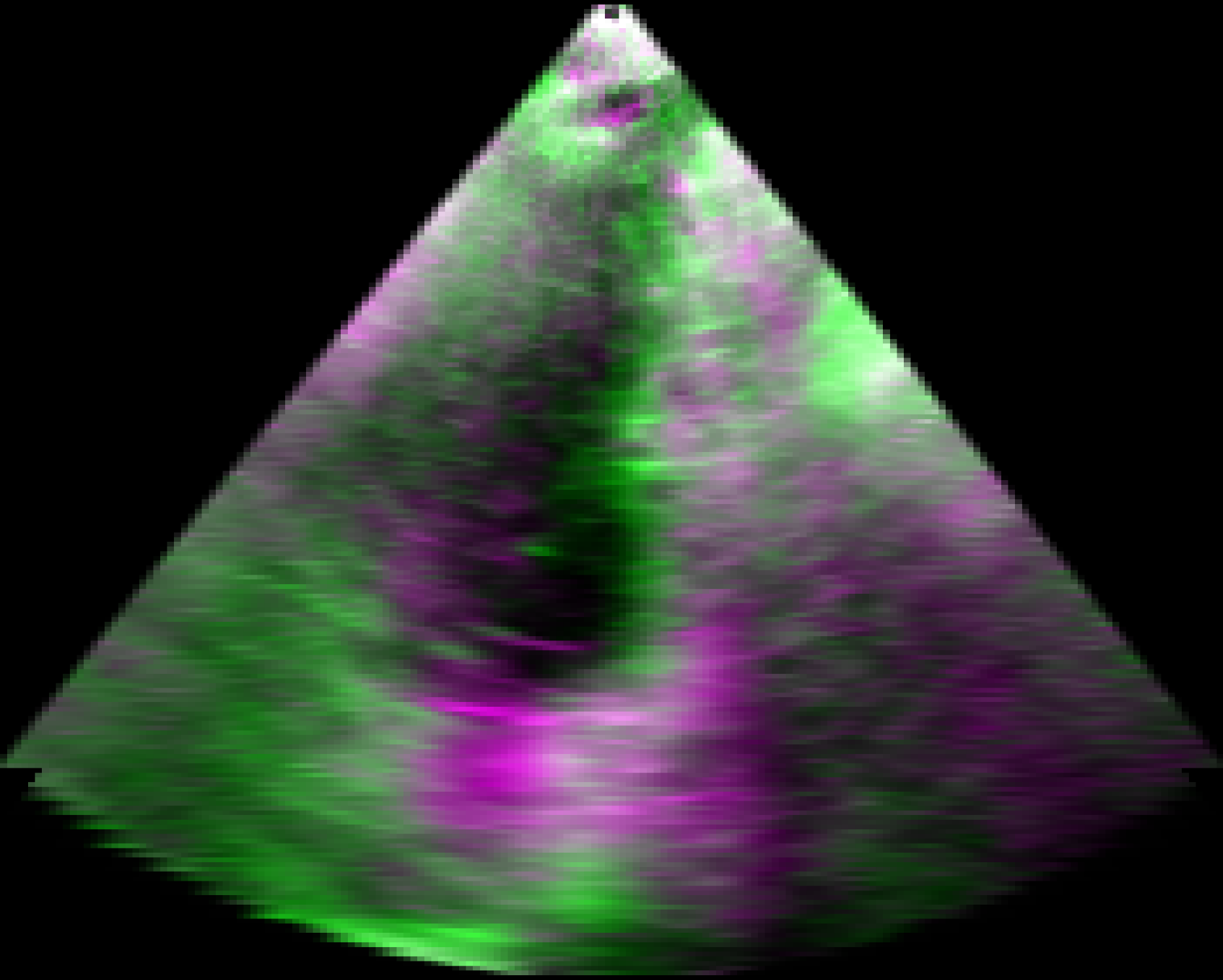}} &
        {\includegraphics[width=0.12\textwidth]{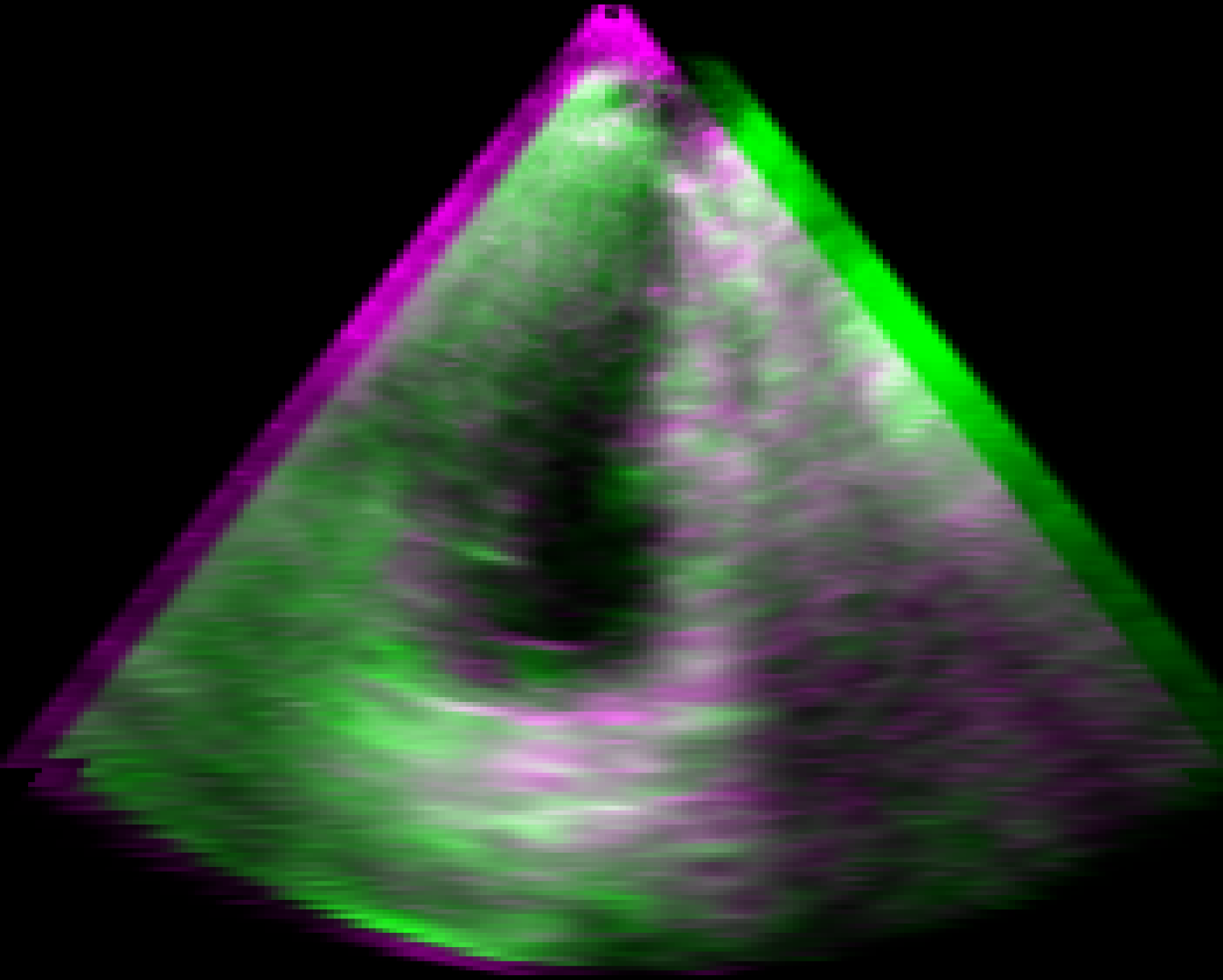}} &
        {\includegraphics[width=0.12\textwidth]{figures/fig_before_Im_013_20240221_143621_3D-Im_004_20240221_143057_3D_Sagittal_0.pdf}} &
        {\includegraphics[width=0.12\textwidth]{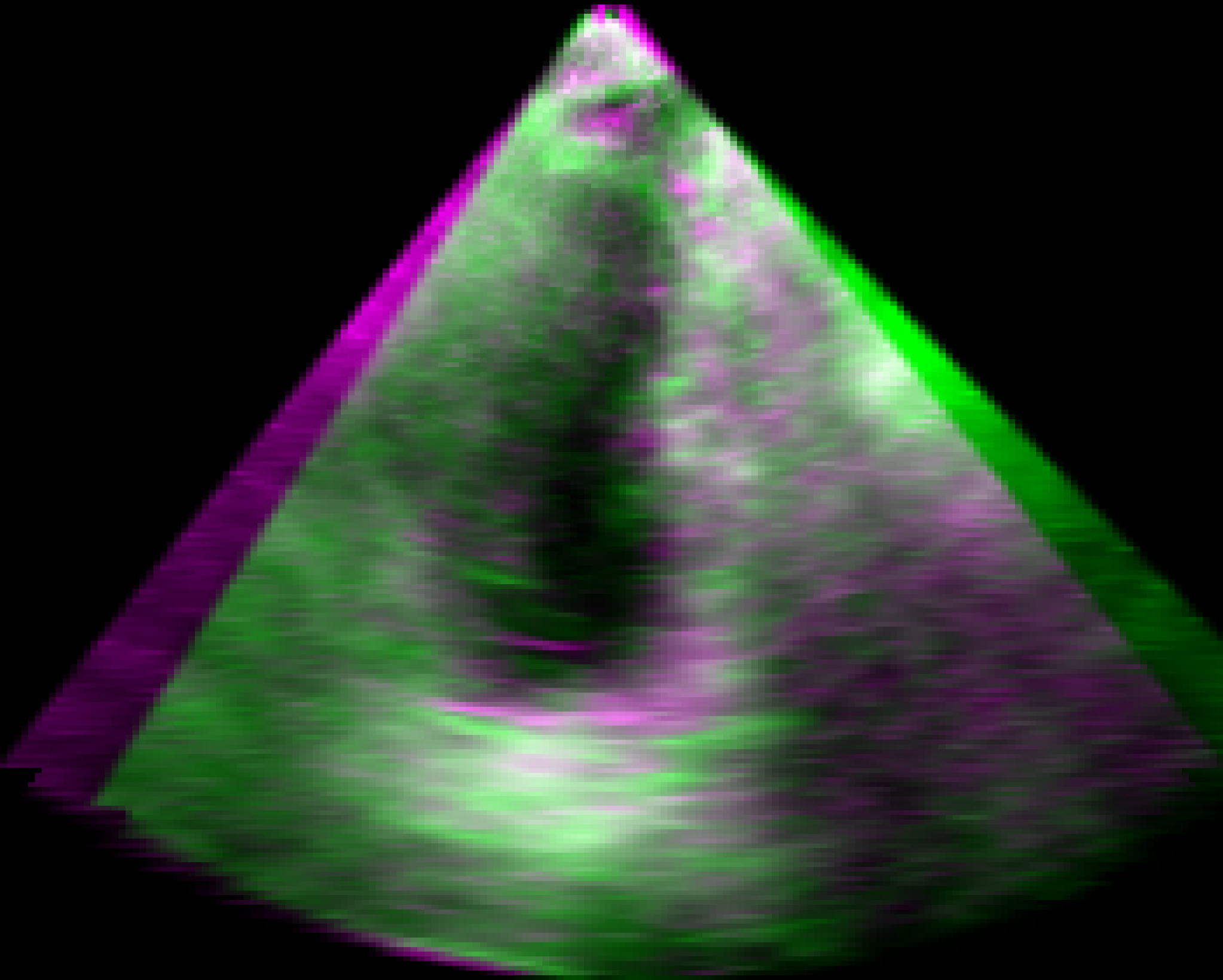}} \\
        \makecell[b]{Max\vspace{15pt}} &
        {\includegraphics[width=0.12\textwidth]{figures/fig_before_Im_003_20240215_151511_3D-Im_010_20240215_151637_3D_Sagittal_0.pdf}} &
        {\includegraphics[width=0.12\textwidth]{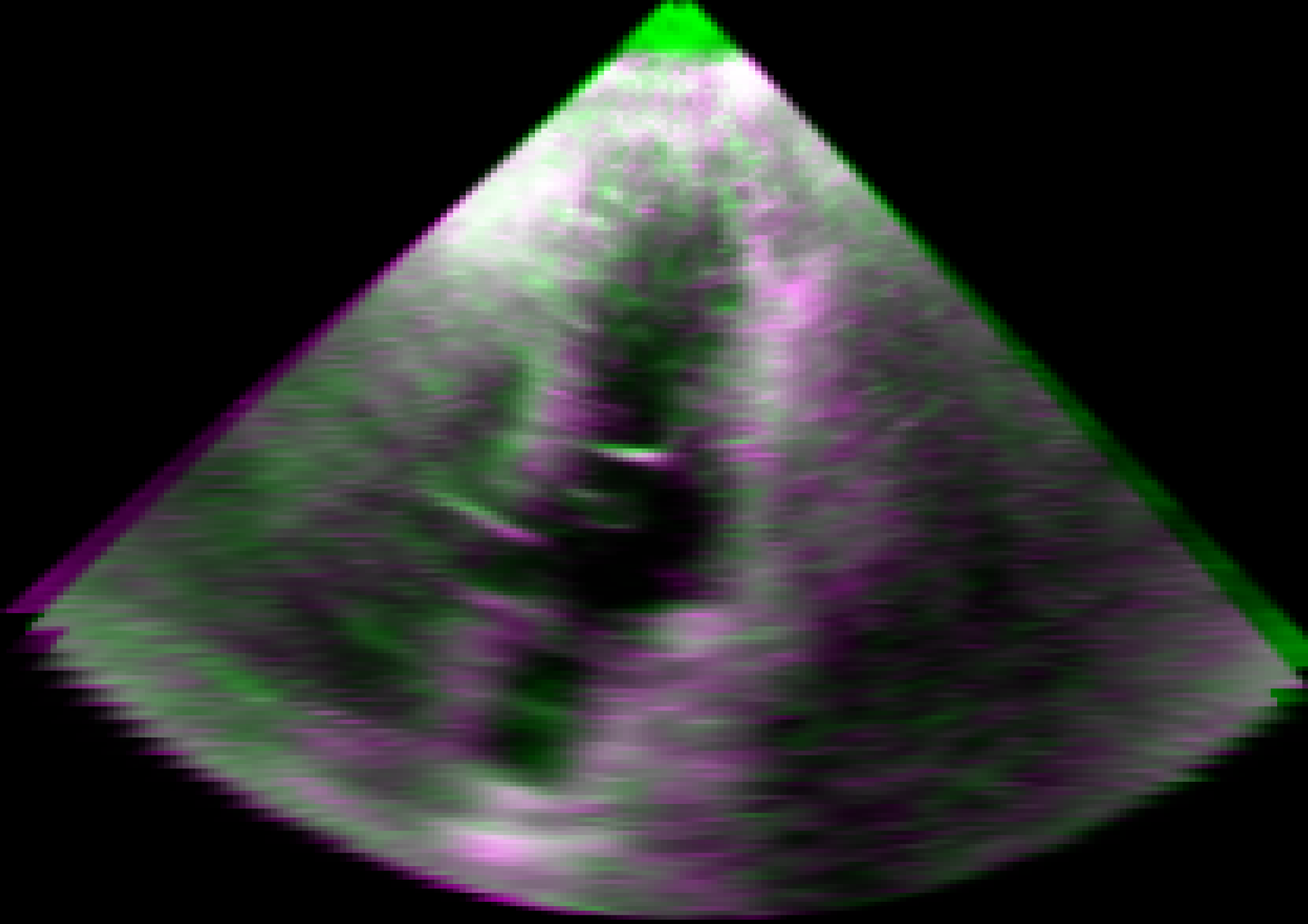}} &
        {\includegraphics[width=0.12\textwidth]{figures/fig_before_Im_013_20240221_143621_3D-Im_002_20240221_143029_3D_Sagittal_0.pdf}} &
        {\includegraphics[width=0.12\textwidth]{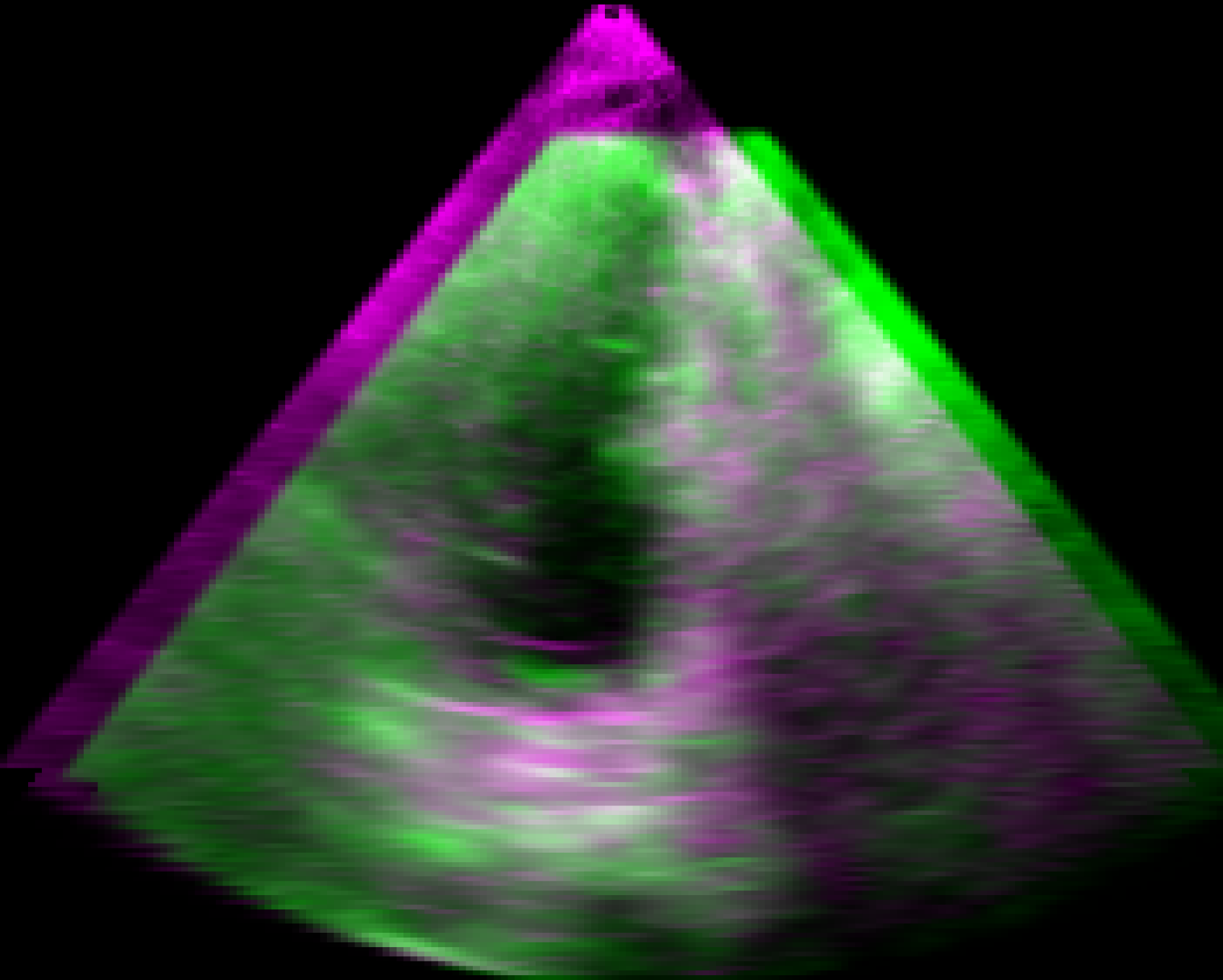}} &
        {\includegraphics[width=0.12\textwidth]{figures/fig_before_Im_004_20240123_153322_3D-Im_010_20240123_153651_3D_Sagittal_0.pdf}} &
        {\includegraphics[width=0.12\textwidth]{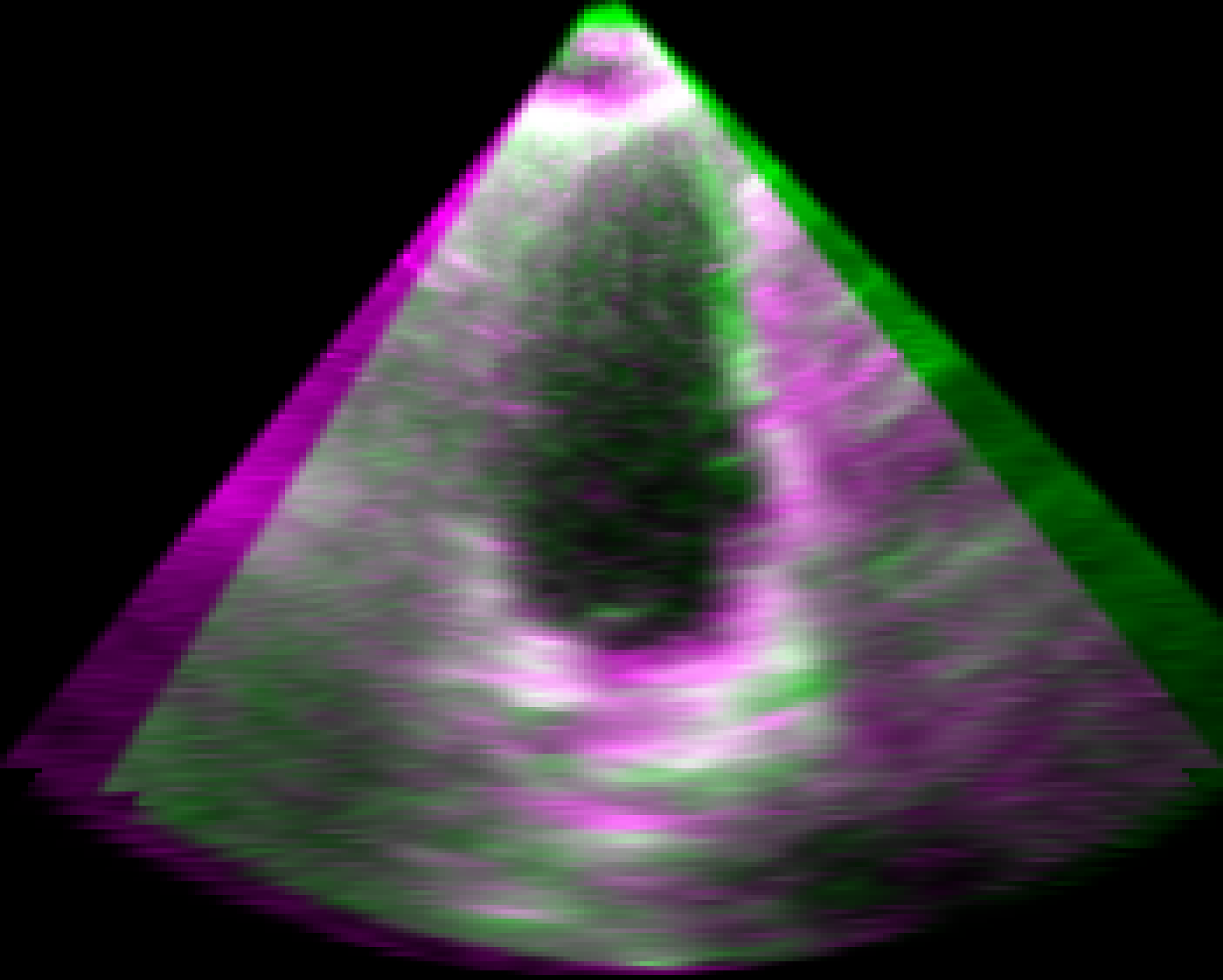}} \\
    \end{tabular}
    \caption{Mask-based rigid (PF and EX) registration results of the ED frame of image pairs for percentile DSC difference values of the sagittal view. Two consecutive columns show images before and after the registration for each category. The source and target images are shown in green and purple colors.}
    \label{fig:fig_perc_images_sagittal_rigid_mask}
\end{figure*}

The DSC improvement of image-based registration for the GPU version of PF is low compared to the CPU version, even though both versions use the exact parameters of PF, except that the GPU version utilizes the PyTorch framework's affine (without scaling) spatial transformer with the trilinear interpolation to compute image coordinates and interpolate voxel values when applying the transformation calculated by the algorithm. In contrast, the CPU version uses the ITK framework's Euler 3D transform and trilinear interpolation. The CPU and GPU versions of PF using mask-based registration have approximately the same average DSC improvement. Figures \ref{fig:fig_perc_images_axial_rigid_mask}, \ref{fig:fig_perc_images_coronal_rigid_mask}, and \ref{fig:fig_perc_images_sagittal_rigid_mask} show the results of rigid registration performed using LV binary masks for axial, coronal, and sagittal views, respectively. It can be seen that rigid registration using binary masks performs better than image intensities for PF(CPU), PF(GPU), and EX methods. The images corresponding to the minimum DSC difference were already aligned (0.828, 0.811, and 0.846), while the maximum DSC difference had limited overlap (0.080, 0.097, and 0.045) before registration for all three methods, respectively.

\begin{table*}[h!]
\caption{The rigid registration results of image pairs using the mean DSC difference for the minimum, maximum, and percentile values. The DSC values after the registration are listed in the next column. IC and IG refer to image-based registration performed on CPU and GPU hardware, respectively. The MC and MG refer to mask-based registration in the same way. The best results are highlighted for each category.}
\label{tab:table_percentile}%
\begin{tabular*}{\textwidth}{NSSSSSSSSSS}
\toprule
Method  & \multicolumn{2}{c}{\normalsize Min} & \multicolumn{2}{c}{\normalsize Q1(25\%)} & \multicolumn{2}{c}{\normalsize Q2 (50\%)} & \multicolumn{2}{c}{\normalsize Q3 (75\%)} & \multicolumn{2}{c}{\normalsize Max}\\
  & Diff & Final & Diff & Final & Diff & Final & Diff & Final & Diff & Final\\
\midrule
PF$_{IC}$ & $-0.005$ & 0.841 & 0.072 & 0.718 & 0.138 & 0.764 & 0.177 & 0.710 & 0.560 & 0.771 \\
PF$_{IG}$ & $-0.018$ & 0.793 & 0.033 & 0.818 & 0.050 & 0.677 & 0.069 & 0.298 & 0.124 & 0.660 \\			
EX$_{IC}$ & $-0.106$ & 0.546 & 0.014 & 0.640 & 0.075 & 0.768 & 0.307 & 0.484 & 0.438 & 0.541 \\
PF$_{MC}$ & $-0.005$ & 0.823 & \textbf{0.151} & 0.883 & 0.204 & 0.781 & 0.490 & 0.631 & 0.730 & 0.810 \\
PF$_{MG}$ & \textbf{0.015} & 0.826 & 0.149 & 0.822 & \textbf{0.216} & 0.841 & \textbf{0.515} & 0.840 & 0.704 & 0.801 \\		
EX$_{MC}$ & $-0.145$ & 0.701 & 0.041 & 0.771 & 0.115 & 0.663 & 0.469 & 0.790 & \textbf{0.750} & 0.795 \\
\bottomrule
\end{tabular*}
\end{table*}

The Kruskal-Wallis test was performed to evaluate the significance of image and mask-based registrations of PF (CPU and GPU versions) and EX methods at a significance level of $0.05$, and it was found that there was a significant difference between these methods, $H(5) = 193.11$, $p = 8.45e-40$. Post-hoc pairwise comparisons using Dunn’s test with Bonferroni correction showed that the CPU and GPU versions of mask-based PF registration have a significant difference with other methods ($p = 0.00$). Further, the image-based PF registration of CPU and GPU versions has a significant difference ($p = 0.00$) between them and between the GPU version of PF and EX registration using the mask ($p = 0.00$).

The deformation of frames within the cardiac cycle of the source and the target images due to volunteer movement and breathing cannot be aligned using rigid registration, and a nonrigid registration is required for that purpose. We used SE registration after rigid registration to improve the accuracy of PF results for the temporal sequence, but SE registration did not improve the overall accuracy for mask-based registration. The DSC improvement for each image pair is shown in Figure \ref{fig:fig_pf_es_se_ncc} along with rigid results, and the discussion part covers more details. The computational time of both the PF and EX algorithms is listed in Table \ref{tab:table_comp_time}. When comparing the execution time of the parallel version of the particle filter with the non-parallel version, $16.2\times$ and $16.7\times$ speedup is gained for image and mask-based registrations, respectively.

\begin{figure}[!ht]
    \centering
    \includegraphics[width=0.5\textwidth]{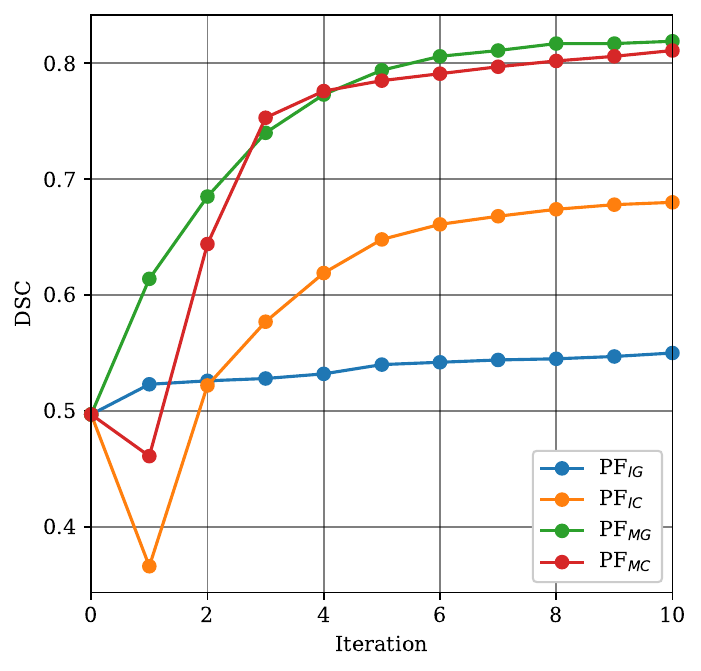}
    \caption{The overlapping score (DSC) value over iterations for the CPU and GPU versions of the PF algorithm. Iteration 0 shows the DSC value before registration. IC and IG refer to image-based registration performed on CPU and GPU hardware, respectively. The MC and MG refer to mask-based registration in the same way.}
    \label{fig:fig_dsc_over_iter}
\end{figure}

\begin{figure}[!ht]
    \centering
    \includegraphics[width=0.5\textwidth]{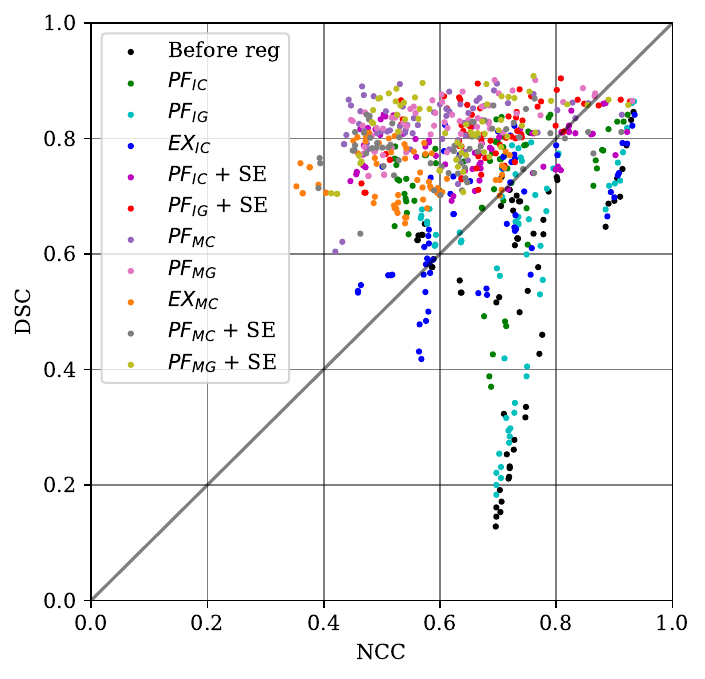}
    \caption{The correlation between overlapping score (DSC) and similarity metric (NCC) of the PF, EX, and PF + SE methods. IC and IG refer to image-based registration performed on CPU and GPU hardware, respectively. The MC and MG refer to mask-based registration in the same way.}
    \label{fig:fig_dsc_ncc}
\end{figure}

Further, the correlation between the similarity metric (NCC) and overlapping score (DSC) was analyzed. Figure \ref{fig:fig_dsc_over_iter} shows DSC changes over iterations, and it is found that the average DSC value drops for the first iteration for the CPU version while it increases smoothly for the GPU version. The DSC value changes over iteration for the EX method are not considered because of the high number of iterations. The correlation between DSC and NCC for approaches listed in Table \ref{tab:table_ncc_dsc} is shown in Figure \ref{fig:fig_dsc_ncc}. It was found that the value was -0.05, and there was no correlation between them.

\renewcommand{\arraystretch}{1.5}
\begin{table}[H]
\caption{The average computational time taken for 3D-3D rigid registration of images on the AMD EPYC 7532 processor for the CPU version of PF and EX methods and on the NVIDIA A100-SXM4-40GB GPU for the GPU version of PF. The speedup is calculated for the GPU version of the PF algorithm with respect to other methods.}
\label{tab:table_comp_time}
\setlength{\tabcolsep}{18pt}
    \begin{tabular}{lcc}
        \toprule
        Method  & Time (Seconds) & Speedup \\
        \midrule
        PF$_{IC}$ & 699.300 & 16.2 \\
        PF$_{IG}$ & 43.076 & \\
        EX$_{IC}$ & 1169.160 & 27.1 \\
        PF$_{MC}$ & 718.125 & 16.7 \\
        PF$_{MG}$ & 42.983 & \\
        EX$_{MC}$ & 1172.651 & 27.3 \\
        \bottomrule
    \end{tabular}
\end{table}
\renewcommand{\arraystretch}{1.0}

\section{Discussion}
This section compares our approach with other registration algorithms used for 3DE as monomodality and their limitations. The study \cite{grau_registration_2007} used three landmarks identified by a cardiologist to perform rigid registration using a phase based metric that considers features and intensities of voxels, and 14 temporal sequences of images were acquired from 7 patients using the 4Q acquisition method that suffers from stitching artifacts. The accuracy of the approach was assessed quantitatively using a 16 segment model, as they did not have the LV masks of images because of poor image quality. The authors claimed that the accuracy of rigid registration was sufficient for spatial registration. Compared to their multiview 3D TTE registration approach, our study uses images from the apical window acquired as a full volume using heart model acquisition (HMQ) with no stitching artifacts. The images that had poor quality and artifacts were ignored. Our method registers standard and nonstandard images obtained from the apical window that need minor probe movement, and we expect that there is a minor change in intensity due to incidence angle and tissue depth. It was noted that mask-based registration yielded better DSC accuracy than image-based registration because the binary masks have an intensity value of 1 in the LV area and 0 everywhere else and will not suffer intensity variations due to attenuation. In addition, the visual results agree with quantitative values. The mask-based registration gave 80\% DSC accuracy for both CPU and GPU versions of PF, and the nonrigid registration did not help to improve the accuracy. Since our method is implemented for GPU hardware, it is faster than their approach implemented for CPU.

Five landmarks (the apex of the LV, the junction of the mitral valve leaflets, and the mitral valve ring) were identified by two observers on apical standard and modified apical views of 3DE images taken from 14 healthy volunteers in the MeVisLab \footnote{\url{https://www.mevislab.de/}} software environment in another study \cite{mulder_registration_2011}. The interobserver variability metric was calculated for annotations, and the ES metric value was low compared to the ED metric due to improved endocardial border visibility. The mean value of the annotations was used for the inter-transformation variability metric. Because more diverse information was used for multiframe registration with the ED and ES frames compared to single frame registration, it yielded the best results, and considering all frames for the multiframe registration approach took more time with low accuracy. When comparing this study with ours, since we have not used the landmarks, the metrics they used for evaluating the accuracy did not apply to the PF algorithm. Both studies agree that the NCC metric can be used to register multiview 3DE images obtained with limited probe movement rigidly. The study \cite{mulder_atlas-based_2017} constructed an atlas manually using 16 patient data with six different views for intrapatient rigid registration and used the manual transformation to test and validate the ABM and regular pairwise registration approaches. The ABM method has the advantage of using the atlas that has a complete FOV to overcome limited FOV and initial overlap while considering multiple candidates to decide the final transformation. Compared to this study, ours uses the same similarity metric for mask-based registration with LV binary masks, and the atlas-based registration can be considered for 3D TTE images to overcome the limited FOV issue.

The study \cite{danudibroto_spatiotemporal_2016} used the apical standard view as the fixed image, and other views were considered moving images obtained from 10 patients, resulting in 4 to 8 views per patient and 50 image pairs in total. Twelve image pairs were considered for temporal registration using the characteristic curve computed with the NCC curve that resembles the LV volume curve over a cardiac cycle. It was reported that the FOV was improved for fused volumes after registration. The average computation time on the CPU was reported as 23.7 ± 17.0 seconds. Similar to their method, we used intensities of images and LV binary masks for the PF without a multiresolution approach for spatial resolution and evaluated the accuracy using the DSC. The computational time for the mask-based PF approach on CPU and GPU is high because the multiresolution approach did not help, and the image dimensions are different. The PF algorithm can be considered for temporal registration in the future with changes to perform deformable registration.

The study \cite{peressutti_registration_2017} used the PCA metric that uses linear dimensionality reduction to describe LV motion in temporal sequences using PCs, and the metric is robust to noise and artifacts while converging to an optimum result quickly without a starting estimate. A temporal interpolation was performed when the number of frames in the source and target sequences was different. The performance of the multiview 4D sequence was evaluated using image pairs that had a minimum initial overlap of 50\%, and it was reported that it outperformed the sum of squared distance (SSD) and phase-based metric. The average execution time of the method was reported as 583 ±  281 seconds in CPU hardware. Our approach uses the NCC similarity metric to align images taken from apical windows with a minimum overlap of 7\%, and the average execution time for mask-based registration is approximately 43 seconds on GPU hardware. The PCA metric can be used with the accelerated version of the PF algorithm in future work.

Comparing our approach with the study \cite{carnahan_multi-view_2022}, multiview 3D TEE image fusion was performed with the help of rigid and nonrigid registration. They did not perform any quantitative validation for the registration results of patient data. It took nearly an hour to perform the fusion, and the authors discussed the importance of using GPU hardware to accelerate the computation for real-time use in surgical planning. The advantage of our approach is that we can perform a pairwise 4D rigid registration in approximately 43 seconds for mask-based registration without the need for a nonrigid registration approach. In addition, we are considering images with limited and significant overlap.

When comparing our approach with \cite{shanmuganathan_two-step_2024}, ours is a training free pairwise registration approach that uses intensities of images and LV masks independently. In contrast, their method uses landmarks identified by a reinforcement learning model in a supervised manner and used Simple ITK’s landmark-based registration to register 53 temporal sequences of multiview images from 7 volunteers. The accuracy of the registration results was reported using the DSC and HD. It can be seen that the DSC value of apical to apical image registration was improved from 65\% to 73\% for rigid registration. At the same time, our method yielded DSC improvement from 50\% to 68\%, 55\%, 81\%, and 82\% for CPU and GPU versions of PF for image and mask-based registrations, respectively. The images considered to evaluate two of these algorithms are different even though they were from the same research study.

\renewcommand{\arraystretch}{1.5}
\begin{table}[h!]
\caption{The change of similarity metric (NCC) and overlapping score (DSC) after rigid (PF) and nonrigid (SE) registration. The best results for nonrigid registration are highlighted. Refer to Table \ref{tab:table_ncc_dsc} for values before and after the PF registration.}
\label{tab:table_ncc_dsc_se}
\setlength{\tabcolsep}{18pt}
    \begin{tabular}{llc}
        \toprule
        Metric & Method  & Value\\
        \midrule
        NCC & PF$_{IC}$ + SE& 0.601 $\pm$ 0.014 \\
         & \textbf{PF}\pmb{$_{IG}$}\textbf{ + SE} & \textbf{0.637} \pmb{$\pm$ 0.013} \\
         & PF$_{MC}$ + SE& 0.567 $\pm$ 0.016 \\
         & PF$_{MG}$ + SE& 0.596 $\pm$ 0.013 \\
        DSC & PF$_{IC}$ + SE& 0.740 $\pm$ 0.035\\
         & PF$_{IG}$ + SE& 0.781 $\pm$ 0.027 \\
         & PF$_{MC}$ + SE& 0.771 $\pm$ 0.005\\
         & \textbf{PF}\pmb{$_{MG}$}\textbf{ + SE} & \textbf{0.816} \pmb{$\pm$ 0.003} \\
        \bottomrule
    \end{tabular}
\end{table}
\renewcommand{\arraystretch}{1.0}

\begin{table*}[h!]
\caption{The nonrigid registration results of image pairs using mean DSC difference for the minimum, maximum, and percentile values. The DSC values after the registration are listed in the next column. IC and IG refer to image-based registration performed on CPU and GPU hardware, respectively. The MC and MG refer to mask-based registration in the same way. The best results are highlighted for each category.}
\label{tab:table_percentile_se}%
\begin{tabular*}{\textwidth}{QSSSSSSSSSS}
\toprule
{\normalsize Method}  & \multicolumn{2}{c}{\normalsize Min} & \multicolumn{2}{c}{\normalsize Q1(25\%)} & \multicolumn{2}{c}{\normalsize Q2 (50\%)} & \multicolumn{2}{c}{\normalsize Q3 (75\%)} & \multicolumn{2}{c}{\normalsize Max}\\
  & Diff & Final & Diff & Final & Diff & Final & Diff & Final & Diff & Final\\
\midrule
PF$_{IC}$ + SE & $-0.061$ & 0.771 & 0.075 & 0.775 & 0.173 & 0.807 & 0.287 & 0.747 & 0.671 & 0.816 \\
PF$_{IG}$ + SE & $-0.011$ & 0.824 & 0.096 & 0.807 & 0.199 & 0.852 & 0.449 & 0.789 & 0.687 & 0.802 \\
PF$_{MC}$ + SE & $-0.058$ & 0.774 & \textbf{0.112} & 0.809 & 0.202 & 0.833 & 0.439 & 0.578 & 0.666 & 0.792 \\
PF$_{MG}$ + SE & $\textbf{-0.008}$ & 0.803 & \textbf{0.112} & 0.859 & \textbf{0.208} & 0.828 & \textbf{0.545} & 0.871 & \textbf{0.741} & 0.886 \\
\bottomrule
\end{tabular*}
\end{table*}

The results of the SE registration are listed in Table \ref{tab:table_ncc_dsc_se}. Due to the minor improvement in alignment between the rigid and nonrigid registration results, figures corresponding to the percentile DSC difference value of the SE registration after the PF algorithm are not shown here. The overall DSC values were improved after applying SE registration on top of images registered rigidly using the image-based PF approach on CPU and GPU hardware. The SE registration failed for 3 and 2 image pairs for the CPU and GPU versions of PF, pre-aligned using image-based rigid registration because there were not enough overlapping sample points. Since the PF registrations performed with masks have already aligned the images well with the overall DSC value of $0.819$, the SE registration could not improve the overall DSC even though it helped to improve the alignment of images with limited overlap. The time taken for SE registration is listed in Table \ref{tab:table_se_comp_time}.

\begin{figure}[h!]
    \centering
    \includegraphics[width=0.5\textwidth]{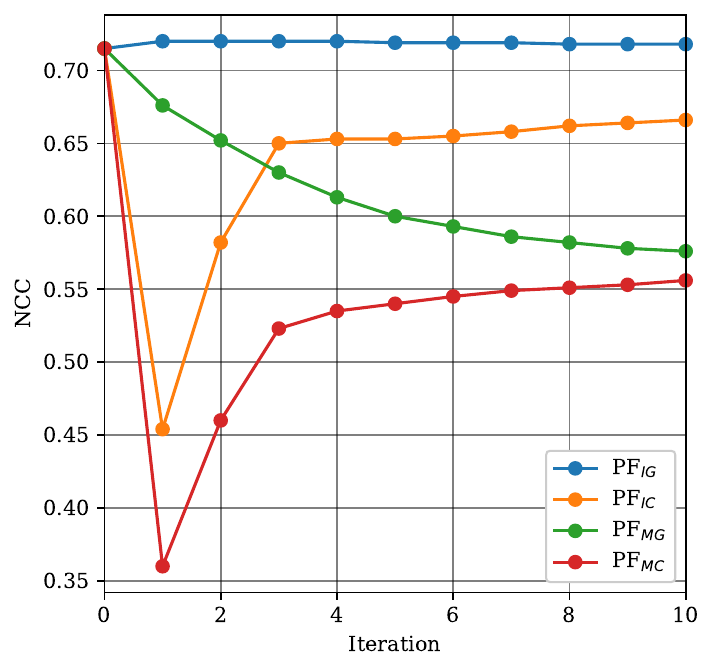}
    \caption{The average similarity metric value over iterations for CPU and GPU versions of PF. Iteration 0 shows the NCC value before registration. IC and IG refer to image-based registration performed on CPU and GPU hardware, respectively. The MC and MG refer to mask-based registration in the same way.}
    \label{fig:fig_ncc_over_iter}
\end{figure}

\begin{figure}[h!]
    \centering
    \includegraphics[width=0.5\textwidth]{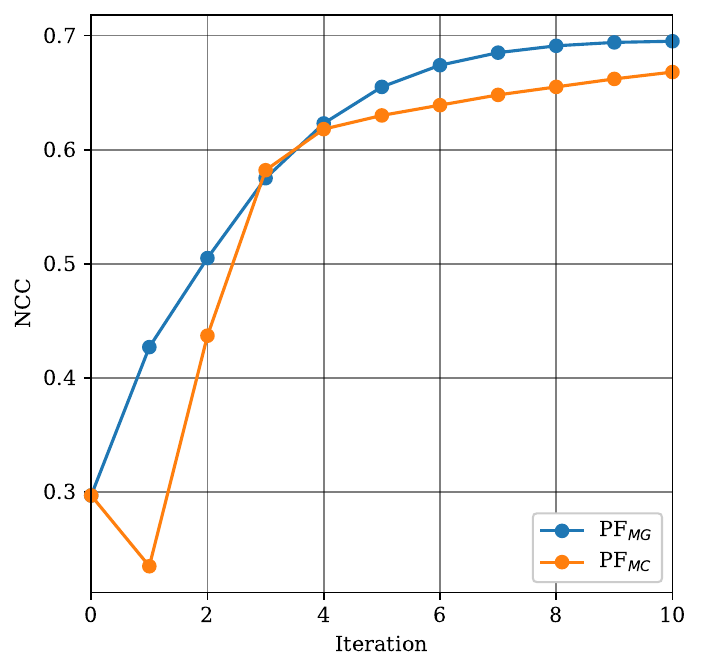}
    \caption{The average similarity metric value of binary masks over iterations for CPU and GPU versions of PF. Iteration 0 shows the NCC value before registration. The MC and MG refer to mask-based registration in the same way.}
    \label{fig:fig_ncc_over_iter_mask}
\end{figure}

The average NCC changes for the CPU and the GPU versions of the image and mask-based PF registration are shown in Figure \ref{fig:fig_ncc_over_iter} where the plotted values were calculated using the voxel intensity values of images. The NCC value increases over iterations even though there was a drop at the first iteration for the CPU version of the PF. However, this trend is not seen for the GPU version because of the difference in the implementation of Euler3D transform and trilinear interpolation. The average NCC value between images decreased gradually when the LV binary masks were used for the mask-based registration of the GPU version in contrast to other methods. However, the estimated NCC values between the binary masks increase over iterations when performing the mask-based registration, as shown in Figure \ref{fig:fig_ncc_over_iter_mask}. The NCC value calculated for images after registration, as listed in Table \ref{tab:table_ncc_dsc}, dropped for all the approaches except for the GPU version of image-based PF registration, even though the DSC values increased.

\renewcommand{\arraystretch}{1.5}
\begin{table}[t!hbp]
\caption{The average computational time taken for 3D-3D nonrigid (SE) registration of images on AMD EPYC 7532 processor for images registered with PF.}
\label{tab:table_se_comp_time}
\setlength{\tabcolsep}{18pt}
    \begin{tabular}{lc}
        \toprule
        Method  & Time (Seconds) \\
        \midrule
        PF$_{IC}$ + SE & 17.979 \\
        PF$_{IG}$ + SE & 18.603 \\
        PF$_{MC}$ + SE & 20.378 \\
        PF$_{MG}$ + SE & 19.486 \\
        \bottomrule
    \end{tabular}
\end{table}
\renewcommand{\arraystretch}{1.0}

Considering the limitations of this study, the LV masks used for the mask-based registration were segmented using Tomtec Arena software. The automatic segmentation of these masks on 4D sequences of TTE images can be considered as preprocessing work for registration. The PF algorithm could not perform the apical to parasternal image registration using image-based registration with the same parameters as both views were obtained by a significant probe movement and the parasternal image needs to be rotated by (90, -180) degrees by y and z axes because of high intensity variation between these views. The mask-based registration could handle more significant probe rotation by up to 60 degrees, and the PF algorithm parameters need to be changed to achieve a perfect alignment of the apical to parasternal image registration. This is a limitation of our method compared to landmark-based approaches of multiview 3D TTE image registration \cite{grau_registration_2007, shanmuganathan_two-step_2024}, and a future study will address this limitation. The entire LV area may not be visible in all views of 3DE images when performing clinical examinations because part of the apex may not be visible in some views. The mask-based registration needs to guess the missing part of the LV area, or the same amount of LV volume should be considered before starting the registration. This is also another limitation of using mask-based registration, and a proper algorithm needs to be defined to do this task in the future.

\section{Conclusion}
The parallel version of the PF algorithm proposed in this study resulted in 55\% and 82\% DSC accuracy for image- and mask-based registration comparable to the non-parallel version of the algorithm, 68\%, and 81\%. The mask-based PF algorithm yields better accuracy than image-based registration, and there is no need for a nonrigid registration algorithm for images with significant overlap. The speedup gained between parallel and non-parallel versions of the PF algorithm is approximately $16 \times$ for both approaches. When comparing the accuracy and speedup of the PF algorithm with the EX, there is a significant difference between these rigid registration approaches at a 0.05 significance level. Post-hoc pairwise comparisons using Dunn’s test with Bonferroni correction showed that image registration performed with mask for both CPU and GPU versions of the PF has a significant difference with other methods (p = 0.00). The automatic LV segmentation of the 4D TTE image sequence can be considered in the future to register images taken from the apical window using mask-based rigid registration by the parallel version of the PF. A machine-learning approach will be implemented to compare the performance of the PF algorithm.

\section{Ethics Approval}
The dataset collection, performed as part of the robot arm assisted 3D multiview fusion echocardiography study conducted at the Mazankowski Alberta Heart Institute, was approved by the Health Research Ethics Board of the University of Alberta on July 9, 2021 (Study ID: Pro00109218). Informed consent was obtained from volunteers who participated in this study, and the data was anonymized.  

\section{Acknowledgement}
The authors thank Bernadette Foster for providing LV segmentation of the images using Tomtec Arena software. We acknowledge research support for Dr. M. Noga from the Alberta Health Services (AHS) Medical Imaging Consultants Research Chair and Dr. K. Punithakumar from the AHS Chair in Diagnostic Imaging positions. This research was enabled in part by support provided by the Digital Research Alliance of Canada (alliancecan.ca).

\section{CRediT authorship contribution statement}
\textbf{Thanuja Uruththirakodeeswaran:} Conceptualization, Data curation, Formal analysis,  Methodology, Software, Validation, Writing – original draft. \textbf{Harald Becher:} Data curation, Funding acquisition, Investigation, Writing – review \& editing. \textbf{Michelle Noga:} Funding acquisition, Writing – review \& editing. \textbf{Lawrence H. Le:} Supervision, Writing – review \& editing. \textbf{Pierre Boulanger:} Funding acquisition, Writing – review \& editing. \textbf{Jonathan Windram:} Investigation, Writing – review \& editing. \textbf{Kumaradevan Punithakumar:} Data curation, Funding acquisition, Investigation, Supervision, Writing – review \& editing.

\section{Funding}
This work was supported by Alberta Innovates for the Accelerating Innovations into CarE (AICE) Concepts grant and the Research Tools and Instruments (RTI) grant (partially) from the Natural Sciences and Engineering Research Council of Canada (NSERC). In addition, the work was supported by the Medical Science Graduate Program (MSGP) scholarship obtained by Thanuja Uruththirakodeeswaran.




\bibliographystyle{elsarticle-num} 
\bibliography{SMC_parallel}


\clearpage



\end{document}